\documentclass[
11pt, 
english, 
onehalfspacing 
nolistspacing, 
liststotoc, 
parskip, 
headsepline, 
oneside]{DoctoralThesis} 
\usepackage[utf8]{inputenc} 
\usepackage[T1]{fontenc} 
\usepackage{verbatim}
\usepackage{amssymb}
\usepackage{amsmath}
\usepackage{mathtools} 
\usepackage{float}
\usepackage[labelfont=bf]{caption} 
\usepackage{commath}
\usepackage{graphicx} 
\usepackage{array}
\usepackage{makecell}
\usepackage{multirow}
\usepackage{hyperref}
\usepackage[nameinlink]{cleveref} 
\usepackage{epigraph}
\usepackage{rotating}
\usepackage{minitoc}
\usepackage{stackrel}
\usepackage[none]{hyphenat} 
\usepackage{dirtytalk} 
\usepackage{slashed}
\usepackage{upgreek}
\usepackage{titlesec} 
\usepackage{lastpage}
\usepackage{subcaption}
\usepackage{float}
\usepackage{datetime}
\usepackage[compat=1.1.0]{tikz-feynman}
\usepackage{physics}
\usepackage{tikz}
\usepackage{simpler-wick}
\usepackage{chngcntr} 

\newdateformat{monthyeardate}{%
  \monthname[\THEMONTH] \THEYEAR}
\usepackage{yfonts} 

  \titleformat{\chapter}
  {\normalfont\Huge\bfseries\color{sectitlecolor}}{\colorbox{orange!60!}{\textcolor{secnumcolor}{\thechapter}}}{1em}{}
\titleformat{\section}
  {\normalfont\Large\bfseries\color{sectitlecolor}}{{\textcolor{secnumcolor}{\thesection}}}{1em}{}
\titleformat{\subsection}{\normalfont\large\bfseries\color{sectitlecolor}}{{\textcolor{secnumcolor}{\thesubsection}}}{1em}{}
\titleformat{\subsubsection}{\normalfont\large\bfseries\color{sectitlecolor}}{{\textcolor{secnumcolor}{\thesubsubsection}}}{1em}{}
\def\beq{\begin{equation}}
\def\eeq{\end{equation}}
\def\Eq#1{Eq.~(\ref{#1})}
\usepackage[backend=bibtex,style=numeric, natbib=true, sorting=none, backref=true]{biblatex} 
\DefineBibliographyStrings{english}{%
  backrefpage = cited on p:,
  backrefpages = cited on pp:,
}

\newbibmacro*{bbx:parunit}{%
  \ifbibliography
    {\setunit{\bibpagerefpunct}\newblock
     \usebibmacro{pageref}%
     \clearlist{pageref}%
     \setunit{\adddot\par\nobreak}}
    {}}

\renewbibmacro*{doi+eprint+url}{%
  \usebibmacro{bbx:parunit}
  \iftoggle{bbx:doi}
    {\printfield{doi}}
    {}%
  \iftoggle{bbx:eprint}
    {\usebibmacro{eprint}}
    {}%
  \iftoggle{bbx:url}
    {\usebibmacro{url+urldate}}
    {}}

\renewbibmacro*{eprint}{%
  \usebibmacro{bbx:parunit}
  \iffieldundef{eprinttype}
    {\printfield{eprint}}
    {\printfield[eprint:\strfield{eprinttype}]{eprint}}}

\renewbibmacro*{url+urldate}{%
  \usebibmacro{bbx:parunit}
  \printfield{url}%
  \iffieldundef{urlyear}
    {}
    {\setunit*{\addspace}%
     \printtext[urldate]{\printurldate}}}

     \renewbibmacro*{pageref}{%
  \addperiod
  \iflistundef{pageref}
    {}
    {\newline\printtext[parens]{
       \ifnumgreater{\value{pageref}}{1}
         {\bibstring{backrefpages}\ppspace}
     {\bibstring{backrefpage}\ppspace}%
       \printlist[pageref][-\value{listtotal}]{pageref}}}}

\usepackage[autostyle=true]{csquotes} 

\usepackage{titlesec}

\titleformat{\paragraph}[runin]{\normalfont\normalsize\bfseries}{}{0pt}{\indent}

\def\id{\boldsymbol I}
\def\ra{\rangle}
\def\la{\langle}

\def\ad#1{{\cal A}_{\rm D}^{(#1)}}

\def\beq{\begin{equation}}
\def\eeq{\end{equation}}
\def\bea{\begin{eqnarray}}
\def\eea{\end{eqnarray}}
\def\nn{\nonumber}
\def\Eq#1{Eq.~(\ref{#1})}
\def\ln#1{\mathrm{log}\left(#1\right)}
\def\id{\boldsymbol I}
\def\ra{\rangle}
\def\la{\langle}
\def\bra#1{\la #1|}
\def\ket#1{|#1\ra}
\def\ii{\imath 0}
\def\uv{{\rm UV}}

\def\r{{\rm R}}
\def\nn{\nonumber}
\def\lb{\boldsymbol{\ell}}
\def\pb{{\bf p}}
\def\qb{{\bf q}}
\def\qon#1{q_{#1,0}^{(+)}}
\def\ad#1{{\cal A}_{\rm D}^{(#1)}}
\def\ps#1{\widetilde \Delta_{#1}}

\def\XXint#1#2#3{{\setbox0=\hbox{$#1{#2#3}{\int}$ }
\vcenter{\hbox{$#2#3$ }}\kern-.6\wd0}}

\newenvironment{dedication}
{
   \cleardoublepage
   \thispagestyle{empty}
   \vspace*{\stretch{1}}
   \hfill\begin{minipage}[t]{0.2\textwidth}
   \raggedright
}%
{
   \end{minipage}
   \vspace*{\stretch{3}}
   \clearpage
}

\def\fussy{%
  \emergencystretch\z@
  \tolerance 2000%
  \hfuzz .1\p@
  \vfuzz\hfuzz}

\thesistitle{Quantum computing applications in High Energy Physics: clustering, integration and generative models} 
\supervisor{Dr.~Germán Vicente Rodrigo García, Dr.~Leandro Javier Cieri} 
\examiner{abcd} 
\degree{Doctor in Physics} 
\author{Jorge Juan Martínez de Lejarza Samper} 
\addresses{} 

\subject{Physical Sciences} 
\keywords{} 
\university{Universitat de València} 
\department{Instituto de Física Corpuscular (UV -- CSIC)} 
\group{LHCPheno group} 
\faculty{Departament de Física Teòrica} 

\AtBeginDocument{
\hypersetup{pdftitle=\ttitle} 
\hypersetup{pdfauthor=\authorname} 
\hypersetup{pdfkeywords=\keywordnames} 
}

\begin{document}
\sloppy 

\frontmatter 

\pagestyle{plain} 


\begin{titlepage}
\begin{center}
\includegraphics[scale=0.5]{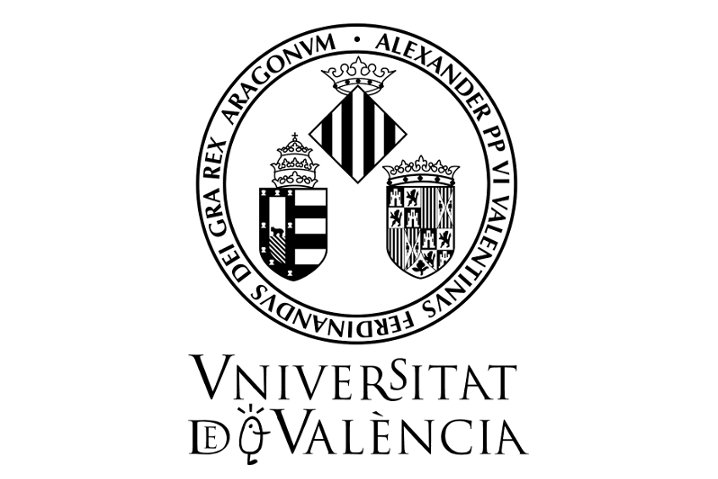}\\
\vspace*{.02\textheight}
\textsc{\Large Doctorat en Física}\\[0.5cm] 

\HRule \\[0.2cm] 
\color{blue}{ \huge \bfseries \ttitle \par}\vspace{0.15cm} 
\color{black}\HRule \\[0.9cm] 
 
\begin{minipage}[t]{0.5\textwidth}
\begin{flushleft} \large
\emph{Author:}\\
\color{black}\large\authorname
\end{flushleft}
\end{minipage}
\color{black}
\begin{minipage}[t]{0.49\textwidth}
\vspace{0.5cm} 
 \begin{flushright} \large
 \emph{Supervisors:} \\
\color{black}\large Dr.~Germán Vicente Rodrigo García\\
\vspace{0.1cm}
\vspace{0.1cm}
\color{black}\large Dr.~Leandro Javier Cieri\\
\end{flushright}
\end{minipage}
\vspace{1.2cm}
 
{\large \facname\\
IFIC, Universitat de València - CSIC}\\
 
\vfill
\vspace*{\fill}
{\large Programa de Doctorat en Física - 3126}\\[1cm]
{\large Valencia, \monthyeardate \today}\\[1cm] 
\end{center}
\newpage

\begin{flushleft}
\emph{Title:}\\
\textbf{\textbf{ \ttitle}}
\end{flushleft}

\vspace{0.3cm}
\begin{flushleft} \large
\emph{Author:} \\
\textbf{\authorname} \\
Departament de Física Teòrica\\
Instituto de Física Corpuscular (IFIC)\\
Universitat de València, Spain\\
Consejo Superior de Investigaciones Científicas (CSIC), Spain\\
\end{flushleft}

\vspace{0.3cm}
 \begin{flushleft}
 \emph{Supervisors:} \\
\textbf{Dr.~Germán Vicente Rodrigo García}\\
Instituto de Física Corpuscular (IFIC)\\
Universitat de València, Spain\\
Consejo Superior de Investigaciones Científicas (CSIC), Spain\\
\vspace{0.1cm}
\textit{and}\\
\vspace{0.1cm}
\textbf{Dr.~Leandro Javier Cieri}\\
Departament de Física Teòrica \\
Instituto de Física Corpuscular (IFIC)\\
Universitat de València, Spain\\
Consejo Superior de Investigaciones Científicas (CSIC), Spain\\
\end{flushleft}  \large


\vspace*{\fill}

Copyright \textcopyright \, 2025 Jorge Juan Martínez de Lejarza Samper

This research has been carried out at Instituto de Física Corpuscular, IFIC, Valencia, which is a joint centre of research between the Spanish Research Council (Consejo Superior de Investigaciones Científicas, CSIC) and the University of Valencia.\\


\end{titlepage}

\newgeometry{
	paper=a4paper, 
	inner=2.5cm, 
	outer=2.5cm, 
	bindingoffset=1cm, 
	top=3cm, 
	bottom=2.5cm, 
}
%
\label{sec:Declaration}
\thispagestyle{empty}

\includegraphics[width=0.8\textwidth]{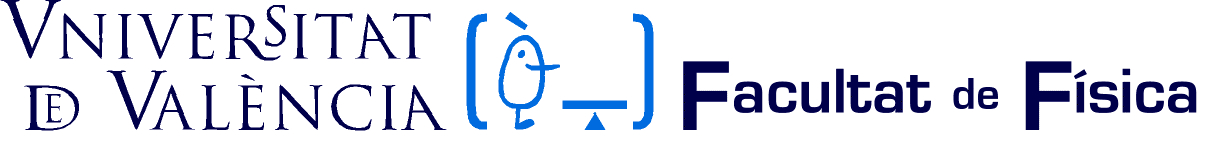}

\vspace{4cm}

\noindent \textbf{Dr.~Germán Vicente Rodrigo García}, Investigador Científico del Consejo Superior de Investigaciones Científicas y  \textbf{Dr.~Leandro Javier Cieri}, Investigador CIDEGENT por la Universidad de Valencia

\bigskip
\noindent CERTIFICAN,
\bigskip

\noindent Que la presente memoria, \enquote{\textbf{\textit{\ttitle}}}, ha sido realizada bajo su dirección en el Instituto de Física Corpuscular (Centro Mixto Universidad de Valencia - CSIC) por \textbf{Jorge Juan Martínez de Lejarza Samper} y constituye su Tesis Doctoral para optar al título de Doctor por la Universitat de València una vez cursados los estudios en el Doctorado en Física.

\noindent Y para que conste, en cumplimiento de la legislación vigente, presenta en la Universitat de València la referida Tesis Doctoral, y firma el presente certificado, en 

\begin{flushright}
Valencia, Agosto 2025
\end{flushright}

\bigskip
\bigskip
\bigskip
\bigskip
\bigskip
\bigskip

\begin{minipage}{0.45\textwidth}
\begin{flushleft}
	\begin{minipage}{6.5cm}
		\rule{\textwidth}{1pt}
		\centering  Dr.~Germán Vicente Rodrigo García
	\end{minipage}
\end{flushleft}
\end{minipage}
\hfill
\begin{minipage}{0.45\textwidth}
\begin{flushright}
	\begin{minipage}{6.5cm}
		\rule{\textwidth}{1pt}
		\centering  Dr.~Leandro Javier Cieri
	\end{minipage}
\end{flushright}
\end{minipage}

\vfill




\cleardoublepage


\begin{quote}
    \textit{We live on an island surrounded by a sea of ignorance. As our island of knowledge grows, so does the shore of our ignorance. 
    - John Archibald Wheeler}
\end{quote}


\newpage

\begin{quote}
    \textit{Life before death. Strength before weakness. Journey before destination. -~Brandon Sanderson}
\end{quote}
\newpage

\begin{dedication}
\textit{A ti, por ser hogar en cualquier lugar}
\end{dedication}
\begin{acknowledgements}
\addchaptertocentry{\acknowledgementname} 
\markboth{Acknowledgements}{Acknowledgements} 
\vspace{1cm}

En estos cuatro años de doctorado he vivido una etapa que me ha enriquecido profundamente, tanto a nivel académico y profesional como a nivel personal. Este camino me ha permitido enfrentarme a retos intelectuales que me han estimulado y formado académicamente, pero también me ha brindado la oportunidad de asistir a conferencias y realizar estancias por todo el mundo, descubriendo otras culturas, otras formas de hacer ciencia, y conociendo a personas increíbles con las que compartir ideas, debatir y de las que aprender. Cada conversación, cada viaje, cada reto superado ha contribuido a darme más perspectiva y a dar forma a la persona que soy hoy. Siento que acabo este doctorado como una persona más madura, con ideas y objetivos vitales más claros, y convencido de que todo este proceso me ha transformado de una forma que difícilmente habría alcanzado por otros caminos. Por todo ello, me siento profundamente agradecido.

En primer lugar, me gustaría agradecer a Germán por ser un excelente supervisor y por encontrar siempre tiempo para mí y mis proyectos. Su infinita sabiduría y su capacidad para transformar proyectos en publicaciones han sido una fuente constante de inspiración. Desde el primer momento me he sentido muy arropado trabajando con él, siempre abierto a que comenzara nuevos proyectos y apoyándome para viajar a escuelas y conferencias para complementar mi aprendizaje en esta etapa. Ahora me toca comenzar una nueva etapa como investigador más independiente pero de verdad espero que nuestros caminos científicos vuelvan a cruzarse en un futuro no muy lejano.

En la misma línea, también quiero agradecer a Leandro por su apoyo como co-director, por sus consejos y por saber transmitir calma y esa característica ``buena onda'' en los momentos más necesarios. Asimismo, deseo agradecer a la Generalitat Valenciana por su apoyo a través del proyecto No. ACIF/2021/219.

I would also like to thank Michele Grossi, for his constant help and interest in keeping a very fruitful collaboration since we started working together at the QT4HEP conference at CERN in 2022. You have always been there for me as an unoffical PhD co-supervisor and I am very grateful for that.

Still from a scientific perspective, although not related to the content of the thesis, I would like to thank Stefano Carrazza from hosting me during my time at University of Milan working on \texttt{
Qibo}. I am also grateful to Fabio Scafirimuto for giving me the opportunity to join IBM Quantum for an internship in Zurich during the final months of my PhD, where I got to experience first-hand how is to work at one of the world leaders in quantum computing. Also to Alvaro Ballon for giving me the amazing opportunity of doing an internship at Xanadu to work with him in Toronto for what it has been one of the best summers of my life so far!

The next lines are for those who might not have affected directly to the content of this thesis but have definitely contributed to my happiness and well-being throughout the journey. 

First, I would like to thank the huge squad of ``El Rincón del IATA'' at IFIC, from those who were there when I started and have already left (Gui, Selo, Fabián, Ele, Drona and Chitra, Ema and Pier, Andrés, Masha) to those who started after and will still be there when I am gone (Baibhab, Androniki, Juliana, Dom, Kostas, David, Muño, Andrea, Aure, Rafa, Arnau, Miguel, Sergio, Juanma). For sharing time over seminars, lunches, picaetas, conferences, trips, geotastic, petanca, M\&M's and more!
Then, I also want to thank all the people I have met during my research stays, internships, conferences that have made a significant impact in my PhD journey. That includes the people from Santander (Celia, Pablo, Carmen) and Oviedo (Andrea, Carlos, Alejandro, Clara), the QTI people at CERN (Matteo, Oriel, Vasilis, Alice, Ema), my fellow ex-residents at Xanadu (Emiliano, Ahmed, Siddhu, Praveen, Kasper, Serene, Juan), and many more names I cannot fit into these lines.

La siguiente línea va para Guillermo Alonso. Recuerdo que empecé el doctorado viendo y aprendiendo de tus vídeos en YouTube sobre computación cuántica, ya que al principio no tenía ni idea. Por eso, poder acabar yendo a Xanadu, conocerte en persona, poder trabajar contigo y aprender de ti y, además, poder considerarte ahora un amigo, ha sido una experiencia muy bonita que me llevo de estos años.

También me gustaría agradecer a mis amigos de la carrera (Pastor, Alicia, Rebeca, Pablet, Zanón, Miró, Ana) por seguir siendo parte de mi vida durante estos años de doctorado. Entre cenas de sushi, pizzas o hamburguesas me transmitíais como era la vida del físico más allá del doctorado y la academia.

Los siguientes en la línea de agradecimientos no pueden ser otros que los integrantes de BOM: Raúl, Jorge, Dani, Adolfo, Iván, Seguí, Papeco y Jaime. Habéis sido, sois y seréis siempre un pilar fundamental en mi vida y lo primero que me viene a la cabeza cuando pienso en el término ``mejores amigos''. Gracias por las risas, nuestras coñas y vocabulario que nadie más entiende, los viajes BOM, y por hacer esta etapa un poco más fácil cada día.

Para acabar, me gustaría dar las gracias a mi familia por esta tesis. A mi padre, por haberme inculcado desde pequeño esa curiosidad y pasión por el conocimiento y la ciencia, y por haberme servido de modelo académico a seguir. A mi hermano, a quien siempre he admirado, por su ayuda a nivel académico desde el colegio hasta la universidad y cuyos pasos he ido siguiendo de una manera casi inconsciente. A mi madre, por vivir con orgullo cada pequeño logro que iba consiguiendo y por creer siempre en mí incluso en los momentos en los que yo mismo dudaba. También a mi familia extendida Salva y Sandra por acogerme como uno más y por su constante implicación y apoyo en este camino.

Finalmente, solo puedo acabar estos agradecimientos nombrando a una persona. Gracias Paula por ser mi apoyo incondicional, mi fan número uno, y la mejor compañera de vida que uno podría pedir. Gracias por ser capaz siempre de sacarme una sonrisa, incluso en los días más complicados. Tu forma de ser y tu energía han sido siempre el mejor antídoto contra el estrés y lo que siempre ha sido capaz de alegrar mis días. Gran parte de lo que he conseguido ha sido gracias a ti, y me siento tremendamente afortunado de tenerte a mi lado y de poder seguir compartiendo contigo cada nueva etapa de mi vida. 

Siempre serás mi hogar.

\end{acknowledgements}


\chapter*{Preface}

\section*{Abstract}

This thesis investigates the potential of quantum computing to address key computational challenges in high-energy physics (HEP), particularly in light of the increasing complexity of modern particle physics experiments and the continued absence of experimental discoveries beyond the Standard Model (SM). Despite its remarkable success, the SM leaves several fundamental questions unresolved, such as the origin of neutrino masses, the nature of dark matter, and the matter-antimatter asymmetry. In this regard, the role of advanced experimental computational tools becomes more crucial than ever, both for analyzing the vast volumes of data produced in collider experiments and for improving the accuracy  and reach of theoretical predictions. 

Quantum computing, inspired by Richard Feynman's vision of simulating quantum systems with quantum machines, offers a paradigm that could outperform classical methods in specific domains. This thesis explores how quantum algorithms can be applied to problems in HEP, which is fundamentally governed by quantum field theory. The contributions span three main research areas: quantum clustering, quantum Monte Carlo integration, and quantum generative modeling for multivariate distributions.

Chapter \ref{chapter:intro} provides a brief historical introduction of quantum mechanics, quantum field theory and quantum computing and outlines the motivation for this work.

Chapter \ref{chap:sm} presents a concise overview of the SM. It introduces the elementary particles and their interactions, and then describes the quantum field theories that form the SM by examining the different terms in its Lagrangian: Quantum Chromodynamics (QCD) for the strong interaction, the Electroweak (EW) theory unifying electromagnetic and weak interactions, the Higgs sector responsible for EW symmetry breaking, and the Yukawa terms generating fermion masses.

Chapter \ref{chap:qc} introduces the basics of QC from a software point of view. First, we highlight the differences between quantum bits (qubits) and classical bits, as well as between quantum gates and their classical counterparts. Then, we explain the structure of quantum algorithms and how they are represented through quantum circuits, illustrating key examples relevant to this thesis, such as the \textit{SwapTest}, Grover's algorithm, and variational algorithms. Finally, it provides an introduction to QML, detailing some of the most widely used methods in the field.

The core results of this work are presented in Chapters \ref{chap:qjets}, \ref{chap:qint}, and \ref{chap:qchebff}, which focus on applications of quantum algorithms to relevant problems in particle physics. First, in Chapter~\ref{chap:qjets}, we introduce two quantum subroutines, one for computing Minkowski-based distances and another for finding the maximum in an unsorted list, which are then applied to a jet clustering problem using simulated LHC data. Chapter \ref{chap:qint} introduces Quantum Fourier Iterative Amplitude Estimation (QFIAE), a quantum Monte Carlo integrator that combines a quantum neural network for Fourier decomposition with Quantum Amplitude Estimation, enabling an end-to-end quantum pipeline without classical preprocessing. This method is validated by computing scattering amplitudes, Feynman loop integrals via the Loop–Tree Duality formalism, and NLO (Next-to Leading Order) decay rates for physical processes on both simulators and real quantum devices. Finally, in Chapter~\ref{chap:qchebff} we propose a quantum Chebyshev probabilistic model designed to learn probability distributions, offering exponential resolution improvements by adding extra qubits. Then, the model is applied to learn and generate fragmentation functions of partons fragmenting into kaons and pions. 

The thesis concludes in Chapter \ref{chapter:outlook} with a discussion of the implications of these results, potential pathways toward finding useful quantum applications in HEP, and future research directions as quantum hardware progresses toward fault-tolerant architectures. Chapter \ref{chapter:resumen} presents a summary of the thesis in Spanish, and the appendices provide further details on selected topics from Chapters \ref{chap:qjets} to \ref{chap:qchebff}.

Overall, these results show how quantum algorithms can already tackle relevant HEP problems on current hardware, while paving the way for future fault-tolerant applications that fully exploit quantum computational advantages.

\newpage

\section*{Resumen}

Esta tesis estudia el potencial de la computación cuántica para abordar desafíos computacionales clave en física de altas energías, en un contexto marcado por la creciente complejidad de los experimentos modernos en física de partículas y la ausencia continuada de descubrimientos experimentales más allá del Modelo Estándar. A pesar de su notable éxito, el Modelo Estándar deja sin resolver cuestiones fundamentales como el origen de las masas de los neutrinos, la naturaleza de la materia oscura o la asimetría entre materia y antimateria. En este sentido, el papel de las herramientas computacionales avanzadas resulta hoy más esencial que nunca, tanto para analizar los enormes volúmenes de datos generados en experimentos de colisionadores como para mejorar la precisión y el alcance de las predicciones teóricas.  

La computación cuántica, inspirada en la visión de Richard Feynman de simular sistemas cuánticos mediante ordenadores cuánticos, ofrece un paradigma con potencial para superar a los métodos clásicos en casos concretos. En esta tesis se explora cómo los algoritmos cuánticos pueden aplicarse a problemas en física de altas energías, gobernados de manera fundamental por la teoría cuántica de campos. Las contribuciones de esta tesis se organizan en tres áreas principales: técnicas de \textit{clustering} cuántico, integración cuántica de Monte Carlo y modelado generativo cuántico de distribuciones multidimensionales.  

El Capítulo~\ref{chapter:intro} presenta una breve introducción histórica a la mecánica cuántica, la teoría cuántica de campos y la computación cuántica, y expone la motivación de este trabajo.  

En el Capítulo~\ref{chap:sm} se ofrece una visión global del Modelo Estándar. Se introducen las partículas elementales y sus interacciones, para después describir las teorías cuánticas de campos que lo conforman, a través de los distintos términos de su lagrangiano: la Cromodinámica Cuántica, que describe la interacción fuerte; la teoría Electrodébil, que unifica las interacciones electromagnética y débil; el sector de Higgs, responsable de la ruptura espontánea de simetría electrodébil; y los términos de Yukawa, que generan las masas de los fermiones mediante su interacción con el campo de Higgs.  

El Capítulo~\ref{chap:qc} introduce los fundamentos de la computación cuántica desde un punto de vista de software. Se comparan los bits cuánticos (\textit{qubits}) con los bits clásicos, así como las puertas cuánticas con sus equivalentes clásicas. A continuación, se describe la estructura de los algoritmos cuánticos y su representación mediante circuitos cuánticos, ilustrando ejemplos relevantes para esta tesis como el \textit{SwapTest}, el algoritmo de Grover y los algoritmos variacionales. El capítulo concluye con una introducción al aprendizaje automático cuántico (QML), repasando algunos de los métodos más empleados en este ámbito.  

Los resultados principales se presentan en los Capítulos~\ref{chap:qjets}, \ref{chap:qint} y \ref{chap:qchebff}, centrados en la aplicación de algoritmos cuánticos a problemas de interés en física de partículas. En el Capítulo~\ref{chap:qjets} se introducen dos subrutinas cuánticas: una para calcular distancias basadas en la métrica de Minkowski y otra para localizar el valor máximo en una lista no ordenada; ambas se aplican al problema de \textit{jet clustering} empleando datos simulados del LHC. El Capítulo~\ref{chap:qint} presenta el integrador cuántico de Monte Carlo denominado \textit{Quantum Fourier Iterative Amplitude Estimation} (QFIAE), que combina una red neuronal cuántica para descomposición en series de Fourier con la técnica de \textit{Quantum Amplitude Estimation}, lo que permite un flujo de trabajo íntegramente cuántico sin preprocesamiento clásico. El método se valida mediante el cálculo de amplitudes de dispersión, integrales de bucle de Feynman con el formalismo \textit{Loop--Tree Duality} (LTD) y tasas de decaimiento a NLO (\textit{Next-to Leading Order}) en teoría cuántica de campos perturbativa, empleando tanto simuladores como dispositivos cuánticos reales. Por último, en el Capítulo~\ref{chap:qchebff} se propone un modelo probabilístico cuántico basado en polinomios de Chebyshev, diseñado para aprender distribuciones de probabilidad y ofrecer mejoras exponenciales de resolución mediante la adición de qubits extra. El modelo se aplica al aprendizaje y generación de funciones de fragmentación de partones en kaones y piones.  

La tesis concluye en el Capítulo~\ref{chapter:outlook} con una discusión sobre el alcance de los resultados, posibles vías para identificar aplicaciones cuánticas útiles en HEP y las líneas de investigación futuras a medida que el hardware cuántico avanza hacia arquitecturas tolerantes a fallos. El Capítulo~\ref{chapter:resumen} presenta un resumen en castellano, mientras que los apéndices recogen detalles adicionales de los temas tratados en los Capítulos~\ref{chap:qjets} a \ref{chap:qchebff}.  

En conjunto, los resultados muestran que, incluso con el hardware actual, los algoritmos cuánticos pueden abordar ya problemas relevantes en HEP, sentando a la vez las bases para aplicaciones futuras tolerantes a fallos que exploten plenamente las ventajas de la computación cuántica.  

\newpage

\section*{Resum}

Aquesta tesi estudia el potencial de la computació quàntica per a abordar desafiaments computacionals clau en física d'altes energies, en un context marcat per la creixent complexitat dels experiments moderns en física de partícules i l'absència continuada de descobriments experimentals més enllà del Model Estàndard. Malgrat el seu notable èxit, el Model Estàndard deixa sense resoldre qüestions fonamentals com l'origen de les masses dels neutrins, la naturalesa de la matèria fosca o l'asimetria entre matèria i antimatèria. En aquest sentit, el paper de les eines computacionals avançades resulta hui més essencial que mai, tant per a analitzar els enormes volums de dades generats en experiments de col·lisionadors com per a millorar la precisió i l'abast de les prediccions teòriques.  

La computació quàntica, inspirada en la visió de Richard Feynman de simular sistemes quàntics mitjançant ordinadors quàntics, ofereix un paradigma amb potencial per a superar els mètodes clàssics en casos concrets. En aquesta tesi s'explora com els algoritmes quàntics poden aplicar-se a problemes en física d'altes energies, governats de manera fonamental per la teoria quàntica de camps. Les contribucions d'aquesta tesi s'organitzen en tres àrees principals: tècniques de \textit{clustering} quàntic, integració quàntica de Monte Carlo i modelatge generatiu quàntic de distribucions multidimensionals.  

El Capítol~\ref{chapter:intro} presenta una breu introducció històrica a la mecànica quàntica, la teoria quàntica de camps i la computació quàntica, i exposa la motivació d'aquest treball.  

En el Capítol~\ref{chap:sm} s'ofereix una visió global del Model Estàndard. S'introdueixen les partícules elementals i les seues interaccions, per a després descriure les teories quàntiques de camps que el conformen, a través dels diferents termes del seu lagrangià: la Cromodinàmica Quàntica, que descriu la interacció forta; la teoria Electrodèbil, que unifica les interaccions electromagnètica i dèbil; el sector de Higgs, responsable de la ruptura espontània de simetria electrodèbil; i els termes de Yukawa, que generen les masses dels fermions mitjançant la seua interacció amb el camp de Higgs.  

El Capítol~\ref{chap:qc} introdueix els fonaments de la computació quàntica des d'un punt de vista de \textit{software}. Es comparen els bits quàntics (\textit{qubits}) amb els bits clàssics, així com les portes quàntiques amb les seues equivalents clàssiques. A continuació, es descriu l'estructura dels algoritmes quàntics i la seua representació mitjançant circuits quàntics, il·lustrant exemples rellevants per a aquesta tesi com el \textit{SwapTest}, l'algoritme de Grover i els algoritmes variacionals. El capítol conclou amb una introducció a l'aprenentatge automàtic quàntic (QML), revisant alguns dels mètodes més emprats en aquest àmbit.  

Els resultats principals es presenten en els Capítols~\ref{chap:qjets}, \ref{chap:qint} i \ref{chap:qchebff}, centrats en l'aplicació d'algoritmes quàntics a problemes d'interés en física de partícules. En el Capítol~\ref{chap:qjets} s'introdueixen dues subrutines quàntiques: una per a calcular distàncies basades en la mètrica de Minkowski i una altra per a localitzar el valor màxim en una llista no ordenada; ambdues s'apliquen al problema de \textit{jet clustering} emprant dades simulades del LHC. El Capítol~\ref{chap:qint} presenta l'integrador quàntic de Monte Carlo denominat \textit{Quantum Fourier Iterative Amplitude Estimation} (QFIAE), que combina una xarxa neuronal quàntica per a descomposició en sèries de Fourier amb la tècnica de \textit{Quantum Amplitude Estimation}, cosa que permet un flux de treball íntegrament quàntic sense preprocessament clàssic. El mètode es valida mitjançant el càlcul d'amplituds de dispersió, integrals de bucle de Feynman amb el formalisme \textit{Loop--Tree Duality} (LTD) i taxes de desintegració a NLO (\textit{Next-to Leading Order}) en teoria quàntica de camps pertorbativa, emprant tant simuladors com dispositius quàntics reals. Finalment, en el Capítol~\ref{chap:qchebff} es proposa un model probabilístic quàntic basat en polinomis de Chebyshev, dissenyat per a aprendre distribucions de probabilitat i oferir millores exponencials de resolució mitjançant l'addició de \textit{qubits} extra. El model s'aplica a l'aprenentatge i generació de funcions de fragmentació de partons en kaons i pions.  

La tesi conclou en el Capítol~\ref{chapter:outlook} amb una discussió sobre l'abast dels resultats, possibles vies per a identificar aplicacions quàntiques útils en HEP i les línies d'investigació futures a mesura que el maquinari quàntic avança cap a arquitectures tolerants a fallades. El Capítol~\ref{chapter:resumen} presenta un resum en castellà, mentre que els apèndixs recullen detalls addicionals dels temes tractats en els Capítols~\ref{chap:qjets} a \ref{chap:qchebff}.  

En conjunt, els resultats mostren que, fins i tot amb el maquinari actual, els algoritmes quàntics poden abordar ja problemes rellevants en HEP, assentant alhora les bases per a aplicacions futures tolerants a fallades que aprofiten plenament els avantatges de la computació quàntica.

\newpage

\addcontentsline{toc}{chapter}{Preface}

\section*{List of publications}

This PhD thesis is based on the following publications:

\subsection*{Articles}
\begin{itemize}
    \item[\cite{deLejarza:2025upd}] \textbf{Jorge J. Martínez de Lejarza}, Hsin-Yu Wu, Oleksandr Kyriienko, Germán Rodrigo, Michele Grossi, \textit{Quantum Chebyshev Probabilistic Models for Fragmentation Functions,   
    \href{https://www.nature.com/articles/s42005-025-02361-1}{Commun. Phys. \textbf{8}, 448}, 
    \href{https://arxiv.org/abs/2503.16073}{		arXiv:2503.16073}} (2025). 
    \item[\cite{deLejarza:2024scm}] \textbf{Jorge J. Martínez de Lejarza}, David F. Rentería-Estrada, Michele Grossi, Germán Rodrigo, \textit{Quantum integration of decay rates at second order in perturbation theory, \href{https://iopscience.iop.org/article/10.1088/2058-9565/ada9c5}{	Quantum Sci. Technol. \textbf{10} 025026}}, \href{https://arxiv.org/abs/2409.12236}{arXiv:2409.12236} (2024). 
    
    \item[\cite{deLejarza:2024pgk}] \textbf{Jorge J. Martínez de Lejarza}, Leandro Cieri, Michele Grossi, Sofia Vallecorsa, Germán Rodrigo, \textit{Loop Feynman integration on a quantum computer,} \href{https://journals.aps.org/prd/abstract/10.1103/PhysRevD.110.074031}{Phys. Rev. D \textbf{110} 074031}, \href{https://arxiv.org/abs/2401.03023}{arXiv:2401.03023} (2024).

    \item[\cite{LTD:2024yrb}] Selomit Ramírez-Uribe, Andrés E. Rentería-Olivo, David F. Rentería-Estrada, \textbf{Jorge J. Martínez de Lejarza}, Prasanna K. Dhani, Leandro Cieri, Roger J. Hernández-Pinto, German F.R. Sborlini, William J. Torres Bobadilla, Germán Rodrigo, \textit{Vacuum amplitudes and time-like causal unitary in the loop-tree duality},  \href{10.1007/JHEP01(2025)103}{JHEP \textbf{01} 103}, \href{https://arxiv.org/abs/2404.05492}{arXiv:2404.05492} (2024).
    
    \item[\cite{deLejarza:2023IEEE}] \textbf{Jorge J. Martínez de Lejarza}, Michele Grossi, Leandro Cieri, Germán Rodrigo, \textit{Quantum Fourier Iterative Amplitude Estimation}, \href{https://ieeexplore.ieee.org/document/10313906/}{2023 IEEE International Conference on Quantum Computing and Engineering (QCE), Bellevue, WA, USA, pp. 571-579}, \href{https://arxiv.org/abs/2305.01686}{arXiv:2305.01686} (2023) .
    \item[\cite{deLejarza:2022bwc}] \textbf{Jorge J. Martínez de Lejarza}, Leandro Cieri, Germán Rodrigo, \textit{Quantum clustering and jet reconstruction at the LHC},  \href{https://doi.org/10.1103/PhysRevD.106.036021}{Phys. Rev. D \textbf{106} 036021}, \href{https://arxiv.org/abs/2204.06496}{arXiv:2204.06496} (2022).
    
    \item[\cite{Delgado:2022tpc}] Andrea Delgado, Kathleen E. Hamilton, Prasanna Date, Jean-Roch Vlimant, Duarte Magano, Yasser Omar, Pedrame Bargassa, Anthony Francis, Alessio Gianelle, Lorenzo Sestini, Donatella Lucchesi, Davide Zuliani, Davide Nicotra, Jacco de Vries, Dominica Dibenedetto, Miriam Lucio Martinez, Eduardo Rodrigues, Carlos Vazquez Sierra, Sofia Vallecorsa, Jesse Thaler, Carlos Bravo-Prieto, Su Yeon Chang, Jeffrey Lazar, Carlos A. Argüelles, \textbf{Jorge J. Martinez de Lejarza}, \textit{Quantum Computing for Data Analysis in High-Energy Physics},   \href{https://arxiv.org/abs/2203.08805}{	arXiv:2203.08805} (2022).
\end{itemize}

\subsection*{Conference proceedings}
\begin{itemize}
\item[\cite{deLejarza:2022vhe}] \textbf{Jorge J. Martínez de Lejarza}, Leandro Cieri, Germán Rodrigo, \textit{Quantum jet clustering with LHC simulated data}, \href{https://pos.sissa.it/414/241}{PoS ICHEP2022 \textbf{241}}, \href{https://arxiv.org/abs/2209.08914}{	arXiv:2209.08914} (2022).

\end{itemize}

\subsection*{Other research works carried out during the course of PhD}
\begin{itemize}
 \item[\cite{Pyretzidis:2025stx}] Konstantinos Pyretzidis, \textbf{Jorge J. Martínez de Lejarza}, Germán Rodrigo, \textit{Unlocking Multi-Dimensional Integration with Quantum Adaptive Importance Sampling \href{https://arxiv.org/abs/2506.19965}{		arXiv:2506.19965}} (2025).
 \end{itemize}
 
 \newpage
 
Different talks about the topic of this thesis were also given at the following conferences and events:

\begin{itemize}
\item Oral presentation of a technical paper in ``Quantum Technologies for High-Energy Physics (QT4HEP25)'': \href{https://indico.cern.ch/event/1433194/contributions/6280035/}{Quantum Chebyshev Generative modeling for fragmentation functions} (2025).
\item Oral presentation during poster session on \textit{Quantum Techniques for Machine Learning }(QTML24): 
\href{https://indico.qtml2024.org/event/1/contributions/75/}{Quantum integration of decay processes at high-energy colliders} (2024).

\item Oral presentation during poster session on \textit{Quantum Techniques for Machine Learning }(QTML23): 
\href{https://indico.cern.ch/event/1288979/page/31312-instructions-for-posters}{Quantum integration of Feynman loop integrals} (2023).
\item Invited speaker to give a seminar at CERN QTI-TH Forum: \href{https://indico.cern.ch/event/1327342/}{Quantum Fourier Iterative Amplitude Estimation} (2023).
\item Oral presentation of a technical paper in \textit{IEEE Quantum Week 2023}: \href{https://ieeexplore.ieee.org/document/10313906/}{Quantum Fourier Iterative Amplitude Estimation} (2023).
\item Oral presentation during poster session on Spring School on \textit{Quantum Information Processing – Applications on Gate-based and Annealing Systems}: \href{https://indico3-jsc.fz-juelich.de/event/44/}{Quantum jet clustering with LHC simulated data}  (2023). 
\item Invited speaker to give a seminar on \textit{2nd workshop of AI Initiative for Science}: 
\href{https://indico.ific.uv.es/event/6792/}{Quantum Machine Learning for Particle Physics} (2022).
\item Oral presentation during poster session on \textit{International Conference on Quantum Technologies for High-Energy Physics} (QT4HEP22): 
\href{https://indico.cern.ch/event/1190278/contributions/5107341/}{Quantum jet clustering with LHC simulated data} (2022).
\item Oral presentation on \textit{International Conference on High Energy Physics} (ICHEP): \href{https://agenda.infn.it/event/28874/contributions/169890/}{Quantum clustering and jet reconstruction at the LHC} (2022).
\end{itemize}

\newpage

\subsection*{Science communication}
Different events or material I have created during my thesis about science communication in quantum computing:
\begin{itemize}
\item \textbf{Jorge J. Martínez de Lejarza} \href{https://github.com/qiskit-community/qgss-2025/blob/main/lab-0/lab0.ipynb}{\textit{Lab 0: Hello Quantum World! - Qiskit Global Summer School 2025}} (2025).
\item \textbf{Jorge J. Martínez de Lejarza}, Alberto Maldonado \href{https://github.com/qiskit-community/qgss-2025/blob/main/lab-2/Lab2.ipynb}{\textit{Lab 2: Cutting through the noise - Qiskit Global Summer School 2025}} (2025).
\item \textbf{Jorge J. Martínez de Lejarza} \href{https://pennylane.ai/codebook/variational-quantum-algorithms}{\textit{Pennylane Codebook: Variational Quantum Algorithms}} (2024).            
\item \textbf{Jorge J. Martínez de Lejarza}, Guillermo Alonso-Linaje  \href{https://pennylane.ai/qml/demos/tutorial_how_to_use_quantum_arithmetic_operators}{\textit{Pennylane Demo: How to use quantum arithmetic operators}} (2024). 
\item \textbf{Jorge J. Martínez de Lejarza}, Serene Shum  \href{https://pennylane.ai/qml/demos/tutorial_qnn_multivariate_regression}{\textit{Pennylane Demo: Multidimensional regression with a variational quantum circuit}} (2024).
\item Assistant in a hands-on tutorial in ``Quantum Machine Learning'' in: \href{https://events.perimeterinstitute.ca/event/75/timetable/}{Navigating Quantum and AI Career Trajectories: A Beginner’s Mini-Course on Computational Methods and their Applications} at Perimeter Institute (2024).         
   
\item \textbf{Jorge J. Martínez de Lejarza},   \href{https://qibo.science/qibo/stable/code-examples/tutorials/qfiae/qfiae_demo.html} {\textit{Tutorial: Quantum Fourier Iterative Amplitude Estimation}} (2023).
\item Outreach presentation with a live demo to a non-technical audience within the Quantum Spain project about {\href{https://www.linkedin.com/posts/gorka-martinez-de-lejarza-samper-51b4a61b6_quantumvalencia-computaciaejncuaerntica-activity-7054515716280004608-oXg7/?utm_source=share&utm_medium=member_android}{Quantum Computing and the protocol BB84}}  (2023).
\end{itemize}

\addcontentsline{toc}{chapter}{Publications}

\pdfbookmark[0]{Contents}{Contents}
\dominitoc
\tableofcontents 
\adjustmtc[2] 


\mainmatter 



\pagestyle{thesis} 

%

\chapter{Introduction}\label{chapter:intro}

\section{Historical introduction}

At the beginning of the 20th century, a series of groundbreaking discoveries made evident something that until that moment was not questionable. Classical theories could not explain certain phenomena and a better theory was needed. 

The first problem that intrigued the physics community was the blackbody radiation problem. Blackbody is a term used to describe objects that absorb all incident electromagnetic radiation and emit radiation across a continuous spectrum that depends only on its temperature. Classical physics failed to explain the radiation that blackbody would emit, predicting infinite energy at high frequencies, a contradiction known as the ultraviolet catastrophe. In 1900, Max Planck proposed that energy is emitted in discrete packets (quanta), leading to a formula that accurately described the observed radiation spectrum and resolved the inconsistency \cite{maxplanck}.

Inspired by Planck's work, in 1905 Albert Einstein proposed his solution to explain the photoelectric effect. This effect consists of the emission of electrons from a material caused by electromagnetic radiation such as ultraviolet light. Classical electromagnetism predicted that light waves transfer energy to electrons continuously, causing emission once enough energy accumulates. However, experiments showed that electrons are emitted only when light exceeds a specific frequency, independent of intensity or exposure time. To explain this, Einstein proposed that light consists of discrete energy packets, or photons, rather than continuous waves, which awarded him the Nobel Prize in 1911.

Years later, in the decade of 1920, a generation of very talented physicists laid the foundation of Quantum Mechanics (QM). In 1924, Louis de Broglie proposed that matter, like light, could present wave-like behavior, introducing the concept of wave-particle duality that would be of vital importance for later discoveries. Then, in 1925, Werner Heisenberg introduced the first mathematical formalism of QM, the Matrix Mechanics, describing particles through probabilities rather than the deterministic laws of classical mechanics. Shortly afterward, Erwin Schrödinger developed Wave Mechanics, describing the quantum behavior in his well-known equation, which describes how particles evolve as waves, a cornerstone of quantum theory.

In 1927, Heisenberg transformed the field again with his Uncertainty Principle \cite{aHeisenberg:1927zz}, which states that one cannot simultaneously determine a particle's position and momentum with complete accuracy. This principle showed a fundamental limit in measurement at the quantum level, revealing a probabilistic nature to reality itself. Around the same time, Niels Bohr and his colleagues developed the Copenhagen interpretation, which suggested that the act of measurement affects quantum systems. Meaning that a quantum system can be in a superposition of different quantum states and the measurement itself forces nature to choose one of the multiple states. This interpretation highlighted the inherent uncertainty in QM, questioning the classical concept of an objective reality.

As QM continued to develop, it had to be reconciled with the theory of Special Relativity, by Albert Einstein \cite{Einstein:1905vqn}. Eventually, this would lead to the formulation of Quantum Field Theory~(QFT) in the 1930s. In particular, Paul Dirac played a critical role by developing the relativistic theory of the electron, predicting antimatter and laying the fundamentals for QFT. By the mid-20th century, quantum electrodynamics~(QED) emerged as the first complete QFT, describing interactions between electrons and photons with extraordinary precision. Over the next few decades, QFT evolved into the Standard Model~(SM) of particle physics, describing all known fundamental particles and their interactions through the strong, weak, and electromagnetic forces. This framework allowed physicists to make precise predictions for high-energy experiments, establishing QFT as the foundation of modern particle physics and one of the theories capable of making the most accurate results ever seen. 

In the meantime, while quantum theory was revolutionizing the physics world, information science was reshaping our world with the advent of the invention of the century, \textit{the machine that changed the world} \cite{Alba_2019}, the computer. In 1936, Alan Turing and Alonzo Church working independently and together, described what an algorithm is and defined the limits of what can be computed. This led to the Church–Turing thesis \cite{church-turing}, which states that any calculation that can be performed can be solved by an algorithm on a computer, as long as there is enough time and memory. That same year, Turing also introduced the concept of Turing Machines \cite{turing-machine}, which are theoretical devices that describe how digital computers work. These machines laid the groundwork for modern computers and introduced the idea of storing programs, which most computers use today. Turing Machines were designed to explore what can and cannot be computed, taking into account the limitations of computational power. 

Since the invention of such machines, scientists of across different fields started using
them to simulate the natural world. In light of the new discoveries, Richard Feynnman in the early 1980s was among the first ones to ponder the idea of simulating physical systems with a system that also follows the same rules, a quantum system. In fact, their words in ``Simulating Physics with Computers'' \cite{Feynman:1981tf} have now become a must-have for papers and seminar talks in the field of quantum computation and quantum information, they read ``nature isn't classical, dammit, and if you want to make a simulation of nature, you'd better make it quantum mechanical, and by golly it's a wonderful
problem, because it doesn't look so easy''.

This idea of simulating nature with quantum systems led to the development of Quantum Computing~(QC). Unlike classical computers, which process information in bits representing either 0 or 1, quantum computers use quantum bits, or qubits, represented by quantum states that can exist in a superposition of both $|0\rangle$ and $|1\rangle$ states. Quantum operations are carried out with quantum gates, fundamental units of quantum circuits that manipulate qubits through carefully controlled operations. These quantum gates enable the use of three main properties of QM that are essential for quantum computation: entanglement, where the state of one qubit directly influences the state of another; superposition, where a qubit can be in a combination of different states simultaneously; and interference, the process by which probabilities of qubit states amplify or cancel each other out.

From a theoretical point of view, the field of QC has made significant progress in understanding quantum algorithms and their potential applications. However, designing these algorithms poses unique challenges. In the notoriously famous book ``Quantum Computing and Quantum Information'' \cite{nielsen}, Nielsen and Chuang highlight two key reasons for this. First, humans naturally have a better intuition for classical computing, making it hard to come up with algorithms well-suited for quantum computers. Second, for a quantum algorithm to be truly valuable, it must outperform classical alternatives by performing the same computations with fewer resources. Some classical problems are theoretically solvable but practically unattainable due to their exponential complexity. The challenge lies in whether efficient quantum algorithms can perform these intractable computations in significantly less time, making them practically solvable. Despite the difficulties, in 1994 Peter Shor introduced a groundbreaking algorithm that efficiently performs prime factorization on quantum computers, with an exponential advantage in time. Soon after in 1996, Lov Grover developed an algorithm that provides a quadratic speedup for searching in unstructured databases. Recently, Variational Quantum Circuits (VQC) have emerged as a promising method for developing new quantum algorithms. These quantum circuits utilize parameterized quantum gates optimized by classical techniques, enabling researchers to tackle complex problems across various fields, including quantum chemistry and machine learning. Quantum Machine Learning (QML), in particular, aims to harness quantum computing's unique properties to enhance data processing and pattern recognition. While these algorithms showcase the capabilities of quantum computing, the range of efficient quantum solutions remains limited. Researchers are actively working to discover more efficient quantum algorithms, requiring creativity due to the absence of a clear roadmap in this emerging and exciting field.

On the other hand, from an experimental point of view, different architectures for quantum computers have been proposed, each leveraging unique technologies to manipulate qubits. Superconducting qubits, used by companies like IBM and Google, rely on circuits cooled to near absolute zero to maintain the coherence in the quantum states. Ion-trap quantum computers, developed by companies such as IonQ and Quantinuum, use electromagnetic fields to trap individual ions and perform quantum operations. Other architectures include photonic quantum computers, pursued by Xanadu and PsiQuantum, which use photons as qubits, and neutral-atom quantum computers, developed by companies like Pasqal and Atom Computing, which trap neutral atoms using lasers. Each approach has its own advantages and challenges, with ongoing development aiming to find the most scalable and stable design for practical QC. It has become evident that a race is taking place in the industry to determine which company can first develop a reliable quantum computer that achieves the long-awaited quantum advantage, potentially transforming the landscape of computational technology as we know it today.

\section{Motivation}
Despite the proliferation of algorithms and new methods in the field of QC, many of these endeavors have fallen short of addressing real-world challenges. Bridging this gap is essential, as tackling real-world problems offers invaluable insights into the limitations of current algorithms and methods. Engaging with such challenges provides a deeper understanding of existing knowledge gaps, facilitating the development of innovative algorithms and solutions. A clear illustration of this phenomenon can be found in the history of the European Council for Nuclear Research (CERN), where the invention of the World Wide Web (WWW) emerged as a solution to the practical challenge of sharing information among scientists across the globe. This historical precedent underscores the importance of addressing problems that pose technical challenges to state-of-the-art classical computers in driving innovation and advancing the field of QC.

Going back to Richard Feynman's quote, the main niche for QC lies in simulating quantum systems. In this direction, several areas have found important applications. For instance, quantum chemistry has seen success with algorithms like the Variational Quantum Eigensolver (VQE) \cite{Peruzzo:2013bzg} which computes the ground state of molecules, a daunting task for classical computers when dealing with large molecules. Another area of great interest is QFT. One of its greatest achievements is the SM of particle physics, one of the most conceptually rich frameworks in physics that provides the most accurate predictions of any scientific theory to date. However, it still presents numerous challenges. High-energy physics collisions, such as those at the Large Hadron Collider (LHC) at CERN, are described by QFT, and represent some of the most complex problems to analyze. Furthermore, inconsistencies observed in their analysis suggest that there may be new physics yet to be discovered or that the SM may need to incorporate new elements to address these new phenomena.

With these considerations in mind, this thesis identifies a promising area of application that combines both concepts. On the one hand, QFT is inherently quantum, making it well-suited for study through quantum systems. On the other hand, it encapsulates some of the most challenging problems in theoretical and computational science. Addressing these challenges with QC tools could yield to new insights into the fundamental particles that constitute the universe and also to advancements for QC itself. This thesis, therefore, aims to deepen our understanding of the universe by inspiring advancements in QC that could reshape modern computational paradigms.

\chapter{Standard Model}\label{chap:sm}

\section{The Standard Model of particle physics in a nutshell}\label{app:nutshell}

The Standard Model (SM) stands as the most widely accepted theory in particle physics. It consists of a Quantum Field Theory (QFT) that merges Quantum Mechanics and Special Relativity into a consistent field theory framework, enabling the classification of all known elementary particles and describing the fundamental interactions among themselves. Specifically, the SM arises from a combination of gauge theories that satisfy Noether's theorem \footnote{Noether's Theorem: If a Lagrangian $\mathcal{L}$ is invariant under a continuous symmetry, then a corresponding conservation law exists.}, such as Quantum Electrodynamics (QED), the Weak theory by Glashow, Weinberg, and Salam, and Quantum Chromodynamics (QCD).

\begin{figure}[th!]
    \centering
    \begin{subfigure}[b]{0.45\textwidth}
        \centering
        \includegraphics[width=\linewidth]{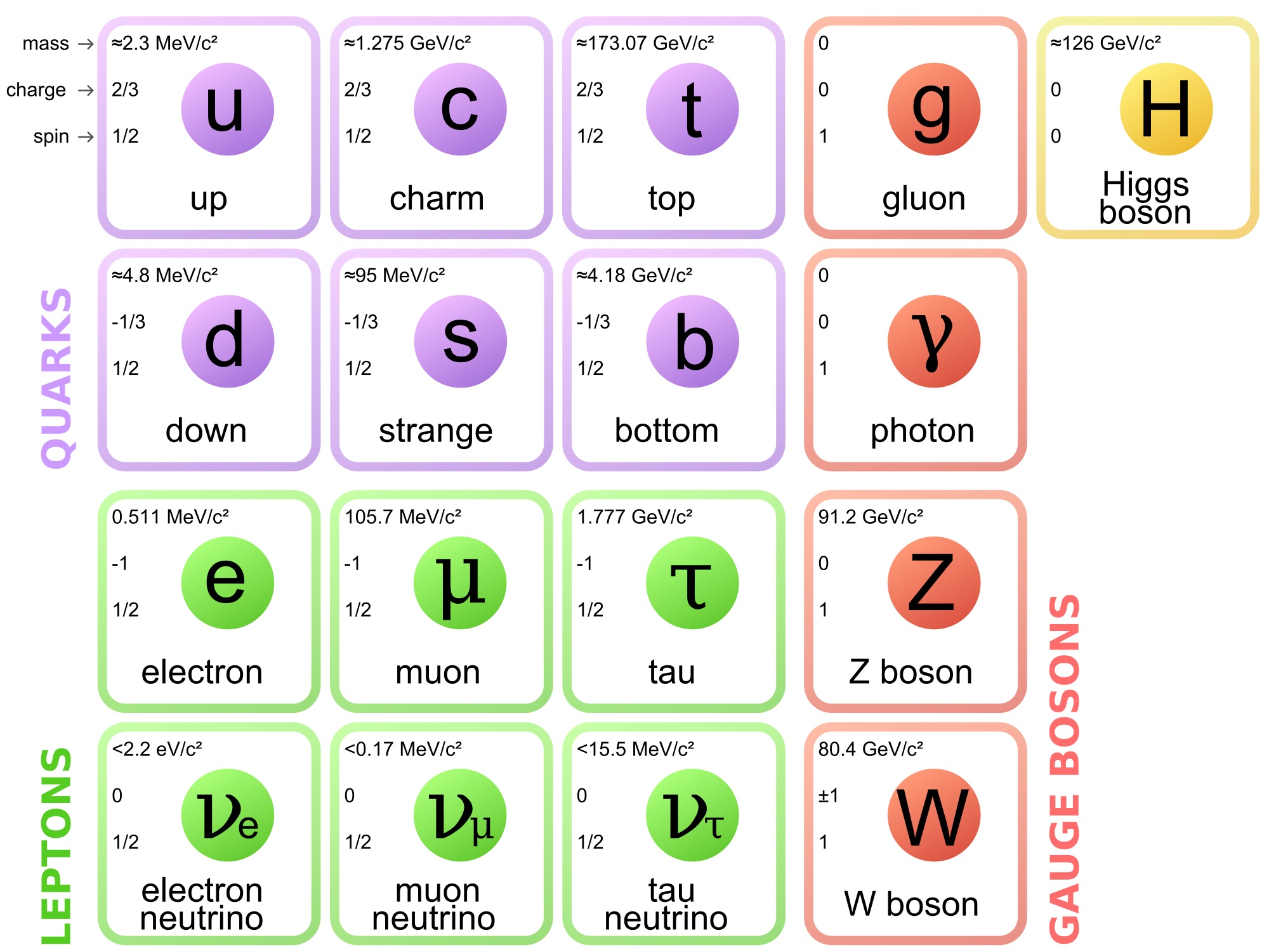}
        \caption{Particles.}
    \end{subfigure}
    \hfill
    \begin{subfigure}[b]{0.45\textwidth}
        \centering
        \includegraphics[width=\linewidth]{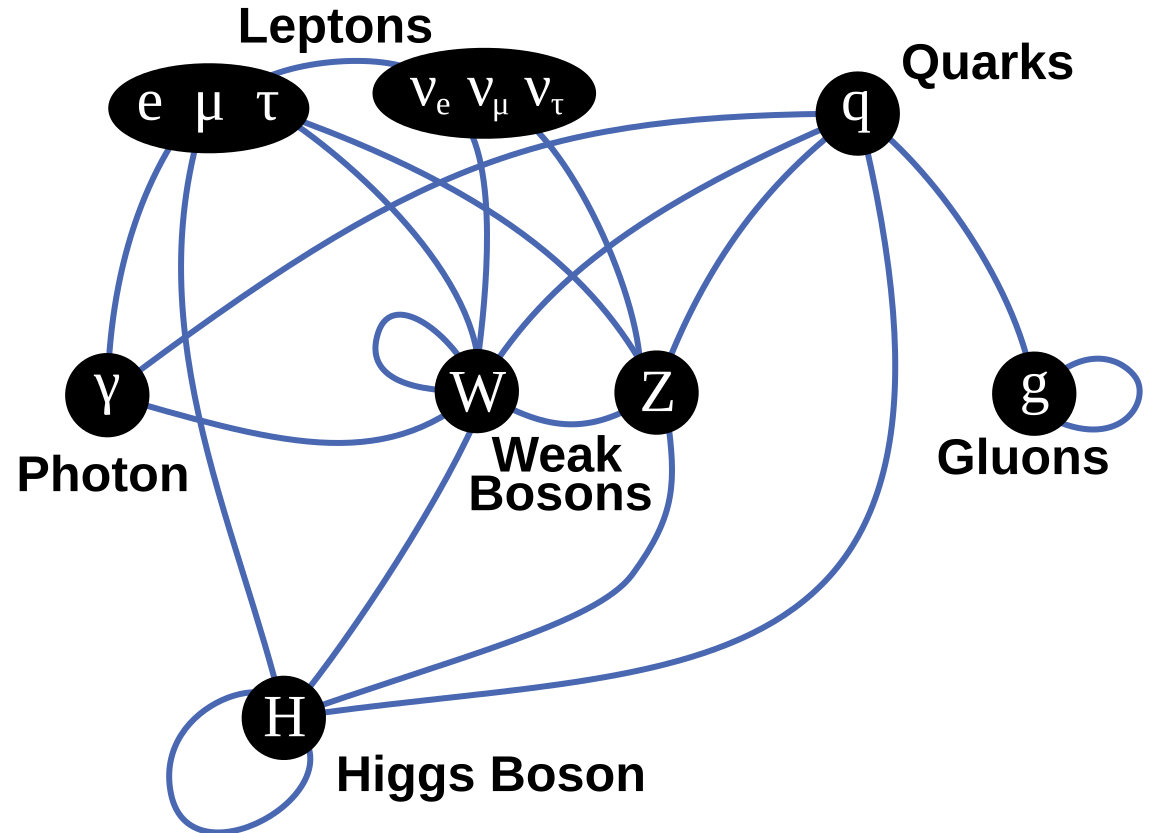}
        \caption{Interactions.}
    \end{subfigure}
    \caption{(a) Classification of the particles in the SM. Picture taken from~\cite{wikiparticles}. (b)~Interactions among particles in the SM. Picture taken from~\cite{wikiinterac}.} 
    \label{fig:sm}
\end{figure}

As shown in Fig.~\ref{fig:sm}, the SM particles can be divided into those constituting matter (fermions) and those mediating interactions (bosons).

The matter particles, or fermions, have half-integer spin and follow the Pauli Exclusion Principle, which states that two identical fermions cannot simultaneously occupy the same quantum state. These particles are named after the Fermi-Dirac statistics that govern their behavior. Each fermion has an associated antiparticle with the same mass but opposite electric charge. Fermions are further classified into quarks and leptons:
\begin{itemize}
    \item \textbf{Leptons}: Elementary particles that can exist freely or bound, do not interact via the strong force, and have nonzero mass. There are three lepton families (or flavors): the electron $e$, muon $\mu$, and tau $\tau$. Each family consists of a pair of particles with opposite lepton numbers. One is the charged, massive particle that gives the family its name, such as the electron, and the other is a neutral, nearly massless particle known as the neutrino.
    \item \textbf{Quarks}: Elementary particles that interact via the strong force and are never found free in nature, instead existing only within hadrons, composite particles made up of quarks, such as protons and neutrons. There are six types of quarks, arranged with leptons into three generations, each ordered by increasing discovery date and mass. The first generation includes the up ($u$) and down ($d$) quarks, which constitute ordinary matter (protons and neutrons). The second generation contains the charm~($c$) and strange ($s$) quarks, which appear in cosmic radiation or decay processes. Finally, the third generation consists of the top ($t$) and bottom ($b$) quarks, observed only under extreme conditions, such as those after the Big Bang or in particle accelerators such as the Large Hadron Collider (LHC) at CERN.
\end{itemize}

The SM describes the fundamental forces as a result of the exchange (emission and absorption) of mediating particles known as bosons. These particles are named after the Bose-Einstein statistics that describe their behavior. In contrast to fermions, bosons have integer spin and are not restricted by the Pauli Exclusion Principle. The different bosons include:
\begin{itemize}
    \item \textbf{Photons ($\gamma$)}: Massless particles mediating the electromagnetic interaction between electrically charged particles.
    \item \textbf{Bosons $W^+$, $W^-$, $Z$}: Massive particles mediating the weak interaction. Together with photons, they are the mediators of the electroweak interaction.
    \item \textbf{Gluons ($g$)}: Massless particles that mediate the strong interaction among particles with color charge, such as quarks and other gluons.
\end{itemize}

In addition to these bosons, the Higgs boson, a massive, spin-zero particle, plays a unique role by providing particles with mass through the Higgs mechanism.

The SM is an extraordinarily complex and elegant theory, where concepts like symmetry play a fundamental role in governing particle dynamics. Its predictive power is remarkable, successfully anticipating major discoveries \cite{Novaes:1999yn}, such as: the second generation of quarks in the 1970s \cite{Bjorken:1964gz}, the third generation of leptons in the 1980s \cite{Perl:1975bf}, the $W^{\pm}$ and $Z$ bosons \cite{UA1:1983crd, UA1:1983mne}, the third generation of quarks in the 1990s \cite{CDF:1995wbb}, and ultimately the Higgs boson in 2012 \cite{higgsdiscovery,CMS:2012qbp}. Not only has the SM predicted previously unknown particles, but it has also achieved unprecedented agreement between theoretical predictions and experimental data, such as the electron's anomalous magnetic moment and the masses of the $W^{\pm}$ and $Z$ bosons, establishing a benchmark for predictive accuracy in physics. For a more detailed review of the SM, see \cite{Pich:2012sx}.

Yet, despite these considerable successes, the SM cannot explain a range of phenomena, preventing it from being a “theory of everything.” Some of its limitations include \cite{apuntesew}:
\begin{itemize}
    \item \textbf{Neutrinos}: The SM is not able to explain why neutrinos are massive. However, experimental evidence confirms that at least two neutrinos possess nonzero mass. Additionally, the SM does not account for the experimentally observed neutrino oscillations.
    \item \textbf{Matter-antimatter asymmetry}: Following the Big Bang, a slight asymmetry between matter and antimatter led to a universe composed predominantly of matter. However, the SM cannot fully explain the magnitude of this asymmetry via baryogenesis.
    \item \textbf{Dark Matter}: Observational cosmology provides indirect evidence of a type of matter that neither emits nor interacts with light, known as dark matter, which the SM does not predict.
    \item \textbf{Gravity}: While the SM unifies three of the four fundamental forces (electromagnetism, weak, and strong forces), it does not incorporate general relativity, leaving the development of a quantum theory of gravity an open challenge.
\end{itemize}

In addition to these questions, the SM raises philosophical ones as well, as certain aspects seem arbitrary: Why are there only three fermion generations? Why is electric charge quantized in multiples of 1/3? Why does the SM contain 19 free parameters?

Analyzing both the strengths and limitations of the SM reveals it to be a highly successful and detailed theory with incredible predictive power. However, as it still cannot account for all observed phenomena, the SM motivates continued research in particle physics, encouraging us to seek answers to the many questions that remain open.

\section{A Lagrangian to rule them all}

The Lagrangian is a mathematical function that is used to describe the dynamics of a physical system. It contains the kinetic and potential energy of a system, allowing us to derive the equations of motion through the Principle of Least Action. In QFT, the Lagrangian provides a compact framework to express the fundamental interactions and symmetries of elementary particles. For the SM, the Lagrangian encodes the behavior of fermions, gauge bosons, and the Higgs field, as well as the interactions that define the electroweak and strong forces. 

The Lagrangian of the SM is built based on fundamental symmetries. These symmetries arise from one of the most fundamental principles of physics: the laws of physics must be independent of the position and speed of a given observer. This translates to the laws of physics must be invariant under Lorentz transformations. As a QFT, the SM uses fields to describe particles, but these fields are not physical observables that can be directly measured. Thus, different field configurations with no physical relevance must produce identical measurable outcomes. The transformations that map one of these configurations to another are known as gauge transformations. 

The symmetry group of the SM is defined by the gauge symmetries of the electroweak and strong interactions:
\begin{equation}
    G_{\textrm{SM}}= SU(3)_C \otimes SU(2)_L \otimes U(1)_Y,
    \label{eq:symmetry}
\end{equation}
where the subgroup $SU(3)_C$ corresponds to the strong interactions, whereas the $SU(2)_L \otimes U(1)_Y$ relates to the electroweak interactions. This connection arises because the mediators~(bosons) of the electroweak and strong interactions are generated by the groups $SU(2)_L \otimes U(1)_Y$ and $SU(3)_C$ respectively.

\subsection{Quantum Chromodynamics}\label{app:qcd}

Quantum Chromodynamics (QCD) is a QFT built to describe the strong interaction between quarks, the fundamental components of hadrons. The discovery of these elementary particles that conform matter marked a significant shift in our understanding of particle physics. Their presence inside the hadrons required the introduction of a new quantum number to satisfy the Fermi-Dirac statistics. This quantum number, the color charge, requires that each quark can have $N_C=3$ different colors. Although this charge needs to be introduced to ensure that Fermi-Dirac statistics is not violated, colored states are not observed in nature. This led to the formulation of the confinement hypothesis, by which all asymptotic states
must be singlets in color space. The main consequence of confinement is that quarks cannot be observed as free particles except at high energies. Instead, they can only be found in color-neutral bound states.

The force that binds the colored quarks within hadronic states is the strong interaction, which is much stronger at long distances than at short distances, unlike other forces such as the electromagnetic force or gravity. This explains how quarks with the same electric charge remain bound together inside a hadron, rather than repelling each other. The QCD theory is built by applying the $SU(3)_C$ gauge symmetry. This requires the introduction of $N^2_C -1 = 8$ new massless boson fields $G_a^{\mu}$ called gluons. With these considerations, the Lagrangian of the QCD theory reads:
\begin{equation}
\begin{split}
\mathcal{L}_{\text{QCD}} = 
& -\frac{1}{4} \left( \partial^\mu G^a_\nu - \partial^\nu G^a_\mu \right) 
\left( \partial_\mu G_a^\nu - \partial_\nu G_a^\mu \right) 
 + \sum_f \bar{q}_f \left( i \gamma^\mu \partial_\mu - m_f \right) q_f \\
& - g_s G^\mu_a \sum_f \bar{q}_f \gamma_\mu \left( \frac{\lambda^a}{2} \right) q_f 
 + \frac{g_s}{2} f^{abc} \left( \partial^\mu G_a^\nu - \partial^\nu G_a^\mu \right) G_\mu^b G_\nu^c \\
& - \frac{g_s^2}{4} f^{abc} f_{ade} G^\mu_b G^\nu_c G_{\mu}^d G_{\nu}^e,
\label{eq:lqcd}
\end{split}
\end{equation}
where $\partial^\mu$ represents a covariant derivative, $q_f,\bar{q}_f$ represent the fields of a quark and an anti-quark, respectively, $\gamma_\mu$ are the Dirac matrices, $m_f$ is the mass of the quark, $g_s$ is the strong coupling, $\lambda^a$ are the generators of the $SU(3)$ group and $f^{abc}$ are the structure constants of $SU(3)$.

The self-interactions, encoded in the cubic and quartic terms of the gluon fields, reveal two key features of QCD, asymptotic freedom and confinement, as they determine the sign of the $\beta$ function that controls the energy evolution of the strong coupling. To understand this concept it is convenient to take a look at the behavior of the strong coupling
\begin{equation}
    g_s(k^2) \sim \frac{1}{\ln{k^2/\Lambda^2}},
\label{eq:strong}
\end{equation}
where $k$ is the energy scale of the interaction and $\Lambda$ is a fundamental parameter in QCD that defines the energy scale at which the strong force becomes non-perturbative. According to \Eq{eq:strong}, the strong coupling decreases as the energy $k$ increases. On the contrary, for $k$ close to $\Lambda$ the strong coupling raises leading to the confinement of quarks and gluons.

\subsection{Electroweak interactions}\label{app:ew}

The Electroweak (EW) sector of the SM unifies the electromagnetic and weak interactions under a single theoretical framework. This unification is achieved by introducing a gauge theory based on the symmetry group $ SU(2)_L \otimes U(1)_Y$. Unlike QCD, which acts only on quarks, the EW interaction involves both quarks and leptons.

The $SU(2)_L$ component corresponds to the weak interactions mediated by the $W^\pm$ and $Z$ bosons, while $U(1)_Y$ describes hypercharge, which is related to the electromagnetic interaction, mediated by photons.

However, in order to define a unified EW theory it makes more sense to talk about four different mediator bosons. Two of these, $W^1_\mu$ and $W^2_\mu$, arise from the $SU(2)_L$ symmetry and are responsible for the weak charged current interactions. These bosons correspond to the $W^\pm$ particles. The other two bosons, $W^3_\mu$ and $B_\mu$, are associated with the neutral current interactions and are related to the $Z$ boson and the photon through the mixing mechanism and the mixing angle $\theta_W$:
\begin{equation}
\begin{pmatrix}
W^3_\mu \\
B_\mu
\end{pmatrix}
=
\begin{pmatrix}
\cos\theta_W & \sin\theta_W \\
-\sin\theta_W & \cos\theta_W
\end{pmatrix}
\begin{pmatrix}
Z_\mu \\
A_\mu
\end{pmatrix},
\end{equation}
where $A_\mu$ and $Z_\mu$ represent the fields of the photon and $Z$ boson, respectively.

The Lagrangian for the EW interaction is expressed as:
\begin{equation}
\mathcal{L}_{\text{EW}} = -\frac{1}{4}B_{\mu \nu}B^{\mu \nu} - \frac{1}{4}W^i_{\mu \nu}W^{i\mu \nu} + \mathcal{L}_{\text{CC}} + \mathcal{L}_{\text{NC}}.
\label{eq:lew}
\end{equation}
where the first two terms correspond to the kinetic terms for the new gauge fields:
$$
B_{\mu \nu} \equiv \partial_\mu B_\nu - \partial_\nu B_\mu, \quad 
W^i_{\mu \nu} \equiv \partial_\mu W^i_\nu - \partial_\nu W^i_\mu - g\epsilon^{ijk}W^j_\mu W^k_\nu~.
$$
The $W^i_{\mu \nu}$ field introduces self-interactions among the gauge fields, whose intensity is quantified by the $g$ coupling.

Regarding the other two terms, $\mathcal{L}_{\text{CC}}$ and $\mathcal{L}_{\text{NC}}$ are the charged and neutral current Lagrangians
\begin{equation}
\mathcal{L}_{\text{CC}} = -\frac{g}{2\sqrt{2}} 
\left( W^\dagger_\mu \left[ 
\sum_{i,j} \bar{u}_i \gamma^\mu (1 - \gamma_5) V_{ij} d_j 
+ \sum_l \bar{\nu}_l \gamma^\mu (1 - \gamma_5) l 
\right] + \text{h.c.} \right),
\label{eq:lcc}
\end{equation}
\begin{equation}
\mathcal{L}_{\text{NC}}  =\mathcal{L}_{\text{QED}}+\mathcal{L}^Z_{\text{NC}}= -eA_\mu \sum_f \bar{f}\gamma^\mu Q_f f - \frac{e}{2\sin\theta_W \cos\theta_W}Z_\mu \sum_f \bar{f}\gamma^\mu \left(v_f - a_f\gamma^5\right)f. \hspace{0.1cm}
\label{eq:nc}
\end{equation}

In the charged-current Lagrangian, $g$ is the $SU(2)$ gauge coupling, $\bar{u}_i$ and $d_j$ denote the different anti-up and down quarks for the different families, $\bar{\nu_l}$ and $l$ represent the antineutrino and lepton for the different families, $\gamma_5=\imath \gamma^0 \gamma^1 \gamma^2 \gamma^3$ is the product of four Dirac matrices and $V_{ij}$ is the  Cabibbo-Kobayashi-Maskawa matrix that describes the  flavor-changing charged currents. In the neutral-current Lagrangian we have separated the Quantum Electrodynamics (QED) Lagrangian $\mathcal{L}_{\text{QED}}$ from the weak interactions in $\mathcal{L}^Z_{\text{NC}}$. In \Eq{eq:nc}, $e$ represents the electric charge of the electron, $\bar{f}$ and $f$ are the fermionic fields, $Q$ is the fermion electromagnetic charge, and $ a_f = T^f_3 $ and $ v_f = T^f_3 \left( 1 - 4|Q_f| \sin^2 \theta_W \right) $, with $T^f_3$ being the weak isospin.

At this point, using gauge symmetry principles we have built a Lagrangian that encodes the dynamics to describe the electromagnetic and weak interactions. Nevertheless, the gauge symmetry employed to construct this theory forbids any mass terms in the EW Lagrangian. Yet, we have experimental proof that both the fermions and the $Z$ and $W^\pm$ bosons are massive, so there has to be a missing piece in our theory. In other words, we need to find a mechanism that gives masses to the particles in the SM.

\subsection{Electroweak Spontaneus Symmetry Breaking}

In order to generate masses in the SM the EW gauge symmetry has to be broken. However, we also need a fully symmetric Lagrangian to ensure renormalizability. This raises an interesting question: how can we reconcile these seemingly contradictory requirements?

To answer this question the Brout-Englert-Higgs mechanism was proposed \cite{PhysRevLett.13.508, PhysRevLett.13.321}. The main idea is to introduce a new particle into the SM which is invariant under transformations of the symmetry group $G_{\rm SM}$, but whose vacuum state is not. Then, we will have a Lagrangian with a degenerate set of states at the minimum energy level. However, only one of those states can be the real vacuum state, and when it is chosen the symmetry is said to be broken. This process, known as Spontaneous Symmetry Breaking~(SSB), leads to the emergence of new degrees of freedom in the form of massless bosons, referred to as Goldstone bosons. An interpretation of these bosons is presented as the Goldstone theorem \cite{Goldstone:1961eq}, in which they are excitations of the field around the vacuum that do not increase the energy of the system. 

This SSB mechanism serves the purpose of giving mass to the $W^\pm$ and $Z$ bosons while leaving the photon massless. Mathematically, let us start by introducing  two additional complex scalar fields which transform under $SU(2)_L$ as a doublet:
\begin{equation}
\phi(x) \equiv 
\begin{pmatrix}
\phi^{(+)}(x) \\
\phi^{(0)}(x)
\end{pmatrix},
\label{eq:doublet}
\end{equation}
and a Lagrangian that describes these fields
\begin{equation}
\mathcal{L}_S = (D_\mu \phi)^\dagger D^\mu \phi - \mu^2 \phi^\dagger \phi - \frac{\lambda}{2} \left( \phi^\dagger \phi \right)^2,
\end{equation}
where $\lambda > 0$ and $\mu^2 < 0$ to ensure that there exists an infinite set of degenerate states that minimize the potential. The covariant derivative is given by
\begin{equation}
D^\mu \phi = \left[ \partial^\mu + i g W^\mu(x) + i g \frac{\tan (\theta_W)}{2} B^\mu(x) \right] \phi.
\end{equation}

Now, the vacuum state has to be chosen among all the infinite degenerate states with minimal energy. This choice of the vacuum state will provoke the SSB of the EW gauge group into the electromagnetic one
\begin{equation}
   SU(2)_L \otimes U(1)_Y \xrightarrow{\text{SSB}} U(1)_{\text{QED}}. 
\end{equation}
When SSB takes place, three of the four fields in the scalar doublet can be reinterpreted as Goldstone bosons, which are absorbed by the $W^\pm$ and $Z$ bosons to provide their longitudinal components. This mechanism explains how the weak mediators acquire mass. The remaining scalar field becomes a fluctuation $H$ around the vacuum value $v$, identified as the EW vacuum expectation value (vev). This introduces a new scalar particle in the SM, known as the Higgs boson.

With these considerations, the scalar Lagrangian becomes:
\begin{equation}
\mathcal{L}_S = \frac{1}{2} \partial_\mu H \partial^\mu H 
+ \left[ \frac{g^2}{4} W_\mu^\dagger W^{\mu } + \frac{g^2}{8 \cos^2 \theta_W} Z_\mu Z^\mu \right] 
(v + H)^2 
- \frac{1}{2} M_H^2 H^2 + \frac{M_H^2}{2v} H^3 - \frac{M_H^2}{8v^2} H^4,
\label{eq:ls}
\end{equation}
where $M_H = \sqrt{\lambda v}$. From \Eq{eq:ls} one can see that the weak bosons acquire masses as they present quadratic terms. In particular, the coefficients associated with these terms represent the boson masses, which are defined as:
\begin{equation}
M_W = M_Z \cos \theta_W =  \frac{1}{2} v g.
\end{equation}

Having successfully generated masses for the weak bosons, the masses of the fermions are still missing. How can this be addressed?

\subsection{Yukawa sector}

The procedure to include the fermion masses into the SM Lagrangian consists of reutilizing the same scalar doublet of \Eq{eq:doublet}. Now, taking into account the symmetry requirements of the SM after SSB, we can write the following gauge-invariant fermion-scalar couplings, known as the Yukawa sector:
\begin{equation}
 \mathcal{L}_Y = -\left(1 + \frac{H}{v}\right) \sum_f \left( m_u \bar{u}_f u_f + m_d \bar{d}_f d_f + m_l \bar{l}_f l_f \right)
 \label{eq:ly}
\end{equation}
and thus provides the SM with mass terms for all fermions besides the neutrinos. Now, we can gather all the pieces  from \Eq{eq:lqcd}, \Eq{eq:lew}, \Eq{eq:ls} and \Eq{eq:ly} to build the full Lagrangian of the SM:
\begin{equation}
\mathcal{L}_{\text{SM}} = \mathcal{L}_{\text{QCD}} + \mathcal{L}_{\text{EW}} + \mathcal{L}_S + \mathcal{L}_Y.
    \label{eq:lsm}
\end{equation}

With this, we have successfully derived the expression of the Lagrangian of the SM. A Lagrangian that unifies the strong, weak, and electromagnetic forces within a single elegant framework, making it one of the most remarkable achievements in modern physics. A Lagrangian that contains the dynamics of all the elementary particles known up to date. A \textit{Lagrangian to rule them all}.




\section{Precision Physics in the Standard Model}\label{app:precision}



The SM has been so far very successful in providing very accurate predictions of the behavior of the particles that conform the universe.
A new generation of particle physics colliders experiments \cite{FCC:2018byv, FCC:2018evy, FCC:2018vvp, FCC:2018bvk, Bambade:2019fyw, Roloff:2018dqu, CEPCStudyGroup:2018rmc, CEPCStudyGroup:2018ghi} is designed to unravel in greater detail the mysteries of the universe. Current and upcoming experiments, such as those conducted at the Large Hadron Collider (LHC) at CERN, are generating unprecedented volumes of data. Just to give a rough idea, when the LHC is active it produces around 1 billion collisions per second. The CERN Data Center\footnote{As part of my PhD journey I carried out a research stay at the CERN Quantum Technology Initiative (CERN-QTI) group and I was sitting in my office every day in the Data Center, just above all this vast amount of data. } is storing more than 30 petabytes of data per year from the LHC experiments, enough to fill about 1.2 million Blu-ray discs, i.e. 250 years of HD video \cite{cern-no-date}. This has increased the pressure on the theoretical physics community to produce more precise predictions, enabling accurate comparisons with experimental results to ensure no subtle details are missed. If new physics exists beyond the SM, these efforts aim to discover it. 

Given the enormous computational challenges of analyzing such vast datasets, advanced and sophisticated methods are essential. Over the past decades, machine learning~(ML) techniques have been extensively applied to address these challenges~\cite{ml4hep}, demonstrating their potential in tasks such as data classification, event reconstruction, and particle identification. ML has been adopted in many analyses of large experimental collaborations such as LHCb, CMS, and ATLAS at the LHC, where it helps in improving the efficiency and precision of the mentioned tasks.

While ML has proven useful in handling a wide range of problems, it also presents its own drawbacks. First, its success depends heavily on the availability of high-quality training data. Second, although their efficiencies often outperform those of the classical methods, they work as black boxes providing little information about their functioning and when doing fundamental science, this might not be what one wants. Third, their accurate predictions rely on the necessity of significant computational resources, which might limit their scalability. On top of the mentioned limitations, despite its widespread use, ML remains a classical computational approach that may struggle with capturing the intrinsic ``quantumness'' of quantum mechanicsal processes. 

In this context, the advent of Quantum Computing (QC) in the past years has opened new possibilities by directly leveraging the principles of quantum mechanics. Quantum algorithms, executed on quantum computers, have the potential to provide computational advantages over problems that are intractable for classical methods. Equally important, they may provide unique insights into the behavior of elementary particles by analyzing quantum phenomena through the lens of QC. Since the SM is a QFT, a theory intrinsically quantum in nature, QC arises as the natural choice of tool to analyze their properties and intricacies. This synergy offers a promising avenue of research for exploring the fundamental building blocks of nature at the microscopic scale.

\chapter{Quantum Computing}\label{chap:qc}

\section{Classical and quantum bits}\label{app:qubits}
The \textit{bit} is the fundamental unit of classical computation and information. It represents the smallest possible piece of information, taking on one of two binary values: 0 or 1. In physical implementations, these binary states are typically represented by the presence (1) or absence (0) of electrical current flowing through a circuit.

The importance of classical bits lies in their simplicity and reliability. They enable the storage and processing of vast amounts of information in a deterministic manner, making them the cornerstone of modern digital technology. Classical bits are used in all digital devices, from the smartphones in our pockets to the supercomputers in data centers, and have been crucial in making technological breakthroughs across different fields.

However, classical bits face several limitations. One significant constraint is their binary nature, which restricts them to representing only one state at a time. This limitation appears when dealing with complex computational problems that require exponential resources as the problem size increases. Another limitation is related to Moore's law, which states that the number of transistors on an integrated circuit doubles approximately every two years while their cost decreases. Although this has enabled technological advancement for decades, we are inevitably approaching a physical limit where further miniaturization of transistors approaches atomic scales, leading to the emergence of quantum mechanical effects. At these dimensions, phenomena such as electron tunneling and the Heisenberg Uncertainty Principle introduce unpredictability in transistor behavior. Additionally, classical bits are unable to accurately simulate certain natural phenomena, particularly those governed by Quantum Mechanics (QM). This limitation restricts our ability to model and understand complex systems in fields such as chemistry, materials science, and particle physics, where quantum effects play a crucial role.

On the other hand, quantum bits, or \textit{qubits}, offer a new approach to addressing some of the inherent limitations of classical bits. They differ to classical bits in the following aspects~\cite{ibmWhatQuantum}:
\begin{itemize}
    \item \textbf{Superposition:} Unlike classical bits, qubits can exist in a superposition of states. This means that a qubit can be in a state $|0\rangle$ or $|1\rangle$, or any linear combination of these states:
    \begin{equation}
        |\psi\rangle = \alpha |0\rangle + \beta |1\rangle,
        \label{eq:qubit}
    \end{equation}
    with $\alpha^2+\beta^2=1$. The state $|\psi\rangle$ of \Eq{eq:qubit} can also be expressed as a vector in a sphere of radius 1 whose position is defined by two angles $\theta$ and $\phi$:
    \begin{equation}
    |\psi\rangle = \cos{\left(\frac{\theta}{2}\right)} |0\rangle + e^{i\phi}\cos{\left(\frac{\theta}{2}\right)} |1\rangle.
    \label{eq:blochsphere}
    \end{equation}
    This way of defining the superposition state of  $|\psi\rangle$ is called the Bloch sphere and can be visualized as in Fig. \ref{fig:blochsphere}
    \begin{figure}[H]
  \centering
  \includegraphics[width=0.4\textwidth]{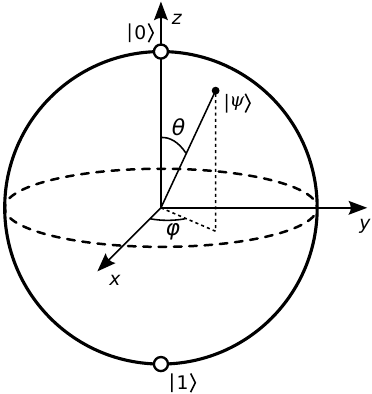}
  \captionsetup{margin={0cm,1cm} }
  
  \caption{Bloch sphere representation of a qubit, showing quantum states as points on a unit sphere. Picture taken from~\cite{wikipediaBlochSphere}.}
  \label{fig:blochsphere}
    \end{figure}
    
    Although a single qubit in superposition has limited utility on its own, groups of qubits in superposition can generate complex, multidimensional computational spaces. These spaces enable new representations and solutions for complex problems.
    \item \textbf{Entanglement:} Qubits can be entangled, a phenomenon where the state of one qubit is completely dependent on the state of another. This property is independent of the distance between them and was difficult to understand by scientists in the early days of QM. Albert Einstein referred to this term as ``spooky action at distance''. From a utility point of view, entanglement can create correlations between qubits that can be exploited to perform complex computations more efficiently than classical systems when combined with other properties such as superposition. For example, in a system of two entangled qubits, the measurement of one qubit immediately determines the state of the other.
    \item \textbf{Decoherence:} Decoherence is the process in which quantum systems lose their ``quantumness'' by collapsing into single states measurable by classical physics. This phenomenon is provoked by interactions with the environment and is one of the key characteristics of qubits. The decoherence time of qubits marks how long they can preserve their quantum properties, hence determining the duration of the quantum algorithms they can execute before decaying into classical states.
    \item \textbf{Interference:} Interference is the phenomenon in which entangled quantum states interact and amplifies certain probabilities while diminishing others. This feature can be leveraged to build algorithms to amplify correct solutions and cancel out incorrect ones. Shor's and Grover's algorithms are examples that leverage quantum interference to achieve  a quantum speedup with respect to their classical counterparts.
    \item \textbf{Measurement:} The measurement process in quantum computing is inherently probabilistic. When a qubit is measured, it collapses from its superposition state into one of its basis states, $|0\rangle$ or $|1\rangle$, with probabilities defined by the squared magnitudes of the coefficients $\alpha$ and $\beta$ of \Eq{eq:qubit}. This contrasts with the deterministic nature of classical bit measurements, presenting unique challenges while also offering new opportunities in the design of quantum algorithms.
\end{itemize}

From an architectural point of view, bits and qubits also present substantial differences. Historically, bits have been represented through various physical mediums, evolving with technology. Early methods included mechanical systems like punched cards, where bits were represented as holes (1) or no holes (0) in a physical card. Later, magnetic storage used the orientation of magnetized regions and transistors, which form today's computers, enabled bits to be encoded as electrical signals.
On the other hand, the architectures for qubits are much more diverse and there is no clear hegemony from one over the others. All of them present their own advantages and challenges. Some of the most used technologies to build qubits include:

\begin{itemize}
    \item \textbf{Superconducting qubits:} This technology leverages superconducting circuits operating at very low temperatures. These qubits present high-speed computation and relatively straightforward fine-tuned control. Major tech companies such as Google, IBM\footnotemark{}, and Intel are building quantum processors based on this architecture. \addtocounter{footnote}{-1}
    \item \textbf{Trapped ion qubits:} They utilize charged atomic particles confined by electromagnetic fields which have long coherence times and high-fidelity measurements. Several companies are developing quantum computers based on this technology, including Quantinuum and IonQ.
    \item \textbf{Neutral atoms:} Laser-cooled neutral atoms are used as qubits, providing scalability and long coherence times. Companies like PASQAL and QuEra are pioneering this technology.
    \item \textbf{Photons:} Photonic quantum computing uses individual photons as qubits, offering advantages like room-temperature operation and compatibility with existing optical infrastructure. Companies developing photonic quantum computers include Xanadu\footnotemark{} and PsiQuantum.
\end{itemize}

\footnotetext{During my PhD, I had the opportunity to do two amazing internships at IBM Research in Zurich and Xanadu in Toronto, where I gained firsthand experience of what it is like to work at and contribute to companies that are leading the development of useful quantum computers.}
\section{Classical and quantum gates}\label{app:gates}

Another major difference between bits and qubits is the kind of operations we use to construct algorithms, or in other words, the logic gates we apply.
In classical computing, every algorithm is decomposed into its fundamental building blocks, the logical gates. From a simple arithmetic operation $0+1$ to the modern and complex transformers that form the Large Language Models (LLMs), such as ChatGPT, all are essentially based on logic gates.

Logic gates are designed using electronic components such as transistors, diodes, and resistors. These gates perform logical operations based on their inputs, producing outputs that are either 0 or 1. Their functionality is based in Boolean algebra, a branch of mathematics that deals with logical operations on binary variables \cite{geeksforgeeksLogicGate}.

\begin{figure}[H]
  \includegraphics[width=\textwidth]{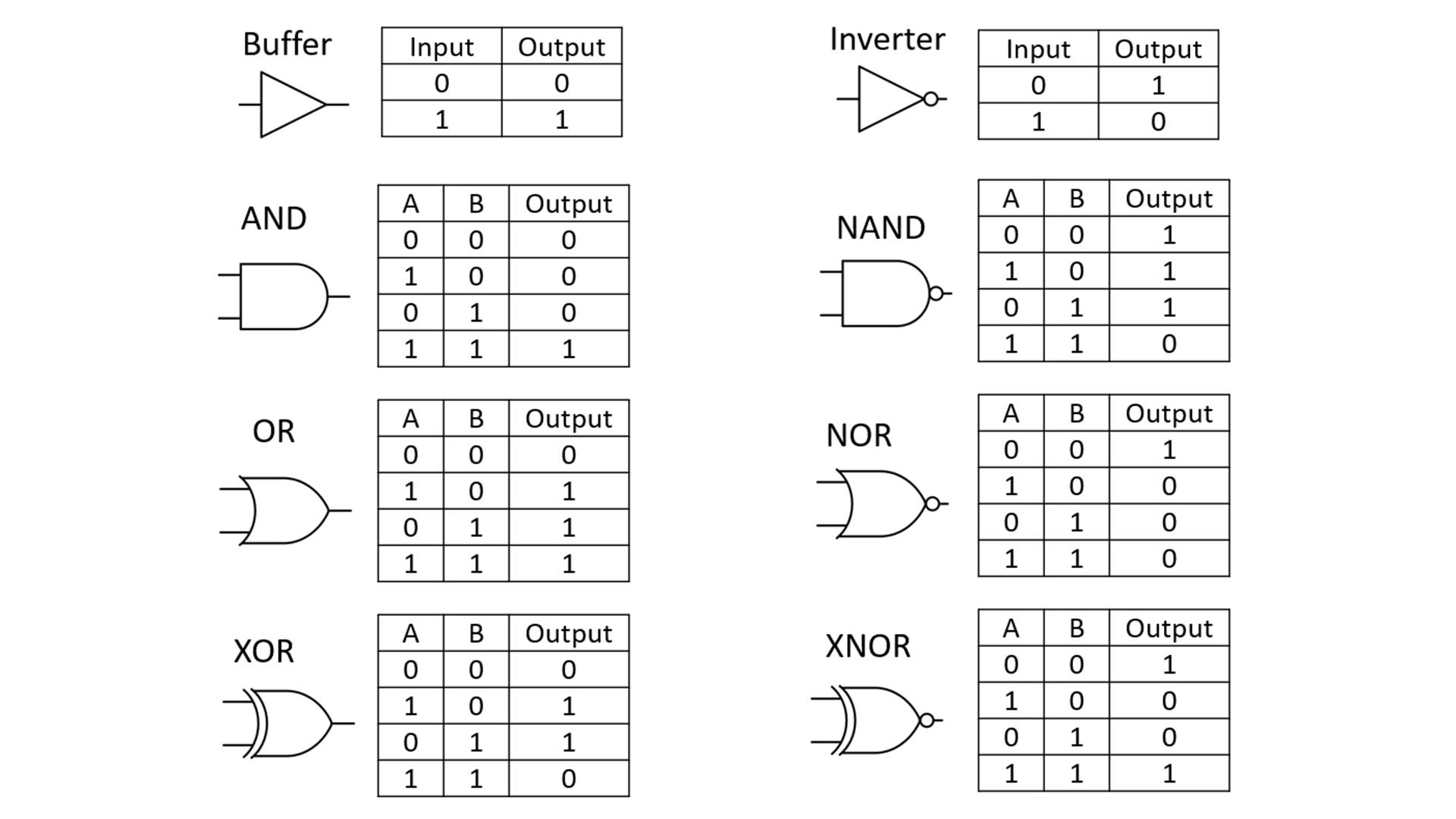}
  \captionsetup{justification=raggedright, singlelinecheck=false} 
  \caption{Truth tables for fundamental logic gates: AND, OR, NOT, NAND, NOR, XOR, and XNOR, and Buffer illustrating input-output relationships for binary operations. Picture taken from~\cite{gsnetworkDigitalLogic}.}
  \label{fig:logic_gates}
\end{figure}

We can classify logic gates into three main categories:
\begin{itemize}
    \item \textbf{AND:} An AND gate is used to perform logical multiplication of binary input. The output of the AND gate will be 1 if both inputs are 1, and 0 otherwise.
    \item \textbf{NOT:} The NOT gate is also known as inverter. It outputs a 1 when the input is 0 and vice versa.
    \item \textbf{OR:} The output of OR gate will be 1 if any of the input states is 1, and 0 otherwise.
\end{itemize}
Then, a combination or negation of these three can build the other 4 logic gates: NOR, NAND, XOR, Buffer and XNOR. The Fig. \ref{fig:logic_gates} show the different logic gates and their truth tables.

Now that we have refreshed how logic gates work on classical bits, let us move on to quantum gates. Unlike many classical logic gates, such as NOR or AND, quantum logic gates are reversible. That means that all quantum operations can be undone. Reversibility is a fundamental property of QM and has important implications for QC. Some of them include that it allows changing from one representation basis to another, such as from the computational basis to the Fourier basis in Shor's algorithm \cite{Shor:1994jg}, and that it enables error correction techniques \cite{Nielsen:1996pv}. 

Mathematically, quantum gates are represented via unitary matrices. A gate acting on $n$ qubits (a quantum register) is described by a $2^n \times 2^n$ unitary matrix. The set of all such gates, combined with matrix multiplication as the group operation, forms the unitary group $U(2^n)$. The quantum states these gates manipulate are unit vectors in a $2^n$-dimensional complex space, utilizing the complex Euclidean norm. The basis vectors of this space, often called eigenstates, represent the possible outcomes when measuring the qubits' states. Any quantum state can be expressed as a linear combination of these basis vectors. Most common quantum gates operate on one or two qubits, analogous to classical logic gates acting on one or two bits. This allows for the construction of more complex quantum circuits and algorithms through the combination of these fundamental operations.

\begin{figure}[H]
  \centering
  \includegraphics[width=0.6\textwidth]{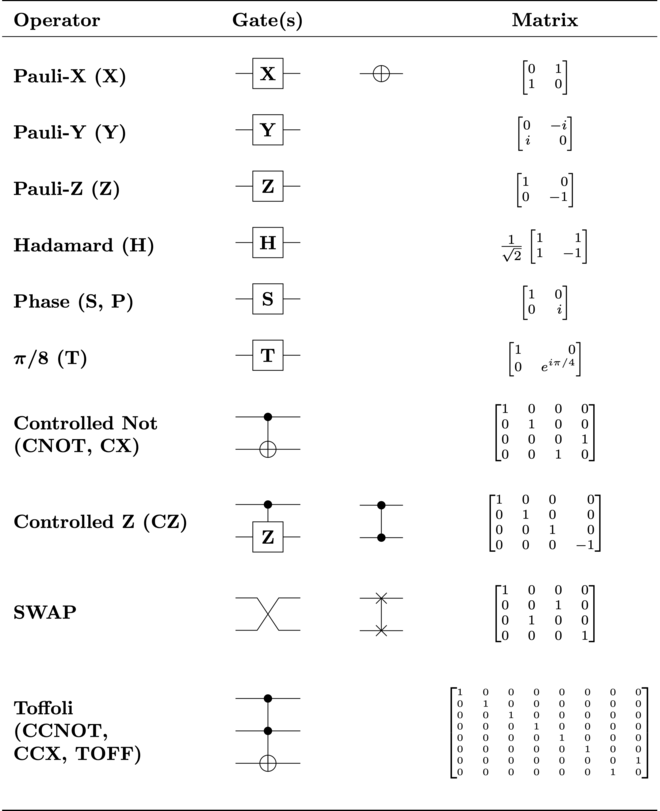}
  \captionsetup{justification=raggedright, singlelinecheck=false} 
  \caption{Common quantum gates, including their names (with abbreviations), circuit representations, and corresponding unitary matrices. Picture taken from~\cite{wikipediaArchivoQuantumLogic}.}
  \label{fig:quantum_gates}
\end{figure}

Common quantum gates include the Hadamard gate ($H$), which creates superposition, the controlled-NOT gate (CNOT) for entanglement operations, and rotation gates like $R_X$, $R_Y$, and $R_Z$ that manipulate the qubit's state on the Bloch sphere. In Fig. \ref{fig:quantum_gates} is depicted a summary of the most common quantum gates, including their mathematical representation.

To illustrate how quantum gates act on a quantum state, let us consider a simple example. We consider the case where the Hadamard gate $H$, which operates on a single qubit, acts on the computational basis state $\vert 0 \rangle$:
\begin{equation}
H \vert 0 \rangle = \frac{1}{\sqrt{2}}
\begin{pmatrix}
1 & 1 \\
1 & -1
\end{pmatrix}
\begin{pmatrix}
1 \\
0
\end{pmatrix}
= \frac{1}{\sqrt{2}}
\begin{pmatrix}
1 \\
1
\end{pmatrix} = \frac{\vert 0 \rangle + \vert 1 \rangle}{\sqrt{2}}.    
\end{equation}
This shows that the Hadamard gate transforms $\vert 0 \rangle$ into an equal superposition of $\vert 0 \rangle$ and~$\vert 1 \rangle$.

\section{Quantum algorithms}\label{app:qalgos}

Not every problem in classical computing can be solved more efficiently using a quantum algorithm. In fact, only a few selected problems can benefit by quantum algorithms, typically those that can exploit quantum phenomena like superposition and entanglement. Problems that require large-scale parallelism or exhibit specific types of structure, such as factoring large numbers (Shor's algorithm \cite{Shor:1994jg}) or solving unstructured search problems (Grover's algorithm \cite{Grover:1996rk}), are where quantum algorithms show the most potential. These problems may lead to a quantum speedup, or \textit{quantum advantage}, where quantum algorithms provide faster solutions than their classical counterparts. For many other computational tasks, classical algorithms remain more efficient and easier to implement. Therefore, the scope of quantum algorithms is currently limited, and their real-world advantages are most evident in specialized areas rather than general-purpose computing.

A quantum algorithm can be understood as a sequence of quantum gates applied to quantum states, followed by measurements, and designed to achieve a specific computational goal. This process is typically represented as a quantum circuit, which serves as a visual and mathematical framework for describing the operations of the algorithm. 

To illustrate how designing a quantum algorithm works, we can show a very simple example that given a two-qubit state in the zero state $|00\rangle$ aims to build a Bell state $\frac{1}{\sqrt{2}} \left( |00\rangle +|11\rangle \right)$. In other words we want to build a quantum circuit that can be represented by the unitary $U_{\rm Bell}$ such that:
\begin{equation}
    U_{\rm Bell} |00\rangle = \frac{1}{\sqrt{2}} \left( |00\rangle +|11\rangle \right).
\end{equation}

Analyzing the form of the Bell state one can realize that we need to perform an operation that incorporates superposition and another one that adds the entanglement to the system. This can be done by the Hadamard gate and the CNOT gate respectively. Therefore the quantum circuit that performs that operation is presented in Fig. \ref{fig:bell_state}:
\begin{figure}[H]
  \centering
  \includegraphics[width=0.6\textwidth]{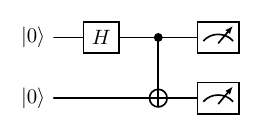}
  \captionsetup{justification=raggedright, singlelinecheck=false} 
  \caption{Quantum circuit that represents the maximally entangled state, the Bell state.}
  \label{fig:bell_state}
\end{figure}
Note that the circuit in Fig.~\ref{fig:bell_state} also includes the measurements of both qubits. The measurements are usually the final part of our quantum algorithm, but sometimes they can be skipped if we plan to insert our quantum algorithm as a subroutine in a more complex quantum algorithm. 

In the next subsections we will explore some quantum algorithms that are relevant for the topics discussed in this thesis.

\subsection{Swap Test}\label{sapp:swaptest}

Measuring the similarity between quantum states is a crucial component in the development of quantum algorithms, as it is a fundamental part of tasks like state verification, clustering, and pattern recognition. The naive solution one can think of to compare the similarity between a state $|\psi\rangle$ and a state $|\phi\rangle$ involves measuring the qubits of each state and reconstructing them from their measurement probabilities. However, this method is not only suboptimal but also ``destructive'', as the act of measurement collapses the quantum states, leaving them unusable for subsequent operations. This limitation highlights the necessity for a non-destructive method to quantify the similarity between quantum states, preserving their coherence and enabling further processing.

One elegant solution to this problem was proposed in \cite{Buhrman:2001rma} and receives the name of the \textit{Swap Test}. A quantum algorithm that measures the overlap between two quantum states~$|\psi\rangle$ and $|\phi\rangle$ in a non-destructive way. The main idea of the \textit{Swap Test} is using an ancillary qubit and a controlled-swap (CSWAP) operation, which exchanges to states if the control qubit is 1, to indirectly infer the overlap between the two states, leaving them unchanged and in their respective quantum registers.

Let us consider $\ket{\psi}$ and $\ket{\phi}$, which are two input real quantum states of $n$ and $m$ qubits, respectively, such that $n \geq m$ (otherwise, we exchange the labels $\psi$ and $\phi$), and an ancillary qubit. The controlled \textit{Swap Test} proceeds in three steps:
\begin{itemize}
    \item \textbf{Initialization:}
    \begin{itemize}
        \item Start with the initial state:
        \begin{equation}
            \ket{\Psi_0} = \ket{0, \psi, \phi}~,
        \end{equation}
        where the ancillary qubit is initialized to $\ket{0}$.
    \end{itemize}

    \item \textbf{Step 1: Apply a Hadamard gate:}
    \begin{itemize}
        \item Apply a Hadamard ($H$) gate to the ancillary qubit, leaving the states $\ket{\psi}$ and $\ket{\phi}$ unchanged:
        \begin{equation}
        \ket{\Psi_1} = \left( H \otimes \id^{\otimes n+m}\right) \ket{\Psi_0}
        =\frac{1}{\sqrt{2}} \left( \ket{0 ,\psi, \phi} + \ket{1 ,\psi, \phi}\right).       
        \end{equation}
        \item Here, $\id^{\otimes n+m}$ is the identity acting on the states $\ket{\psi}$ and $\ket{\phi}$, and the tensor product $\otimes$ is omitted in the composed states (e.g $\ket{0} \otimes \ket{\psi} \otimes \ket{ \phi} = \ket{0 ,\phi, \psi}$).
    \end{itemize}

    \item \textbf{Step 2: Apply the controlled swap gate (CSWAP):}
    \begin{itemize}
        \item Apply the CSWAP gate, which swaps the $m$ qubits of $\ket{\phi}$ with the first $m$ qubits of $\ket{\psi}$:
        \begin{equation}
        \ket{\Psi_2} = {\rm CSWAP} \ket{\Psi_1}
        =\frac{1}{\sqrt{2}} \left( \ket{0, \psi, \phi} + \ket{1 ,\phi ,\psi'}\right)~.            
        \end{equation}
        \item Here, $\ket{\psi'}$ is the swapped version of $\ket{\psi}$, where the $m$ first qubits of $\psi$ are swapped with the rest $n-m$ qubits.
    \end{itemize}

    \item \textbf{Step 3: Apply another Hadamard gate:}
    \begin{itemize}
        \item Apply a Hadamard gate to the ancillary qubit again:
        \begin{equation}
        \ket{\Psi_3} = \left( H \otimes \id^{\otimes n+m}\right) \ket{\Psi_2}
        =\frac{1}{2} \left( \ket{0} \otimes \left(\ket{\psi ,\phi} + \ket{\phi, \psi'}\right)
        +\ket{1} \otimes \left( \ket{\psi, \phi} - \ket{\phi, \psi'} \right)
        \right)~.
        \end{equation}
    \end{itemize}
\end{itemize}

The resulting probability of measuring the ancillary qubit in the state $\ket{0}$ is given by:  
\begin{equation}
\begin{split}
P_{\Psi_3}(\ket{0}) &= \left|\langle 0\ket{\Psi_3} \right|^2 = 
\frac{1}{4} \left| \ket{\psi, \phi} + \ket{\phi, \psi'} \right|^2 
=\frac{1}{2}+\frac{1}{2} \rm{Re} 
\left[ \la \phi, \psi' \ket{\psi, \phi} \right] \\
&=\frac{1}{2}+\frac{1}{2} \la \psi'| \phi \ra \la \phi| \psi \ra ~,
\end{split}
\end{equation}

\begin{figure}[H]  
    \centering  
    \includegraphics[width=0.45\textwidth]{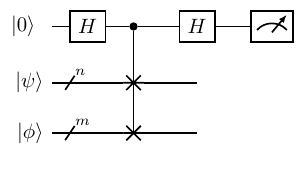}  
    \caption{Quantum circuit \textit{Swap Test}.}  
    \label{fig:swaptest0}  
\end{figure}

If $m = n$, the swapped state $\ket{\psi'}$ simplifies to $\ket{\psi}$, and the probability becomes:  
\begin{equation}
P_{\Psi_3}(\ket{0}) =
\frac{1}{2}+\frac{1}{2} \left|\la \psi| \phi \ra \right|^2~.
\end{equation}
This provides the squared inner product between the two states, with an uncertainty of ${\cal O}(\epsilon)$ after ${\cal O}(\epsilon^{-2})$ shots. The corresponding quantum circuit for the \textit{Swap Test} method is shown in Fig.~\ref{fig:swaptest0}.  

\subsection{Grover's Algorithm}\label{sapp:swaptest}

Grover's algorithm is a quantum search algorithm initially proposed by Lov K. Grover \cite{Grover:1996rk} that provides a quadratic speedup over classical methods for searching in unstructured databases. Consider the case of searching for a specific phone number in a randomly-ordered catalogue containing $N$ entries. Using classical methods, one would need to check each entry one by one until the desired number is found. That takes an average of $N/2$ steps and in the worst case one would need to check all the $N$ numbers. That means, the complexity of the classical algorithm to solve this problem is $\mathcal{O}(N)$. However, in~\cite{Grover:1996rk} Grover presents a quantum algorithm that solves this problem in $\mathcal{O}(\sqrt{N})$ steps, hence providing a quadratic speedup.  For example, if the catalogue contained $10000$ items, a classical search might require up to $10000$ checks, while Grover's quantum algorithm would need only around 100 quantum operations.

Before digging into Grover's algorithm workflow, let us first introduce one concept that constitutes the algorithm's core, the \textit{oracle}. In Grover's algorithm, the quantum \textit{oracle} is a black box that can recognize the solutions to the search problem. This oracle is represented as a unitary operator $O$ whose action over the index $x$ of an element of the list is:
\begin{equation}
    O |x\rangle = (-1)^{f(x)}|x\rangle,
    \label{eq:grovers_oracle}
\end{equation}
where $|x\rangle$ is the index register and by definition $f(x)=1$ if $x$ is a solution to the search problem, and $f(x)=0$ otherwise. Looking at \Eq{eq:grovers_oracle}, we can say that the oracle marks the solutions to the search problem by introducing a $-1$ phase.

Another crucial component of Grover's algorithm is the \textit{diffuser} or diffusion operator. The intuition behind this operator is that it will amplify the amplitudes of the states that are marked by the oracle. Mathematically, for a problem with $N = 2^n $ entries, where the indices are represented by $n$ qubits, the diffusion operator, $D$, is defined as:
\begin{equation}
    D=H^{\otimes n} (2 \ket{0}\bra{0} - I) H^{\otimes n} = 2 \ket{\psi}\bra{\psi} - I,
    \label{eq:grovers_diffusion}
\end{equation}
where $\ket{\psi}$ is the uniform superposition state:
\begin{equation}
\ket{\psi}= \frac{1}{\sqrt{N}}\sum_{x=0}^{N-1}\ket{x}.
    \label{eq:grovers_init}
\end{equation}

By joining the oracle and the diffuser, one builds the Grover operator, $G=DO$, which will mark and amplify the probability of measuring a solution to the problem. Moreover, the Grover operator has also a particularly interesting geometrical interpretation. It can be understood as a rotation in the two-dimensional space constituted by the initial state $|\psi\rangle$ and the state consisting of a uniform superposition of all the $M$ solutions to the search problem $\ket{\beta}$. To do so, let us define $\ket{\alpha}$ and  $\ket{\beta}$ as:
\begin{equation}
|\alpha\rangle \equiv \frac{1}{\sqrt{N-M}} \sum_x^{\prime \prime}|x\rangle,
\end{equation}
\begin{equation}
|\beta\rangle \equiv \frac{1}{\sqrt{M}} \sum_x^{\prime}|x\rangle,
\end{equation}
where the convention $\sum_x^{\prime}$ indicates a sum over all $x$ which are solutions and $\sum_x^{\prime \prime}$ indicates a sum over the rest $x$ which are not solutions. Now, it is convenient to rewrite the initial state $\ket{\psi}$ in a new basis:
\begin{equation}
|\psi\rangle=\sqrt{\frac{N-M}{N}}|\alpha\rangle+\sqrt{\frac{M}{N}}|\beta\rangle.
\label{eq:grover_initial_state}
\end{equation}
In the orthogonal basis defined by $\ket{\alpha}$ and $\ket{\beta}$, the operator $G$ can be understood as a combination of two reflections. The first one is a reflection of the oracle $O$ about the state $\ket{\alpha}$ in the plane defined by $\ket{\alpha}$ and $\ket{\beta}$. Since $O(a\ket{\alpha} + b\ket{\beta}) = a\ket{\alpha} - b\ket{\beta}$. The second one is performed by the diffusion operator $D$  about the state $\ket{\psi}$. This combination of two reflections results in a rotation. It also explains why the state $G^k\ket{\psi}$ remains in the subspace defined by $\ket{\alpha}$ and $\ket{\beta}$ for all $k$. 

\begin{figure}[h]
  \centering
  \includegraphics[width=0.5\textwidth]{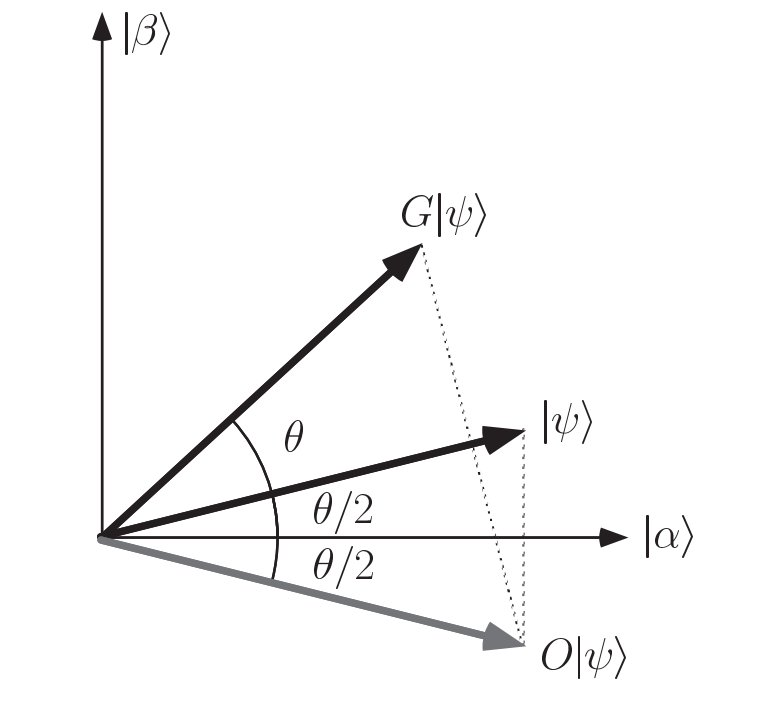}
  \caption{Geometrical interpretation of the action of Grover operator $G$ over the initial state $\ket{\psi}$. Picture taken from~\cite{nielsen}.}
  \label{fig:rotation_grover}
\end{figure}

Furthermore, this process also gives us the rotation angle. Let $\cos \theta/2 = \sqrt{(N - M)/N}$, so that $\ket{\psi} = \cos \theta/2 \ket{\alpha} + \sin \theta/2 \ket{\beta}$. As illustrated in Fig. \ref{fig:rotation_grover}, the two reflections that form $G$, transform $\ket{\psi}$ into the following state:
\begin{equation}
    G |\psi\rangle = \cos\frac{3\theta}{2} |\alpha\rangle + \sin\frac{3\theta}{2} |\beta\rangle,
\end{equation}
so the rotation angle is $\theta$. It can be demonstrated that applying $G$ a number $k$ of times gives us:
\begin{equation}
    G^k |\psi\rangle = \cos\left(\frac{2k+1}{2} \theta\right) |\alpha\rangle + \sin\left(\frac{2k+1}{2} \theta\right) |\beta\rangle. 
    \label{eq:grover_k}
\end{equation}

From \Eq{eq:grover_k}, it can be understood that the operator $G$ is a rotation that is making the initial state to get closer to the state $\ket{\beta}$ which represents the $M$ solutions of our problem. However, the amplitude of the state $\ket{\beta}$ also depends on the number of iterations $k$ of Grover's operator. Hence, there exists an optimal number of iterations, $R$, such that applying $G^R$ results in $G^R\ket{\psi} \approx \ket{\beta}$. Beyond this point, for $k > R$, the state begins to deviate from $\ket{\beta}$. Thus, Grover's algorithm can also be viewed as a quantum preparation subroutine, where the initial state $\ket{\psi}$ evolves to approximate the solution $\ket{\beta}$.

At this point, we should ponder the question: how can we determine the number of rotations required to bring the state $ \ket{\psi}$ close to the target state $\ket{\beta}$? In other words, how do we estimate $R$? To answer this, it is helpful to first recognize that if the initial state is given by \Eq{eq:grover_initial_state}, a rotation by an angle $\theta' = \arccos \sqrt{\frac{M}{N}} $ will bring the state to $ \ket{\beta}$. That can be verified by applying a rotation of angle $\theta'$ to $\ket{\psi}$ in the $\ket{\alpha}$, $\ket{\beta}$ basis:
\begin{equation}
R(\theta')|\psi\rangle = \begin{pmatrix}
\cos\theta' & -\sin\theta' \\
\sin\theta' & \cos\theta'
\end{pmatrix}
\begin{pmatrix}
\sqrt{\frac{N-M}{N}} \\
\sqrt{\frac{M}{N}}
\end{pmatrix}
= \begin{pmatrix}
\cos\theta' \sqrt{\frac{N-M}{N}} - \sin\theta' \sqrt{\frac{M}{N}} \\
\sin\theta' \sqrt{\frac{N-M}{N}} + \cos\theta' \sqrt{\frac{M}{N}}
\end{pmatrix},
\end{equation}
applying $\cos\theta'=\sqrt{M/N}$ and $\sin\theta'=\sqrt{(N-M)/N}$ we obtain:
\begin{equation}
R(\theta')|\psi\rangle = \begin{pmatrix}
\sqrt{\frac{M}{N}} \sqrt{\frac{N-M}{N}} - \sqrt{\frac{N-M}{N}} \sqrt{\frac{M}{N}} \\
\sqrt{\frac{N-M}{N}}  \sqrt{\frac{N-M}{N}} + \sqrt{\frac{M}{N}} \sqrt{\frac{M}{N}}
\end{pmatrix}
= \begin{pmatrix}
0\\
1
\end{pmatrix}=\ket{\beta}.
\end{equation}

However, we cannot reach the state $|\beta\rangle$ in a single step. It will require several iterations of the Grover operator. This is exactly what we aim to determine: the number of required iterations $R$. Essentially, we can decompose this large rotation of angle $\theta'$ into $R$ smaller rotations of angle $\theta$, which means that $R\theta = \theta'$. Hence, we can obtain $R$ as:
\begin{equation}
    R=\textrm{CI} \left(\frac{\arccos \sqrt{M/N}} {\theta}\right),
    \label{eq:grover_R}
\end{equation}
where $\textrm{CI}(x)$ denotes the integer closest to the real number $x$. The expression in \Eq{eq:grover_R} provides an exact formula for the number of oracle calls required to perform Grover's algorithm. However, in some cases,, it will be more convenient to have a simpler expression that captures the essential behavior of $R$. In particular, from \Eq{eq:grover_R} we see that $R \leq \lceil \pi/2\theta \rceil$, which means that a lower bound on $\theta$ will give an upper bound on $R$. Assuming that $M\leq N/2$ we have:
\begin{equation}
   \sqrt{\frac{M}{N}} = \sin \theta/2 = \frac{\theta}{2}.
\end{equation}

Combining this with the upper bound we already had on $R$ we obtain this elegant upper bound:
\begin{equation}
   R \leq \left\lceil \frac{\pi}{4}\sqrt{\frac{N}{M}} \right\rceil.
\end{equation}
That is, $R =  \cal O \left(\sqrt{\frac{N}{M}}\right)$ Grover iterations (and thus oracle calls) are needed to obtain a solution to the search problem with high probability, representing a quadratic improvement over the $\cal O\left(\frac{N}{M}\right)$ oracle calls classically required.

\begin{figure}[h]
  \centering
  \includegraphics[width=0.75\textwidth]{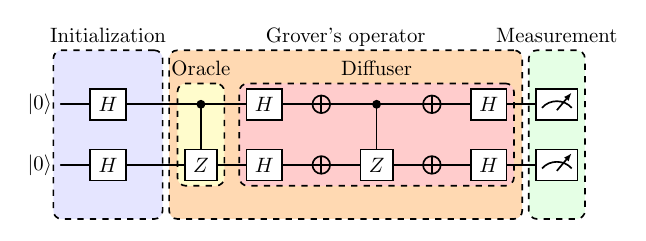}
  \caption{Quantum circuit representing Grover's algorithm for two qubits.}
  \label{fig:grover_2q}
\end{figure}

We now illustrate Grover's algorithm with a simple example. Let us consider the case where we are looking for one item in a list of four elements, whose indices are represented as: $|00\rangle$, $|01\rangle$, $|10\rangle$, $|11\rangle$. Without lose of generality we will consider $|x_0\rangle=|11\rangle$ as the solution. The quantum circuit that represents Grover's search is depicted in Fig.~\ref{fig:grover_2q}.

Note that the oracle of Fig. \ref{fig:grover_2q} marks the state $|11\rangle$ as the $CZ$ will add a $-1$ phase if and only if both qubits are in the state $|1\rangle$. The quantum circuit of Fig.~\ref{fig:grover_2q} performs the following operations:

\begin{itemize}
    \item \textbf{Initialization:}
        \begin{equation}
        \ket{\psi}= \frac{1}{2} \left( |00\rangle+ |01\rangle+ |10\rangle+ |11\rangle \right).
            \label{eq:grovers_init_ex}
        \end{equation}

    \item \textbf{Apply Grover operator $R\approx \left\lceil\pi \sqrt{4} / 4\right\rceil \approx 1$ times:}
        \begin{equation}
        \begin{split}
        G|\psi\rangle &= D O   |\psi\rangle = D\left[ \frac{1}{2} \left( |00\rangle+ |01\rangle+ |10\rangle - |11\rangle \right) \right] \\
        &= \left(2 \ket{\psi}\bra{\psi} - I\right) \left[ \frac{1}{2} \left( |00\rangle+ |01\rangle+ |10\rangle - |11\rangle \right) \right]  \\
        &= \left(2 \ket{\psi}\frac{1}{4} \left( 1+ 1+ 1 - 1 \right) 
        - \frac{1}{2} \left( |00\rangle+ |01\rangle+ |10\rangle - |11\rangle \right) \right) \\
        &=\ket{\psi} - \frac{1}{2} \left( |00\rangle+ |01\rangle+ |10\rangle - |11\rangle \right) \\
        &= \frac{1}{2} \left( |00\rangle+ |01\rangle+ |10\rangle + |11\rangle \right) - \frac{1}{2} \left( |00\rangle+ |01\rangle+ |10\rangle - |11\rangle \right) \\
        &=  \frac{1}{2} \left( |00\rangle-|00\rangle+ |01\rangle-|01\rangle + |10\rangle-|10\rangle + |11\rangle+|11\rangle \right)\\
        &= |11\rangle
        \end{split}   
        \end{equation}

    \item \textbf{Measurement:}
    
         We measure the $n=2$ qubits and obtain $x_0=|11\rangle$ with probability 1.

\end{itemize}

In this example, we showed how Grover's algorithm solves a relatively simple unstructured search problem. However, its applications extend far beyond this. Grover's algorithm can speed up tasks such as finding the minimum of a function \cite{Durr:1996nx}, addressing encryption problems involving prime factorization~\cite{pennylaneBasicArithmetic}, and one of its most fascinating uses: quantum Monte Carlo integration~\cite{Montanaro_2015}, which is achieved through the Quantum Amplitude Estimation (QAE) algorithm \cite{Brassard:2000xvp}. This application of Grover's algorithm in quantum integration will be explored in detail in Chapter~\ref{chap:qint}.

\subsection{Variational Quantum Algorithms}\label{sapp:varalgos}
The NISQ (Noisy Intermediate-Scale Quantum)~\cite{Preskill:2018jim} era refers to the current stage of quantum computing\footnote{Part of the quantum computing community is starting to discuss the transition from the NISQ era to the early fault-tolerant era, or ISQ era \cite{pennylaneFromNISQ}, where noise becomes less of a problem due to advancements in error correction. This would enable the implementation of quantum algorithms with significantly deeper quantum circuits.}, where quantum processors have tens to a few hundred qubits, which are enough to explore interesting problems but are still subject to noise and errors. These devices are not yet capable of performing fault-tolerant quantum computations, which are not restricted by the number of qubits or gates, but they offer a unique opportunity to experiment with quantum algorithms and explore quantum advantage in real-world applications. In this context, Variational Quantum Algorithms (VQAs) are an incredibly popular approach for quantum algorithms on near-term quantum devices.

VQAs rely on Parameterized Quantum Circuits (PQC), which are quantum circuits that contain gates with free parameters that can be tuned. Hence, the goal of a VQA is to find the optimal parameters that achieve the desired task. This parameter tuning is handled by a classical computer, making VQA a quantum-classical hybrid algorithm, with the PQC representing the quantum component, while the tuning of parameters is the classical component.
The next question is, how to find the correct set of parameters that solve a particular problem? To do so, the so-called \textit{cost function} quantifies how far the PQC is from achieving its goal. Therefore, the cost function is a function of the parameters that, when minimized, indicates that we have performed the task well enough.

In most cases, VQAs is divided into three main steps:

\begin{enumerate}
    \item A PQC, often referred to as an \textit{Ansatz}, that is defined with a set of free parameters.
    \item Measurement of an observable and computation of a cost function based on these measurements.
    \item Optimization of the free parameters using a classical computer, which interacts with the quantum device to refine the parameters iteratively. This optimization is usually performed by a gradient descent method, an optimization algorithm to find a local minimum of a differentiable function.
\end{enumerate}

\subsubsection{Ansatz}

In physics and mathematics, the German word ``Ansatz'' or in plural ``Ansätze'' (``Ansatzes'' in English), refers to an educated guess for solving a problem. In the context of VQAs, the Ansatz refers to the quantum circuit $U(\vec{\theta})$ we design to solve a specific task. But how do we choose such a circuit Ansatz? There are two main approaches. One approach is to use prior knowledge about the problem or physical intuition to design a tailored Ansatz, often referred to as a ``problem-inspired Ansatz''.

What if we lack sufficient understanding of the problem to guide our design? In such cases, we can use universal, ``problem-agnostic'' Ansatzes that are applied without specific information about the problem. These Ansatzes are generally larger, containing many gates and parameters, which makes the optimization process more challenging. The better an Ansatz is tailored to the specific task, the easier it becomes for the optimizer to find the right parameters, increasing the likelihood of solving the problem in a reasonable amount of time.

There are two key factors to consider when evaluating how well an Ansatz is suited to a particular problem: \textit{expressibility} and \textit{trainability}. \textit{Expressibility} measures the range of functions the PQC can represent, which is often related to the amount of entanglement and the number of parameters in the circuit. While greater expressibility allows the PQC to describe more complex functions, it also makes the optimization process, and hence the \textit{trainability}, more difficult. This is because a larger space of functions increases the complexity of finding the optimal parameters. \textit{Trainability} also depends on the choice of the cost function and optimizer.

Typically, there is a trade-off between \textit{expressibility} and \textit{trainability}, since larger and more expressive PQCs are harder to train. Finding an Ansatz that balances enough \textit{expressibility} with feasible \textit{trainability} is essential for designing effective quantum circuits.

\subsubsection{Cost functions}

Once we have chosen the Ansatz circuit for our VQA, the next step is to determine the parameters required to perform the desired task. This task is addressed by defining a cost function, which is a mathematical function that quantifies how close the output of the PQC is to the desired result. 

However, that does not mean that we necessarily know the answer. In supervised problems, we often know the solution for some specific inputs, even if not for all. For these known cases, we optimize the parameters to match the expected outcomes, assuming that the same parameters will work for unknown inputs.

In other cases, the goal could simply be to minimize a cost function without having prior knowledge of the solution. VQAs are well-suited for this task, as they iteratively adjust the parameters to find the minimum value of the cost function. The cost function is typically defined as the expectation value of a chosen observable $\hat{M}$:
\begin{equation}
C(\vec{\theta}) = \langle 0 | U^\dagger(\vec{\theta}) \hat{M} U(\vec{\theta}) | 0 \rangle.
\end{equation}

However, constructing an effective cost function $C(\vec{\theta})$ can be challenging. As explained in~\cite{Cerezo:2020jpv}, some points can help us make the right choice:
\begin{itemize}
    \item \textbf{Faithfulness}: The minimum of the cost function $C(\vec{\theta})$ should correspond to the solution of the problem.
    \item \textbf{Efficient Estimation}: It must be possible to estimate $C(\vec{\theta})$ efficiently through quantum measurements and potentially classical post-processing. Additionally, to maintain a quantum advantage, $C(\vec{\theta})$ should not be efficiently computable classically.
    \item \textbf{Operational Meaningfulness}: Lower values of $C(\vec{\theta})$ should correspond to better solutions.
    \item \textbf{Trainability}: It should be feasible to optimize the parameters efficiently.
\end{itemize}
By carefully considering these criteria, we can design cost functions that enhance the effectiveness and performance of VQAs.

\subsubsection{Optimizer and Gradient Descent}

Once the Ansatz and the cost function are defined, the next step is to optimize the trainable parameters. Mathematically, this optimization task is expressed as finding the set of parameters $\boldsymbol{\theta}^*$ that minimize the cost function $C(\boldsymbol{\theta})$:
\begin{equation}
    \boldsymbol{\theta}^* = \underset{\boldsymbol{\theta}}{\text{argmin}} \, C(\boldsymbol{\theta}).
\end{equation}
There are numerous methods to perform this optimization, but gradient-based techniques are particularly advantageous. These methods leverage information about the gradient (or higher-order derivatives) of the cost function to iteratively update the parameters, driving the cost function towards a minimum. 

\begin{figure}[h]
    \centering
    \includegraphics[width=0.6\textwidth]{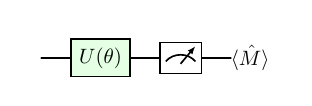}
    \caption{Quantum circuit measuring the expectation value of the observable $\hat{M}$.}
    \label{fig:utheta.pdf}
\end{figure}

At this point, a natural question arises: How can we compute the gradient of a PQC? At first glance, this might seem challenging due to the quantum nature of the system. However, thanks to the Parameter Shift Rule (PSR), it is possible to compute these gradients analytically, even for quantum circuits. The PSR is a powerful mathematical tool that enables gradient computation by evaluating the expectation value of the cost function at two specific points. While PQCs typically consist of multiple gates, we focus here on computing the gradient of a single gate for simplicity and pedagogical clarity. For a more general treatment of PQCs with multiple gates, we refer the reader to~\cite{Schuld:2018aiz}. That said, let us consider a unitary gate $U(\theta)$ of the form:
\begin{equation}
U(\theta) = e^{-i\frac{\pi}{2}\theta \hat{G}_{j}},
\end{equation}
where $\hat{G}_{i}$ is the Hermitian generator of $U(\theta)$ and corresponds to a Pauli operator. By using Taylor expansion of the matrix exponential, the gradient of $U(\theta)$ is expressed as:
\begin{equation}
\nabla U(\theta) = -\frac{i}{2}\hat{G}_{j}U(\theta) = -\frac{i}{2}U(\theta)\hat{G}_{j}.
\end{equation}
Then, we aim to compute the gradient of the following quantum circuit in Fig.~\ref{fig:utheta.pdf}, which corresponds to the function:
\begin{equation}
f(\theta) = \langle\hat{M}\rangle = \langle 0|U^{\dagger}(\theta)\hat{M}U(\theta)|0\rangle.
\end{equation}

The gradient of $f(\theta)$ is given by:
\begin{equation}
\begin{aligned}
\nabla f(\theta) &=\nabla \langle  \hat{M} \rangle = \langle 0| (\nabla U^\dagger(\theta) \hat{M} U(\theta) + U^\dagger(\theta) \hat{M} \nabla U(\theta)) |0\rangle \\
&= \frac{i}{2} \langle 0| U^\dagger(\theta) (\hat{G}_j \hat{M} - \hat{M} \hat{G}_j) U(\theta) |0\rangle \\
&= \frac{i}{2} \langle 0| U^\dagger(\theta) [\hat{G}_j, \hat{M}] U(\theta) |0\rangle~.
\end{aligned}
\end{equation}

This can be simplified using the following expression for commutators involving Pauli operators:
\begin{equation}
   [\hat{G}_j, \hat{M}] = -i \left( U^\dagger \left(\frac{\pi}{2}\right) \hat{M} U\left(\frac{\pi}{2}\right) - U^\dagger\left(-\frac{\pi}{2}\right) \hat{M} U\left(-\frac{\pi}{2}\right) \right),
\end{equation}
and substituting in previous expression we obtain the gradient of the circuit:
\begin{equation}
   \nabla f(\theta) = \frac{1}{2} \langle 0 | U^\dagger\left(\theta + \frac{\pi}{2}\right) \hat{M} U\left(\theta + \frac{\pi}{2}\right) | 0 \rangle - \frac{1}{2} \langle 0 | U^\dagger\left(\theta - \frac{\pi}{2}\right) \hat{M} U\left(\theta - \frac{\pi}{2}\right) | 0 \rangle.
\end{equation}

Finally, rewriting this in terms of $f(\theta)$, the gradient of the quantum circuit is:
\begin{equation}
\nabla f(\theta) = \frac{1}{2}\left[f\left(\theta + \frac{\pi}{2}\right) - f\left(\theta - \frac{\pi}{2}\right)\right].
\end{equation}

The PSR unlocks the possibility of using gradient-based optimization algorithms, such as Gradient Descent (GD), for variational quantum circuits. (GD) is a fundamental optimization technique used to iteratively minimize a function. It is widely employed in various fields, including Machine Learning, numerical optimization, and Quantum Computing. The goal of GD is to find the minimum of a cost function by iteratively adjusting parameters in the direction of the steepest descent of the function.

The GD algorithm can be summarized in the following steps:
\begin{enumerate}
    \item \textbf{Initialization:}
    Start with an initial guess for the parameters $\boldsymbol{\theta}$, denoted as $\boldsymbol{\theta}^{(0)}$. This initial guess can be random or based on some heuristic.

    \item \textbf{Compute gradient:}
    Calculate the gradient $\nabla f(\boldsymbol{\theta}^{(k)})$, which represents the vector of partial derivatives of $f(\boldsymbol{\theta})$ with respect to each parameter $\theta_k$. For each parameter:
    \begin{equation}
    (\nabla f(\theta))_k = \frac{\partial f(\theta)}{\partial \theta_k}.
    \end{equation}
    In practice, libraries like \texttt{PennyLane}~\cite{pennylane} use efficient methods such as the PSR to compute these gradients for quantum circuits.

    \item \textbf{Update parameters:}
    Adjust the parameters in the opposite direction of the gradient to minimize the function. The update rule is given by:
    \begin{equation}
           \boldsymbol{\theta}^{(kt+1)} = \boldsymbol{\theta}^{(k)} - \eta \nabla f(\boldsymbol{\theta}^{(kt)}), 
    \end{equation}

    where $\eta$ is the learning rate, a small positive scalar that controls the size of the steps taken towards the minimum. Choosing an appropriate learning rate is crucial: too small learning rate can result in slow convergence, while too large learning rate can cause oscillations or even divergence.

    \item \textbf{Iterate until convergence:}
    Repeat steps 2 and 3 until a stopping criterion is met. This criterion can be based on the change in the function value or the gradient magnitude falling below a threshold, indicating that further updates are not significantly improving the result. Alternatively, steps 2 and 3 can be executed for a fixed number of iterations.
\end{enumerate}

\begin{figure}[h]
    \centering
    \includegraphics[width=0.6\textwidth]{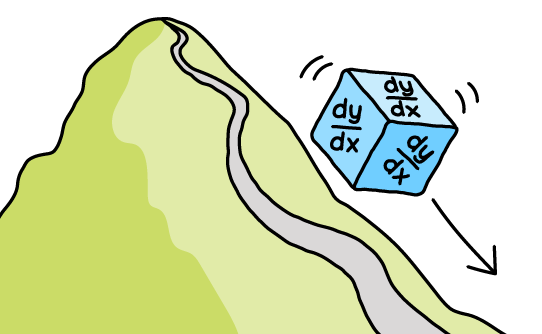}
    \caption{Graphical representation of Gradient Descent. The algorithm iteratively moves towards the minimum of the cost function by following the direction of the negative gradient. Picture taken from~\cite{pennylaneDoublyStochastic}}
    \label{fig:gradient_descent}
\end{figure}

A graphical interpretation of GD is shown in Figure~\ref{fig:gradient_descent}. Essentially, the gradient $\nabla f(\boldsymbol{\theta})$ indicates the direction of the steepest increase of the function $f(\boldsymbol{\theta})$. Moving in the opposite direction ensures movement towards the function's minimum. The gradient is a vector of partial derivatives, representing how much the function value changes with respect to changes in its parameters. The learning rate $\eta$ controls the size of the step taken in the direction of the gradient, balancing the trade-off between convergence speed and stability. A well-chosen learning rate ensures efficient and stable convergence to the function's minimum.

In Quantum Computing, GD is used to optimize parameters in PQCs. By tuning the variational parameters of a PQC, GD minimizes the cost function. The combination of the Parameter Shift Rule and GD provides a powerful framework for training VQAs, enabling efficient optimization of PQCs.

In summary, the PSR provides a practical and efficient way to compute gradients of PQCs, enabling the use of gradient-based optimization techniques. GD's ability to iteratively refine parameters towards a function's minimum, coupled with its efficiency and versatility, makes it indispensable for optimization tasks in modern computational problems. Whether optimizing classical Machine Learning models or fine-tuning quantum circuits, GD constitutes a fundamental technique for achieving effective and efficient optimization.

\newpage

\section{Quantum Machine Learning}\label{app:qml}

\begin{figure}[H]
    \centering
    \includegraphics[width=0.7\linewidth]{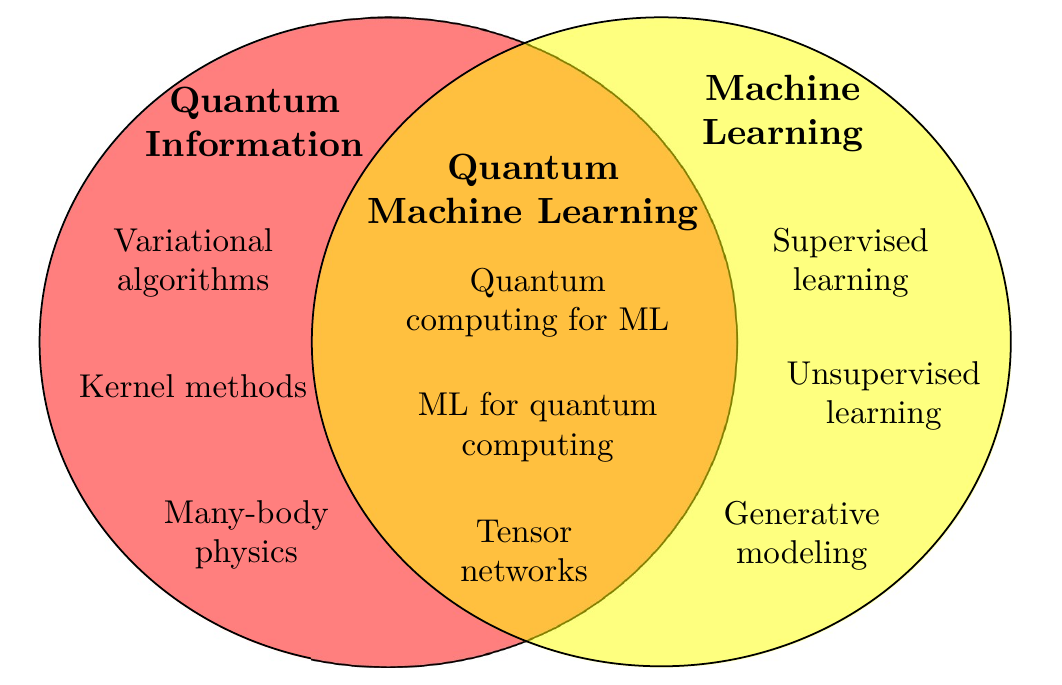}
    \caption{Venn diagram of the intersection between quantum information and Machine Learning (ML).}
    \label{fig:qml_venn}
\end{figure}

Over the past years, the term `Quantum Machine Learning' (QML) has become ubiquitous, appearing in various contexts and usually referring to the intersection of quantum information theory and machine learning, see Fig.~\ref{fig:qml_venn}.

This broad field includes Machine Learning (ML) techniques applied to quantum information processing \cite{Carrasquilla_2017,PhysRevLett.104.063603,Biamonte_2017} and classical computational tools developed within the quantum community, such as the use of tensor networks from many-body physics to train neural networks \cite{stoudenmire2017supervisedlearningquantuminspiredtensor}, and the use of quantum algorithms to tackle ML tasks \cite{Schuld_2014, Schuld2017QuantumML}. 

\begin{figure}[h]
    \centering
    \includegraphics[width=0.4\linewidth]{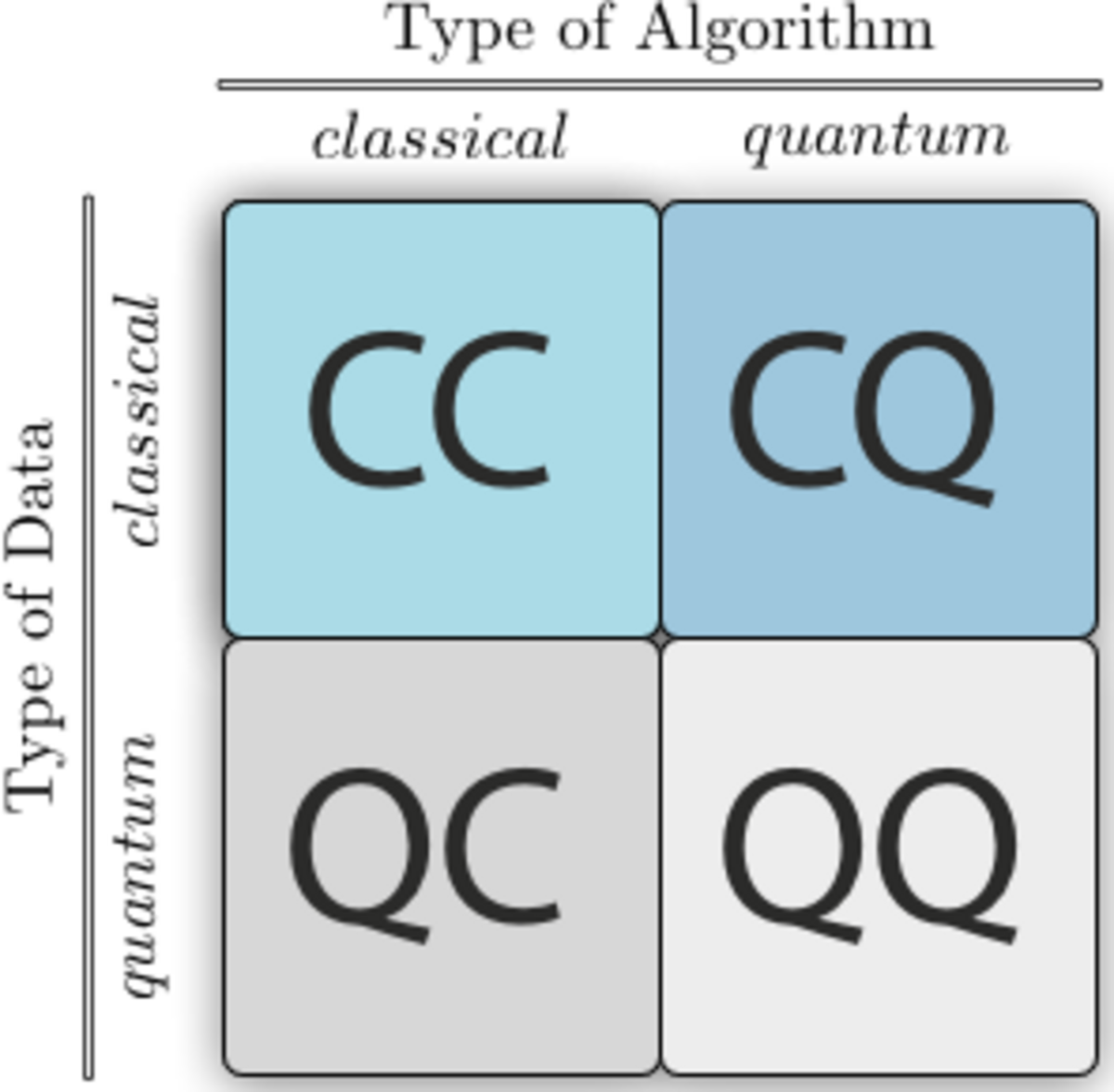}
    \caption{Quantum Machine Learning (QML) approaches by algorithm and data nature. Picture taken from~\cite{wikipediaQuantumMachine}.}
    \label{fig:qml_data}
\end{figure}

Another interesting point of view to analyze the different ways of doing QML is to analyze the nature of the algorithms and the data, as depicted in Fig.~\ref{fig:qml_data}

In this thesis, we will primarily focus on quantum algorithms for tackling ML problems with classical data (first item on the overlapping area in Fig.~\ref{fig:qml_venn} and top right corner of Fig.~\ref{fig:qml_data}). Within this category, we consider three different categories that have been widely explored: quantum kernels, quantum neural networks and quantum generative models. Nevertheless, before diving into these categories it is convenient to consider the  important question of how to encode data in a quantum computer in the context of QML.

\subsection{Encoding} \label{sapp:encoding}
Data is the core component of ML. Its efficient processing allows models to make predictions that might be otherwise unreachable. In QML, this role also remains crucial. However, the way data is processed and represented varies significantly. Quantum computers offer unique computational advantages, but to leverage their properties effectively, data must first be mapped onto quantum states. Encoding classical data into a quantum computer, is a key challenge in designing QML algorithms, as the choice of encoding method impacts efficiency, expressiveness, and the feasibility of the potential quantum advantage. Moreover, the best encoding strategy often depends on the specific problem at hand. The most common approaches to load data into quantum computers include basis encoding, amplitude encoding, angle encoding and dynamical encoding~\cite{Schuld2017QuantumML}.

\subsubsection{Basis encoding}

Basis encoding (or basis embedding) maps a classical $n$-bit string (e.g., $0011$) to a computational basis state of an $n$-qubit system (e.g., $| \psi \rangle = \lvert 0011 \rangle$). This approach is relatively straightforward, as each classical bit is directly replaced by a qubit. Like classical computing, basis encoding relies on a binary representation of numbers. A quantum state~$| x \rangle$, where $x \in \mathbb{R}$, refers to a binary representation of $x$ using $\log x $ qubits. 

Let us now consider a classical dataset $D$ that we want to load into a quantum computer. For basis embedding, each item $x^{(m)}$ has to be a $N$-bit binary string; $x^{(m)} = (b_1, \ldots, b_N)$ with $b_i \in \{0,1\}$ for $i = 1, \ldots, N$. Assuming all features are represented with unit binary precision (one bit), each input example $x^{(m)}$ can be directly mapped to the quantum state $|x^{(m)}\rangle$. This means that the number of quantum subsystems, $n$, must be at least equal to~$N$. An entire dataset is represented in superpositions of computational basis states as
\begin{equation}
\ket{D} = \frac{1}{\sqrt{M}} \sum_{m=1}^M |x^{(m)}\rangle.
\end{equation}
To illustrate this, consider the set $D=\{2,3,7,9\}$, which corresponds to $D=\{0010,0011,0111,1001\}$ in binary. Therefore, the quantum state is given by:
\begin{equation}
\ket{D} = \frac{1}{\sqrt{4}} \left( |0010\rangle+|0011\rangle+|0111\rangle+|1001\rangle \right).
\end{equation}
This encoding technique is particularly useful in binary classification, quantum feature mapping, and quantum error correction, but it might not be ideal for other tasks such as regression. 

\subsubsection{Amplitude encoding}

Amplitude encoding represents classical data by encoding each value in the amplitudes of a quantum state. Specifically, a classical dataset $D = \{x^{(1)}, x^{(2)}, \dots, x^{(M)}\}$, where each $x^{(m)}$ is a scalar, is mapped to a quantum state $\ket{D}$ such that each classical value $x^{(m)}$ appears as the amplitude of the corresponding computational basis state. The quantum state $\ket{D}$ is a superposition of basis states, where the amplitudes are directly proportional to the values in the dataset. Mathematically, the quantum state is written as:
\begin{equation}
\ket{D} = \frac{1}{\sqrt{\sum_{m=1}^{M} |x^{(m)}|^2}} \sum_{m=1}^M x^{(m)} |x^{(m)}\rangle.
\end{equation}
Let us now, consider the same dataset as before $D = \{2, 3, 7, 9\}$. The corresponding quantum state will now be:
\begin{equation}
\ket{D} = \frac{1}{\sqrt{2^2 + 3^2 + 7^2 + 9^2}} \left( 2\ket{00} + 3\ket{01} + 7\ket{10} + 9\ket{11} \right).
\end{equation}
This approach could be more efficient than basis encoding, as it allows the quantum system to represent a larger range of data with fewer qubits. Amplitude encoding is particularly useful for representing compactly and efficiently large datasets in quantum computing. It is commonly used in tasks like multivariable classification.

\subsubsection{Angle encoding}

Angle encoding maps classical data to a quantum computer by representing each value as the angle of a quantum rotation. This encoding technique is implemented encoding each angle in a different qubit via a rotation gate such as $R_X, R_Y \text{ or } R_Z$. Specifically, a classical dataset $D = \{\theta^{(1)}, \theta^{(2)}, \dots, \theta^{(M)}\}$, where each $\theta^{(m)}$ is a scalar, is mapped to a quantum state $\ket{D}$ such that each qubit codifies one angle. Therefore, if we consider rotations around the $X$ axis the quantum state is written as:
\begin{equation}
\ket{D} =  \bigotimes_{m=1}^M R_X|0\rangle =\cos\left(\frac{\theta_m}{2}\right) \ket{0} -i \sin\left(\frac{\theta_m}{2}\right) \ket{1},
\end{equation}
Now, if we want to encode the vector of the dataset $D = \{2, 3, 7, 9\}$  using $R_X$ rotations, the resulting quantum state would be:
\begin{equation}
    \begin{split}
            \ket{D} &= \left( R_X(2)\ket{0} \otimes R_X(3)\ket{0} \otimes R_X(7)\ket{0} \otimes R_X(9)\ket{0} \right)\\ &=  \bigg(  
        \big( \cos(1)\ket{0} - i\sin(1)\ket{1} \big) \otimes 
        \big( \cos(1.5)\ket{0} - i\sin(1.5)\ket{1} \big) \otimes \\ 
        & \big( \cos(3.5)\ket{0} - i\sin(3.5)\ket{1} \big) \otimes 
        \big( \cos(4.5)\ket{0} - i\sin(4.5)\ket{1} \big) 
        \bigg).
    \end{split}
\end{equation}
Angle encoding is particularly efficient for tasks where the data needs to be represented in a way that is compatible with quantum operations. It can provide a more compact representation of data, offering an advantage in certain problems such as regression.

\subsubsection{Dynamic encoding}
 
Dynamic encoding is a technique that embeds classical or quantum information into the evolution of a quantum system. Instead of encoding data into static quantum states, this approach encodes information into unitary operators that govern the system's time evolution. A common strategy is to associate a given square matrix $A$ with a Hamiltonian~$H$, enabling the representation of $A$ through quantum evolution. If $A$ is non-Hermitian, one defines an extended Hermitian matrix  as
\begin{equation}  
\tilde{A} =  
\begin{pmatrix}  
0 & A \\  
A^\dagger & 0  
\end{pmatrix},  
\end{equation}  
and use only part of the quantum system's output. This method allows quantum algorithms to manipulate matrices efficiently, including operations such as matrix-vector multiplication and matrix inversion, particularly in amplitude encoding schemes. Another perspective on dynamic encoding emerges when analyzing the evolution of a subsystem within a larger unitary system. This encoding strategy provides a flexible way to load information into a quantum computer, allowing for applications in quantum algorithms and quantum simulations.

\subsection{Quantum Kernels}\label{sapp:qkernels}

Quantum kernels have emerged as a promising tool in QML, offering potential advantages over classical kernel methods. Kernel methods work by mapping data into a higher-dimensional space, where patterns that may be complex in the original space become more easily separable. These techniques are widely used in supervised learning to transform nonlinear problems into linear ones by leveraging high-dimensional feature representations.

\begin{figure}[h]
    \centering
    \includegraphics[width=0.5\linewidth]{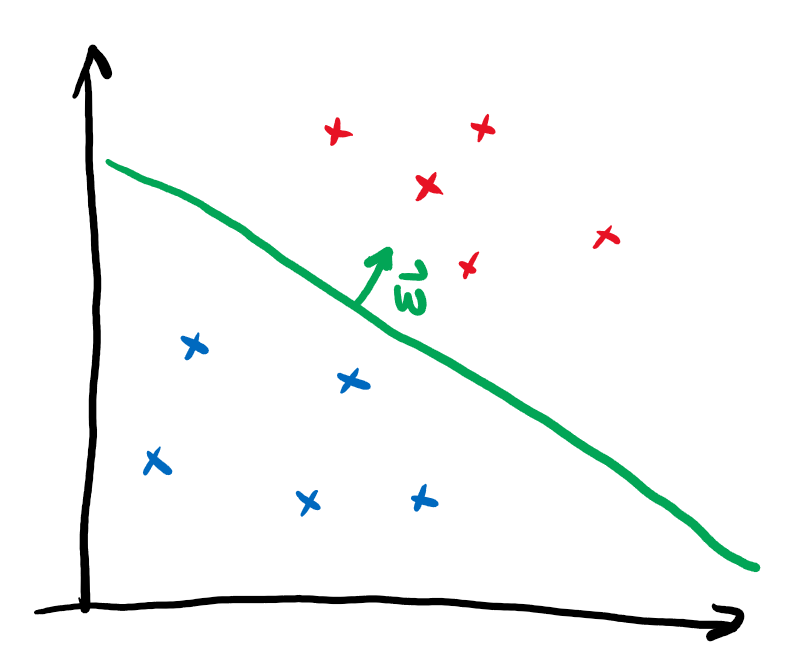}
    \caption{Binary classification problem using a Support Vector Machine (SVM). Picture taken from~\cite{pennylaneTrainingEvaluating}.}
    \label{fig:linear_classification}
\end{figure}

To understand how kernel methods work we follow \cite{Hubregtsen:2021lqn} approach of considering the example of linear classification. One of the most well-known kernel methods for solving such problems is the support vector machine (SVM) \cite{Cortes1995SupportVectorN}, a supervised binary classifier that aims to determine the hyperplane that best divides data in two groups. The SVM algorithm will perform the binary classification of a datapoint according to the following expression:
\begin{equation}
    y(\vec{x})=sign\left(\langle\vec{w}, \vec{x}\rangle+b\right),
\end{equation}
where $\langle\vec{w}, \vec{x}\rangle$ denotes the inner product of the vector $\vec{w}$, which is perpendicular to the hyperplane, and the input $\vec{x}$, while $b$ determines the hyperplane's position. A sketch of this binary classification problem is depicted in Fig. \ref{fig:linear_classification}.

Then, following the kernel's definition, one remaps $\vec{x}$ into a larger feature space: $\vec{x} \rightarrow \phi(\vec{x})$. Also, one rewrites $\vec{w}$ as a linear combination of the remapped datapoints: $\vec{w}=\sum_i \alpha_i \phi(\vec{x}_i)$. Putting this all together, the previous formula reads:
\begin{equation}
    y(\vec{x})=sign\left(\sum_i \alpha_i \langle\phi(\vec{x_i}), \phi(\vec{x})\rangle+b\right).
\end{equation}

At this point, the prediction only depends on inner products in the higher-dimensional feature space. These inner products, which measure the similarity between two datapoints, are what we will call kernel functions:
\begin{equation}
k(\vec{x}_i,\vec{x}_j)=\langle\phi(\vec{x_i}), \phi(\vec{x_j})\rangle.
\end{equation}

\begin{figure}[h]
    \centering
    \includegraphics[width=0.7\linewidth]{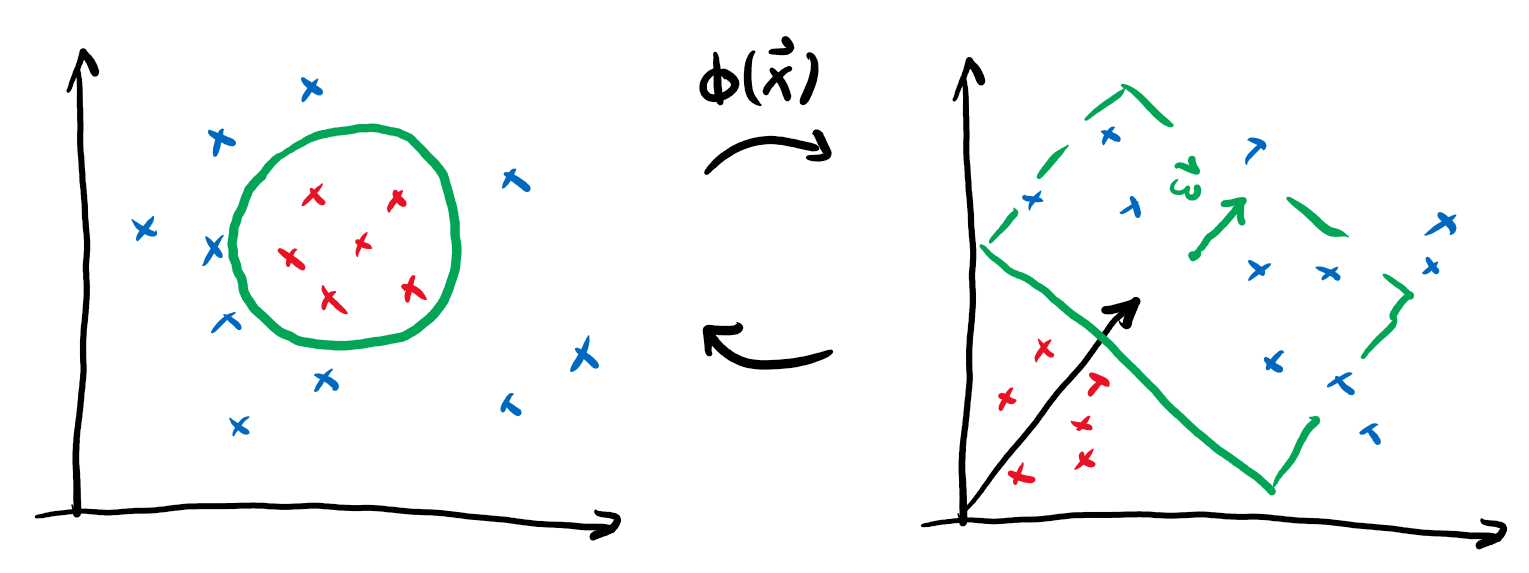}
    \caption{A nonlinear embedding used to enhance the capabilities of a linear classifier. Linear classification in the embedding space translated to nonlinear decision boundaries in the original space. Picture taken from~\cite{pennylaneTrainingEvaluating}}
    \label{fig:enter-label}
\end{figure}

It is now relatively straightforward to generalize this kernel concept to a more general framework, introducing the notion of a \textit{quantum} kernel. This connection to quantum computing naturally arises when we interpret the feature space as a Hilbert space and the mapping $\phi(\vec{x})$ as a quantum state within it:  
\begin{equation}  
|\phi(\vec{x})\rangle = S(\vec{x})|0\rangle,  
\end{equation}  
where $S(\vec{x})$ represents the embedding routine used to prepare this state.  With this perspective, we can define the \textit{quantum} kernel as:  
\begin{equation}  
k(\vec{x}_i, \vec{x}_j) = |\langle \phi(\vec{x_i}) | \phi(\vec{x_j}) \rangle|^2.  
\end{equation}  
This quantum kernel is then be used as a subroutine to efficiently compute similarities between datapoints in a quantum-enhanced SVM, leading to the so-called Quantum Support Vector Machine (QSVM)~\cite{Rebentrost:2013bin}.

\begin{figure}[h]
    \centering
    \includegraphics[width=0.6\linewidth]{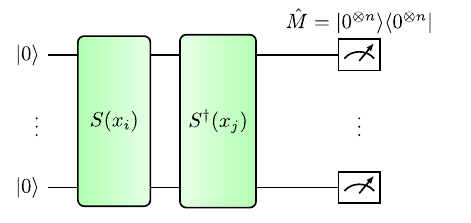}
    \caption{ Quantum kernel circuit for states encoded by $S(x_i)$ and $S(x_j)$. }
    \label{fig:kernel_circuit}
\end{figure}

To implement a quantum kernel in a quantum circuit, we prepare the two states  
$|\phi(x_i)\rangle$ and $|\phi(x_j)\rangle$  
on different sets of qubits using angle-embedding routines $S(x_i)$ and $S(x_j)$, and measure their overlap with the SWAP test subroutine explained in Section \ref{sapp:swaptest}. However, we only need half the number of qubits $n$ if we prepare  $|\phi(x_i)\rangle$  and then apply the inverse embedding with $x_j$ on the same qubits. We then measure the projector onto the initial state $|0^{\otimes n}\rangle \langle 0^{\otimes n}|$. This corresponds to the quantum circuit of Fig. \ref{fig:kernel_circuit}. 

We can mathematically verify that this quantum circuit, when measuring the observable $\hat{M}=|0^{\otimes n}\rangle \langle 0 ^{\otimes n}|$, produces the desired quantum kernel:
\begin{equation}
    \begin{split}
    \langle 0^{\otimes n} | S(x_j) S(x_i)^\dagger \hat{M} S(x_j)^\dagger S(x_i) | 0^{\otimes n} \rangle 
    &= \langle 0^{\otimes n} | S(x_j) S(x_i)^\dagger | 0^{\otimes n} \rangle  
    \langle 0^{\otimes n} | S(x_j)^\dagger S(x_i) | 0^{\otimes n}\rangle  \\
    &= |\langle 0^{\otimes n} | S(x_j)^\dagger S(x_i) | 0^{\otimes n} \rangle |^2  \\
    &= |\langle \phi(x_j) | \phi(x_i) \rangle |^2  \\
    &= \kappa(x_i, x_j).
\end{split}
\end{equation}
As expected, the quantum circuit computes the quantum kernel, which is inserted to enhance ML algorithms. Thus, we have shown how quantum kernels can be leveraged in QML.

\subsection{Quantum Neural Networks}\label{sapp:qnn}

The term Quantum Neural Network (QNN) has been widely used in recent years~\cite{andrecut,panella,Cao:2017tnw,Farhi2018ClassificationWQ,Tacchino_2020} to describe ML models that combine elements from quantum computing and artificial neural networks. The term originates from the fact that these models mimic, to some extent, the behavior of Neural Networks (NN). They share with their classical analogues a modular structure, with quantum gates functioning similarly to layers in classical neural networks, and the optimization of trainable parameters using techniques such as GD. Nevertheless, their internal functioning is fundamentally different. To provide a comprehensive comparison and explanation of these topics, let us begin by introducing how neural networks work.

\begin{figure}[h]
\centering
\includegraphics[width=0.8\textwidth]{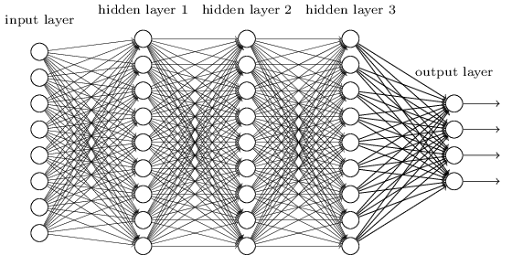}
\caption{Diagram of a classical Neural Network (NN) with three hidden layers. Picture taken from~\cite{nielsenneural}.}
\label{fig:cnn}
\end{figure}

A NN consists of a set of nodes or neurons that communicate with each other by transmitting information~\cite{GmezRamos2013ARO}. Input data passes through the network, where it undergoes various transformations and produces the algorithm's predictions.
The architecture of a classification NN is as follows (see Fig.~\ref{fig:cnn}): an input layer with as many neurons as the number of input variables, one (simple NN) or several (deep NN) hidden layers with $n$ neurons, where $n$ is a hyperparameter of the model introduced by the user and an output layer with as many neurons as classes.

The NN works as follows. First, the neurons in the first layer process the value corresponding to the input variables. Then, in the rest of the layers, each neuron has a weighted sum of the values of the neurons in the previous layer \cite{SCHMIDHUBER201585}. The weights of this weighted sum are initially randomized, but through backpropagation learning, they are modified in successive iterations of the training process to make predictions increasingly accurate \cite{Rojas1996TheBA}.

This type of algorithm has several hyperparameters introduced by the user that affect the functioning and learning process of ML algorithms. Some of the hyperparameters of NN algorithms are~\cite{kerasKerasDocumentation}:
\begin{itemize}
\item \textbf{hidden layers}: number of hidden layers.
\item \textbf{neurons}: number of neurons in each layer.
\item \textbf{dropout}: percentage of neurons randomly eliminated in each hidden layer during each iteration of the training process. This procedure prevents overfitting, as patterns are not established between subsets of neurons that memorize training data instead of learning.
\item \textbf{batch size}: number of training data samples used to train in each iteration of the process.
\item \textbf{validation split}: fraction of training data used as validation data. The model separates this fraction of data for the training process and subsequently performs a validation of the trained model to adjust the weights of the neurons, obtaining a more accurate model.
\item \textbf{epochs}: number of times the entire training set is used for training.
\end{itemize}

\begin{figure}[h]
    \centering
    \includegraphics[width=0.7\linewidth]{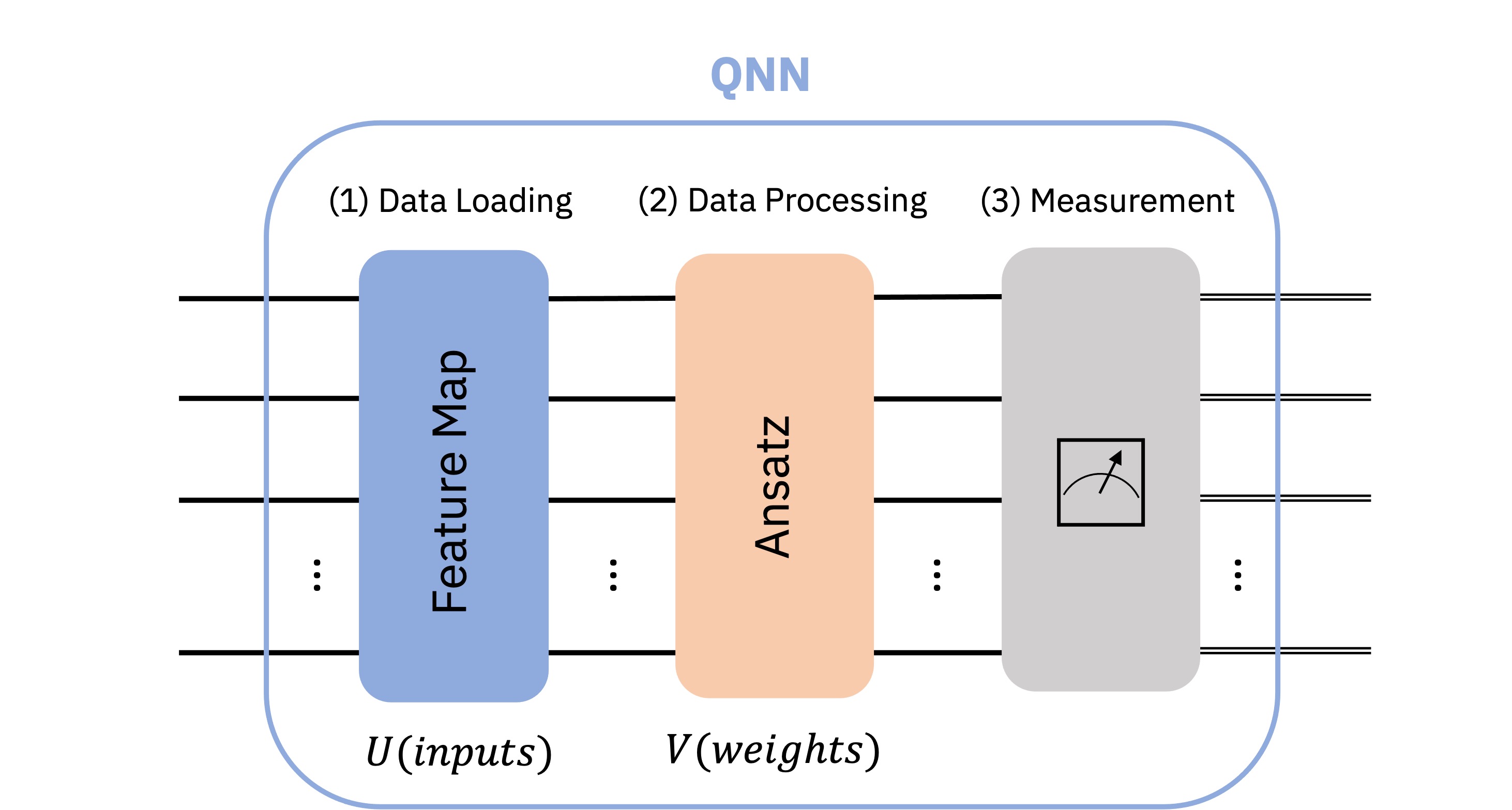}
    \caption{Generic Quantum Neural Network (QNN) structure. Picture taken from~\cite{qiskit2024}.}
    \label{fig:qnn_qiskit}
\end{figure}

On the other hand,  QNN have a workflow that presents similarities and differences. QNNs can load classical data (inputs) into quantum states, which can be later processed by using quantum gates parametrized by trainable parameters~(weights). Fig.~\ref{fig:qnn_qiskit} depicts a generic QNN example including the data loading and processing steps. After this processing, the quantum state is measured and the result is employed to compute a loss function. This loss function serves as the metric to be minimized during the training process. The weights are adjusted through optimization routines, which can be either gradient-free or gradient-based, such as GD. These optimization routines are usually the same as those used in classical NNs.

While the QNN structure depicted in Fig.~\ref{fig:qnn_qiskit} provides a good general representation, it is worth noting that other architectures have also demonstrated effectiveness, sometimes surpassing this generic model in specific applications. One remarkable example, which will be explored further in this thesis, is the data-reuploading scheme~\cite{P_rez_Salinas_2020}.
The data-reuploading approach to QNNs addresses a significant challenge in implementing these algorithms on NISQ devices. Traditional QNN schemes, as illustrated in Fig.~\ref{fig:qnn_qiskit}, often require a large number of qubits to effectively learn and represent the intricate underlying patterns in complex high-dimensional datasets \cite{Perez-Salinas:2021blv}. The data-reuploading technique offers an elegant solution to this limitation.
By repeatedly "re-uploading" input data into the quantum system, this approach enhances the flexibility and expressivity of quantum models. This enables the approximation of complex functions without necessitating large quantum systems or sophisticated encoding strategies. The data-reuploading QNN architecture consists of a sequence of alternating trainable and encoding blocks. In particular, the QNN consists of $L$ layers. Each containing a data-encoding block $S(\vec{x})$ and a trainable block $\mathcal{A}(\vec{\theta})$. The quantum unitary that represents the quantum circuit is expressed as:
\begin{equation}
U(\vec{x},\vec{\theta}) = \mathcal{A}(\vec{\theta}_L)S(\vec{x})\mathcal{A}(\vec{\theta}_{L-1})\ldots \mathcal{A}(\vec{\theta}_1)S(\vec{x})\mathcal{A}(\vec{\theta}_0).
\end{equation}

\begin{figure}[h]
    \centering
    \includegraphics[width=0.85\linewidth]{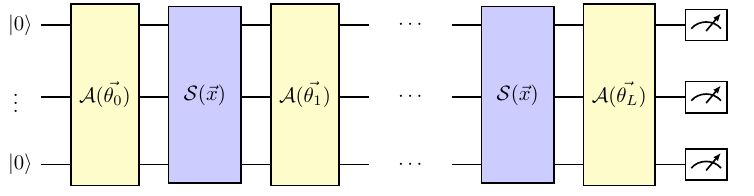}
    \caption{Quantum neural network following a data-reuploading scheme.}
    \label{fig:qnn_reuploading}
\end{figure}

The training blocks $\mathcal{A}(\vec{\theta})$ depend on the parameters $\vec{\theta}$ that are optimized classically. A sketch of this method is depicted in Fig.~\ref{fig:qnn_reuploading}.
Another key motivation behind data re-uploading is overcoming the limitations imposed by the quantum no-cloning theorem, which prevents the duplication of quantum states. Unlike classical NNs, which can reuse input data freely, quantum circuits face restrictions on copying quantum information. The re-uploading scheme overcomes this issue by repeatedly feeding the data into the quantum circuit at different layers alongside additional quantum gates with trainable parameters. This approach draws a parallel with classical NNs, where the same input is passed through multiple layers to extract hierarchical features. As a result, the re-uploading framework provides a natural bridge between quantum circuits and classical NNs, reinforcing the relevance of quantum models in ML tasks.

Furthermore, the re-uploading technique has implications beyond QML algorithms. This strategy allows quantum information stored in registers to be efficiently processed through controlled operations. This flexibility makes it a valuable tool for designing larger quantum algorithms where classical data needs to be integrated with quantum processing, such as loading a function into a quantum state. 

Another relevant aspect of QNNs is how the chosen encoding scheme influences the set of functions they can represent. In this section, we demonstrate how a specific encoding enables QNNs to represent Fourier series \cite{Schuld:2020enb, Ostaszewski:2019vnn}. For simplicity, we focus on the case of univariate functions with inputs $x \in \mathbb{R}$, although this can be generalized as outlined in Appendix A of \cite{Schuld:2020enb}. We define a quantum model $f_\theta(x)$ as the expectation value of an observable, where the state is prepared through a PQC:
\begin{equation}
    f_\theta(x) = \langle 0 | U^{\dagger}(x, \theta) \hat{M} U(x, \theta) | 0 \rangle,
    \label{eq:qnn_model}
\end{equation}
where $|0\rangle$ is the $n$-th qubit initial state of the quantum computer, $U(x, \theta)$ is a quantum circuit that depends on the input $x$ and a set of parameters $\theta$, and $\hat{M}$ is some observable. 

Such quantum circuit is constructed from $L$ layers, each consisting of a data encoding circuit block $S(x)$ and a trainable circuit block $W(\theta)$. The data encoding block is the same in every layer and consists of single or multiple qubit gates of the form $\mathcal{G}(x) = e^{-ixH}$, where $H$ is a Hamiltonian that generates the ``time evolution'' used to encode the data.

Since we want to focus on the role of the data encoding, and to avoid further assumptions on how the trainable circuit blocks are parametrized, we view the trainable circuit blocks as arbitrary unitary operations, $W(\theta) = W$, and drop the subscript $\theta$ of $f_\theta$ from here on. With this assumption, the overall quantum circuit has the form:
\begin{equation}
    U(x) = W^{(L+1)} S(x) W^{(L)} \dots W^{(2)} S(x) W^{(1)}.
\end{equation}
Our goal is to prove that $f(x)$ can be written as a partial Fourier series:
\begin{equation}
    f(x) = \sum_{n \in \Omega} c_n e^{i n x},
    \label{eq:qnn_fourier}
\end{equation}
with integer-valued frequencies. 
The first step is to note that one can always find an eigenvalue decomposition of the generator Hamiltonian  
\begin{equation}
    H = V^\dagger \Sigma V
\end{equation} where $\Sigma$ is a diagonal operator that contains the eigenvalues of $H$, $\lambda_1, \dots, \lambda_d$ on its diagonal.  
Now for simplicity, we consider the case where $S(x)= \mathcal{G}(x)$, although it can easily be generalized to the more general case where $S(x)= \mathcal{G}_1(x)\otimes \ldots \otimes \mathcal{G}_N(x)$. In this case, the data encoding unitary becomes  
\begin{equation}
    S(x) = V^\dagger e^{-i x \Sigma} V~,
\end{equation}
and we can ``absorb'' $V$ and $V^\dagger$ into the arbitrary unitaries  $ W' = V W V^\dagger$.   Hence, without loss of generality, we will assume that $H$ is diagonal.  This allows us to separate the data-dependent expressions from the rest of the circuit in each component $i$ of the quantum state $U(x) \ket{0}$,
\begin{equation}
[U(x) \ket{0}]_i = \sum_{j_1,\dots,j_L=1}^{d} e^{-i(\lambda_{j_1} + \dots + \lambda_{j_L})x} 
 W^{(L+1)}_{i j_L} \dots W^{(2)}_{j_2 j_1} W^{(1)}_{j_1 1}. 
\end{equation}
To ease the notation, we introduce the multi-index $j = \{ j_1, \dots, j_L \} \in [d]^L$, where $[d]^L$ refers to the set of $L$ integers between $1, \dots, d$. We then rewrite the sum of eigenvalues for a given $j$ as $\Lambda_j = \lambda_{j_1} + \dots + \lambda_{j_L}$
\begin{equation}
[U(x) \ket{0}]_i = \sum_{j \in [d]^L} e^{-i \Lambda_j x} W^{(L+1)}_{i j_L} \dots W^{(2)}_{j_2 j_1} W^{(1)}_{j_1 1}.
\end{equation}

Now, computing the adjoint of this expression we rewrite the \Eq{eq:qnn_model} that describes the quantum model: 
\begin{equation}
f(x) = \sum_{k,j \in [d]^L} e^{i(\Lambda_k - \Lambda_j)x} a_{k,j},
\label{eq:qnn_lambdas}
\end{equation}
where $a_{k,j}$ contain the terms coming from the parametrized unitaries and the observable,
\begin{equation}
a_{k,j} = \sum_{i,i'} \left(W^{*}\right)_{1k_1}^{(1)} \left(W^{*}\right)_{j_1j_2}^{(2)} \ldots \left(W^{*}\right)_{j_Li}^{(L+1)} M_{i,i'}
 W^{(L+1)}_{i' j_L} \ldots W^{(2)}_{j_2 j_1} W^{(1)}_{j_11}.
\end{equation}

At this point \Eq{eq:qnn_lambdas} almost look like a Fourier series. What is left now is to put together all the terms in \Eq{eq:qnn_lambdas} whose basis function $e^{ (\Lambda_k - \Lambda_j) x}$ have the same frequency $\omega = \Lambda_k - \Lambda_j.$  All frequencies accessible to the quantum model are contained in its frequency spectrum  
\begin{equation}
   \Omega = \{ \Lambda_k - \Lambda_j, \quad k,j \in [d] \}. 
\end{equation} 
With that, we finally obtain  
\begin{equation}
    f(x) = \sum_{\omega \in \Omega} c_{\omega} e^{i \omega x}  ,
\end{equation}
where the coefficients are obtained by summing over all  
$a_{k,j}$ 
contributing to the same frequency  
\begin{equation}
c_{\omega} = \sum_{\substack{k, j \in [d]^L \\ \Lambda_k - \Lambda_j = \omega}} a_{k, j}.
\end{equation}

The frequency spectrum $\Omega$ has key properties: $0 \in \Omega$, for every $\omega \in \Omega$, $-\omega \in \Omega$, and since $c_{\omega} = c_{-\omega}^*$, \Eq{eq:qnn_model} is a real-valued function. The size of the spectrum is denoted by $K = (|\Omega| - 1)/2,$ indicating the number of independent nonzero frequencies and the largest frequency $D = \max(\Omega)$ is the spectrum's degree. The coefficients $c_{\omega}$ are determined by gates $W^{(1)} \dots W^{(L+1)}$ and the measurement observable $\hat{M}$. Thus, the frequency spectrum depends on the eigenvalues of the data encoding gates, while Fourier coefficients depend on the entire quantum circuit. For integer-valued eigenvalues $\lambda_1, \dots, \lambda_d$, frequencies in $\Omega$ are integer-valued, yielding a partial Fourier series as in \Eq{eq:qnn_fourier}.

The implications of interpreting QNNs as Fourier series will be explored further in Section~\ref{chap:qint}, where we demonstrate how to utilize this property to build a Quantum Monte Carlo integrator.

\subsection{Quantum Generative Models}\label{sapp:qgm}
Generative models have emerged as a transformative tool to revolutionize how we handle information in the XXI century by creating new data that mimic real-world distributions~\cite{sengar2024generativeartificialintelligencesystematic,Bond_Taylor_2022}. These models have significantly advanced artificial intelligence by enabling the creation of datasets that are often indistinguishable from real data. Their applications range from generating text, images, videos or music. Large Language Models (LLMs), such as GPT are a good example of generative models that leveraging transformer architectures, are capable of capturing complex patterns in language and reproducing it to a disturbing level of accuracy.

In this regard, quantum generative models aim to leverage the strengths of quantum computing to enhance data generation tasks \cite{Riofrio:2023ncy,Zoufal:2021pbi}. Generally, these models learn a target distribution from training data and generate new samples upon measurement. This aligns well with one of the tasks that quantum computers are better at: sampling. In fact, one of the few demonstrations of quantum supremacy, where a quantum device performs a task faster than any classical counterpart, was Google's 2019 experiment \cite{Arute:2019zxq}, which was fundamentally a sampling problem, showcasing the potential of quantum computers to outperform classical systems. Given that generative modeling is a cornerstone of classical ML and sampling is one of the strongest capabilities of quantum computers, the intersection of these fields has gained significant attention within the quantum computing community.

Several quantum generative models have shown particular promise, following a general workflow based on variational quantum circuits. These circuits contain tunable parameters that are optimized to minimize a cost function that encodes the problem that we aim to solve. Typically, this cost function is designed to ensure that the output distribution obtained from measuring the quantum circuit matches the target distribution used for training. By exploring an exponentially larger Hilbert space, these models offer a different approach to generative learning, potentially leading to more expressive and efficient representations compared to classical methods. 

There exist different type of learning strategies that allows us to differentiate into different groups of quantum generative models. Two relevant examples are Quantum Circuit Born Machines (QCBMs) and Quantum Generative Adversarial Networks (QGANs), each employing distinct mechanisms for training and sample generation.

\begin{figure}[h]
    \centering
    \includegraphics[width=0.75\linewidth]{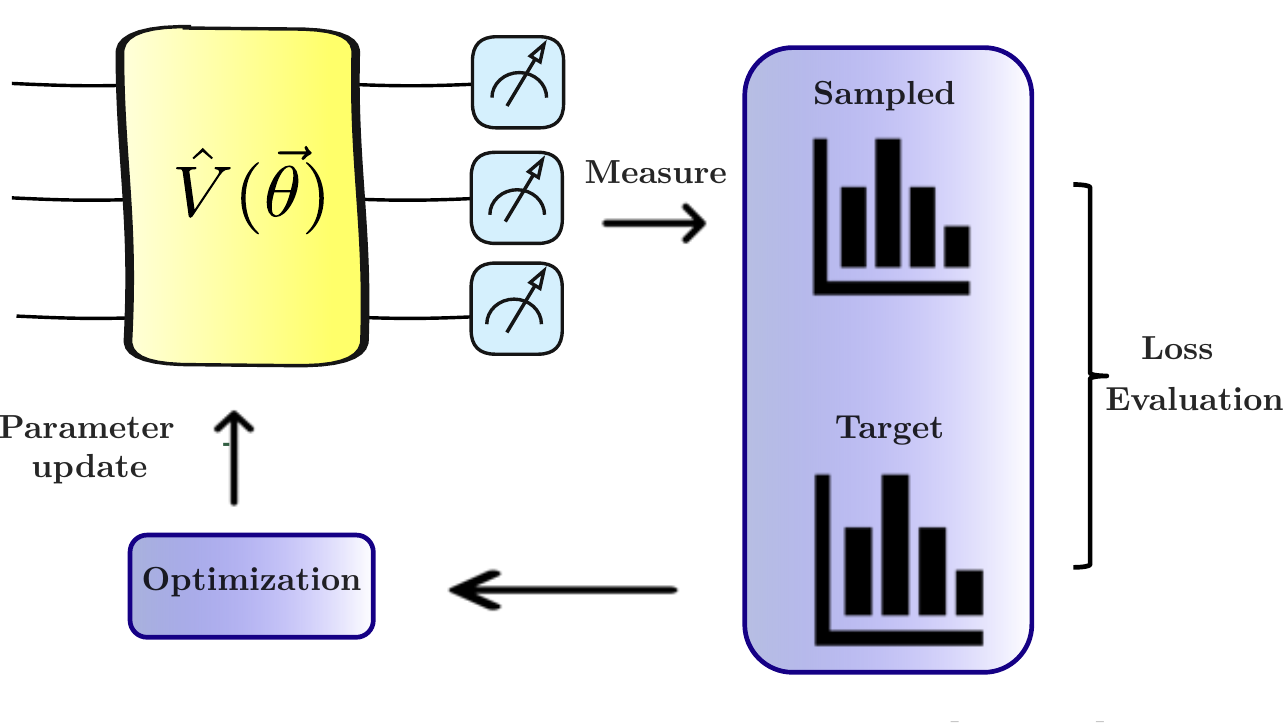}
    \caption{Quantum Circuit Born Machine (QCBM) workflow.}
    \label{fig:qgm_sketch}
\end{figure}

QCBMs aim to directly learn a target probability distribution. They achieve this by optimizing a cost function, such as the negative log-likelihood, to fit the probability of measurement outcomes from a quantum circuit to the given data distribution. Given a dataset $\mathcal{D} = \{x\}$ consisting of independent and identically distributed samples from an unknown target distribution $\pi(x)$, a QCBM is employed to generate new samples that resemble the target distribution. The QCBM maps an initial product state $\ket{0}$ to a parameterized quantum state $\ket{\psi_{\theta}}$. Measuring this state in the computational basis produces a sample of bits, following the distribution $x \sim p_{\theta}(x)$ \cite{pennylaneQuantumCircuit}.  
\begin{equation}
    p_{\theta}(\mathbf{x}) = |\braket{x | \psi_{\theta}}|^2.
\end{equation}
The goal is to optimize the model so that the generated probability distribution $p_{\theta}$ approximates the target distribution $\pi$ as closely as possible. A general worfklow of the QCBM is depicted in Fig. \ref{fig:qgm_sketch}

\begin{figure}[h]
    \centering
    \includegraphics[width=0.75\linewidth]{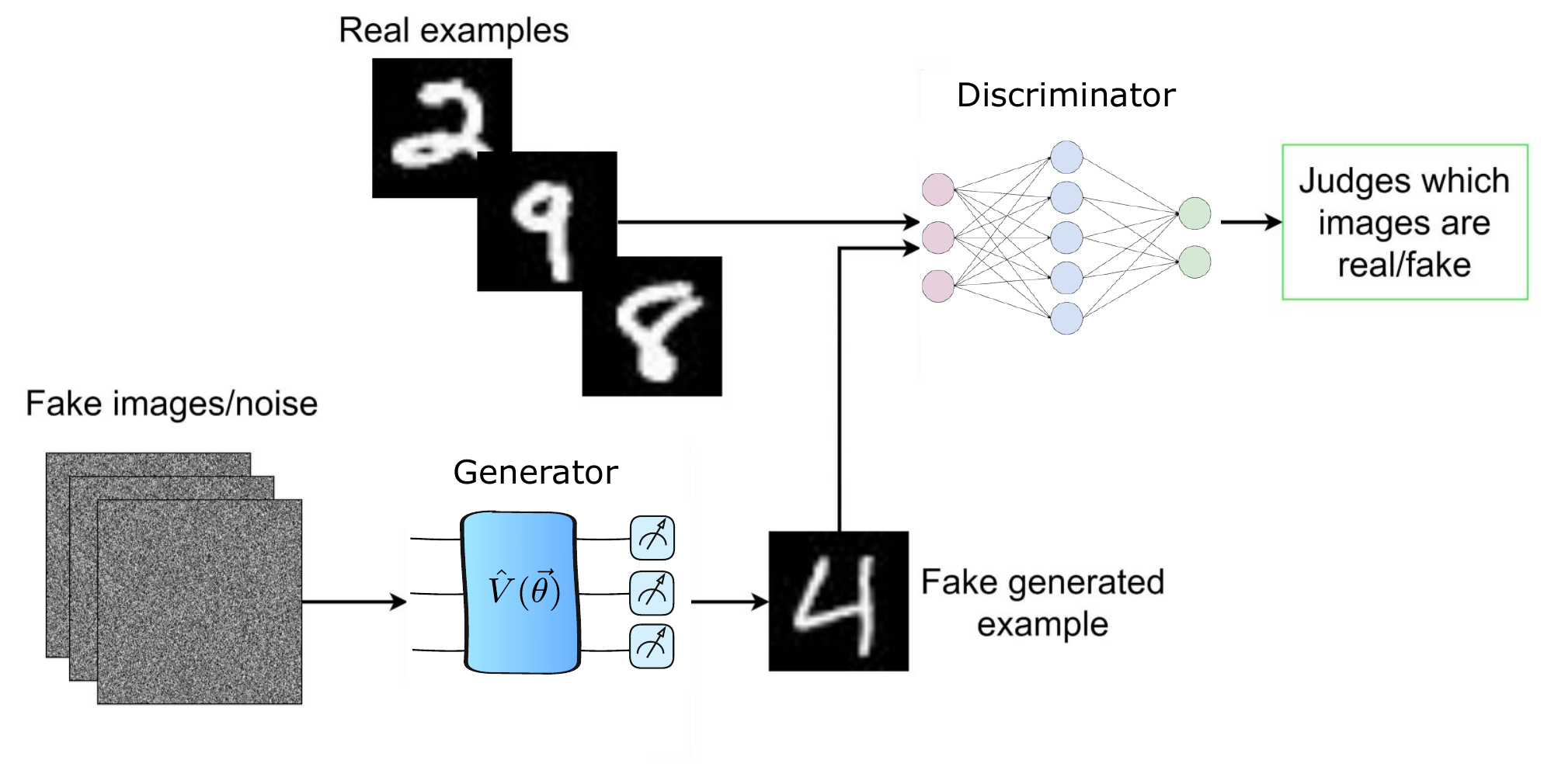}
    \caption{Quantum Generative Adversarial Network (QGAN) workflow.}
    \label{fig:qgan_sketch}
\end{figure}

On the other hand, QGANs are especially good at generating data with multiple distinct patterns that follow different behaviors. This is due to its learning structure, which consists of two pieces:
\begin{itemize}
    \item[] 1. A generator (quantum) creates fake data samples, trying to make them look real.
    \item[] 2. A discriminator (usually classical) checks whether a sample is real (from actual data) or fake (generated).
\end{itemize}
These two components compete, which helps the generator learn to focus on different patterns in the data rather than just averaging everything out, like sometimes other models such as QCBM tend to do. This makes QGANs optimal for capturing complex, multi-modal distributions, meaning they can generate diverse and structured data instead of blurry or unrealistic results. This ability to separate different modes of data makes them very useful for tasks like realistic image synthesis, financial modeling, and anomaly detection. A general worfklow of the QGAN is depicted in Fig.~\ref{fig:qgan_sketch}

\chapter{Quantum clustering and jet reconstruction at the LHC}\label{chap:qjets}

\section{Introduction}
Quantum computing devices, based on the principles of Quantum Mechanics, offer the potential to solve certain problems efficiently that become intractable for classical computers due to exponential or super-polynomial scaling. Quantum algorithms leverage superposition, entanglement and interference to achieve a computational advantage over their classical counterparts. In recent years, these algorithms have rapidly advanced, finding applications in diverse fields such as optimization problems in fintech~\cite{Orus:2019xrh}, quantum chemistry~\cite{Liu:2020eoa}, nuclear physics and Monte Carlo simulations~\cite{Holland:2019zju,Lynn:2019rdt,montanaro:2015}, combinatorial optimization~\cite{Kokail:2018eiw}, and state diagonalization~\cite{Larose:2019,PhysRevA.101.062310}. More recently, they have also been applied to challenges in high-energy physics (HEP)~\cite{Delgado:2022tpc,DiMeglio:2023nsa}. The Large Hadron Collider (LHC) at CERN, along with its Run 3 and the future high-luminosity phase (HL-LHC), generates vast amounts of data requiring intensive analysis and precise theoretical predictions~\cite{EPPPG:2019qin}. The computational demands will only grow with planned future colliders~\cite{FCC:2018byv,Roloff:2018dqu,CEPCStudyGroup:2018ghi}.

Quantum algorithms have been explored for various HEP subfields, including jet clustering~\cite{Wei:2019rqy,Pires:2021fka,Pires:2020urc}, jet quenching~\cite{Barata:2021yri}, parton density determination~\cite{Perez-Salinas:2020nem}, parton shower simulation~\cite{Bauer:2019qxa,Williams:2021lvr,Bepari:2020xqi}, heavy-ion collisions~\cite{DeJong:2020riy}, quantum machine learning~\cite{Guan:2020bdl,Wu:2020cye,Felser:2020mka,Abel:2022lqr,Araz:2022haf,Ngairangbam:2021yma,Araz:2021zwu,Blance:2020nhl}, lattice gauge theories~\cite{Jordan:2012xnu,Banuls:2019bmf,Zohar:2015hwa,Byrnes:2005qx}, and multi-loop Feynman integrals~\cite{Ramirez-Uribe:2021ubp,Clemente:2022nll,Ramirez-Uribe:2024wua,Ochoa-Oregon:2025opz,deLejarza:2024scm,deLejarza:2024pgk,Pyretzidis:2025stx}.

In this chapter, we focus on the problem of clustering and jet reconstruction from collider data, a computationally expensive task involving optimization over large sets of final-state particles. Current state-of-the-art jet clustering algorithms require months to process one year's worth of LHC data~\cite{Evans2009TheLH}. The upcoming HL-LHC will further increase event rates and pile-up~\cite{HEPSoftwareFoundation:2017ggl,Collaboration:2802918}, increasing the computational challenge. Without improvements, processing times could last to several decades. This evidences the need for faster and more efficient jet clustering algorithms.

In this chapter, we explore the potential of quantum algorithms to accelerate jet identification, focusing on three well-established classical methods: \texttt{K-means} clustering~\cite{macqueen1967some,ball1967clustering}, Affinity Propagation (\texttt{AP})~\cite{Frey2007ClusteringBP}, and the $k_T$-jet clustering algorithm in its various forms~\cite{Catani:1991hj,Catani:1993hr,Ellis:1993tq,Cacciari:2008gp,Dokshitzer:1997in}. In particular, we propose~\cite{deLejarza:2022bwc,deLejarza:2022vhe} quantum analogs for each: quantum \texttt{K-means}, quantum \texttt{AP}, and quantum $k_T$-based algorithms.

Clustering is a fundamental problem in ML and computational geometry, widely used in marketing, data mining, bioinformatics, image processing, and pattern recognition, as well as in HEP. The \texttt{K-means} method~\cite{ball1967clustering,macqueen1967some} groups $N$ datapoints into $K$ clusters, where each points belongs to the nearest mean. Finding the optimal solution is NP-hard~\cite{Drineas:2004fa}, but the widely used Lloyd's algorithm~\cite{Lloyd:1982ls} provides an iterative heuristic approach. The \texttt{K-means++} variant~\cite{K-Means++} improves performance by selecting better initial centers. In HEP, \texttt{K-means} has been applied to improve mass resolution in top quark and $W$ boson studies~\cite{Chekanov:2005cq}, boosted object tagging~\cite{Thaler:2011gf}, and minijet identification at low $p_T$~\cite{Wong:2018frb}.

The \texttt{AP} algorithm clusters data by exchanging messages between points to identify exemplars. It has been applied in face recognition, gene identification~\cite{Leone_2007, Sumedha_2008,Bailly_Bechet_2009}, astrophysics~\cite{GonzlezMartn2017}, and parton density studies in HEP~\cite{Carrazza:2016sgh}. Unlike \texttt{K-means}, it does not require to specify the number of clusters beforehand, making it interesting for jet clustering.

Hierarchical clustering builds on a recursive approach without fixing the number of clusters in advance. The $k_T$-based algorithms~\cite{Cacciari:2011ma} fall into this category, relying on a distance metric to iteratively merge clusters. These methods are the most widely used for jet reconstruction at the LHC.

Quantum versions of \texttt{K-means} clustering have been introduced in Refs.~\cite{Blance:2020ktp,Pires:2021fka} for HEP, following earlier studies~\cite{Abhi:2020}. These implementations use Euclidean distance for clustering. In this chapter, we present a novel quantum \texttt{K-means} algorithm based on Minkowski distance, making it the first quantum clustering method adapted to relativistic kinematics. For the quantum \texttt{AP} algorithm, we define a similarity metric based on invariant sum squared and compute it using a quantum subroutine identical to quantum \texttt{K-means}. Additionally, we introduce the first quantum $k_T$-based algorithm, featuring a novel quantum search method for finding minimum distances. This new quantum method is generalizable to other applications, which we discuss further in the chapter.

This chapter is organized as follows. In Section \ref{subsubsec:qinvmass} we introduce our notation and we define the Euclidean and Minkowskian quantum distances. In Section \ref{subsec:qsearching} we present our new quantum algorithm in order to search the approximate maximum in a set of a given number of elements. We consider the classical versions of the \texttt{K-means} clustering, \texttt{AP} and $k_T$-based algorithms in Section \ref{app:clusteringalgos}. In Section \ref{app:qclusteringalgos} we present our results considering the quantum simulations of these algorithms and a proof-of-concept implementation with Gaussian datasets as well as with simulated LHC physical events. We also compare their performance in detail. We discuss their differences and conceptual similarities and we compare them with their classical counterparts. A brief summary of our results is presented in Section \ref{sec:qjets_conclusions}. 

The code to reproduce the results of this chapter is available in \cite{githubGitHubGmlejarzaQuantumjetclustering}.

\section{Jets}\label{app:jets}

\begin{figure}[h]
    \centering
    \includegraphics[width=0.7\linewidth]{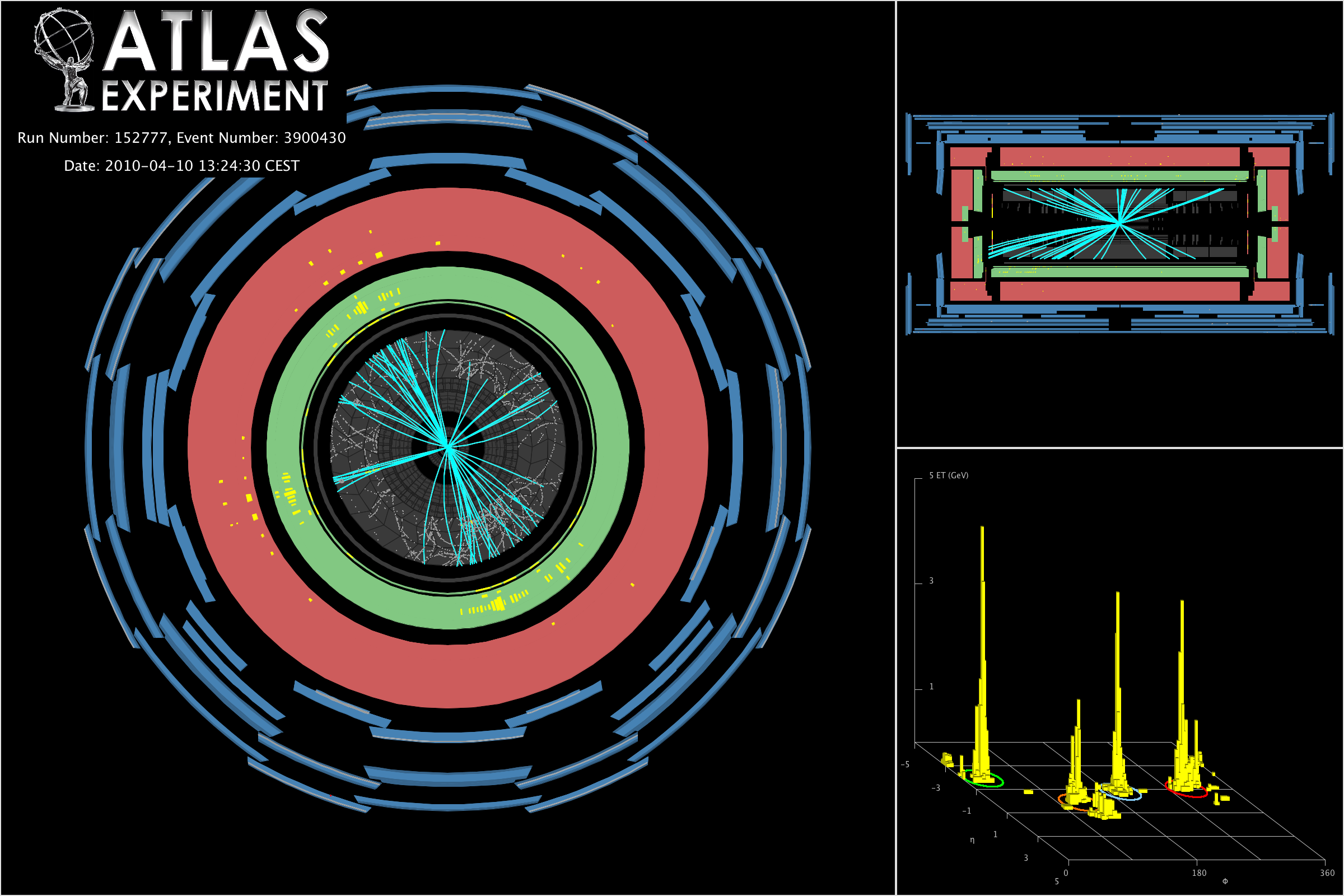}
    \caption{Event with four reconstructed hadronic jets observed by the ATLAS detector at CERN. Picture taken from the ATLAS public event display repository~\cite{cernEventDisplayStandAlonelt}}
    \label{fig:atlas_jets}
\end{figure}

The main goal of particle colliders such as the Large Hadron Collider (LHC) at CERN is to study the fundamental interactions of elementary particles with the greatest detail and to search for new physics, including the discovery of unknown particles. Among the richest processes that include more relevant information are those involving quarks and gluons, collectively known as partons, as they constitute other particles known as hadrons that are color-neutral states. When partons are produced in high-energy collisions, they undergo a cascade of radiation, emitting secondary quarks and gluons mostly collinear to the parent parton through a process called parton showering. This is a prediction of perturbative QCD at high energies. Because quarks and gluons cannot exist as free particles due to QCD's color confinement they finally hadronize into color-neutral bound states by combining with ``sea'' partons spontaneously created from the quantum vacuum. The avalanche of hadrons created is observed in the detector as highly collimated bunches of particles see Fig.~\ref{fig:atlas_jets}. These bunches of particles are known as hadronic jets, as their constituents are hadrons. If we look inside a jet we might find protons and neutrons, the constituents of atomic nuclei, and other less known hadrons such as pions and kaons~\cite{Kar:2015nxu,banfi}. 

At this point, it has been clear that the partons produced in the collisions cannot be directly observed since they are not free particles. However, jets provide a relevant information on the underlying quarks and gluons produced in high-energy collisions. Analyzing individual hadrons would obscure the original parton dynamics, as hadronization is a non-perturbative process that introduces significant complexity. On the other hand, jets preserve key properties of the originating parton due to the preference for the production of secondary particles in the original direction, as predicted by QCD. This allows for meaningful comparisons with QCD predictions and facilitates searches for new physics, where deviations from expected jet properties could indicate the presence of beyond the Standard Model (BSM) physics~\cite{Salam:2010nqg}.

Jets are fundamental objects in HEP analysis  \cite{Salam:2010nqg,ATLAS:2013bqs,Gras:2017jty,Cacciari:2008gp}. The transverse momentum ($p_T$) of jets, measured perpendicular to the beam axis, distinguishes hard-scattering processes (high $p_T$) from background interactions (low $p_T$). Also the jet mass, calculated as the invariant mass of all constituent particles, helps differentiate jets produced by quarks (typically lighter) from those produced by gluons or boosted jets. Furthermore, the angular substructure, reveals radiation patterns critical for identifying QCD splittings or resonant decays. Jets originated from bottom quarks are specifically tagged via displaced secondary vertices. Finally, the jet radius parameter ($R$), defining the angular scale $\Delta R = \sqrt{(\Delta\eta)^2 + (\Delta\phi)^2}$ for particle clustering, balances resolution and noise: smaller radius ($R\leq 0.4$) mitigates pileup in proton-proton collisions, while larger radius $R$ ($R\sim 1.0$) captures the broad substructure of boosted decays. Together, these properties enable jets to serve as effective tools to identify and describe the partons that participate in the hard scattering at very short distances.

Accurately identifying jets in high-energy collisions is a daunting task that is essential for understanding the fundamental processes at particle colliders. Since jets are the primary experimental signatures of quarks and gluons, their correct reconstruction directly impacts precision tests of QCD and searches for new physics. A well-designed jet identification algorithm must efficiently group the particles originating from a parent parton while minimizing contamination from unrelated radiation or detector noise. However, this is challenging due to the complex and dynamic nature of hadronization, the presence of overlapping jets, and fluctuations in the detector response. Poor jet reconstruction can lead to noisy measurements of key observables, such as jet energy and angular distributions, reducing the sensitivity to deviations from the SM. In searches for new physics, subtle modifications in jet substructure could provide hints of novel phenomena, making robust jet algorithms a critical component of high-energy collider experiments.

\section{Clustering algorithms}\label{app:clusteringalgos}
\subsection{\texttt{K-means} algorithm}
The \texttt{K-means} algorithm is an unsupervised learning method that classifies the elements of a dataset into $K$ groups called clusters \cite{ball1967clustering,macqueen1967some}. The goal of this algorithm is to maximize similarity within clusters while maximizing dissimilarity between them. The algorithm inputs and outputs are:
\begin{itemize}
    \item Input: $N$ $d$-dimensional data points and cluster count $K$ ($K \leq N$)
    \item Output: $K$ centroids computed as cluster means, and the $N$ points assigned to their respective centroids.
\end{itemize}

\begin{figure}[h]
    \centering
    \includegraphics[width=0.5\linewidth]{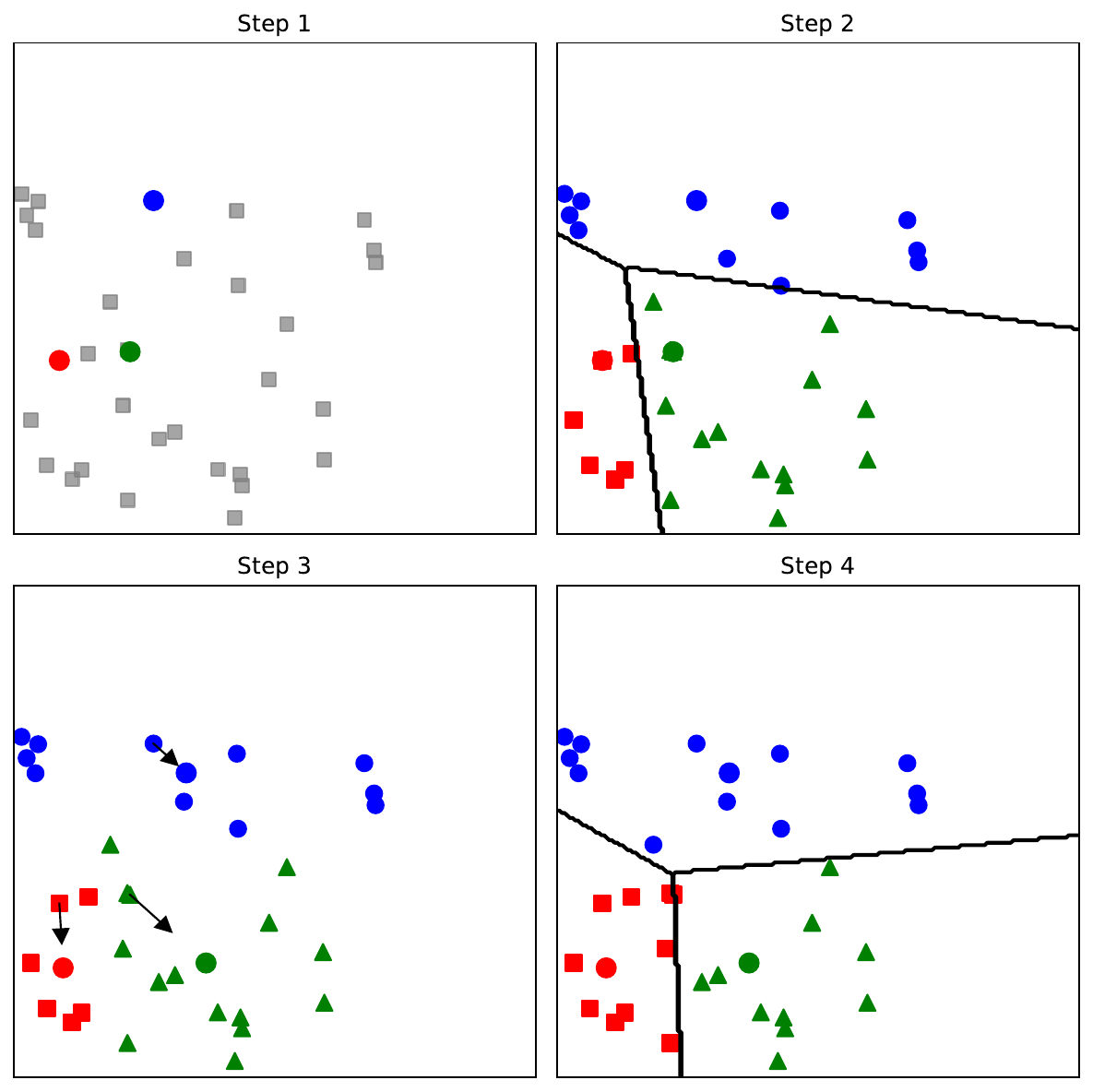}
    \caption{Workflow of the \texttt{K-means} algorithm.}
    \label{fig:kmeans}
\end{figure}

The iterative procedure follows these steps, and is also depicted in Fig.~\ref{fig:kmeans}:
\begin{enumerate}
    \item \textbf{Initialization}: Select $K$ initial centroids, either randomly or via smart initialization, such as \texttt{k-means++}~\cite{K-Means++}. 
    \item \textbf{Assignment}: Associate each point to its nearest centroid using a predefined distance metric, such as the Euclidean distance.
    \item \textbf{Update}: Recompute centroids as the mean of all points in each cluster.
    \item \textbf{Termination}: Repeat steps 2 and 3 until centroids stabilize and convergence is achieved.
\end{enumerate}

The time complexity of this algorithm is estimated by analysing the time complexity of its components. The number of distances that have to be calculated required $\mathcal{O}(N)$ operations, the search of a minimum distance for every data point with respect to the centroids would be $\mathcal{O}(K)$, and the calculation of each distance itself would require $\mathcal{O}(d)$, this results in a complexity of $\mathcal{O}(NKd)$.

\subsection{Affinity Propagation algorithm}
Although \texttt{K-means} is an effective algorithm for clustering data, it requires the number of clusters, $K$, to be predetermined, which is often not practical in HEP applications. In contrast, the Affinity Propagation (\texttt{AP}) algorithm \cite{Frey2007ClusteringBP}, another unsupervised machine learning technique, does not require the number of clusters as an input. Instead, \texttt{AP} only requires the data points that need to be classified. Let $x_1, \ldots ,x_N$ represent a set of data points. A similarity function $s$ is then computed to quantify the relationship between points, such that $s(i,j)\geq s(i,k)$ if and only if $x_i$ is more similar to $x_j$ than to $x_k$. A common similarity metric is the negative squared distance between two points: $s(i,j)=-|x_i-x_j|$.

The diagonal elements of the similarity matrix, $s(i,i)$, are particularly important as they represent ``preferences'' that indicate the likelihood of a point becoming an exemplar, or a cluster center, similar to \texttt{K-means} centroids. During the first iteration, each diagonal element, $s(i,i)$, is initialized to a fixed value, typically the median similarity of all point pairs. 

\begin{figure}
    \centering
    \includegraphics[width=0.9\linewidth]{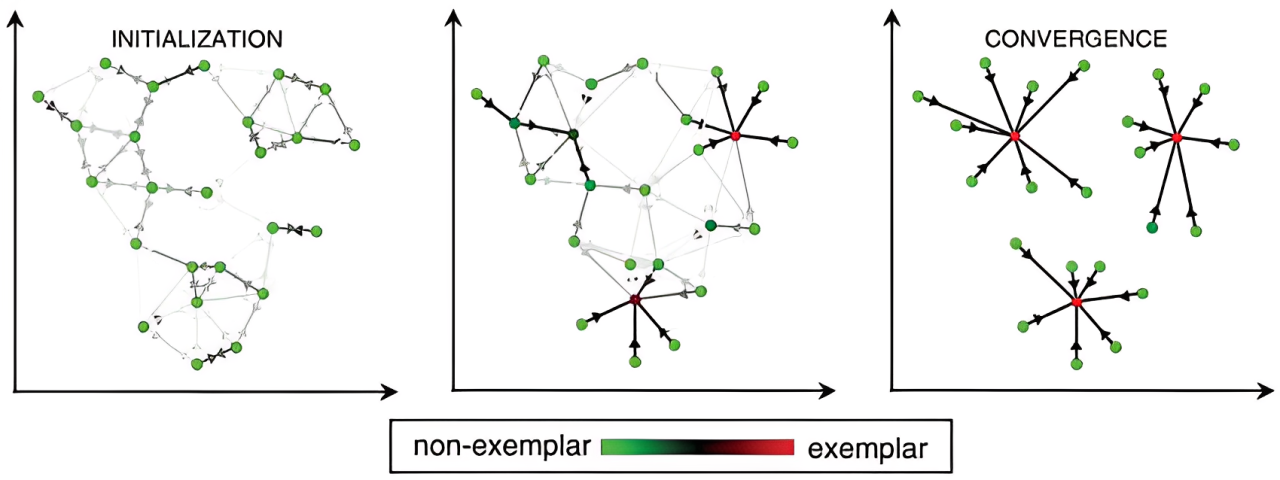}
    \caption{Workflow of the \texttt{AP} algorithm. Picture taken from \cite{applot}}
    \label{fig:ap}
\end{figure}

Next, two matrices are computed, representing message exchanges between data points~\cite{Frey2007ClusteringBP}. The responsibility matrix $R$ contains values $r(i,k)$, which measure the suitability of point $k$ as the exemplar for point $i$, compared to other candidate exemplars. The availability matrix $A$ contains elements $a(i,k)$, which indicate how suitable it would be for point $i$ to select point $k$ as its exemplar, based on the preferences of other points. Both matrices can be interpreted as log-probability ratios. The steps of the \texttt{AP} algorithm are as follows and are also depicted in Fig.~\ref{fig:ap}:
\begin{enumerate}
    \item The matrices $R$ and $A$ are initialized to zero.
    \item The responsibility matrix is computed:
    \beq
    r(i,k)= s(i,k)-\max_{q\neq k}\lbrace a(i,q)+s(i,q)\rbrace.
    \eeq
    
    \item The availability matrix is computed:
    \beq
    a(i,k)= \min \left( 0, r(k,k)+\sum_{q\notin \lbrace i,k \rbrace}\max(0,r(q,i))\right) \mathrm{for} \, i \neq k, \, \mathrm{and}
    \eeq
    \beq
    a(k,k)= \sum_{q\neq k} \max(0,r(q,k).
    \eeq
    \item Steps 2 and 3 are repeated until either the cluster boundaries remain unchanged for several iterations, or a predetermined number (of iterations) is reached.
\end{enumerate}

After convergence, the exemplars (clusters) are identified from the final matrices as those for which $r(i,i) + a(i,i) > 0$.

This algorithm requires $\mathcal{O}(N^2)$ operations to fill the similarity matrix, and computing each element takes $\mathcal{O}(d)$ time, as it involves calculating the distance between two $d$-dimensional points. Furthermore, steps 2 and 3 are repeated $T$ times, resulting in an overall complexity of $\mathcal{O}(N^2 Td)$.

\subsection{Generalised $k_T$-jet algorithm}

The inclusive variant of the generalised $k_T$-jet algorithm is defined as follows~\cite{Cacciari:2011ma}:
\begin{enumerate}
\item For each pair of particles $i$ and $j$, a distance measure is calculated according to:
\begin{equation}
d_{ij}=\mathrm{min} (p_{T,i}^{2p},p_{T,j}^{2p}) \Delta R_{ij}^2/R^2,
\label{eq:kt}
\end{equation}
where $\Delta R_{ij}^2=(y_i-y_j)^2+(\phi_i-\phi_j)^2$. Here, $p_{T,i}$, $y_i$, and $\phi_i$ are the transverse momentum (relative to the beam direction), rapidity, and azimuth angle of particle~$i$, respectively. The parameter $R$ defines the characteristic jet radius and is typically set to a value around~$1$. In addition, a distance to the beam is defined for each particle~$i$ as $d_{iB}=p_{T,i}^{2p}$.
\item The smallest distance, $d_{min}$, is then identified among all $d_{ij}$ and $d_{iB}$. If $d_{min}$ corresponds to a $d_{ij}$, particles $i$ and $j$ are merged by summing their four-momenta (using the E-scheme recombination). Alternatively, if $d_{min}$ is a $d_{iB}$, particle $i$ is designated as a final jet and removed from further consideration.
\item Step 1 is repeated until no particles remain.
\end{enumerate}

It is worth noting that for specific choices of the parameter $p$ in \Eq{eq:kt}, the generalised $k_T$-jet algorithm simplifies to well-known jet algorithms: the $k_T$ algorithm for $p=1$, the Cambridge/Aachen algorithm for $p=0$, and the anti-$k_T$ algorithm for $p=-1$.

As mentioned in Ref.~\cite{2006}, the traditional implementation of the $k_T$-jet algorithm has a computational complexity of $\mathcal{O}(N^3)$. This stems from the fact that the algorithm's performance is limited by the need to scan a table of size $\mathcal{O}(N^2)$ containing all distances $d_{ij}$ and $d_{iB}$, an operation that must be performed $N$ times.

However, the \texttt{FastJet}~\cite{Cacciari:2011ma} algorithm achieves a reduced complexity of ${\cal O}(N^2)$. This improvement relies on identifying the geometrical nearest neighbor for each particle, which avoids the need to construct the full size-$N^2$ table of $d_{ij}$. Instead, it is sufficient to consider only the size-$N$ array $d_{i\mathcal{G}_i}$, where $\mathcal{G}_i$ represents $i$'s geometrical nearest neighbor. Furthermore, \texttt{FastJet} offers opportunities for additional optimization using Voronoi diagrams, leading to a further reduction in time complexity from $\mathcal{O}(N^2)$ to $\mathcal{O}(N \log N)$.

\section{Quantum  algorithms}\label{app:qclusteringalgos}

In this section we design specific quantum algorithms intended to be inserted as subroutines in the clustering algorithms mentioned in the previous section with the goal of speeding up some bottleneck tasks. 

\subsection{Quantum distance}

In quantum computing, it is crucial to have efficient tools to measure quantum overlap between two states, as in many cases it determines the possibility of obtaining a quantum advantage~\cite{Foulds:2020ajt}.  One approach to efficiently measure similarity between quantum states is the \textit{SwapTest} method~\cite{Buhrman:2001} (see \Eq{eq:p0swaptest} in Appendix~\ref{app:swaptest} for more details). The quantum distances (Euclidean distance or Minkowski invariant sum squared) presented in this subsection, make use of the \textit{SwapTest} procedure.

\subsubsection{Euclidean quantum distance}

Let us start by considering $N$ data points or vectors
in an Euclidean $d$-dimensional space, 
$\{{\bf x}_i\}_{i=1,\ldots, N}$, which
are encoded as quantum states of the form 
\beq
\ket{x_i} = |{\bf x}_i|^{-1} \sum_{\mu=1}^{d} x_{i,\mu} \, \ket{\mu}~,
\eeq
where $|{\bf x}_i| = \sqrt{\sum_{\mu=1}^d (x_{i,\mu})^2}$ is the modulus 
of the vector ${\bf x}_i$, and $x_{i,\mu}$ are its components.  
Since we rely on amplitude encoding, each vector requires $n \ge \log_2 d$ qubits to be encoded, i.e. for $d=3$ we need two entangled qubits where one of its states remains free and is not used. The Euclidean distance between two vectors ${\bf x}_i$ and ${\bf x}_j$ is defined classically as
\beq
\label{eq:EuclidDist}
d_E^{\rm (C)}({\bf x}_i,{\bf x}_j) = |{\bf x}_i - {\bf x}_j|~,
\eeq
where $E$ is a subscript indicating Euclidean distance, and the superscript ${\rm C}$ specifies that it is the classical version.

To obtain the quantum counterpart of Eq.~\eqref{eq:EuclidDist}, we employ the controlled \textit{SwapTest} approach. In order to define the Euclidean quantum distance connecting the $d$-dimensional vectors ${\bf x}_i$ and ${\bf x}_j$, we entangle their corresponding quantum states, $\ket{x_i}$ and $\ket{x_j}$. Auxiliary states are then defined as
\begin{equation}
\ket{\psi_1} =\frac{1}{\sqrt{2}} \left( \ket{0, x_i} + \ket{1, x_j} \right), \qquad
\ket{\psi_2} =\frac{1}{\sqrt{Z_{ij}}} \left( |{\bf x}_i| \ket{0} -|{\bf x}_j| \ket{1} \right),
\label{eq:varphi}
\end{equation}
where $Z_{ij}=|{\bf x}_i|^2+|{\bf x}_j|^2$
acts as a normalization constant, and $\ket{0}$ and $\ket{1}$ are the states of an auxiliary qubit. Furthermore, let us define the swapped state $\ket{\psi'_1}$ as
\begin{equation}
\ket{\psi_1'} =\frac{1}{\sqrt{2}} \left( \ket{x_i, 0} + \ket{x_j, 1}  \right).
\label{eq:psiprime}
\end{equation}
The inner products involving the quantum states defined in Eqs. \eqref{eq:varphi} and \eqref{eq:psiprime} is expressed as
\begin{equation}
\langle\psi_1'|\psi_2\rangle = \frac{1}{\sqrt{2Z_{ij}}} 
\left(|{\bf x}_i| \langle x_i| - |{\bf x}_j| \langle x_j| \right), \qquad
\langle\psi_2|\psi_1\rangle =\frac{1}{\sqrt{2Z_{ij}}} 
\left(|{\bf x}_i| \ket{x_i} - |{\bf x}_j| \ket{x_j} \right).
\label{eq:Edistance}
\end{equation}
From this, it follows that
\begin{equation}
\langle\psi_1'|\psi_2\rangle \langle\psi_2|\psi_1\rangle 
= \frac{1}{2Z_{ij}}|{\bf x}_i-{\bf x}_j|^2.
\label{eq:overlap}
\end{equation}
Thus, the quantum Euclidean distance (as detailed in \Eq{eq:p0swaptest} within Appendix~\ref{app:swaptest}) is given by
\begin{equation}
d_E^{\rm (Q)}({\bf x}_i, {\bf x}_j) = \sqrt{2Z_{ij}(2P_{\Psi_3}(\ket{0})-1)},
\label{eq:qdistance}
\end{equation}
where the superscript $Q$ indicates the quantum formulation of the distance $d_E$, and the subscript $\Psi_3$ of the probability $P$ refers to the probability of measuring the ancillary qubit in state $\ket{0}$ at the last stage of the \textit{SwapTest} method.

\subsubsection{Quantum invariant sum squared in Minkowski space}
\label{subsubsec:qinvmass}

In HEP, vectors are typically defined within a four-dimensional spacetime in the Minkowski metric. These vectors take the form $x_i = (x_{i,0}, {\bf x}_i)$, where $x_{i,0}$ is the temporal component and ${\bf x}_i$ represent the three spatial components. We will subsequently assume a spacetime dimension of $d$, with $d-1$ indicating the number of spatial components. The analogue of the Euclidean classical distance, in the Minkowski space is the invariant sum squared~$s_{ij}^{\rm (C)}$, often known as the invariant mass squared for particle four-momenta:
\beq
s_{ij}^{\rm (C)} = (x_{0,i}+x_{0,j})^2 
- |{\bf x}_i + {\bf x}_j|^2~. 
\eeq

This Lorentz-invariant quantity serves as a similarity test. It also corresponds to the distance employed in certain traditional jet-clustering algorithms at $e^+e^-$ colliders~\cite{JADE:1982ttq,Bethke:1991wk,Rodrigo:1999qg}. Computing this Minkowski-type distance via a quantum algorithm necessitates two applications of the \textit{SwapTest} subroutine (detailed in Appendix A): one for the spatial and one for the temporal components.

The spatial distance is calculated following the method outlined in the previous section, with a minor adjustment to \Eq{eq:Edistance} (a sign change in the term multiplying the qubit $\ket{1}$):
\beq
\ket{\psi_2} \longrightarrow \ket{\psi_2}  =\frac{1}{\sqrt{Z_{ij}}} \left( |{\bf x}_i| \ket{0} +|{\bf x}_j| \ket{1} \right)~,
\eeq
The temporal distance is obtained by calculating the overlap between these states:
\begin{equation}
\ket{\varphi_1} = H \ket{0} = \frac{1}{\sqrt{2}} \left( \ket{0} + \ket{1} \right)~, \qquad 
\ket{\varphi_2} =\frac{1}{\sqrt{Z_0}} \left(x_{0,i} \ket{0} +x_{0,j} \ket{1} \right)~,
\label{eq:minkstates}
\end{equation}
where $Z_{0}=x_{0,i}^2+x_{0,j}^2$. Applying the \textit{SwapTest} to these states results in:
\begin{equation}
P(|0\rangle|_{time})=\frac{1}{2}+\frac{1}{2} |\langle \varphi_1| \varphi_2 \rangle|^2 \ ,
\label{eq:p0time}
\end{equation}
where the overlap $|\langle \varphi_1| \varphi_2 \rangle|^2$ is trivially given by
\begin{equation}
|\langle \varphi_1| \varphi_2 \rangle|^2 = \frac{1}{2Z_0}(x_{0,i} +x_{0,j})^2.
\label{eq:hphioverlap}
\end{equation}
Therefore:
\begin{equation}
(x_{0,i}+x_{0,j})^2 =2 Z_{0}(2P_{\Psi_3}(|0\rangle|_{time})-1)~.
\label{eq:parttime}
\end{equation}
At this juncture, the quantum version of the invariant sum squared emerges from the combination of Eqs.~\eqref{eq:qdistance} and \eqref{eq:parttime}:
\beq
s_{ij}^{\rm (Q)} =2\big( 
Z_0(2P_{\Psi_3}(|0\rangle|_{time})-1)-Z_{ij}(2P_{\Psi_3}(|0\rangle|_{spatial})-1)\big).
\label{eq:qdistancemink}
\eeq

\begin{figure}[th!]
    \centering
    \includegraphics[width=0.6\textwidth]{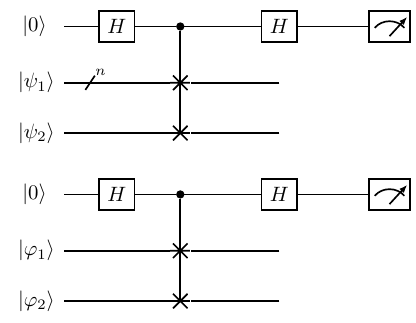}
    \caption{Quantum circuit to obtain the invariant sum squared between two $d$-dimensional vectors in Minkowski space.}  
    \label{fig:qsmassinvariant}
\end{figure}

The quantum circuit used to implement the invariant sum-squared distance is presented in Fig.~\ref{fig:qsmassinvariant}. 
For the first three wires, the \textit{SwapTest} is executed on the spatial components, where we assume that the states $\psi_1$, $\psi_2$ have been loaded from a quantum Random Access Memory (qRAM) in $\mathcal{O}(\log (d-1))$, since the state $\psi_1$ is encoded in $\log_2 (d-1)$ qubits. On the other hand, from the fourth wire onward, the \textit{SwapTest} is applied to the temporal components. In this case, it takes $\mathcal{O}(1)$, since we only have 1-dimensional qubit states.

\subsection{Quantum maximum searching algorithm}
\label{subsec:qsearching}

Finding a specific element within a dataset is a common and computationally demanding challenge in data analysis. However, quantum computing provides efficient tools to accelerate data queries. In particular, Grover's algorithm~\cite{Grover:1997fa} is well known for achieving a quadratic speedup.

In this section, we introduce a significantly simpler algorithm designed only for identifying the maximum value in a list. Despite its simplicity, this algorithm is sufficiently precise for the applications that we will addres in Sectio~n\ref{app:qclusteringalgos}. To the best of our knowledge, this method has not been previously reported in the literature.

\begin{figure}[th!]
    \centering
    \includegraphics[width=0.9\textwidth]{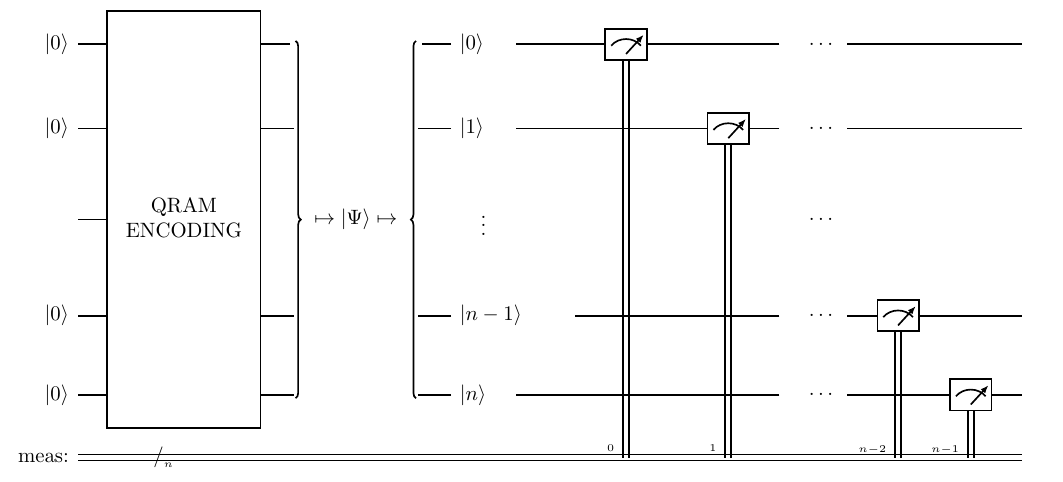}
    \caption{Quantum circuit for quantum maximum searching by qRAM amplitude encoding.}  
    \label{fig:qsearch}
\end{figure}

Let $L[0, \ldots ,N-1]$ be an unsorted list of $N$ elements. The problem of finding the maximum requires determining the index $y$ such that $L\left[y\right]$ is the largest value. The quantum algorithm leveraging amplitude encoding follows two steps:
\begin{enumerate}
    \item  The list of $N$ elements is encoded into a $\log(N)$-qubit state as follows:
    \begin{equation}
    \ket{\Psi} = \frac{1}{\sqrt{L_{sum}}}\sum_{j=0}^{N-1}L\left[j\right] \, \ket{j}~,
        \label{eq:amplitude encoding}
    \end{equation}
    where $L_{sum}=\sum_{j=0}^{N-1}L[j]^2 $ is a normalization constant.
    This amplitude encoding is implemented using qRAM.
    \item  The final state is measured multiple times to mitigate statistical uncertainty. The most frequently observed state corresponds to the maximum value.
\end{enumerate}
The quantum circuits that represents this algorithm is shown in Fig.~\ref{fig:qsearch}, where $n=\log(N)$ qubits are needed to encode a list of $N$ (real) elements. 

The bottleneck of this procedure underlies in encoding data into a quantum state. Assuming data is stored in a qRAM, as would be the case on a true universal quantum computer, encoding takes $\mathcal{O}(\log(N))$ steps~\cite{lloyd2013quantum,2008QRAM,2008Arch,demartini2009experimental,2010Mag,PhysRevLett.105.140501,PhysRevLett.105.140502,PhysRevLett.105.140503}. The best classical algorithm used to obtain the minimum of an unsorted list of $N$ items is of order $\mathcal{O}(N)$. Therefore, under these assumptions, the improvement introduced by this quantum algorithm is exponential.

The well-known quantum minimum search algorithm by Dürr and Høyer~\cite{Durr:1996nx} achieves a complexity of $\mathcal{O}(\sqrt{N})$. Following its theoretical proposal~\cite{Durr:1996nx}, it has been studied and implemented in quantum simulators (see Ref.~\cite{DBLP:journals/corr/abs-1804-03719}). However, prior implementations~\cite{DBLP:journals/corr/abs-1804-03719} indicate potential areas for improvement, including the large number of qubits required, the infeasibility of hardcoding an oracle for each element, the high number of \textit{shots} needed, and, in some cases, suboptimal performance. The quantum maximum search algorithm presented here, based on amplitude encoding through qRAM, aims to address these challenges and enhance the efficiency of quantum search methods.

Nonetheless, the new algorithm introduced in this paper and the Dürr and Høyer quantum method share a probabilistic nature that could lead to misidentification of the absolute maximum or minimum. These challenges typically arise when the dataset has a very low standard deviation, or when the largest and smallest values are nearly indistinguishable. In such cases, the probability of measuring multiple candidates becomes almost identical, making it difficult to select the correct maximum/minimum.

However, as we will show in Section \ref{app:qclusteringalgos}, these potential challenges do not significantly affect the practical implementation of our quantum algorithm in the context of jet clustering.

Beyond HEP and jet clustering, our quantum algorithm could also be valuable in other fields, particularly in Extreme Value Theory (EVT)\cite{smith1990extreme}. According to Gumbel (1958)\cite{Gumbel1958}, EVT focuses on analyzing the probability distributions of extreme values, aiming to predict rare events based on historical data. Since predictive models require identifying extreme values within large datasets, our algorithm could be well-suited for statistical analysis in this domain. This includes applications in actuarial and financial sciences, meteorology, material sciences, engineering, climatology, geology, hydrology, and highway traffic analysis~\cite{Reiss2007,Coles2001,castillo2004}.

\section{Quantum clustering algorithms}

In this Section, we use the open-source IBM Quantum software \textit{Qiskit}~\cite{qiskit2024} to build the quantum algorithms presented in Section~\ref{app:qclusteringalgos}. Specifically we design the quantum circuits to calculate the invariant sum squared described in Section \ref{subsubsec:qinvmass} for the \texttt{K-means} and the \texttt{AP} algorithm, and to finding the minimum distance in the \texttt{K-means} and the $k_T$-jet algorithm. Then, these quantum subroutines have been introduced into their respective classical algorithm substituting the classical part they are replacing. The  algorithms presented here have been executed on a quantum simulator that offers an unrestricted and noise-free environment.
A quantum implementation in an existing quantum device taking advantage of the claimed speed-up is today not possible, as a qRAM architecture does not exist yet. Nonetheless, the quantum simulations in this section show a satisfactory performance and clustering efficiencies comparable to those of their classical counterparts.

\subsection{Quantum \texttt{K-means} with Minkowski-type distance}
\label{subsec:qkmeans}

The \texttt{K-means} algorithm has a precedent of a quantum version that differs from its classical counterpart~\cite{kmeans} in two points~\cite{kopczyk2018quantum}. First, the quantum \texttt{K-means} introduces a quantum method to calculate the distance between data points. Second, the quantum version also includes a procedure for obtaining the minimum distance of each data point with respect to the $K$ centroids, which is achieved by Dürr and Høyer's algorithm~\cite{Durr:1996nx}.

We focus on a new quantum version of the \texttt{K-means} algorithm, where the calculation of distances is made quantumly and the minimum distance of each data point to the centroids is obtained with the quantum maximum searching algorithm\footnote{We may apply this algorithm for finding the minimum since obtaining the minimum amongst the distances is equivalent to obtaining the maximum of their inverses: $s_{ij}^{-1}$.} explained in Section~\ref{subsec:qsearching}. 
Other quantum versions of the \texttt{K-means} algorithm have been studied in Refs.~\cite{Blance:2020ktp,Pires:2021fka} and~\cite{Abhi:2020}, where an Euclidean distance was used to separate the particles from each other. Instead, we analyse for the first time an implementation of the \texttt{K-means} algorithm with a Minkowski-type quantum distance, as defined in Section~\ref{subsubsec:qinvmass}.

The time complexity of this quantum algorithm is estimated by analysing the time complexity of its components. The distances that have to be calculated are $\mathcal{O}(N)$, the search of a minimum distance for every data point with respect to the centroids would be $\mathcal{O}(\log K)$, and the calculation of each distance itself would require $\mathcal{O}(\log (d-1))$ qubits assuming the data is stored in a qRAM. This results in a speedup from $\mathcal{O}(NKd)$ in the classical version to $\mathcal{O}(N\log K \log(d-1))$ in our quantum version. Therefore an exponential speed-up in the  number of clusters and in the vector dimensionality would be achieved. 

We now present our implementation of the quantum \texttt{K-means} algorithm using the invariant sum squared as a distance metric, along with a maximum searching algorithm, and compare their performance with classical counterparts. To do so, we generate 15 Gaussian-clustered datasets of $N=300$ three-dimensional vectors~\footnote{In general, it is possible to relate this generated set of three-dimensional vectors, to a physical event at the LHC. It is enough to consider the set of $n$ three-dimensional vectors as massless partons recoiling against a small number of tagged particles.} with different noise and clustering levels using \textit{Scikit-learn}'s \textit{make\_blobs} function, which provides \textit{true\_labels} for validation. These labels allow us to assess clustering accuracy by computing the true efficiency, $\varepsilon_t$, defined as the fraction of data points correctly classified by the algorithm.

\begin{figure}[h]
       \centering
\begin{subfigure}{0.49\textwidth}
  \centering
  \includegraphics[width=.99\linewidth]{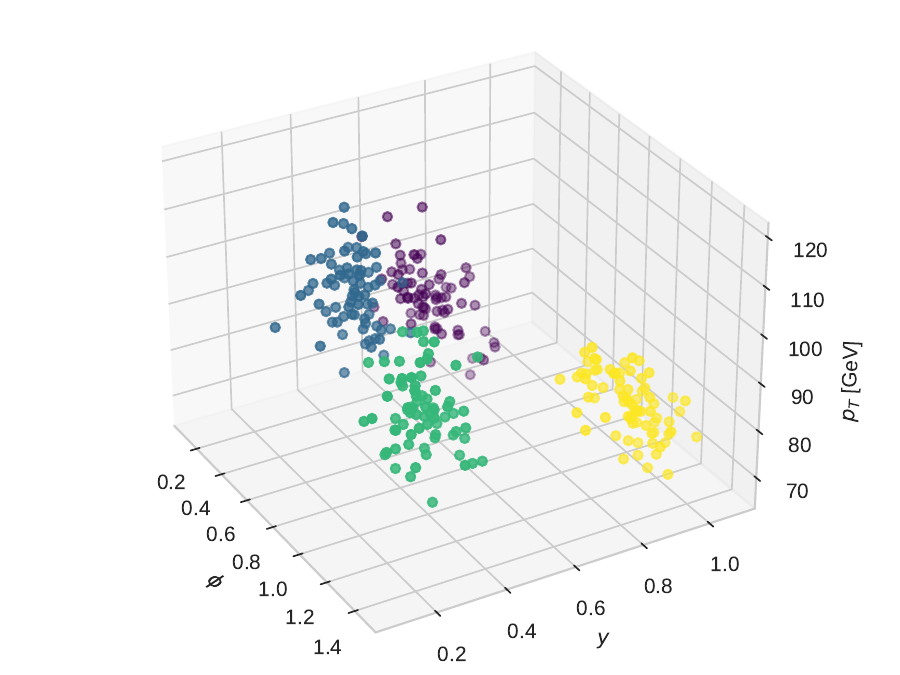}  
  \caption{ Classical \texttt{K-means} clustering, $\varepsilon_t=1.00$.}
  \label{fig:sub-first}
\end{subfigure}
\begin{subfigure}{0.49\textwidth}
  \centering
  \includegraphics[width=0.99\linewidth]{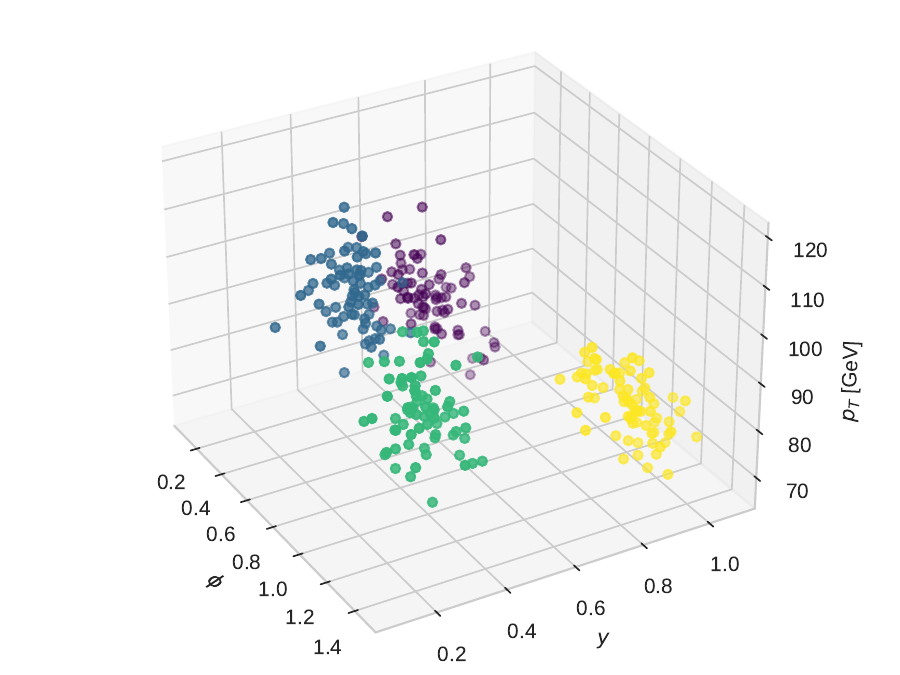}  
  \caption{ Quantum \texttt{K-means} clustering, $\varepsilon_t=1.00$.}
  \label{fig:sub-second}
\end{subfigure}
        \caption{Clusters identified after five iterations by the classical and quantum versions of the \texttt{K-means} algorithm are shown in different colors. The dataset consists of Gaussian-distributed points generated with a random seed and a standard deviation of 2.0 from the cluster centroids. Clustering was performed using a Minkowski-type distance, assuming all particles are massless and on-shell, with both algorithms achieving an efficiency of $\varepsilon_t = 1.00$. }
        \label{fig:qandckmeans}
\end{figure}

\begin{figure}[h]
       \centering
\begin{subfigure}{0.49\textwidth}
  \centering
  \includegraphics[width=.99\linewidth]{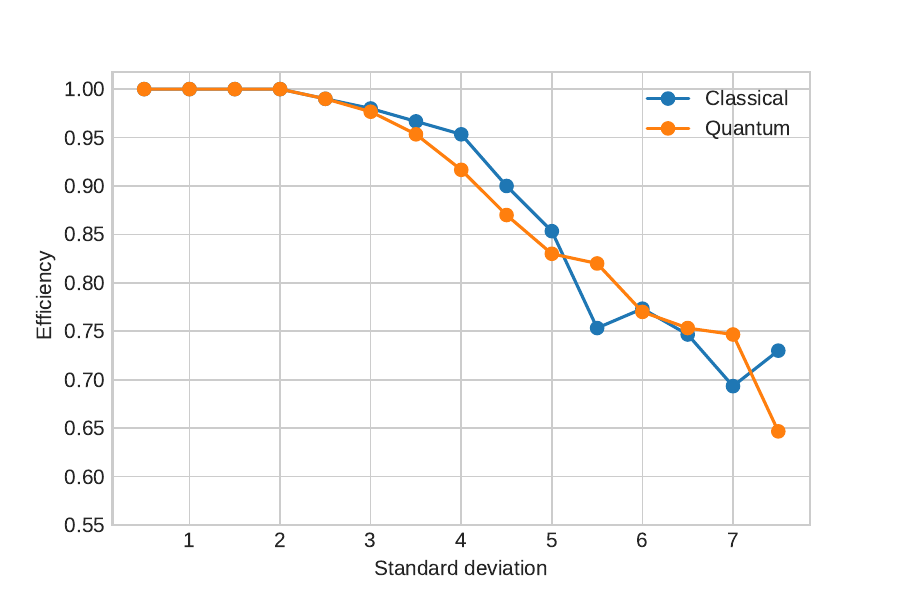}  
  \caption{Random seed.}
  \label{fig:efvsdeviation_sub-first}
\end{subfigure}
\begin{subfigure}{0.49\textwidth}
  \centering
  \includegraphics[width=0.99\linewidth]{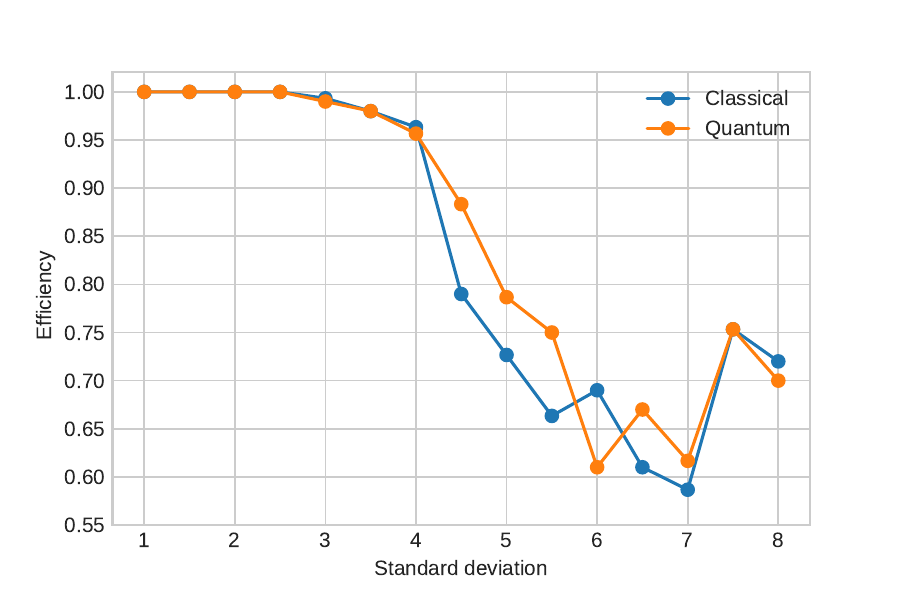}  
  \caption{\texttt{K-means}$++$ seed.}
  \label{fig:efvsdeviation_sub-second}
\end{subfigure}
        \caption{Cluster efficiency of the \texttt{K-means} algorithm as a function of the data's standard deviation from the centroids. Both the classical and quantum versions were tested on 15 datasets with standard deviations ranging from 0.5 to 7.5.}
        \label{fig:efvsdeviation}
\end{figure}

Both the hybrid and classical \texttt{K-means} algorithms were applied to each dataset. The analyzed data represent particle four-momenta, with the three-dimensional vectors corresponding to spatial components, while temporal components are inferred under the assumption that all particles are massless and on-shell. Results are shown in Figs.~\ref{fig:qandckmeans} and~\ref{fig:efvsdeviation}.

\begin{figure}[h]
       \centering
\begin{subfigure}{0.49\textwidth}
  \centering
  \includegraphics[width=0.99\linewidth]{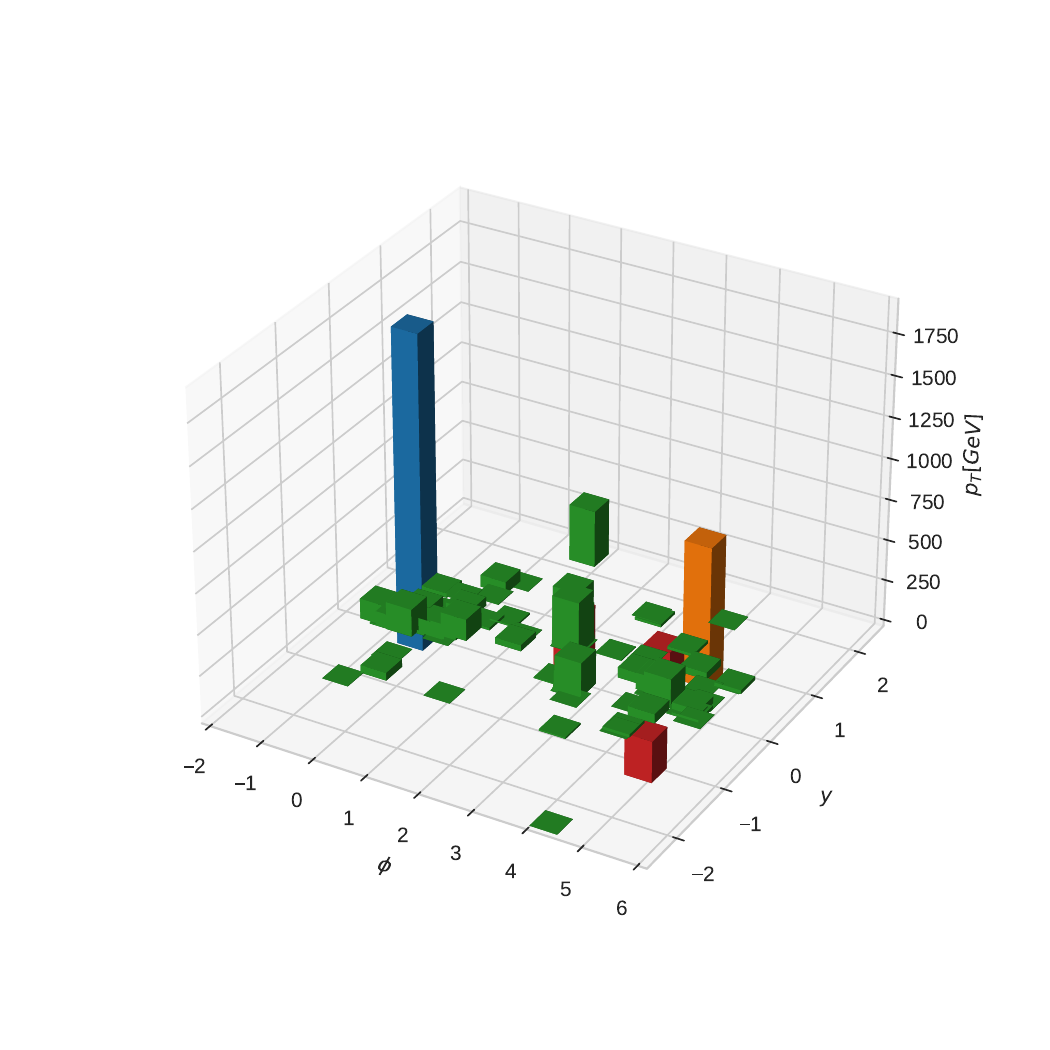}  
  \caption{\centering{Classical \texttt{K-means} applied to \hspace{\textwidth} LHC physical events.}}
  \label{fig:sub-first}
\end{subfigure}
\begin{subfigure}{0.49\textwidth}
  \centering
  \includegraphics[width=0.99\linewidth]{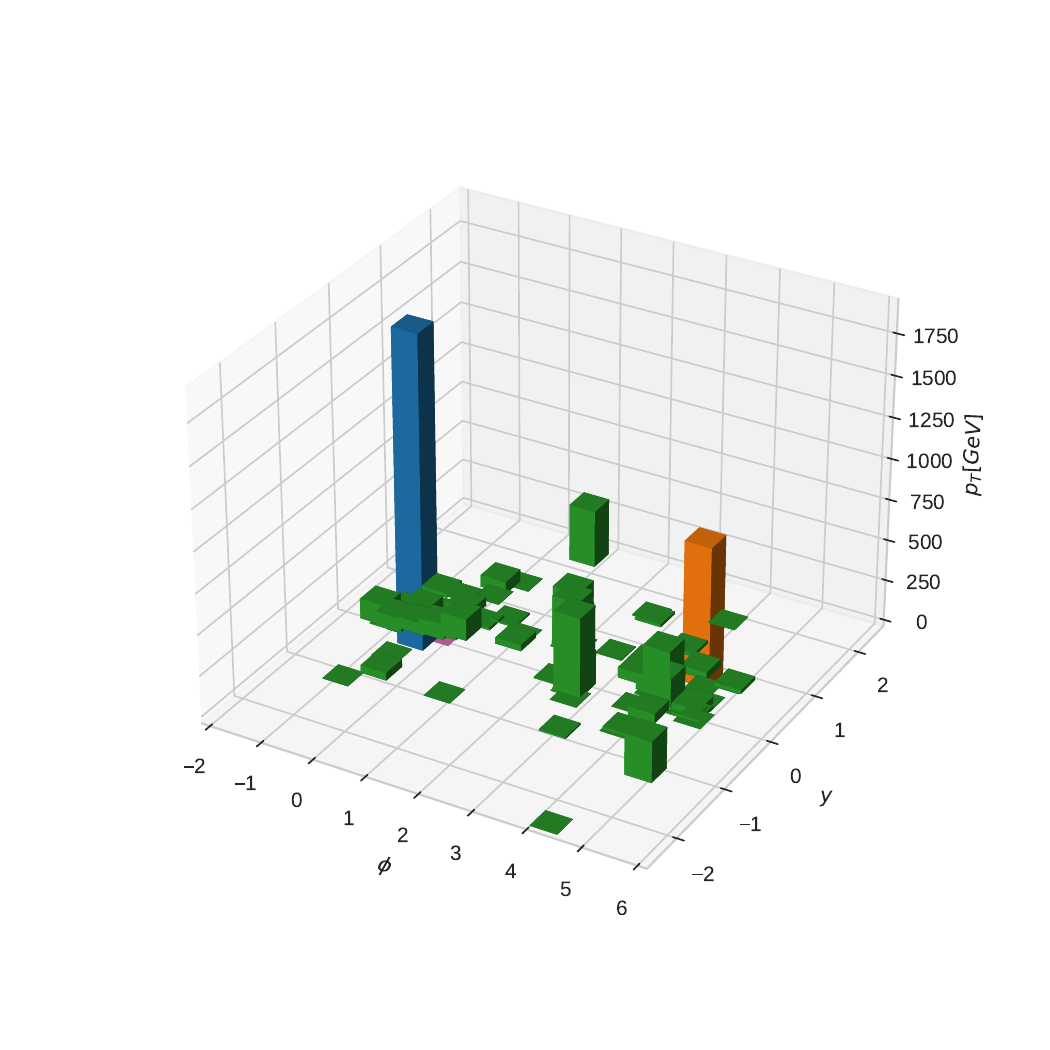}  
  \caption{\centering{Quantum \texttt{K-means} applied to \hspace{\textwidth} LHC physical events, $\varepsilon_c=0.94$.}}
  \label{fig:qkmeansrealdata_sub-second}
\end{subfigure}
        \caption{A sample parton-level event generated as described in the text and clustered with the classical and the quantum version of the \texttt{K-means++} algorithm, taking $K=8$.}
        \label{fig:qkmeansrealdata}
\end{figure}

Fig.~\ref{fig:qandckmeans} illustrates that both the classical and quantum versions of the \texttt{K-means} algorithm cluster data similarly in the three-dimensional space of transverse momentum~($p_T$), rapidity~($y$), and azimuthal angle~($\phi$). On the other hand, Fig.~\ref{fig:efvsdeviation} presents the clustering efficiency as a function of the standard deviation used to generate the data, assessing whether clustering occurs as expected. For small standard deviations, both algorithms perform well, achieving efficiencies close to one. However, as the standard deviation increases—introducing more noise—efficiencies decline for both approaches.

The performance of \texttt{K-means} is further studied by comparing two centroid initialization strategies: a random seed (Fig.~\ref{fig:efvsdeviation_sub-first}) and the \texttt{K-means}$++$ method (Fig.~\ref{fig:efvsdeviation_sub-second}), which selects initial centroids to be as far as possible. The random seed approach presents a linear efficiency decrease with increasing standard deviation, with classical and quantum versions performing similarly. In contrast, the \texttt{K-means}$++$ variant shows distinct behavior: from a standard deviation of 4 onward, the quantum version generally outperforms the classical one. In this case, both algorithms experience a drop in efficiency from a standard deviation of 4 to 7, followed by a slight recovery from 7 to 8. Comparing both methods, \texttt{K-means}$++$ achieves higher efficiencies for low standard deviations ($<4$), whereas the random seed approach performs better for larger standard deviations.

We now apply our quantum \texttt{K-means} method to LHC physical events. A key challenge arises from the fact that a negative vector $-\vec{x}$ represents the same quantum state $|x\rangle$ as its positive counterpart $\vec{x}$, differing only by a global phase. To address this, we preprocess the data by rescaling each component of every data point to the range \{1,10\}, ensuring all values remain positive\footnote{The value 0 is excluded to prevent numerical and statistical fluctuations}. Unlike the previous analysis, LHC event data lack \textit{true\_labels}, making it impossible to compute $\varepsilon_t$. Instead, we introduce the efficiency metric $\varepsilon_c$, defined as the quotient of the number of particles clustered in the same way as their classical counterpart and the total number of particles to be classified.

We simulate an $n$-particle event at the LHC using a custom phase-space event generator implemented in \texttt{C++} and based on \texttt{ROOT} \cite{Brun:1997pa}. This generator can handle $n$-particle events with $n$ reaching tens of thousands, allowing for a flexible choice of final-state configurations, including massless QCD partons, QCD partons with photons, massive vector bosons, or top quarks.
The accuracy of each generated event is validated by enforcing kinematical constraints between the initial state and the $n$-particle final state. The required precision\footnote{The kinematical constraint is evaluated over the three-momentum space vector, assuming all momenta are outgoing. The test is performed at the highest multiplicity, ensuring that the precision is at least $10^{-2}$, with further improvements as the number of final-state particles decreases.} is consistently better than $10^{-2}$.

We study the production of $n$-particle massless final states in proton-proton~\footnote{Since we consider unweighted events, our study is also applicable to $e^{+}e^{-}$ colliders.} collisions at a center-of-mass energy of $\sqrt{s} = 14$ TeV. The final-state selection follows these criteria: jets are identified using the $k_T$-jet algorithm with a minimum transverse momentum of $p_{T, \text{min}} \geq 10$ GeV and a radius of $R = 1$. We focus on events with $n = 128$ massless particles in the final state.

The application of the quantum \texttt{K-means++} algorithm to LHC events is shown in Fig.~\ref{fig:qkmeansrealdata}. Despite setting $K=8$ clusters initially, Fig.~\ref{fig:qkmeansrealdata} reveals that the algorithm consistently identifies only 3 or 4 distinct clusters (jets). This behavior arises because, although the algorithm starts with $K$ centroids, it converges to a local minimum where some clusters remain empty.
Fig.~\ref{fig:qkmeansrealdata} also illustrates that both the classical and quantum algorithms classify the data similarly, with the quantum algorithm achieving an efficiency close to one. These results suggest that the quantum approach performs successfully when applied to physical LHC data.

\subsection{Quantum Affinity Propagation algorithm}
\label{subsec:qapalgorithm}

The quantum (hybrid) Affinity Propagation (\texttt{AP}) algorithm~\cite{Frey2007ClusteringBP} uses the invariant sum squared as a metric in the similarity matrix and calculates it through a quantum subroutine, as the quantum \texttt{K-means} algorithm just described. Then, a speedup would be achieved, since computing the distances only requires $\mathcal{O}(\log(d-1))$ qubits. So, the quantum \texttt{AP} algorithm, which is as far as we know completely original, would have a time complexity of $\mathcal{O}(N^2T\log(d-1))$ instead of the classical $\mathcal{O}(N^2Td)$.

We present a simulation of the quantum \texttt{AP} algorithm. First, we apply the algorithm to Gaussian datasets with varying numbers of clusters, generated using a standard deviation of 0.6. This particular value for the standard deviation was chosen for convenience. The resulting efficiencies for both the classical and quantum versions are shown in Table~\ref{tab:qapclusters}. Table~\ref{tab:qapclusters} illustrates that both the classical \texttt{AP} algorithm and its quantum counterpart successfully clustered the low-noise Gaussian datasets.

\begin{table}[th!]
\begin{longtable}{| p{3cm} | p{3cm} | p{3cm}|}
\hline
\centering {Number of clusters $K$}& \centering{ Efficiency classical \texttt{AP}} ($\varepsilon_t)$ & \centering {Efficiency quantum \texttt{AP} ($\varepsilon_t)$} \cr   \hline 
 \centering 4 &\centering 1.00 &\centering 0.99 \cr   \hline
\centering 5 &\centering 1.00 &\centering 1.00 \cr   \hline
\centering 6 &\centering 0.99 &\centering 0.98 \cr   \hline
\centering 7 &\centering 1.00 &\centering 0.98  \cr   \hline
\centering 8 &\centering 0.98 &\centering 0.94  \cr    \hline 
\omit
    \\    
\caption{Efficiencies of \texttt{AP} algorithms for Gaussian datasets with different number of clusters.}
\label{tab:qapclusters}
\end{longtable}
\end{table}

\begin{figure}[ht!]
       \centering
\begin{subfigure}{.49\textwidth}
  \centering
  \includegraphics[width=.99\linewidth]{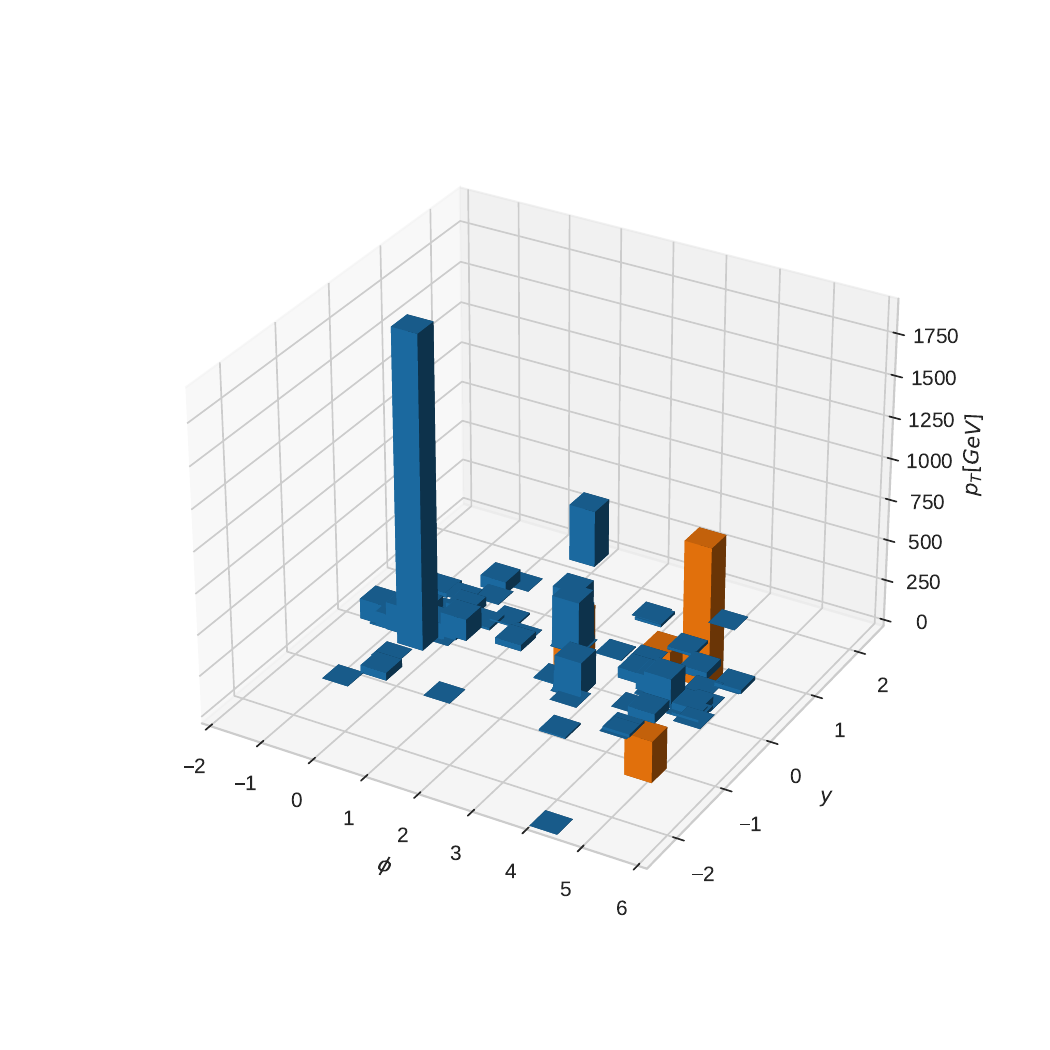}  
  \caption{\centering Classical \texttt{AP} algorithm applied to \hspace{\textwidth} LHC physical events.}
  \label{fig:applots_sub-first}
\end{subfigure}
\begin{subfigure}{.49\textwidth}
  \centering
  \includegraphics[width=.99\linewidth]{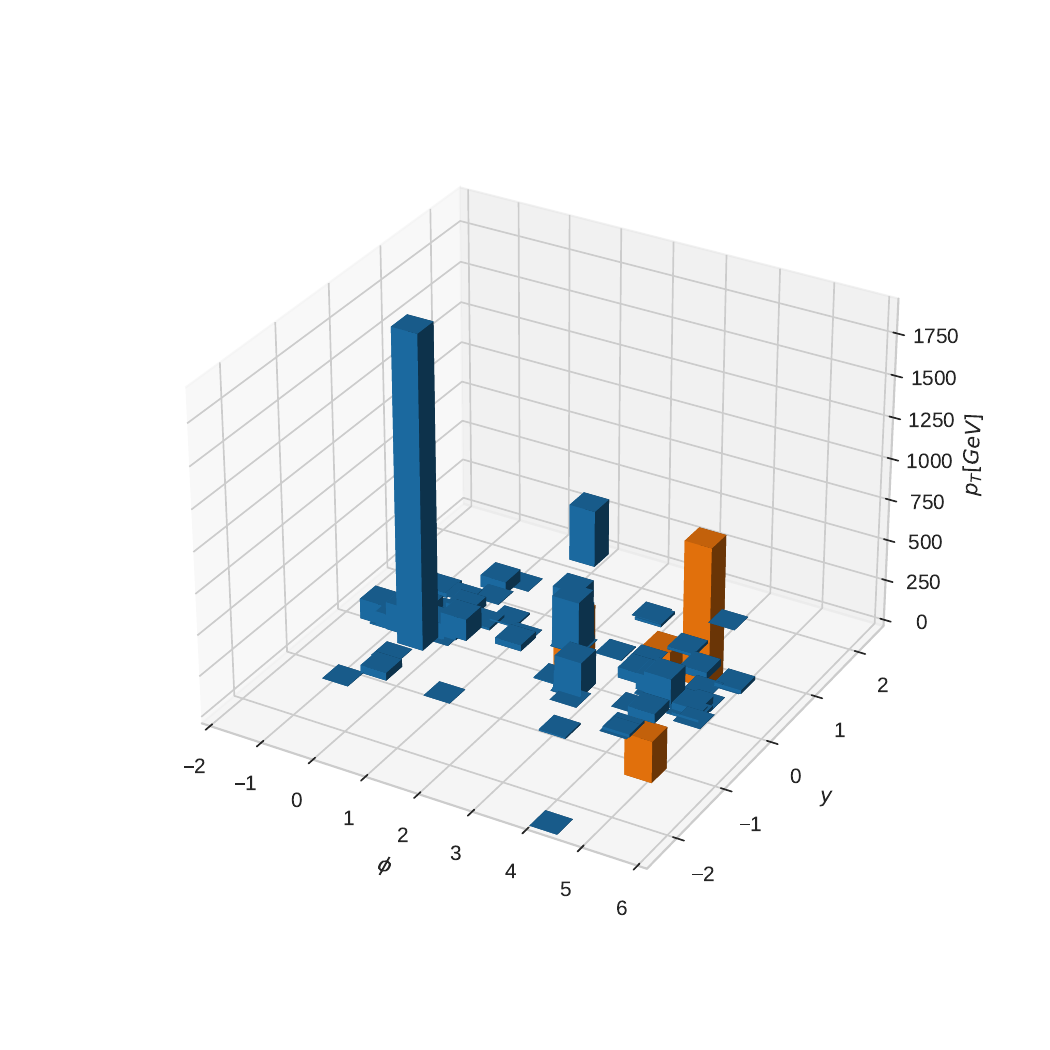}  \caption{\centering Quantum \texttt{AP} algorithm applied to \hspace{\textwidth} LHC physical events, $\varepsilon_c=1.00$.}
  \label{fig:applots_sub-second}
\end{subfigure}
        \caption{A sample parton-level event, generated as described in the text, is clustered into $K=2$ distinct clusters using both the classical and quantum versions of the \texttt{AP} algorithm. }
        \label{fig:applots}
\end{figure}

Now, we apply the quantum \texttt{AP} algorithm to the physical dataset described in Section~\ref{subsec:qkmeans}, which has been preprocessed as explained there. The results are presented in Fig.~\ref{fig:applots}. In Fig.~\ref{fig:applots_sub-second}, the same clustering is performed as in Fig.~\ref{fig:applots_sub-first} (note that the efficiency of the quantum version is $\varepsilon_c=1.00$). Although both algorithms show similar results, the \texttt{AP} algorithm identifies only 2 clusters, whereas the \texttt{K-means} algorithm finds 3 or 4 clusters (see Fig.~\ref{fig:qkmeansrealdata}). Nevertheless, both algorithms correctly identify the most energetic jets of the event (the blue and orange clusters), while the majority of the remaining particles are not classified in the same way, probably because they are soft particles.

\subsection{Quantum $k_T$ jet algorithm}
\label{subsec:qktalgorithm}

The quantum maximum search algorithm discussed in Section~\ref{subsec:qsearching} can be effectively applied to the $k_T$-jet algorithm. Even if the quantum algorithm does not select the absolute maximum during one of the multiple iterations, it is unlikely to impact the overall jet clustering process. This is because QCD predicts that secondary radiation is emitted mostly collinear to the parent parton. Therefore, an small deviation in finding the true maximum would alter the order in which two particles are merged, but in many cases, the final result remains unaffected by this permutation.

It is also worth noting that the quantum version of the $k_T$-jet algorithm would have a complexity of $\mathcal{O}(N^2\log(N))$. This is because computing all the distances requires $\mathcal{O}(N^2)$, while finding the minimum takes $\mathcal{O}(\log(N))$, compared to the classical version, which has a complexity of $\mathcal{O}(N^3)$ \cite{2006}. Additionally, the quantum minimum search could be integrated into the \texttt{FastJet}~\cite{Cacciari:2008gp} algorithm, which has a complexity of ${\cal O}(N^2)$. In this case, the resulting quantum algorithm would have a complexity of $\mathcal{O}(N\log(N))$, matching the efficiency of the \texttt{FastJet} algorithm with Voronoi diagrams, the most efficient clustering algorithm known to date.

\begin{figure}[H]
\begin{subfigure}{.5\textwidth}
  \centering
  \includegraphics[width=.8\linewidth]{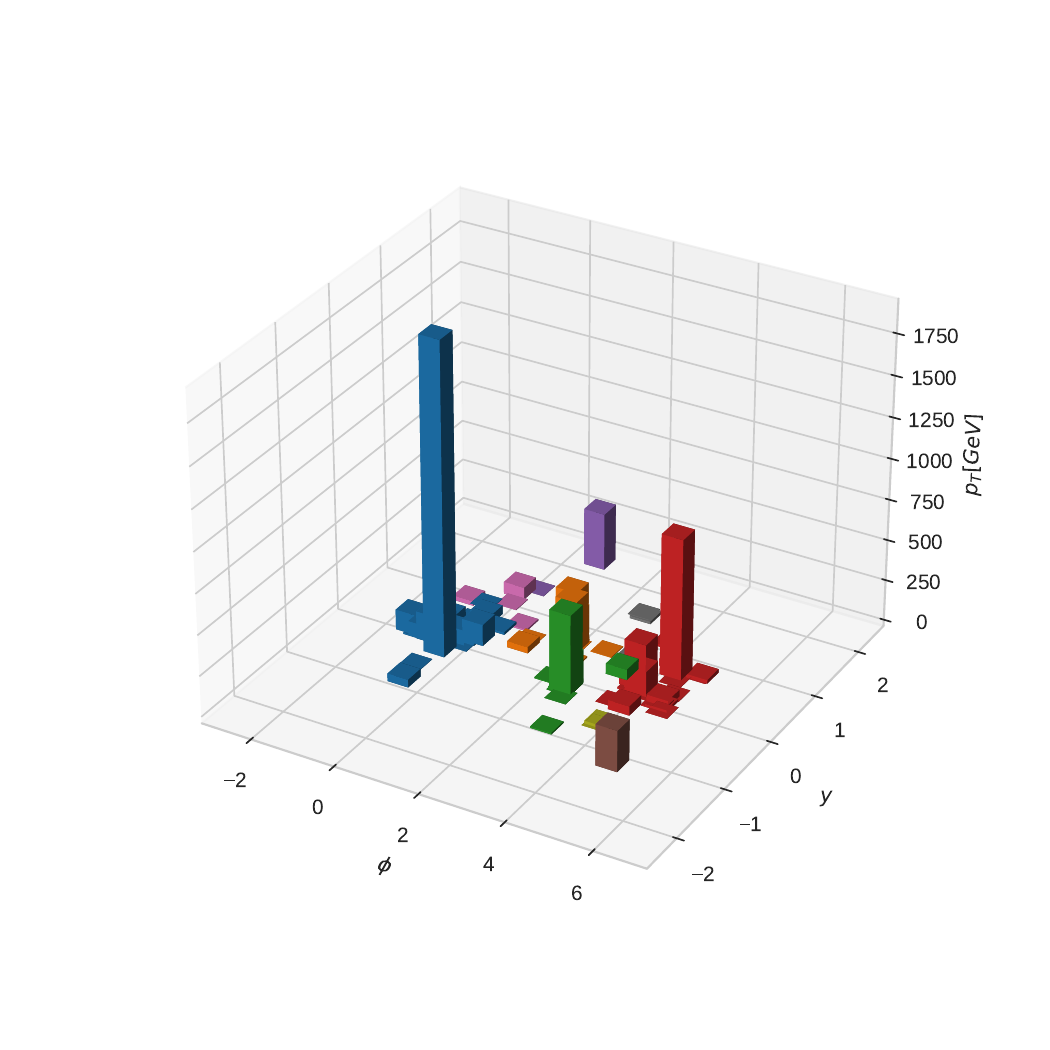}  
  \caption{Classical anti-$k_T$, $p=-1$, $R=1$.}
  \label{fig:sub-first}
\end{subfigure}
\begin{subfigure}{.5\textwidth}
  \centering
  \includegraphics[width=.8\linewidth]{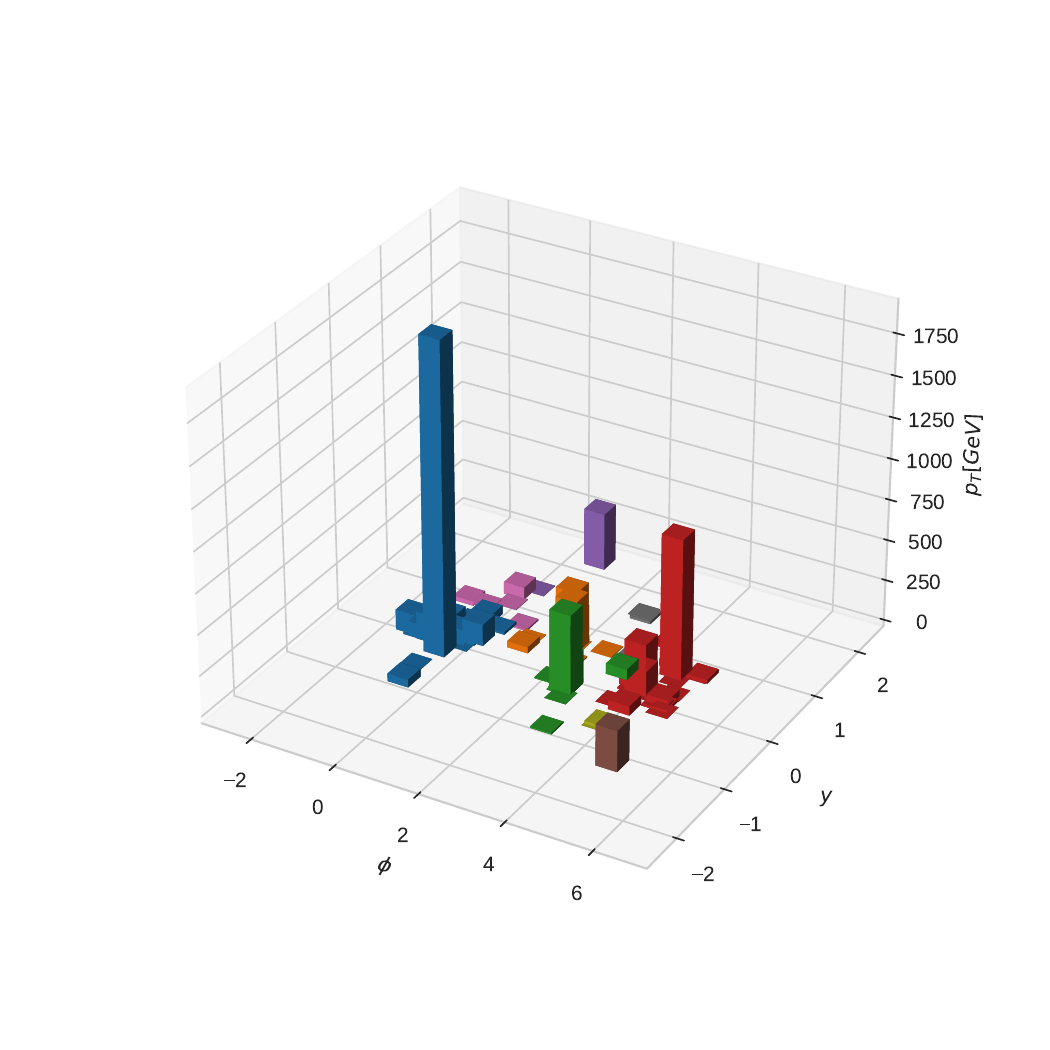}  
  \caption{Quantum anti-$k_T$, $p=-1$, $R=1$, $\epsilon_c =0.99$.}
  \label{fig:sub-second}
\end{subfigure}
\newline
\begin{subfigure}{.5\textwidth}
  \centering
  \includegraphics[width=.8\linewidth]{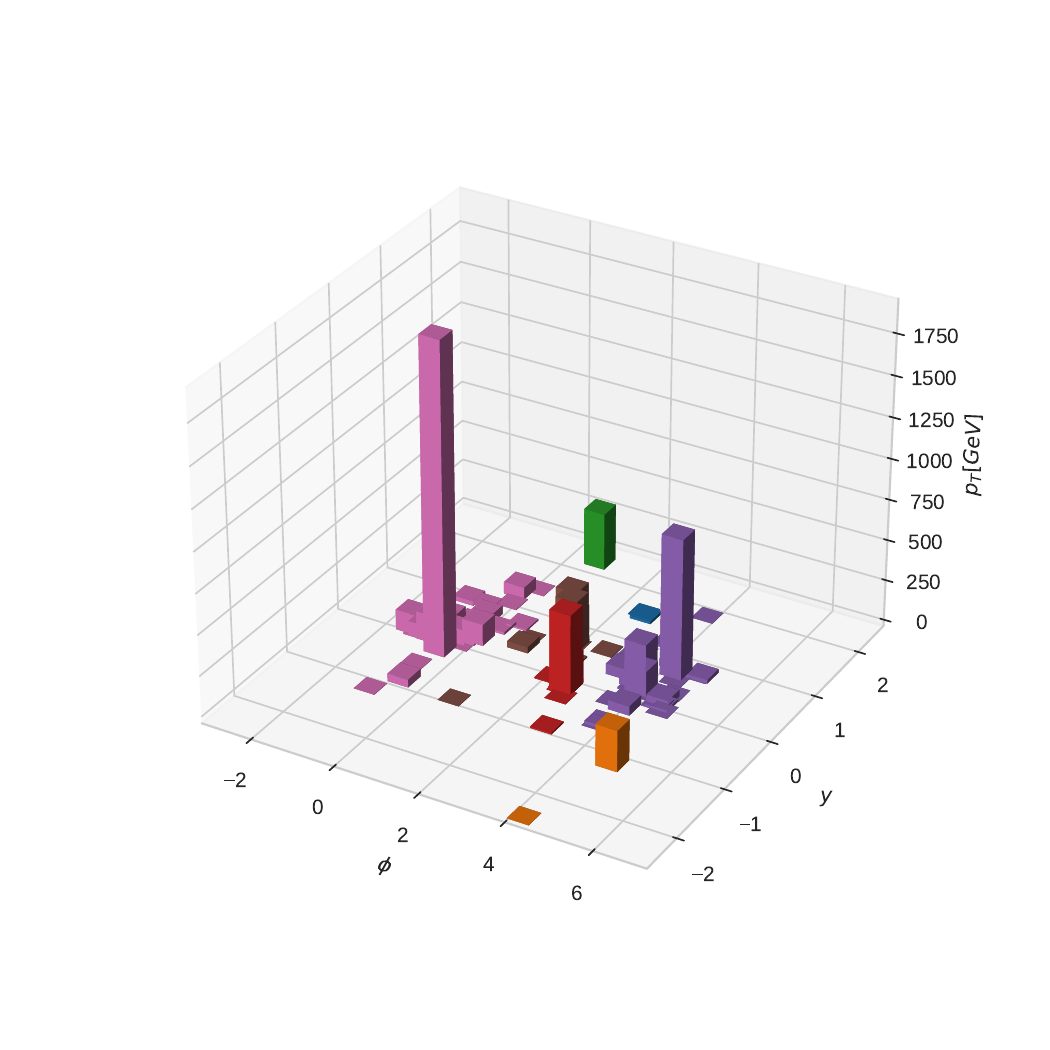}  
  \caption{Classical $k_T$, $p=1$, $R=1$.}
  \label{fig:sub-third}
\end{subfigure}
\begin{subfigure}{.5\textwidth}
  \centering
  \includegraphics[width=.8\linewidth]{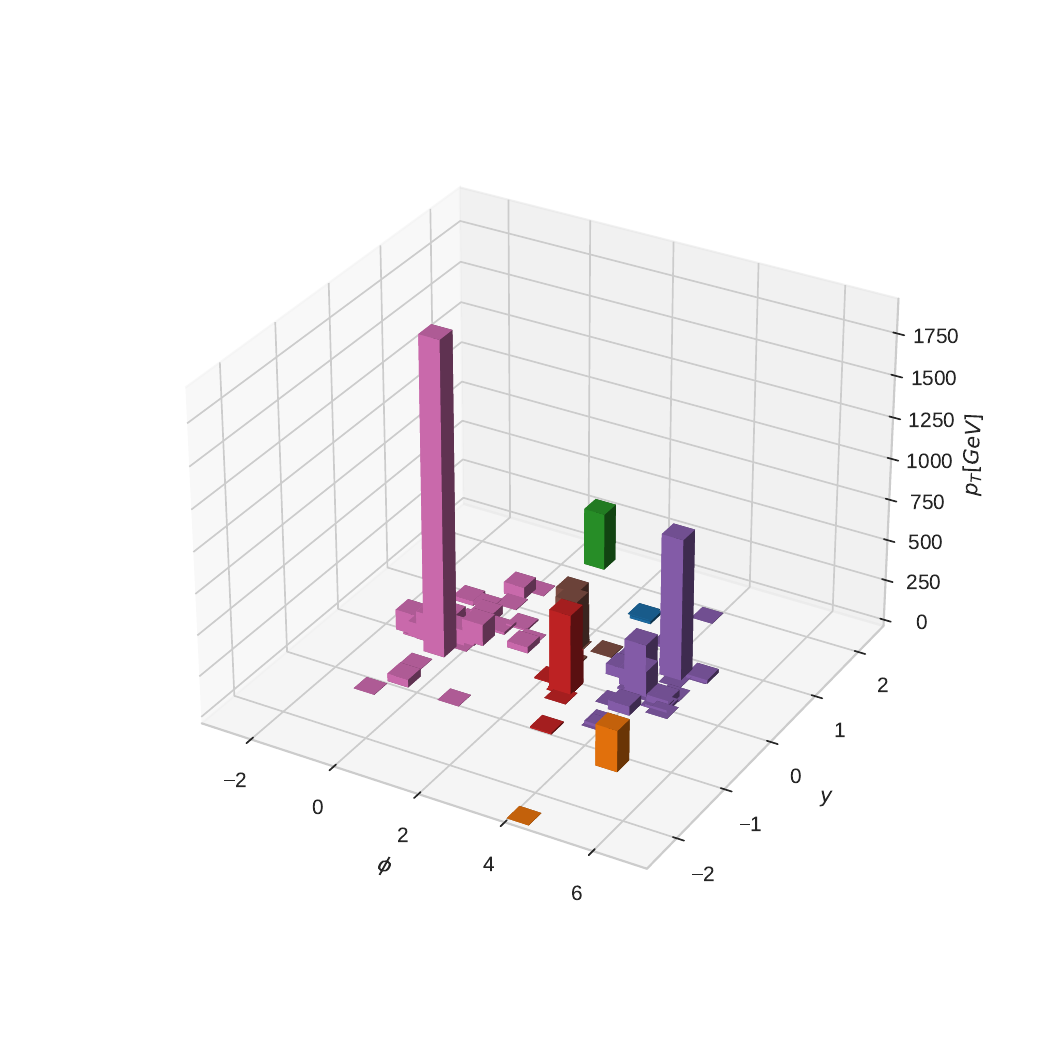}  
  \caption{Quantum $k_T$, $p=1$, $R=1$, $\epsilon_c =0.98$.}
  \label{fig:sub-fourth}
\end{subfigure}
\newline
\begin{subfigure}{.5\textwidth}
  \centering
  \includegraphics[width=.8\linewidth]{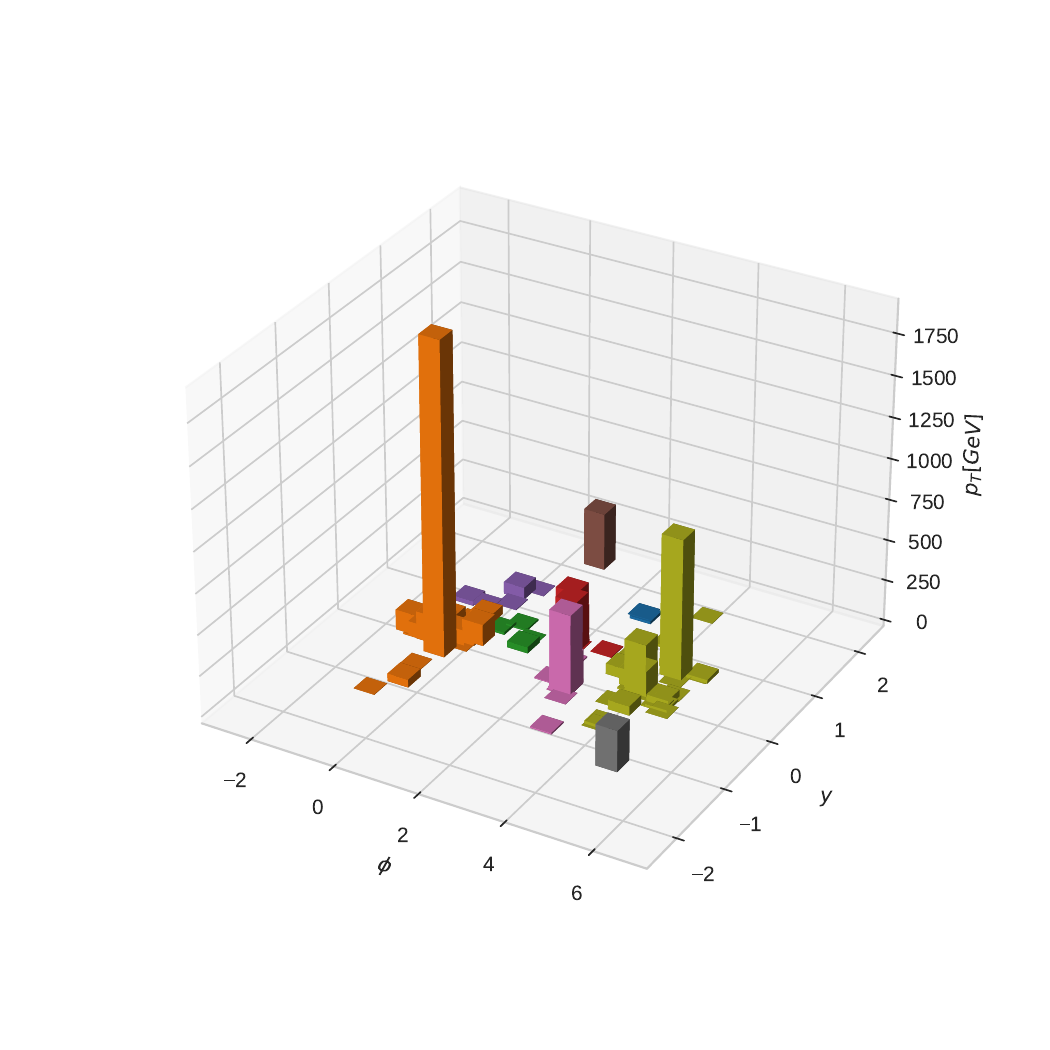}  
  \caption{Classical Cam/Aachen, $p=0$, $R=1$.}
  \label{fig:sub-fifth}
\end{subfigure}
\begin{subfigure}{.5\textwidth}
  \centering
  \includegraphics[width=.8\linewidth]{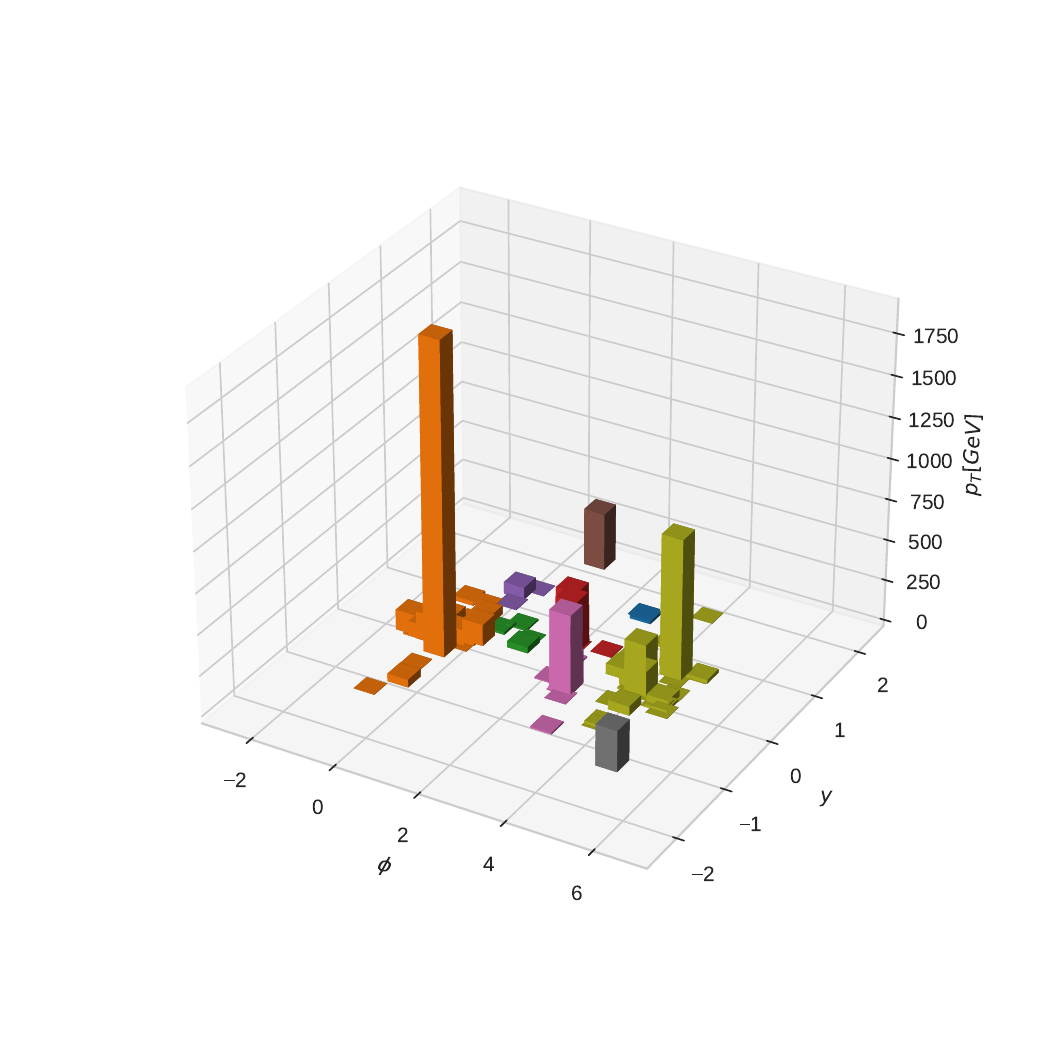}  
  \caption{Quantum Cam/Aachen, $p=0$, $R=1$, $\epsilon_c =0.98$.}
  \label{fig:sub-sixth}
\end{subfigure}
\caption{A sample parton-level event generated as described in the text and clustered with three different $k_T$-jets algorithms as well as its quantum versions.} 
\label{fig:ktclassicandquantum}
\end{figure}

We apply the quantum version of \texttt{FastJet} with the LHC physical datasets of the two previous sections. Figure~\ref{fig:ktclassicandquantum} shows the performance of both classical and quantum $k_T$ jet algorithms, illustrating the jet clustering process for each of the $k_T$ algorithm variants: anti-$k_T$, $k_T$, and Cambridge/Aachen. Both the classical and quantum versions yield identical jet clusterings.

When comparing Figs.~\ref{fig:qkmeansrealdata}, \ref{fig:applots}, and \ref{fig:ktclassicandquantum}, we observe that the $k_T$ algorithm performs a cleaner clustering with a larger number of jets. This difference is primarily due to the fact that jet clustering is represented graphically in three dimensions, which aligns with the dimensionality of the $k_T$ metric. In contrast, the \texttt{K-means} and \texttt{AP} algorithms use a four-dimensional Minkowski distance, which leads to a different clustering representation.

Furthermore, we analyze the efficiencies and the number of \textit{shots} required for all quantum versions as a function of the $a$ parameter. These results are presented in Table~\ref{tab:effsandshots}. Table~\ref{tab:effsandshots} demonstrates that the efficiencies of the quantum algorithms are close to one, meaning they classify particles nearly identically to their classical counterparts. Additionally, it can be observed that as the parameter $a$ increases, the number of \textit{shots} required to achieve a successful efficiency decreases. Specifically, by setting $a$ to 5, we attain the desired efficiencies with at most 10 \textit{shots}. For other problems involving larger datasets, using a higher value of $a$ can further separate the data points and maximize efficiency while minimizing the number of \textit{shots} needed.
\begin{table}[h]
\begin{longtable}{| p{0.5cm} | p{1.7cm} | p{1.7cm} | p{1.7cm} | p{1.7cm} | p{2.3cm} | p{2.3cm} |}
\hline
\centering {$a$}& \centering{ Efficiency anti-$k_T$} & \centering {\textit{Shots} anti-$k_T$} & \centering {Efficiency $k_T$} & \centering{\textit{Shots} $k_T$} & \centering {Efficiency Cam/Aachen} &  \centering{  \textit{Shots} Cam/Aachen} \cr   \hline 
 \centering 1 &\centering 0.96 &\centering 50 &\centering 0.98 &\centering 50 &\centering 0.96 &\centering 70\cr   \hline
\centering 2 &\centering 0.99 &\centering 40 &\centering 0.99 &\centering 45 &\centering 0.98 &\centering 60 \cr   \hline
\centering 3 &\centering 1.00 &\centering 25 &\centering 0.98 &\centering 20 &\centering 0.97 &\centering 40\cr   \hline
\centering 4 &\centering 1.00 &\centering 15 &\centering 0.95 &\centering 15 &\centering 1.00 &\centering 20 \cr   \hline
\centering 5 &\centering 0.99 &\centering 5 &\centering 1.00 &\centering 8 &\centering 0.98 & \centering 10 \cr    \hline 
\omit
    \\    
\caption{Efficiencies and number of \textit{shots} of the quantum $k_T$-jet algorithms as a function of the parameter $a$.}
\label{tab:effsandshots}
\end{longtable}
\end{table}

A summary of the time complexity of all the jet clusterings we consider alongside their quantum versions is presented in Table \ref{tab:qclusters_complexity}.

\begin{longtable}{|  p{3.7cm} | p{2.7cm}| p{2.7cm} | p{4.2cm} | }
\hline
\centering{Jet clustering algorithm}  & \centering{ Quantum subroutine} & \centering{ Classical version}& \centering {Quantum \\ version}     \cr   \hline 
 \centering \texttt{K-means}  & \centering{Both}&\centering $\mathcal{O}(NKd)  $&\centering $\mathcal{O}(N\log K\log(d-1)) $  \cr   \hline
 \centering \texttt{AP}  & \centering{ Distance} &\centering $\mathcal{O}(N^2Td) $&\centering $ \mathcal{O}(N^2T\log(d-1)) $  \cr   \hline
 \centering anti-$k_T$ \texttt{Jet} & \centering{ Maximum}&\centering$\mathcal{O}(N^2) $&\centering $\mathcal{O}(N\log N) $  \cr   \hline
 \centering anti-$k_T$ \texttt{FastJet} & \centering{ Maximum}&\centering $\mathcal{O}(N\log N) $&\centering $\mathcal{O}(N\log N) $  \cr   \hline 
 \omit
    \\  
   \caption{Time complexity of quantum and classical jet clustering algorithms when one or both subroutines are applied.}
\label{tab:qclusters_complexity}
\end{longtable}

\section{Conclusions}\label{sec:qjets_conclusions}
In this chapter, we have developed quantum versions of three well-known clustering algorithms~\cite{deLejarza:2022bwc,deLejarza:2022vhe}: \texttt{K-means}, Affinity Propagation, and $k_T$-jet clustering.
These quantum algorithms are built upon two novel quantum techniques. The first is a quantum subroutine designed to compute distances that satisfy the Minkowski metric, and the second is a quantum circuit used to track the maximum value from a list of unsorted data.

For the \texttt{K-means} algorithm, the quantum version builds upon the classical algorithm by incorporating a quantum procedure to compute Minkowski distances and a quantum circuit for assigning each particle to the closest centroid. We found that the quantum \texttt{K-means} algorithm achieves clustering efficiency on par with its classical counterpart, while also offering an exponential speed-up in computational time, particularly as the vector dimensionality $d$ and the number of clusters $K$ increase, when executed on a quantum device with qRAM.

Next, we examined the quantum version of the Affinity Propagation algorithm, an unsupervised machine learning method. Similar to the \texttt{K-means} case, the quantum procedure computes the similarity using the same quantum technique for Minkowski distance. This results in an exponential speed-up over the classical version in terms of the vector dimensionality $d$, while preserving clustering efficiency.

Finally, we presented the quantum adaptations of the $k_T$-jet clustering algorithms. On a true universal quantum device, these algorithms would demonstrate an exponential speed-up in finding the minimum distance. Specifically, while the classical algorithm requires $\mathcal{O}(N^3)$ computational time, where $N$ is the number of particles, the quantum version would only require $\mathcal{O}(N^2\log(N))$. It is important to note that this comparison is made between the classical \textit{non-optimal} and unoptimized version and its quantum counterpart. Further improvements can be made by applying a geometrical nearest-neighbor optimization procedure to the quantum algorithm, similar to that used in \texttt{FastJet}. This would result in a quantum version with a complexity of $\mathcal{O}(N\log(N))$, matching the fully optimized \texttt{FastJet} version.

In summary, the quantum simulations presented in this chapter demonstrate excellent performance and clustering efficiencies across all considered algorithms in comparison with their classical counterparts, while providing theoretical speedups when executed on quantum devices.
\def\ii{\imath 0}
\chapter{Quantum integration}\label{chap:qint}

\section{Introduction}\label{app:intro_qint}

Quantum Field Theories (QFT) have emerged as highly successful for understanding the behavior of nature at sub-atomic scales.
This achievement necessarily requires the development of robust computational methods to extract reliable predictions from QFTs.
The perturbative approach has become a cornerstone technique in high-energy particle physics, since higher-order contributions within the perturbative expansion lead systematically to more precise and accurate theoretical predictions.  
However, the computation of higher-order contributions in perturbative QFT is not straightforward. The primary challenge lies in the virtual quantum fluctuations, which involve dealing with multiloop-multileg Feynman integrals. Loop Feynman integrals are an essential component in the calculation of quantum contributions to scattering and decay processes of elementary particles at high-energy colliders~\cite{Heinrich:2020ybq}. Their accurate evaluation is crucial for making precise theoretical predictions for experimental measurements. The difficulty of computing these integrals stems from their multi-dimensional nature, the dependence on multiple scales, and the emergence of ultraviolet (UV), infrared (IR) and threshold singularities, which often requires the use of regularization and re\-nor\-ma\-li\-za\-tion techniques. Traditional methods can be computationally demanding, making it a challenging task.

The standard approach to computing multiloop scattering matrix elements starts with the corresponding Feynman diagrams and their associated scattering amplitudes. The corresponding Feynman integrals appearing in the expression are written in terms of bases of master integrals (MIs) using integration-by-parts identities (IBPs) relations~\cite{Chetyrkin:1979bj,Chetyrkin:1981qh}. These MIs can be obtained analytically or numerically. One of the most widely used methods for calculating MIs is the so-called differential equation method \cite{Kotikov:1990kg,Remiddi:1997ny}. The success in completing the analytical formulae is based, roughly, on our knowledge of the \textit{Multiple Polylogarith\-mic} functions (MPLs)~\cite{Goncharov:2001iea}. More difficult cases involving elliptic functions \cite{Adams:2017tga} are challenging. Recently, much effort has been devoted to understanding the analytic structure of Feynman integrals that do not admit expressions in terms of MPLs~\cite{Broedel:2014vla,Broedel:2017kkb,Adams:2014vja,Frellesvig:2023iwr,Gorges:2023zgv,Duhr:2022pch,Duhr:2022dxb,Bourjaily:2022bwx}. However, our understanding of the analytic structure of Feynman integrals beyond the MPLs class is not the same as for the MPLs case. Even when an analytic solution is available, the numerical evaluation of the functions associated with such a solution can be extremely challenging for phenomenological applications~\cite{Chawdhry:2019bji,Czakon:2021mjy,Abreu:2022vei,Abreu:2022cco,Badger:2023mgf}.
To overcome the difficulties described above, the generalized power series method~\cite{Lee:2017qql,Mandal:2018cdj,Moriello:2019yhu} has recently been developed to solve the system of differential equations associated with the MIs. This technique has attracted much interest due to its wide range of applicability, and it has been successfully exploited in several phenomenological applications~\cite{Becchetti:2020wof,Bonciani:2021zzf,Armadillo:2022bgm,Becchetti:tab}.

Both approaches to multiloop scattering amplitudes (analytical or numerical/semi-analytical) produce results that are very slow to evaluate under phase-space integration. In the case of the two-loop scattering amplitude for three-photon production~\cite{Chawdhry:2019bji} the CPU cost is 29 years in a single core. Another example is the case of three-jet production~\cite{Czakon:2021mjy}, the time needed to complete the calculation at the second order in strong coupling is 115 CPU years\footnote{Due to the high gluon multiplicity in the final state (up to 5 gluons), the most time-consuming contribution in this case is the double real emission.}. 
The preceding processes represent the current two-loop technical frontier, which refers to $2 \rightarrow 3$ particle collisions. It is important to note that all precedent CPU costs represent the time required for a single phenomenological study at the LHC. Each of them requires multi-cluster computers running continuously for several months (or even years, as in the case of three-jet production~\cite{Czakon:2021mjy}). It is clear that our current state-of-the-art methodologies will not be able to cope with the huge demand for phenomenological studies that the LHC will require in the current Run~3 and future phases.

This evidences the need for computing Feynman integrals more efficiently, surpassing the current limitations imposed by hardware capabilities. In this regard, there is a growing interest in the development of innovative strategies based on the new paradigm brought by quantum computing with the potential to tackle traditionally challenging problems across various domains. The potential speedup offered by quantum computers has sparked numerous ideas, such as emplo\-ying Grover's algorithm for efficient database querying~\cite{Grover:1997fa}, Shor's algorithm for factorization of large integers~\cite{shor}, or quantum annealing for Hamiltonian minimization~\cite{qannealing}. In the realm of particle physics, quantum algorithms have found applications in different areas, such as solving problems associated with lattice gauge theories~\cite{Zohar_2016,preskill}.
Furthermore, quantum algorithms have been applied to various tasks in the context of high-energy colliders \cite{Delgado:2022tpc}. These include jet identification and clustering~\cite{thaler,delgado_jets,deLejarza:2022bwc,deLejarza:2022vhe}, jet properties in a medium~\cite{Barata:2023clv,Barata:2022wim,Barata:2021yri}, determining parton densities (PDFs)~\cite{carrazza}, simulating parton showers~\cite{Bepari:2021kwv}, detecting anomalies~\cite{Belis:2023atb,Schuhmacher:2023pro, Bermot:2023kvh}, unveiling the causal structure of multiloop Feynman diagrams~\cite{Ramirez-Uribe:2021ubp,Clemente:2022nll,Ramirez-Uribe:2024wua,Ochoa-Oregon:2025opz}, integrating elementary particle processes at tree level~\cite{deLejarza:2023IEEE,AGLIARDI2022137228}, and optimizing existing particle accelerator beam lines~\cite{Schenk:2022pgo}. The list of applications continues to expand rapidly as quantum algorithms demonstrate their suitability for a range of purposes. A very recent status of quantum computations in high-energy physics, split into theoretical and experimental target benchmark applications that can be addressed in the near future, can be found in~\cite{DiMeglio:2023nsa}.

In view of these recent successes, the aim of the work presented in this Chapter is to investigate the potential of quantum algorithms for the efficient computation of loop Feynman integrals~\cite{,deLejarza:2024scm,deLejarza:2024pgk}. 
In particular, we apply a novel quantum Monte Carlo integration algorithm called Quantum Fourier Iterative Amplitude Estimation~(QFIAE)~\cite{deLejarza:2023IEEE} to numerically evaluate several benchmark one-loop Feynman integrals and decay rates at second order in perturbation theory in the Loop-Tree Duality~(LTD)~\cite{Aguilera-Verdugo:2020set,Catani:2008xa,Bierenbaum:2010cy,Bierenbaum:2012th}. We use the LTD representation of the loop integrals because of its several advantages compared to other more conventional representations, as will be explained below.

The underlying basis of QFIAE are other well-known quantum algorithms, upon which QFIAE introduces a suitable solution to successfully solve some of the bottlenecks arising from these methods.
Quantum Amplitude Estimation~(QAE)~\cite{Brassard_2002} is a quantum algorithm that estimates the amplitude value of a quantum state. It uses amplitude amplification, a generalization of Grover's searching algorithm~\cite{Grover:1997fa}, to enhance the likelihood of measuring the desired state over the non-desired state. The QAE method finds practical applications in various domains, including our field of interest, i.e. numerical integration~\cite{Montanaro2015, Pooja}. However, the dependence of the original QAE algorithm on the resource-intensive Quantum Phase Estimation~(QPE) subroutine~\cite{9781107002173}, involving operations considered expensive for current Noisy Intermediate Scale Quantum (NISQ) devices, potentially undermines the speedup provided by Grover's algorithm. To cope with this problem~\cite{Intallura:2023yvu}, several solutions have been suggested~\cite{Suzuki2020,wie2019simpler,Plekhanov:2021kir,Kitaev2002ClassicalAQ,aaronson,Ghosh:2023qze,Grinko_2021}. Among them, the Iterative Quantum Amplitude Estimation (IQAE) variant proposed in~\cite{Grinko_2021} stands as a successful candidate that maintains the speedup.

In the context of numerical integration, a Quantum Monte Carlo Integration~(QMCI) method that utilizes QAE, or any of its variants, to achieve a quadratic speedup over the number of queries compared to its classical counterpart has been presented in~\cite{Montanaro2015}. However, this proposal does not address properly the challenge of preparing the quantum initial state, and in certain scenarios, it requires an extensive amount of quantum arithmetic operations, which may potentially diminish the quantum advantage. In view of this, 
the Fourier Quantum Monte Carlo Integration (FQMCI) method~\cite{Herbert_2022} introduces a unique approach to QMCI that harnesses the power of quantum computing without relying on arithmetic or phase estimation. This method stands out by accom\-plishing multiple goals simultaneously, which was not seen in previous proposals. The key idea is to use Fourier series decomposition to approximate the integrand and then estimate each component separately using QAE. Nevertheless, FQMCI relies on certain assumptions regarding the acquisition of Fourier coefficients, which may not hold in general. When these assumptions are not met, the quantum speedup might be wiped out.

The main novelty introduced by QFIAE~\cite{deLejarza:2023IEEE} with respect to FQMCI~\cite{Herbert_2022} involves decomposing the target function into its Fourier series using a Quantum Neural Network (QNN) and subsequently integrating each trigonometric component using IQAE. As a result, this approach constitutes an end-to-end quantum algorithm that offers a viable strategy to maintain the quadratic speedup achieved by IQAE.

This chapter is organized as follows. First, in Section~\ref{app:feynmanintegrals} the loop Feynman integrals in LTD are introduced. Then in Section~\ref{app:qfiae}, the QFIAE method is introduced and described in detail. In Section~\ref{app:lfintqc}, we motivate the application of quantum algorithms to specific loop Feynman diagrams, and present numerical results in simulators and hardware of the quantum integration of different loop topologies: tadpole, bubble, triangle and pentagon. Then, in Section~\ref{app:qdecaysint}, we present the decay rates of Higgs bosons, photons, and scalars computed using QFIAE on both simulators and real quantum devices. Finally, Section~\ref{app:qintconclusions} provides a brief summary and discussion of the results.

\section{Loop Feynman integrals in LTD}\label{app:feynmanintegrals}

QFT stands as one of the most accurate theories for producing theoretical predictions at high-energy colliders. However, its inherent complexity makes analytical calculations of physical processes exceptionally challenging. The perturbative approach in QFT expresses physical observables (such as cross sections or decay rates) as power series in the interaction couplings, with each term in the expansion corresponding to tree-level and loop Feynman integrals of increasing complexity. These mathematical expressions involve sophisticated multi-dimensional integrations over internal momenta, creating significant computational challenges due to UV and IR divergences that require careful regularization and renormalization. Regarding their physical interpretation, terms at high perturbative orders are elegantly represented through Feynman diagrams, where loop diagrams specifically capture quantum fluctuations arising from virtual particles circulating in closed paths \cite{Feynman:1963ax}.

Despite the technical difficulties in estimating their value, loop Feynman integrals are essential to deliver precision calculations in particle physics. They enable theoretical predictions for scattering amplitudes, decay rates, and cross-sections that can be directly compared with experimental measurements.

\subsection{Loop-Tree Duality}\label{sec:ltd}
In this context, the Loop-Tree Duality (LTD)~\cite{Aguilera-Verdugo:2020set,Catani:2008xa,Bierenbaum:2010cy,Bierenbaum:2012th} is an innovative methodology to deal with multiloop Feynman integrals and scattering amplitudes. Specifically, LTD transforms loops defined in the Minkowski space of the loop four-momenta into trees defined in the Euclidean space of their spatial components, and reinterprets virtual states as configurations that resemble real-radiation processes. Among other advantages, this transformation provides a more intuitive understanding of the singular structure of loop integrals~\cite{Buchta:2014dfa,Aguilera-Verdugo:2019kbz}. In particular, the most remarkable property of LTD is the existence of a manifestly causal representation~\cite{Aguilera-Verdugo:2020set,Aguilera-Verdugo:2020kzc,Ramirez-Uribe:2020hes,JesusAguilera-Verdugo:2020fsn,Sborlini:2021owe,TorresBobadilla:2021ivx}, i.e., an integrand representation where certain nonphysical singularities are absent and therefore yields integrands that are numerically more stable. 

The fact that the integration domain in LTD is Euclidean and not Minkowski also brings additional advantages, both for analytic applications such as asymptotic expansions~\cite{Driencourt-Mangin:2017gop,Plenter:2020lop}, where the hierarchy of scales is well defined, and numerical applications~\cite{Buchta:2015wna,Driencourt-Mangin:2019yhu} because the number of loop integration variables is independent of the number of external particles. For example, at one loop the number of independent integration variables is always three, the number of spatial components of the loop momentum, although for certain kinematic configurations this number can be reduced when the dependence on any of these variables is trivial. This is the case for tadpole and bubble diagrams at one loop, as we will see in more detail in Section~\ref{app:lfintqc}. 

Moreover, LTD offers a unified framework for cross-section calculations, since the dual represen\-tation of loop integrals in Euclidean domains allows a direct combination of virtual and real contributions at the integrand level, resulting in a fully local cancellation of IR singularities, the so-called Four-Dimensional Unsubtraction (FDU)~\cite{Hernandez-Pinto:2015ysa,Sborlini:2016gbr,Sborlini:2016hat,Prisco:2020kyb,deJesusAguilera-Verdugo:2021mvg} and UV singularities by appropriate UV local counterterms~\cite{Driencourt-Mangin:2017gop,Driencourt-Mangin:2019aix}, without the need for additional regulators, such as Dimensional Regularization (DREG)~\cite{Bollini:1972ui,tHooft:1972tcz}.
Also, it is worth to remark that quantum algorithm have already been used to bootstrap the causal structure of multiloop Feynman diagrams in LTD~\cite{Ramirez-Uribe:2021ubp,Clemente:2022nll,Ramirez-Uribe:2024wua,Ochoa-Oregon:2025opz}.

To illustrate mathematically how the LTD formalism work let us consider a generic $L$-loop scattering amplitude with $N$ external legs $\{p_j\}^N$ and $n$ sets of internal lines, each defined by specific dependencies on loop momenta, can be expressed as \cite{deJesusAguilera-Verdugo:2021mvg}
\begin{equation}
\mathcal{A}_N^{(L)}(1, \ldots, n) = \int_{\ell_1,\ldots,\ell_L} \mathcal{A}_F^{(L)}(1, \ldots, n) ,
\label{eq:lloopint}
\end{equation}
where
\begin{equation}
\mathcal{A}_F^{(L)}(1, \ldots, n) = \mathcal{N}(\{\ell_s\}^L, \{p_j\}^N) \times G_F(1, \ldots, n)~.
\end{equation}
This integral in the Minkowski space involves the $L$-loop momenta $\{\ell_s\}^L$, the product of Feynman propagators $G_F(q_i) = (q_i^2 - m_i^2 + i0)^{-1}$, and numerators $\mathcal{N}(\{\ell_s\}^L, \{p_j\}^N)$ derived from the Feynman rules of the theory. In Dimensional Regularization (DREG), the $d$-dimensional integration measure is defined as
\begin{equation}
\int_{\ell_s} = -i \mu^{4-d} \int \frac{d^d\ell_s}{(2\pi)^d} .
\end{equation}
The standard Feynman propagator for a single particle reads
\begin{equation}
G_F(q_{i_s}) = \frac{1}{q_{i_s,0}^2 - \left(q_{i_s,0}^{(+)}\right)^2} ,
\end{equation}
where
\begin{equation}
q_{i_s,0}^{(+)} = \sqrt{\vec{q}_{i_s}^{,2} + m_{i_s}^2 - \ii}
\end{equation}
represents the positive on-shell energy of the loop momentum $q_{i_s}$ in terms of its spatial components $\vec{q}_{i_s}$, mass $m_{i_s}$, and the infinitesimal Feynman complex prescription $\ii$.
Also, $s$ denotes the set of internal propagators with momenta $q_{i_s} = \ell_s + k_{i_s}$, depending on the loop momentum $\ell_s$ or specific linear combinations of loop momenta, along with external momenta $k_{i_s}$, where $i_s \in s$. The Feynman propagators can be raised to arbitrary powers.

The LTD representation is derived through iterative application of Cauchy's residue theorem. This process integrates out one degree of freedom for each loop momentum, with the Cauchy contour closed below the real axis to select poles with negative imaginary components in the complex plane. This procedure modifies the infinitesimal complex prescription of the Feynman propagators, requiring careful treatment to preserve causality.
Starting from Eq.~\ref{eq:lloopint}, we set on-shell the propagators dependent on the first loop momentum $q_{i_1}$ and define
\begin{equation}
\mathcal{A}_D^{(L)}(1; 2, \ldots, n) \equiv \sum_{i_1 \in 1} \text{Res}\left(\mathcal{A}_F^{(L)}(1, \ldots, n), \text{Im}(q_{i_1,0} < 0)\right) ,
\end{equation}
where taking the residue effectively integrates out the energy component of the loop momenta. We then construct the nested residue, iterating until the $r$-th set as
\begin{equation}
\mathcal{A}_D^{(L)}(1, \ldots, r ; r+1, \ldots, n) = \sum_{i_s \in s} \text{Res}\left(\mathcal{A}_D^{(L)}(1, \ldots, r-1 ; r, \ldots, n), \text{Im}(q_{i_s,0} < 0)\right) .
\end{equation}
In this representation, all sets preceding the semicolon contain one propagator set on-shell and are linearly independent, while remaining propagators stay off-shell. This transforms the loop amplitude into non-disjoint trees. Then, after integration of the energy component, the integration measure becomes
\begin{equation}
\int_{\vec{\ell}_s} \equiv -\mu^{d-4} \int \frac{d^{d-1}\ell_s}{(2\pi)^{d-1}} ,
\end{equation}
converting the $d$-dimensional Minkowski space into a $(d-1)$-dimensional Euclidean space.

\section{Quantum Fourier Iterative Amplitude Estimation}\label{app:qfiae}

This section is based on \cite{deLejarza:2023IEEE} where we present a Quantum Monte Carlo Integrator dubbed Quantum Fourier Iterative Amplitude Estimation (QFIAE). The QFIAE algorithm is build upon Quantum Amplitude Estimation and uses a Quantum Neural Network to decompose the target function to be integrated into a Fourier series to ensure efficient encoding into a quantum circuit. A tutorial on using this quantum integrator, and the source code used to produce the results in this section, are available in \cite{tutorial_qfiae} and \cite{githubGitHubGmlejarzaQuantumFourierIterativeAmplitudeEstimation}, respectively.
\subsection{Iterative Quantum Amplitude Estimation}
\label{sec:iqae}
Quantum Amplitude Estimation (QAE) is a quantum algorithm that relies on Grover's amplitude amplification subroutine to efficiently estimate the amplitude of a quantum state. One of their applications include integration, as long as one can encode the desired integral as the amplitude of what we will call the ``good state''. However, the general implementation \cite{Brassard_2002} relies on Quantum Phase Estimation (QPE) which involves many quantum gates and makes it difficult to be implemented in current NISQ devices. In this context the Iterative Quantum Amplitude Estimation~(IQAE) algorithm proposed in~\cite{Grinko_2021} is an alternative implementation of QAE that relies exclusively on Grover's algorithm. This variant replaces the QPE subroutine with an efficient classical post-processing methodology, thereby reducing the necessary quantum resources in terms of qubits and gate operations.
The IQAE algorithm incorporates two essential components: quantum amplitude amplification and a methodical iterative search to identify the optimal number $k$ of amplifications required for accurate estimation.
The amplitude amplification~\cite{Brassard,PhysRevLett.80.4329,Brassard_2002} technique, which extends the Grover's quantum search algorithm~\cite{10.1145/237814.237866}, achieves a quadratic enhancement in performance relative to classical approaches.
Let us consider a unitary operator $\mathcal{A}$ operating on $n+1$ qubits such that:
\begin{equation}
|\psi\rangle=\mathcal{A}|0\rangle_{n+1}=\sqrt{a}|\tilde{\psi}_1\rangle|1\rangle +\sqrt{1-a}|\tilde{\psi}_0\rangle|0\rangle,
\label{eq:qae}
\end{equation}
where $a\in[0,1]$ represents the parameter we aim to estimate, while $|\tilde{\psi}_1\rangle$ and $|\tilde{\psi}_0\rangle$ denote the $n$-qubit good and bad states, respectively. Through Eq.~\ref{eq:qae}, parameter $a$ could be determined from the ratio of good to bad states, but this approach provides no computational advantage as its query complexity would match that of classical methods.
The quantum advantage materializes through amplitude amplification, enabling a quadratic reduction in query complexity through the application of the amplification operator:
\begin{equation}
\mathcal{Q}=-\mathcal{A}S_0\mathcal{A}^{-1}S_{\chi},
\label{eq:grover_operator}
\end{equation}
where $S_0$ marks the $|0\rangle_{n+1}$ state with a negative sign while preserving other states, and $S_\chi$ inverts the sign of the good state, specifically $S_\chi |\tilde{\psi}_1\rangle|1\rangle=-|\tilde{\psi}_1\rangle|1\rangle$.
By defining a parameter $\theta_a\in [0,\pi/2]$ where $\sin^2{\theta_a}=a$, we can express:
\begin{equation}
|\psi\rangle=\mathcal{A}|0\rangle_{n+1}=\sin{\theta_a}|\tilde{\psi}_1\rangle|1\rangle +\cos{\theta_a}|\tilde{\psi}_0\rangle|0\rangle.
\label{eq:psi_A}
\end{equation}
Brassard et al. demonstrated in~\cite{Brassard_2002} that applying $\mathcal{Q}$ repeatedly $k$ times to $\psi$ yields:
\begin{equation}
\mathcal{Q}^k|\psi\rangle=\sin{((2k+1)\theta_a)}|\tilde{\psi}_1\rangle|1\rangle +\cos{((2k+1)\theta_a)}|\tilde{\psi}_0\rangle|0\rangle.
\label{eq:grover_k_operator}
\end{equation}
Eq. \ref{eq:grover_k_operator} indicates that after $k$ applications of operator $\mathcal{Q}$, the probability of obtaining the good state increases by approximately $4k^2$ times compared to the probability from Eq.~\ref{eq:psi_A} for sufficiently small values of~$a$. This represents the quadratic acceleration achieved through amplitude amplification, as $2k$ measurements from $\mathcal{A} |0\rangle_{n+1}$ would only enhance the probability of obtaining the good state by a factor of $2k$. Thus, if we can determine the proportion of good states following amplitude amplification, we can estimate $a$ based on the query count required to achieve this proportion.
The procedure for identifying the optimal value of $k$ constitutes the innovative contribution of the IQAE algorithm. This process involves a confidence interval $[\theta_l,\theta_u] \subseteq [0,\pi/2]$ with $\theta_l < \theta_a < \theta_u$, a power $k$ of $\mathcal{Q}$, and an estimate for $\sin^2((2k+1)\theta_a)$. Applying the trigonometric identity $\sin^2(x)=(1-\cos(2x))/2$, our estimates for $\sin^2 ((2k + 1)\theta_a)$ are converted to estimates for $\cos((4k + 2)\theta_a)$, considering whether the argument falls within $[0,\pi]$ or $[\pi, 2\pi]$. The objective is to identify the maximum $k$ for which the interval $[(4k+ 2)\theta_l,(4k+ 2)\theta_u]_{\mod 2\pi}$ remains entirely within either the upper or lower half-plane. This establishes an upper bound for $k$, and the core innovation of this algorithm lies in the methodology used to determine the next value of $k$ given the interval $[\theta_l,\theta_u]$.

\subsection{Quantum Monte Carlo Integration}\label{sec:qmci}

Monte Carlo integration is a powerful technique widely used in fields such as particle physics, finance, and cosmology. It estimates definite integrals using random sampling:
\begin{equation}
     I=\int_{x_{min}}^{x_{max}} p(x)f(x) dx.
\end{equation}
To compute this integral using the Monte Carlo method, we draw $2n$ independent and identically distributed (i.i.d.) samples $x_i \in \{0,1\}^n$, with $i=0,\ldots, 2n-1$, from the probability distribution $p(x)$ over the interval $[x_{min},x_{max}]$. The integral is then approximated by the mean value of $f(x)$ over these samples.

The main objective of discrete Monte Carlo integration is to compute the expected value of a real function $0\leq f(x) \leq 1$ defined for $n$-bit inputs $x \in \{0,1\}^n$ with probability $p(x)$:
\begin{equation}
    \mathbb{E}[f(x)]=\sum_{x=0}^{2^n-1} p(x)f(x).
\end{equation}

Although Monte Carlo integration has been successfully applied across various fields, certain complex integrals remain highly challenging. This is particularly true for multivariate functions with sparse probability distributions, which are common in HEP. These challenges in Monte Carlo integration methods have motivated the development of a quantum counterpart \cite{Montanaro2015}. The key idea behind this quantum approach is to encode the integral as the amplitude of a quantum state and then apply QAE to estimate this amplitude, and thus the integral.

The convergence of Quantum Monte Carlo Integration (QMCI), measured by the Mean Square Error (MSE), offers a quadratic advantage in the number of samples drawn from the probability distribution. This speedup comes from the use of the Grover's quantum amplitude amplification subroutine, which underlies QAE.

As discussed in~\cite{Suzuki2020}, the QMCI algorithm prepares the operator $\mathcal{A}$ from two components:
\begin{equation}
    \mathcal{P}|0\rangle_n=\sum_{x=0}^{2^n-1} \sqrt{p(x)}|x\rangle_n ,
\end{equation}
which encodes the probability distribution $p(x)$ into the state $|0\rangle_n$, and
\begin{equation}
    \mathcal{R}|x\rangle_n|0\rangle=|x\rangle_n \left(  \sqrt{f(x)}|1\rangle +\sqrt{1-f(x)}|0\rangle \right),
\end{equation}
which encodes the function $f(x)$ into an ancillary qubit. Applying $\mathcal{A}$ to the $(n+1)$-qubit initial state results in:
\begin{equation}
\begin{split}
|\psi\rangle &= \mathcal{A}|0\rangle_{n+1}=\mathcal{R}(\mathcal{P}\otimes\mathbb{I}^1)|0\rangle_{n+1}\\
&=\sum_{x=0}^{2^n-1} \sqrt{p(x)}|x\rangle_n \left( \sqrt{f(x)}|1\rangle +\sqrt{1-f(x)}|0\rangle \right),
\end{split}
\end{equation}
where $\mathbb{I}^1$ is the identity operator acting on the ancillary qubit. We now define:
\begin{equation}
    a=\sum_{x=0}^{2^n-1} p(x)f(x)= \mathbb{E}[f(x)],
\end{equation}
\begin{equation}
    |\tilde{\psi}_1\rangle=\frac{1}{\sqrt{a}}\sum_{x=0}^{2^n-1} \sqrt{p(x)}\sqrt{f(x)}|x\rangle_n,
\end{equation}
\begin{equation}
    |\tilde{\psi}_0\rangle=\frac{1}{\sqrt{1-a}}\sum_{x=0}^{2^n-1} \sqrt{p(x)}\sqrt{1-f(x)}|x\rangle_n,
\end{equation}
and express $|\psi\rangle$ as:
\begin{equation}
    |\psi\rangle=\sqrt{a}|\tilde{\psi}_1\rangle|1\rangle+\sqrt{1-a}|\tilde{\psi}_0\rangle|0\rangle.
    \label{eq:psi}
\end{equation}
which looks identical to Eq. \ref{eq:qae}.

\begin{figure}[ht!]
\centering
\includegraphics[width=.69\linewidth]{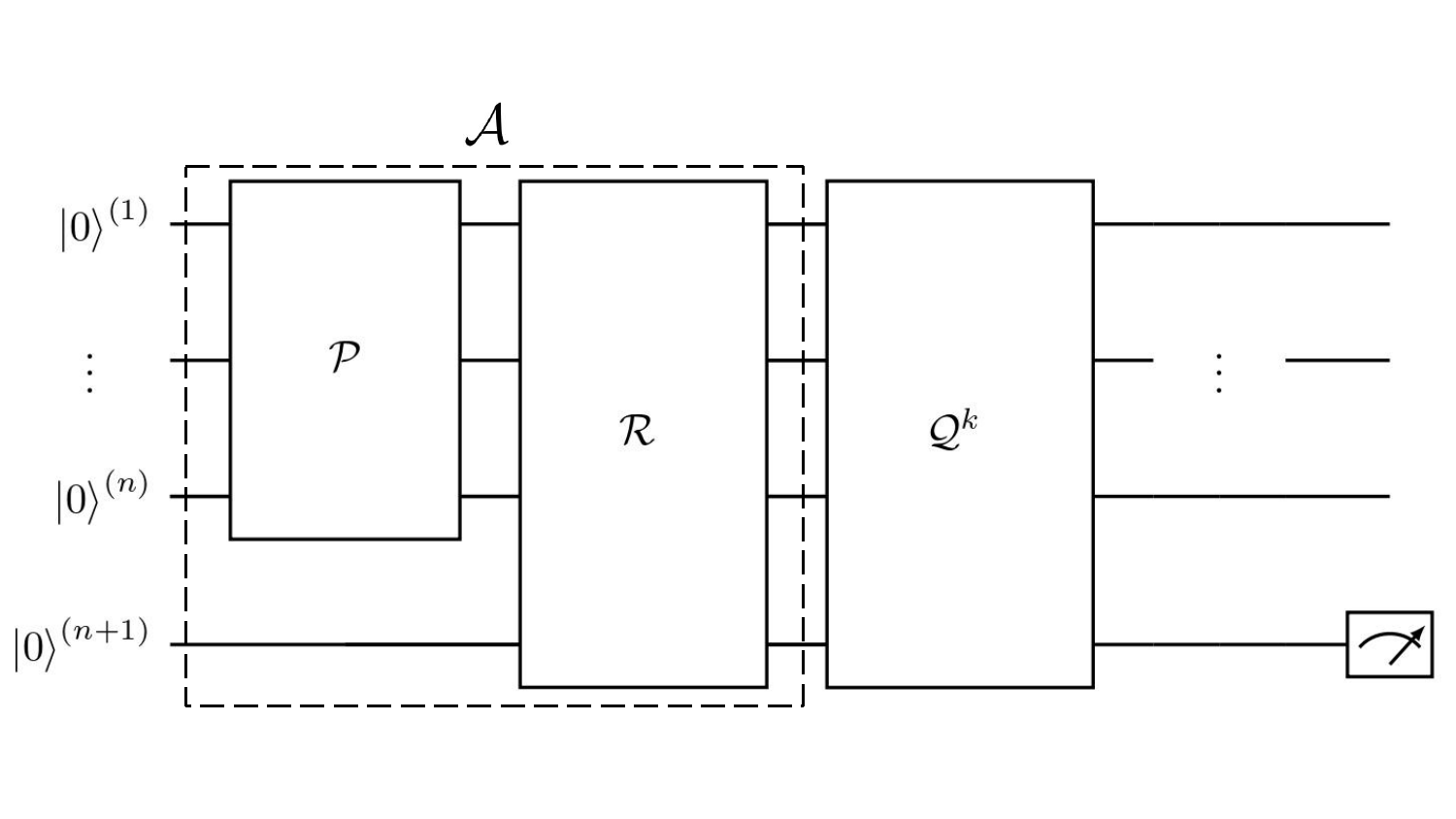}
\caption{Quantum circuit for amplitude amplification and estimation in Monte Carlo integration.}
\label{fig:qae_circuit}
\end{figure}

As motivated before, we can now interpret Monte Carlo integration as an amplitude estimation problem, where the probability of the target state, $a$, corresponds to the integral to be computed, $\mathbb{E}[f(x)]$. What is left now is the operator $\mathcal{Q}$, which is constructed using $U_{\psi}$ and $U_{\psi_0}$:
\begin{equation}
\mathcal{Q} = U_{\psi} U_{\psi_0},
\end{equation}
where
\begin{equation}
\begin{split}
& U_{\psi_0}=\mathbb{I}_{n+1}-  2\mathbb{I}_n|0\rangle \langle 0|, \,  \hspace{0.4cm} \, \qquad U_{\psi}=\mathbb{I}_{n+1}-2|\psi\rangle  \langle\psi|,  \qquad \mathbb{I}_n\equiv \mathbb{I}^{\otimes n}.
\end{split}
\end{equation}
Defining $a = \sin^2{\theta_a}$ and applying Eq. \ref{eq:psi}, the IQAE algorithm described earlier can be used for Monte Carlo integration. The quantum circuit implementing this procedure is shown in Fig.~\ref{fig:qae_circuit}.

\subsection{Fourier Quantum Monte Carlo Integration}\label{sec:fqmci}

Although it is possible to integrate any 
function $f(x)$ that follows a probability distribution $p(x)$ using the QMCI described in Sec.~\ref{sec:qmci}, this comes at the cost of creating the quantum circuit $\mathcal{R}$, which encodes the function $f(x)$. If one wants to consider the most general scenario, where $f(x)$ has a complicated behavior, the computational cost of building such circuit $\mathcal{R}$ could be prohibitively complex, potentially killing the quantum speedup provided by Grover's amplitude amplification subroutine.

Even the trivial case where $f(x) = x$ (i.e., just finding the mean of $p(x)$) would imply significant amounts of arithmetic to be performed quantumly. However, an important exception is the specific case where $f(x) = \sin^2 (mx + c)$ for some constants $m$ and $c$, which can be encoded by a bank of $R_y$ rotation gates.

With this in mind, the author in~\cite{Herbert_2022} introduced the Fourier Quantum Monte Carlo Integration (FQMCI) method that is based on the idea of decomposing the desired function $f(x)$ into sines and cosines. The procedure starts by extending $f(x)$ as a periodic piecewise function. Let us consider \textbf{f}$(x)$, which repeats with period $x_{\tilde{u}}$--$x_l$:
\begin{equation}
\textbf{f}(x)=
    \begin{cases}
         f(x) & \text{if } x_l\leq x \leq x_u \\
        \tilde{f}(x) & \text{if } x_u\leq x \leq x_{\tilde{u}},
    \end{cases}
    \label{eq:fx}
\end{equation}
where $\tilde{f}(x)$ is itself sufficiently smooth and chosen such that $f(x_l) =\tilde{f}(x_{\tilde{u}})$ and $f(x_u) =\tilde{f}(x_u)$, which holds true for their derivatives w.r.t to $x$. Since \textbf{f}$(x)$ is periodic, it has a Fourier series:
\begin{equation}
\textbf{f}(x)=c+ \sum_{n=1}^\infty \left( a_n\cos(n\omega x)+ b_n\sin(n\omega x) \right),
    \label{eq:ffourier}
\end{equation}
where $\omega = 2\pi/T$ and $T$ is the period of the periodic piecewise function. Those various trigonometric components of $\textbf{f}(x)$ can be estimated individually, using IQAE, and then all the individual integrals are collected to obtain the final result.

Furthermore, it is shown that the quadratic quantum advantage is indeed retained with this method under certain conditions. Those include the ability to prepare a specific probability distribution as a shallow-depth quantum state and apply a function to random samples such that the mean cannot be calculated analytically. Additionally, it is important to be cautious when evaluating Fourier coefficients as numerical integration may shift the computational load rather than reduce the complexity. In~\cite{Herbert_2022}, the author claims that for commonly used functions such as the mean ($f(x)=x$), Fourier coefficients can be calculated symbolically. Moreover, he adds that if the coefficients cannot be found symbolically, for commonly used functions it may be reasonable to assume that they have been pre-computed and stored in advance. Therefore they would not add to the complexity of the overall cost. 

Nevertheless, in the most general possible scenario, the Fourier coefficients need to be calculated and we cannot rely on them being symbolically computable or stored in advance. Hence, to attain the desired speedup for the whole computation, a new procedure to estimate the Fourier coefficients without relying on numerical integration must be introduced. This leads us to propose in Sec.~\ref{subsec:qfiae} a novel method for obtaining the Fourier series of a function using a QNN.

\subsection{Quantum Fourier Series}\label{subsec:qfourierseries}

As explained in Sec.~\ref{sapp:qnn}, a QNN typically consists of a layered architecture that encodes input data into quantum states and processes them in a high-dimensional feature space. The encoding strategy and circuit Ansatz play a key role in outperforming classical NNs. While complex data encoding with feature maps that are hard to simulate may enable a quantum advantage, too expressive circuits can lead to flat cost landscapes, making the model difficult to train~\cite{holmes2022connecting}.

With this in mind, it is possible to implement a QNN to fit a multivariate function $f(\vec{x})$ and extract its Fourier series from the trained quantum circuit. The architecture considered as introduced in~\cite{P_rez_Salinas_2020}, consists of interleaved encoding and trainable circuit blocks.  
The general form of the Ansatz is:  
\begin{equation}
    U_0 \equiv \mathcal{A}(\vec{\theta_0})~, \qquad U_l \equiv \mathcal{A}(\vec{\theta_l})\mathcal{S}(\vec{x})~.
\end{equation}
More specifically, the circuit consists of three main components: an initial operator, $\mathcal{A}(\vec{\theta_0})$, that prepares a superposition state; an encoding layer, $\mathcal{S}(\vec{x})$, that maps the input data; and a trainable layer, $\mathcal{A}(\vec{\theta_l})$, optimized according to a chosen metric. The cost function is defined as the squared difference between the circuit's output (expected value in the computational basis) and the target function.

\begin{figure}[h]
       \centering
  \begin{subfigure}[b]{0.8\textwidth}
  
  \includegraphics[width=.99\linewidth]{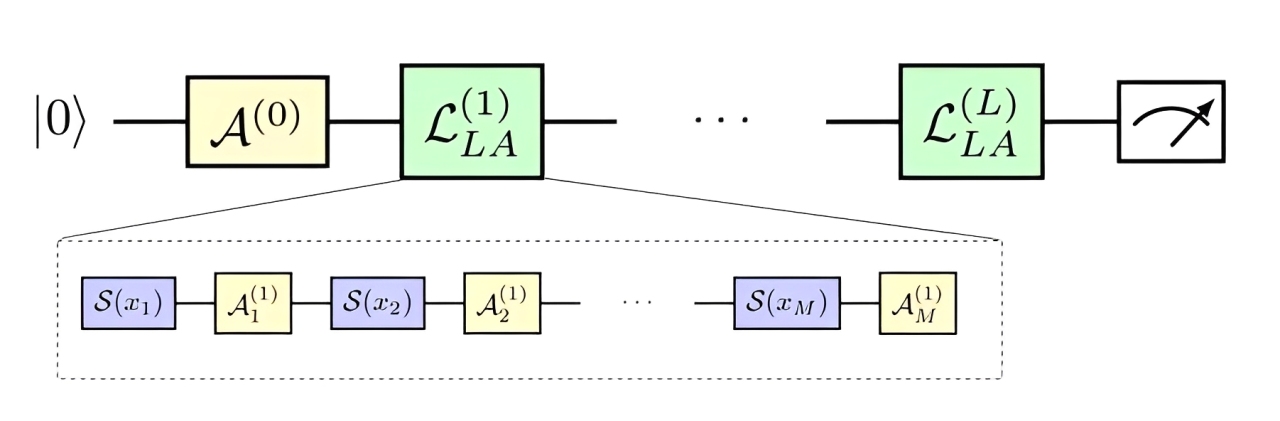} 
  \caption{Linear Ansatz}
  \end{subfigure}
  \begin{subfigure}[b]{0.8\textwidth}

  \includegraphics[width=.99\linewidth]{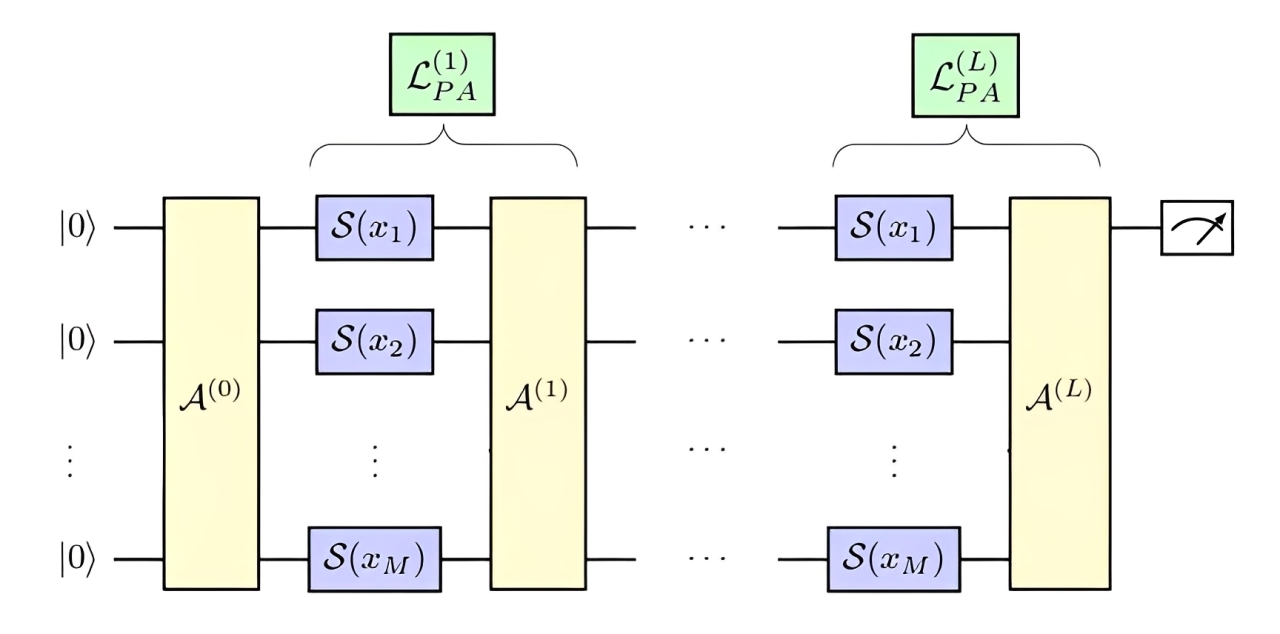} 
  \caption{Parallel Ansatz}
  \end{subfigure}
        \caption{Quantum circuit Ans\"atze of the QNN. The Linear Ansatz (a) encodes each data feature in a single qubit. Thus the circuit depth grows linearly with the total number of variables~$M$. The Parallel Ansatz (b) encodes the~$M$ variables in~$M$ qubits. In both Ans\"atze the circuit depth grows linearly with the number of layers $L$. Note that for $M=1$, both the Linear and Parallel Ansatz converge to the same linear circuit.}
        \label{fig:ansatzes}
\end{figure}

We consider two strategies to encode the data sampled from the target function.  
In the first approach, the circuit follows a linear Ansatz, where each feature, i.e., each $x$ value, is encoded in a single qubit, as shown in Fig.~\ref{fig:ansatzes}(a). The second approach, the circuit follows a parallel Ansatz, where each feature, i.e. each $x$ value, is encoded in a different qubit as shown in Fig.~\ref{fig:ansatzes}(b).

Following the explanation in Sec.~\ref{sapp:qnn}, the expectation value of this quantum model corresponds to a universal multidimensional Fourier series representation~\cite{Schuld_2021,Casas:2023ure,Atchade-Adelomou:2023mjf}:
\begin{equation}
    \langle \textit{M} \rangle (\vec{x}, {\vec{\theta}}) = \sum_{\vec{w} \in \Omega} c_w e^{i\vec{x}\cdot\vec{w}}~,
\end{equation}
where the Fourier components arise from the circuit elements. The encoding gate for each variable $\mathcal{S}(x) = e^{xH}$ determines the frequency $w$, given a Hamiltonian $H$. For a single qubit with $H = \frac{1}{2} \sigma_z$, the frequency spectrum corresponds to the eigenvalues of $H$. The Fourier coefficients $c_w$ depend on the trainable parameters $\vec{\theta_l}$ of the unitary gates $\mathcal{A}(\vec{\theta_l})$. 

The choice of data encoding influences the maximum frequency in the output Fourier series, which scales with the number of layers $L$ and the computational space dimension $D = \max(\Omega) = (d-1)L$. In the qubit setup considered in this Section, where $d=2$, increasing $L$ enables the circuit to represent more complex functions.

\subsection{Quantum Fourier Iterative Amplitude Estimation}\label{subsec:qfiae}

\begin{figure}[ht!]
       \centering
  \includegraphics[width=.991\linewidth]{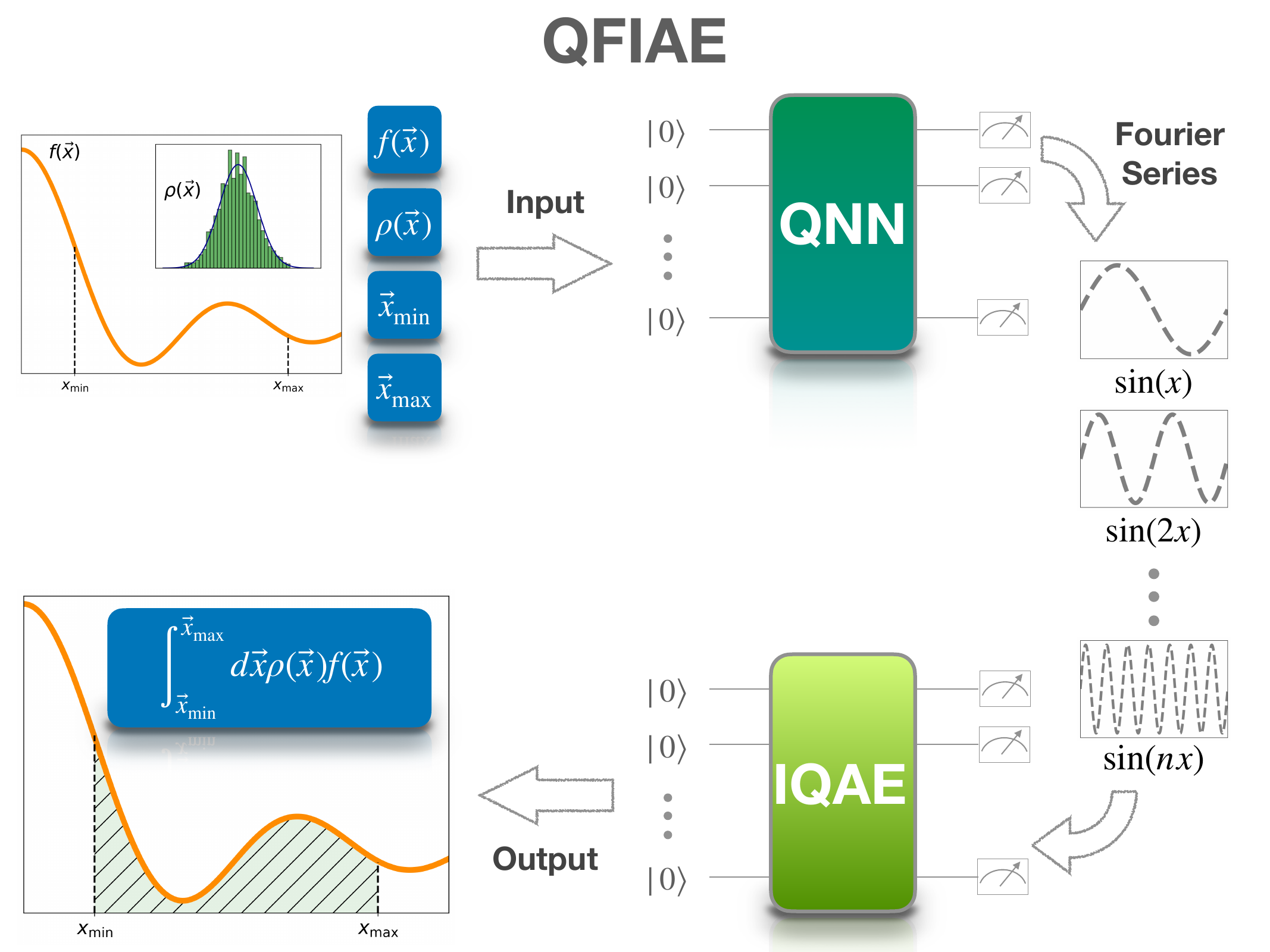}  

        \caption{Workflow of QFIAE. The input consists of the target function $f(\vec{x})$, the probability distribution $\rho(\vec{x})$, and the integration domain $\{\vec{x}_{min},\vec{x}_{max} \}$. The QNN fits $f(\vec{x})$ and extracts its Fourier series from the quantum circuit. Next, IQAE estimates the integral for each trigonometric term in the Fourier series. Finally, these integrals are added with their corresponding coefficients to obtain the final integral result. }
        \label{fig:sketch qfiae}
\end{figure}

The quantum algorithm we are introducing in this section presents an elegant solution to the problem presented in the FQCMI method. The core idea of the Quantum Fourier Iterative Amplitude Estimation (QFIAE) algorithm is to exploit an appropiate choice of the encoding to design a QNN that represent a truncated Fourier series, as demonstrated in~\cite{Schuld:2020enb, Ostaszewski:2019vnn} and explained in Sections~\ref{sapp:qnn} and~\ref{subsec:qfourierseries}.  Then, if a quantum subroutine is used to obtain the Fourier series, this QNN can be integrated with the FQCMI algorithm to establish a fully quantum pipeline for Monte Carlo integration. The workflow of the QFIAE algorithm is presented in Fig. \ref{fig:sketch qfiae}, and a tutorial using the open-source software for quantum computation \texttt{Qibo}~\cite{qibo_paper} is available in~\cite{tutorial_qfiae}.

Let us see how this algorithm performs in practice with a simple example.

\subsection{Benchmark example: integral of $1+x^2$}

To benchmark the QFIAE method for integration, we consider the following integral:
\beq
I=\int_0^1 \left(1+x^2\right) dx. \label{eq:integral_example}
\eeq

This choice is motivated by its simplicity, which provides a clear pedagogical demonstration of the method while being relevant to a well-known particle physics case. The integrand, $f(x)=1+x^2$, represents a simple yet non-trivial form of the differential cross-section for the scattering process $e^+e^- \rightarrow q \bar{q}$ at the Leading Order (LO).  In quantum electrodynamics (QED), this process is well described by parametrizing the phase space with two angles, leading to the total cross-section expression\footnote{Note that \Eq{eq:integral_example} and \Eq{eq:Xsection22} differ in their integration intervals. However, since the function is even, they are equivalent up to a factor of~$2$.}:
\begin{equation}
  \sigma \sim \int^{1}_{-1} \int^{2\pi}_0 \mathrm d \cos \theta \mathrm d \phi \left( 1+\cos^2 \theta\right) 
  \label{eq:Xsection22}
 \end{equation}

This integral was previously computed using the IQAE algorithm with a different approach in~\cite{AGLIARDI2022137228}.  
There, a quantum Generative Adversarial Network (QGAN) was employed to load the normalized distribution $1+x^2$. The QGAN was trained using classically generated samples to encode the function $f(x)$ into a quantum state, after which the integral was estimated using IQAE. While~\cite{AGLIARDI2022137228} demonstrated a quadratic speedup over classical Monte Carlo integration in terms of query complexity, the total circuit depth, given by the complexity of the QGAN
architecture and the IQAE part makes this approach unsuitable for NISQ devices. In contrast, the QFIAE method we propose maintains a feasible circuit depth for NISQ hardware, making it more practical for real-world applications in near-term devices.

Starting from the mentioned example, we evaluate the performance of our quantum algorithm in computing the same integral. Since the QNN can only fit functions with values in the range $[-1,1]$, we first normalize the function $f(x) = 1 + x^2$ before applying the QFIAE method.  

\begin{figure}[ht!]
       \centering
  \includegraphics[width=0.69\linewidth]{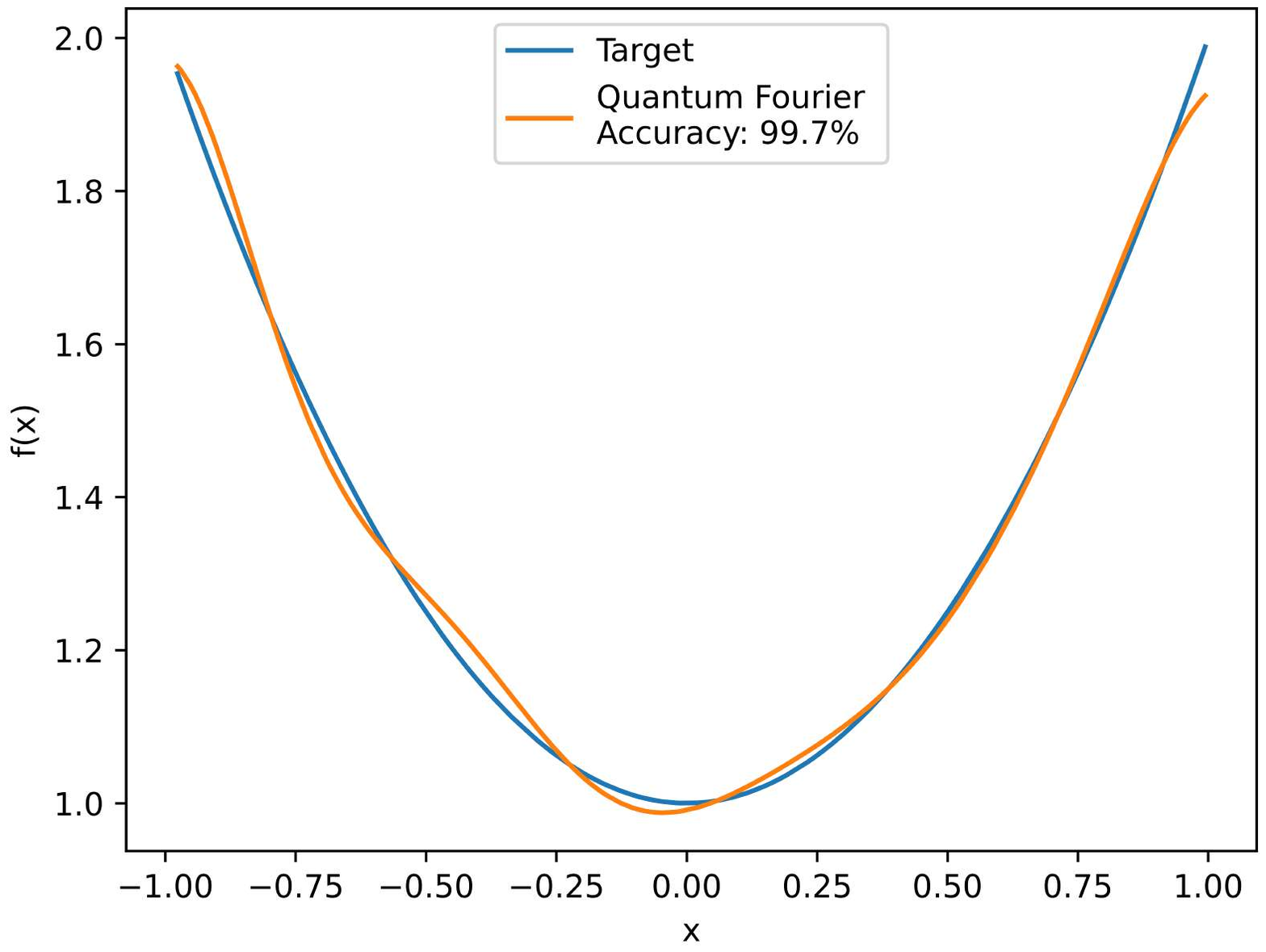}  
  \caption{Quantum simulation using a Linear Ansatz in the QNN to fit the function $1 + x^2$. The setup uses $n_{\text{qubits,QF}} = 1$ and $L = 10$, resulting in a Fourier series with $n_{\text{Fourier}} = 10$ coefficients. A gradient descent method, the Adam optimizer, is employed with 200 data points in the range $[-1,1]$. The optimizer hyperparameters are \textit{learning\_rate} = 0.05 and \textit{nepochs} = 100.}
        \label{fig:qf_example}
\end{figure}

The first step of the quantum integration algorithm involves fitting the function using a QNN, as shown in Fig.~\ref{fig:qf_example}. The QML model achieves remarkable accuracy, exceeding 99\%, demonstrating the reliability of our method for function fitting. The model's performance can be optimized by tuning hyperparameters such as the number of layers, which is related to the number of Fourier terms $n_{\text{Fourier}}$, the size of training and test datasets, and the number of optimizer iterations, \textit{nepochs}. Adjusting these parameters allows balancing between learning time and accuracy.  

The next step extracts the Fourier coefficients from the quantum circuit and reconstructs the Fourier series:  
\beq
\begin{split}
f(x) &\approx 0.476 + 1.169 \cos(x) - 0.263\cos(2x) + \ldots  \\ 
     &\quad - 0.017\cos(9x) + 0.004\cos(10x) \\ 
     &\quad - 0.125\sin(x) - 0.278\sin(2x) + \ldots  \\ 
     &\quad - 0.029 \sin(9x) - 0.004\sin(10x).
\label{eq:fourierexample}
\end{split}
\eeq
Once the function is expressed as a Fourier series, as shown in Eq.~\eqref{eq:fourierexample}, the trigonometric components are transformed into the $\sin^2(ax+b)$ form and integrated using IQAE. The probability distribution considered in this example is $p(x) = 1/2^n$, which is prepared as a shallow-depth quantum state to preserve the quadratic quantum advantage. Specifically, it is generated by applying an $n$-dimensional Hadamard gate, $\mathcal{H}^{\otimes n}$.

At this point, one can compute the integrals, which are weighted by their corresponding coefficients, and summed them to estimate the integral in \Eq{eq:integral_example}. The results, along with a comparison to the analytical value, are presented in Table \ref{table:integral_example}.

\begin{table}[h!]
\begin{center}

\begin{tabular}{ |p{3.8cm} |p{2.3cm}|p{2.3cm}| }
 \hline
 \centering method = QFIAE & \centering $I_{est}$ &  \centering $\epsilon$  \cr 
 \hline
 \centering $n_{Fourier}=5$, $shots=100$ &
  \centering  \vspace{0.01cm} 1.32 $\pm$ 0.05 &    \centering \vspace{0.01cm} 0.991 \cr 
 \hline
 \centering $n_{Fourier}=10$, $shots=100$ &
  \centering  \vspace{0.01cm} 1.34 $\pm$ 0.06&    \centering \vspace{0.01cm} 1.006 \cr 
 \hline
 \centering $n_{Fourier}=5$, $shots=1000$ &
  \centering  \vspace{0.01cm} 1.33 $\pm$ 0.05&    \centering \vspace{0.01cm} 0.998 \cr 
 \hline
 
 \centering $n_{Fourier}=10$, $shots=1000$ &
  \centering  \vspace{0.01cm} 1.33 $\pm$ 0.04&    \centering \vspace{0.01cm} 0.999 \cr 
 \hline
\centering method = FQMCI & \centering $I_{est}$ &  \centering $\epsilon$  \cr 
 \hline
 
 \centering $n_{Fourier}=5$, $shots=100$ &
  \centering  \vspace{0.01cm} 1.35 $\pm$ 0.07&    \centering \vspace{0.01cm} 1.010 \cr 
 \hline
 \centering $n_{Fourier}=10$, $shots=100$ &
  \centering  \vspace{0.01cm} 1.34 $\pm$ 0.07&    \centering \vspace{0.01cm} 1.003 \cr 
 \hline
 \centering $n_{Fourier}=5$, $shots=1000$ &
  \centering  \vspace{0.01cm} 1.34 $\pm$ 0.07&    \centering \vspace{0.01cm} 1.007 \cr 
 \hline

 \centering $n_{Fourier}=10$, $shots=1000$ &
  \centering  \vspace{0.01cm} 1.34 $\pm$ 0.06&    \centering \vspace{0.01cm} 1.002 \cr 
 \hline

\end{tabular}
\caption{Integration of $1+x^2$ from $[0,1]$ using QFIAE and FQMCI. With $\epsilon=I_{est}/I_{exact}$.}
\label{table:integral_example}

\end{center}
\end{table}

In Table~\ref{table:integral_example}, we present an estimate of the integral of the function $1+x^2$ over the interval $[0,1]$. The integral is calculated using two methods: the QFIAE method and the FQMCI method introduced in~\cite{Herbert_2022}. The key distinction between these two algorithms lies in the calculation of the Fourier series. In QFIAE, the Fourier series is estimated using a QNN, while in FQMCI, it is computed classically through numerical integration. The integrals are evaluated for different values of the number of Fourier coefficients, $n_{Fourier}$, and the number of shots used in the IQAE algorithm, $shots$. In all cases, the IQAE algorithm employs $n_{qubits,\mathrm{IQAE}} = 4$ qubits, with a confidence interval of $\alpha = 0.05$ and an estimated error of $\epsilon = 0.01$. The third column of Table~\ref{table:integral_example} compares the obtained values with the exact result ($I_{exact} = 4/3$). Each integral is computed 50 times to minimize statistical fluctuations, and the results are averaged to compare the method's performance and identify the optimal set of parameters. The errors reported by the IQAE method in \texttt{Qibo} could be reduced by adjusting the $\epsilon$ parameter. From the analysis in the third column, we observe that QFIAE performs at least as well as FQMCI, demonstrating that estimating the Fourier series with a QNN does not produce any loss in accuracy. Additionally, the best performance is achieved with $n_{Fourier} = 10$ and $shots = 1000$, confirming the intuition that a higher number of Fourier coefficients increases the expressibility of the Fourier series, and also a larger number of shots improves precision.

\begin{table}[h!]
\begin{center}

\begin{tabular}{ |p{1.5cm}|p{1.5cm}|p{1.5cm}||p{1.5cm}|}
 \hline
 \multicolumn{4}{|c|}{${\rm QF}_{depth}=\mathcal{A}_{depth}+layers(\mathcal{A}_{depth}+\mathcal{S}_{depth})$} \\
 \hline
  \centering $\mathcal{A}_{depth}$& \centering $\mathcal{S}_{depth}$ &  \centering $layers$ &  \centering ${\rm  \,QF}_{depth}$ \cr 
 \hline
  \centering 3&  \centering 1 &  \centering 10 & \centering 43 \cr 
 \hline
 \multicolumn{4}{|c|}{${\rm IQAE}_{depth}=\mathcal{A}_{depth}+k\mathcal{Q}_{depth}$} \\
 \hline
   \centering $\mathcal{A}_{depth}$& \centering $\mathcal{Q}_{depth}$ &  \centering $k$ &  \centering ${\rm IQAE}_{depth}$ \cr 
 \hline
  \centering 4&  \centering 12 &  \centering 9\footnotemark{}& \centering 112 \cr 
 \hline

\end{tabular}

\caption{Depths of the Quantum Fourier (QF) and Iterative Quantum Amplitude Estimation (IQAE) parts of the QFIAE method.}
\label{table:depth}
\end{center}
\end{table}
\footnotetext{The value of $k$ is calculated by the algorithm and it changes in every iteration and integral. Hence the different values have been averaged, and the mean value is presented in the table.}

To evaluate the feasibility of our algorithm on current NISQ devices, we consider the Quantum Volume (QV) metric introduced by IBM in~\cite{IBMQV}. This metric quantifies the maximum size of square quantum circuits that can be successfully executed by a quantum computer. However, it is important to note that the QV metric may not accurately predict the performance of circuits that are deep and narrow or wide and shallow~\cite{ionqarticle}. This limitation is particularly relevant to our analysis, as shown in Table~\ref{table:depth}, where two circuits are much deeper than they are wide, with $n_{qubits, \mathrm{QF}}=1$ and $n_{qubits,\mathrm{IQAE}}=4$. Consequently, the suitability of our circuits is further assessed by referencing a recent study conducted by IonQ~\cite{ionqarticle}, which is detailed in~\cite{ionqnews}.

In their study, a set of quantum algorithms was tested and compared using QED-C benchmarks. The results indicate that our algorithm, with a quantum depth of 112 and a low qubit count ($\leq 4$), can be executed with a high probability of success on IonQ and Quantinuum devices. This suggests that running our proposed algorithm on current NISQ processors, is not only feasible but also a promising endeavor.

In the following sections, Sec.~\ref{app:lfintqc} and Sec.~\ref{app:qdecaysint}, we demonstrate how the QFIAE algorithm is fully or partially implementated on quantum computers and successfully computes integrals that are highly relevant to HEP processes.

\section{Loop Feynman integration on a quantum computer}\label{app:lfintqc}

In this section, we apply the QFIAE method to estimate a selection of loop Feynman integrals. While quantum algorithms have been previously applied to loop integrals~\cite{Ramirez-Uribe:2021ubp,Clemente:2022nll,Ramirez-Uribe:2024wua,Ochoa-Oregon:2025opz} to unveil their causal structure, and quantum integration methods have been employed in elementary particle physics processes at tree level~\cite{AGLIARDI2022137228}, this represents the first application of a quantum algorithm that computes Feynman loop integrals. This achievement holds great significance as these integrals present both technical and conceptual challenges. From a technical standpoint, they involve cumbersome multi-dimensional functions with UV, IR and threshold singularities. In terms of their physical importance, loop Feynman integrals serve as fundamental components in precision physics. Therefore, developing an efficient method to calculate them is crucial for facilitating rapid computations that have the potential to drive new discoveries and enhance our understanding of the fundamental principles governing the universe. 
\subsection{QFIAE implementation}\label{subsec:qfiae_implementation}
In this subsection we will explain the details of how we will implement the QFIAE method introduced in Sec.\ref{app:qfiae} in hardware and simulators.
\subsubsection{Simulators implementation}\label{subsubsec:simulators_imp}

At this point, we can implement the QFIAE algorithm using two different simulation frameworks. \texttt{Pennylane}~\cite{pennylane} will be used for QNN implementation, whereas \texttt{Qibo}~\cite{qibo_paper} will be used for applying  IQAE to the Fourier series.

In particular, the linear Ansatz corresponding to each layer $\mathcal{L}_{LA}^{(l)}(\vec{x},\vec{\theta})$ for training the QNN to fit a $M$-dimensional function is the following \cite{Casas:2023ure}:
\begin{equation}
    \mathcal{L}_{LA}^{(l)}(\vec{x},\vec{\theta})=\prod_{i=1}^M\mathcal{S}(x_i)\mathcal{A}_i^{(l)}(\vec{\theta}_{l,i})~,
    \label{eq:linear_ansatz}
\end{equation}
where $\mathcal{S}(x_i)$ and $\mathcal{A}_i^{(l)}$ are chosen as:
\begin{equation}
    \mathcal{S}(x_i)=R_z(x_i)~, \quad \quad \mathcal{A}_i^{(l)}(\vec{\theta}_{l,i})=R_z(\theta_{l,i,1})R_y(\theta_{l,i,2})R_z(\theta_{l,i,3})~.
    \label{eq:SandA}
\end{equation}

Once the Fourier coefficients are obtained from the QNN, we implement the IQAE algorithm. To uphold the claimed quantum advantage provided by Grover's amplitude amplification, certain conditions must be fulfilled. First, the probability distribution of the functions to be integrated should be encodable into a shallow quantum circuit. In view of this requirement, we will use the distribution $p(x_i)=1/2^n$ generated by applying an $n$-dimensional Hadamard gate, denoted as $\mathcal{H}^{\otimes n}$, which corresponds to a quantum circuit of depth 1. The second condition is that the target function has to be encodable with a minimum number of quantum arithmetic operations. That will be achieved selecting the target function as a $\sin^2(x_i)$ to be integrated in $[x_{i,min},x_{i,max}]$. Then the integrals of the Fourier terms are obtained from the integral of the sine function.

\begin{figure}[th]
       \centering
       \begin{subfigure}[b]{0.49\textwidth}
       \centering
           \includegraphics[width=.99\linewidth]{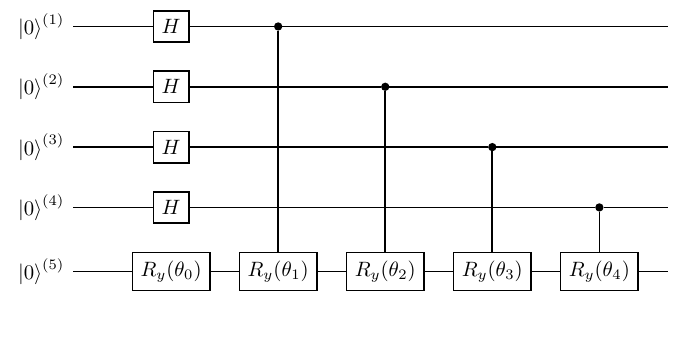}  
           \caption{Amplitude operator $\mathcal{A}$}
       \end{subfigure}
       \hfill
         \begin{subfigure}[b]{0.49\textwidth}
         \centering
           \includegraphics[width=.99\linewidth]{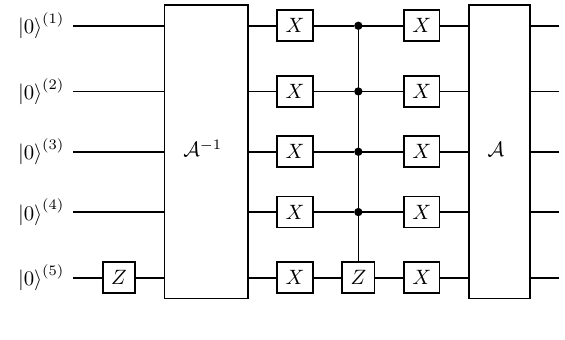} 
           \caption{Amplification operator $\mathcal{Q}$}
       \end{subfigure}
        \caption{Quantum circuits of the operators for the IQAE component of the QFIAE algorithm.
        }
        \label{fig:qaecircuits}
\end{figure}

Under these considerations and choosing $n_{qubits}=5$, the quantum circuits corresponding to the $\mathcal{A}$ and $\mathcal{Q}$ operators are shown in Fig.~\ref{fig:qaecircuits}. Note that in Fig.~\ref{fig:qaecircuits}(a) the rotation angles encode the information about the limits of integration $x_{i,min}$ and $x_{i,max}$. In particular, they are defined as:
\beq
\theta_0=(x_{max} - x_{min}) / 2 ^ n + 2x_{min}, \quad \theta_i=2^{(i+ 1)} ( x_{max} - x_{min}) / 2 ^ n, \quad n=n_{qubits}.
\eeq
For more information, a tutorial on QFIAE implementation can be found at \cite{tutorial_qfiae}.

\subsubsection{Hardware implementation}
\label{subsubsec:hardware_imp}

Although the implementation of Monte Carlo integration in quantum simulators is of great interest for proof of concept purposes, the potential quantum advantage will only be materialized when the quantum algorithm is run on a real quantum device. To this aim, we have addressed the challenge of implementing an end-to-end quantum integrator method into two different quantum devices.

First, the QNN has been trained using an updated version of the Adam gradient descent method first presented in~\cite{Robbiati:2022dkg} and recently improved in~\cite{rtqem}. In this new version of the Adam algorithm, the authors propose a Real-Time Quantum Error Mitigation~(RTQEM) procedure, that allows to mitigate the noise in the QNN parameters during training. We use the full-stack \texttt{Qibo}~\cite{qibo_paper} framework. The high-level algorithm has been written using \texttt{Qibo}, while \texttt{Qibolab}~\cite{qibolab} and \texttt{Qibocal}~\cite{qibocal} are used to respectively control and calibrate the 5-qubit superconducting quantum device hosted in the
Quantum Research Centre~(QRC) of the Technology Innovation Institute~(TII).

In this case, a more hardware-friendly linear Ansatz has been chosen to construct the QNN in one qubit. In particular each layer $\mathcal{L}_{LA}^{(l)}(\vec{x},\vec{\theta})$ is defined as:
\begin{equation}
    \mathcal{L}_{LA}^{(l)}(\vec{x},\vec{\theta})=\prod_{i=1}^MR_z(\theta_3 x_i+\kappa\theta_4)R_y(\theta_1 x_i+\theta_2), \quad \textrm{with} 
        \begin{cases}
        \kappa=1 & \text{if } l \text{ is the last layer,} \\
        \kappa=0 & \text{otherwise.}
    \end{cases}
    \label{eq:linear_ansatz_hw}
\end{equation}

On the other hand, the IQAE has been executed using \texttt{Qiskit} \cite{Qiskit} on the IBM Quantum 27-qubits device \textit{ibmq\_mumbai}. To mitigate quantum noise during the execution of this algorithm, we employed a pulse-efficient transpilation technique \cite{Earnest_2021}. This technique effectively reduces the number of two-qubit gate operations by harnessing the hardware-native cross-resonance interaction, potentially leading to a reduction in quantum noise. Furthermore, we also applied two more error mitigation techniques, Dynamical Decoupling (DD) and Zero Noise Extrapolation (ZNE), which are automatized within the \texttt{Qiskit} Runtime Estimator primitive \cite{estimator}.

\subsection{Tadpole loop integrals}\label{subsec:tadpole}

\begin{figure}[h]
       \centering
  \includegraphics[width=.491\linewidth]{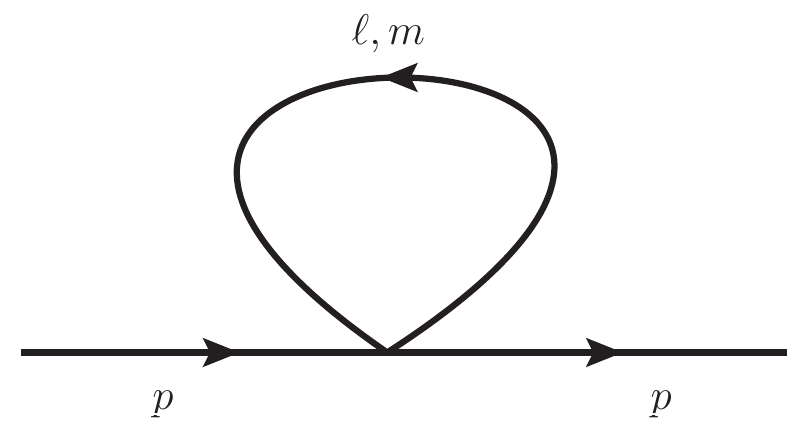}  

        \caption{Tapdole Feynman diagram of an external particle with momentum $p$ and an internal loop momentum $\ell$ with mass $m$. }
        \label{fig:tadpole_draw}
\end{figure}

Let us start with the simplest one-loop Feynman integral, a tadpole-like, see Fig. \ref{fig:tadpole_draw}, where the Feynman propagator is raised to a given power, 
\beq
{\cal A}^{(1)}_a(m) = \int_\ell \left(G_F(\ell) \right)^a = 
\int_\ell \frac{1}{\left(\ell^2-m^2+\ii \right)^a}~, \qquad a\in \mathbb{N}~.
\label{tadpole}
\eeq

For $a\in \{1,2\}$, this integral is singular in the UV. So, we renormalize it by introducing a local UV counterterm 
\beq
{\cal A}^{(1,\r)}_a(m;\mu_\uv) =
 {\cal A}^{(1)}_a(m) - {\cal A}^{(1)}_{\uv,a}(\mu_\uv)~,
\qquad a=\{1,2\}~,
\label{tadpoleR}
\eeq
where $\mu_\uv$ is the renormalization scale.
Regarding the integration measure, we have in $d$ spacetime dimensions
\beq
\int_{\ell} \equiv -\imath \, \mu^{4-d} \int \frac{d^d \ell}{(2\pi)^d}~.
\eeq

We work with the LTD representation of the integrals in~\Eq{tadpole} and~\Eq{tadpoleR}
\beq
{\cal A}^{(1)}_a(m) = \int_{\lb} \frac{c_a}{(2 \qon{1})^{2a-1}}~, \qquad
c_a= \{-1,2,-6,20,-70, \ldots \}~,
\eeq
where the on-shell energy is $\qon{1}=\sqrt{\lb^2+m^2-\ii}$,
and $\qon{\uv}=\sqrt{\lb^2+\mu_\uv^2-\ii}$ for the UV counterterm.
The UV counterterms are 
\beq
{\cal A}^{(1)}_{\uv,1}(\mu_\uv) = \int_{\lb} 
\frac{c_1}{2 \qon{\uv}}\left(1+ \frac{\mu_\uv^2-m^2}{2 (\qon{\uv})^{2}} \right)~,  
\label{eq:A1_5}
\eeq
\beq
{\cal A}^{(1)}_{\uv,2}(\mu_\uv) 
= \int_{\lb} \frac{c_2}{(2 \qon{\uv})^{3}}~.
\label{eq:A1UV2}
\eeq
Notice that ${\cal A}^{(1)}_{1}(m)$  is quadratically singular in the UV, so we need to subtract up to logarithmic order. Therefore, the extra term in the UV expansion in \Eq{eq:A1_5}. 
The integration measure involves now only the spatial components of the loop momentum. In $d=4-2\epsilon$ spacetime dimensions
\beq
\int_{\lb} \equiv \mu^{4-d} \int \frac{d^{d-1} \ell}{(2\pi)^{d-1}}
= \frac{\mu^{2\epsilon}}{(2\pi)^{3-2\epsilon}}\int (\lb^2)^{1-\epsilon} d|\lb| d\Omega^{(2-2\epsilon)}~.
\eeq
As the integrals we consider are renormalized, we can fix the spacetime dimensions to $d=4$, or $\epsilon=0$. The integral does not depend on the solid angle, so the angular integration is straightforward, $\Omega^{(2)}=4\pi$, and we are left with an integral in one variable, the modulus of the loop three-momentum. 

This change of variable remaps the integration domain to the interval $[0,1)$~,
\beq
|\lb|= \frac{m \,z}{1-z}~, \qquad z=[0,1)~.
\label{eq:changevar_5}
\eeq
These tadpole integrals can be computed using the QFIAE method. The integrals that are particularly interesting are those with $a=1,2$, since we have to deal with UV singularities by introducing appropriate UV counterterms. The corresponding analytic integrated expressions are 
\beq
{\cal A}^{(1,\r)}_1(m;\mu_\uv) = \frac{1}{16\pi^2} 
\left( - m^2 \ln{\frac{m^2}{\mu_\uv^2}} + m^2 - \mu_\uv^2
\right)~, 
\eeq
\beq
{\cal A}^{(1,\r)}_2(m;\mu_\uv) = \frac{1}{16\pi^2} 
\left( - \ln{\frac{m^2}{\mu_\uv^2}}
\right)~.
\eeq

\subsubsection{Tadpole integration on simulator}
\begin{figure}[!ht]
       \centering
  \includegraphics[width=.49\linewidth]{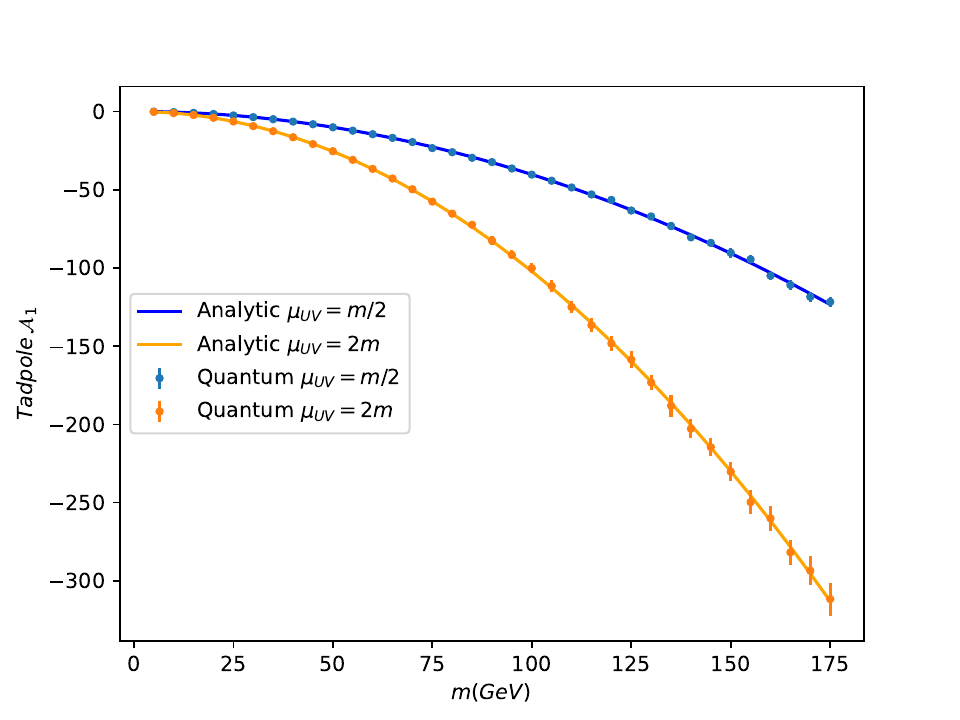}  
  \includegraphics[width=.49\linewidth]{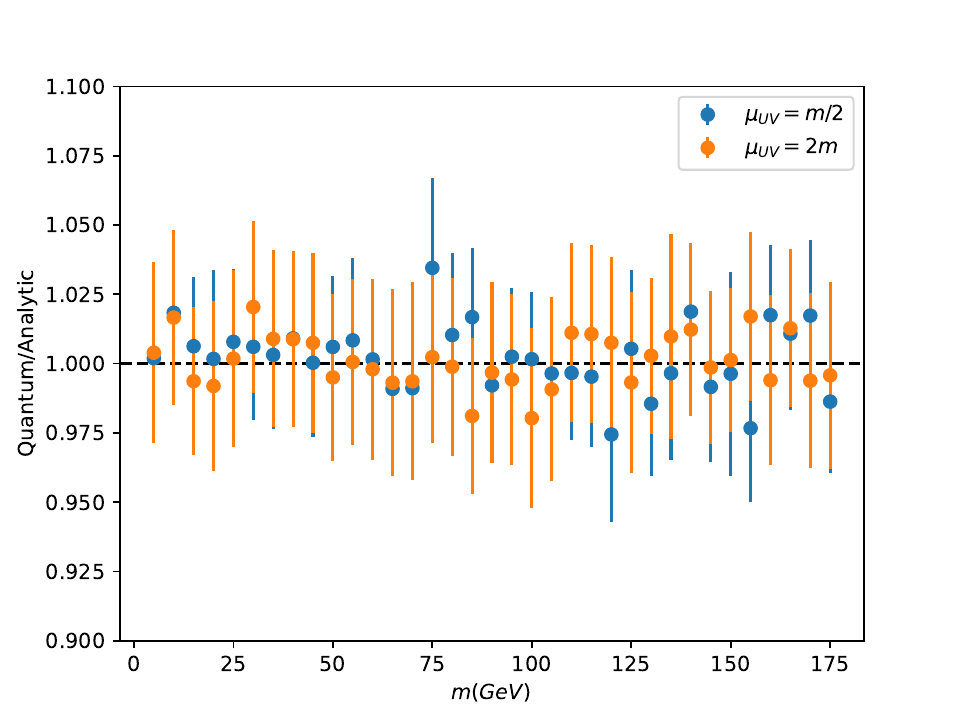}  
        \caption{Quantum integration using QFIAE of the renormalized tadpole integral ${\cal A}^{(1,\r)}_{1}(m; \mu_\uv)$ as a function of the mass $m$ and the renormalization scale $\mu_\uv$ (left) and comparison with the analytical result (right). The parameters used in the quantum implementation are: $max\_steps=300$, $step\_size=0.06$, $layers=n_{Fourier}=5$ for the quantum Fourier part and $n_{qubits}=5$, $n_{shots}=10^3$, $\epsilon=0.01$, $\alpha=0.05$ for the IQAE part. 
        }
        \label{fig:a1tadpole}
\end{figure}

Applying the QFIAE on the quantum simulators of \texttt{Pennylane} and \texttt{Qibo} for the QNN and the IQAE respectively, we obtain the results presented in Fig. \ref{fig:a1tadpole}.

The Fig.~\ref{fig:a1tadpole}~(left) shows the result of the quantum integration with QFIAE of the tadpole loop integral with $a=1$ for different values of the mass $m$ and the renormalization scale $\mu_\uv$. We do not show results for $a=2$ because the integral is constant when we fix $\mu_{\uv}$ as a multiple of $m$.

The uncertainties of the integrals have been calculated as the quadratic sum of the individual uncertainties provided by the IQAE algorithm when integrating every trigonometric piece separately. Regarding the uncertainty of each integral, it has been set up beforehand to be smaller than $\epsilon=0.01$ for each contribution with a confidence interval of $\alpha=0.05$. These uncertainties are statistical. There will be also systematic errors in each integral due to the quantum estimation of the Fourier coefficients. In Fig.~\ref{fig:a1tadpole}~(right) the accuracy of the results obtained in comparison to the analytical expression of the integral is depicted. Showing a relatively small deviation ($\sim 2$\%), which is in concordance with the uncertainties we have estimated, from the ideal scenario $\mathcal{A}_{quantum}=\mathcal{A}_{analytical}$. 

\subsubsection{Tadpole integration on hardware}
Since the results obtained in the simulation are quite successful, we can go one step further and run QFIAE on hardware to compute for the first time a loop Feynman integral into a quantum computer.

\begin{figure}[ht!]
  \begin{subfigure}{0.49\textwidth} 
    \centering
    \includegraphics[width=0.8\linewidth]{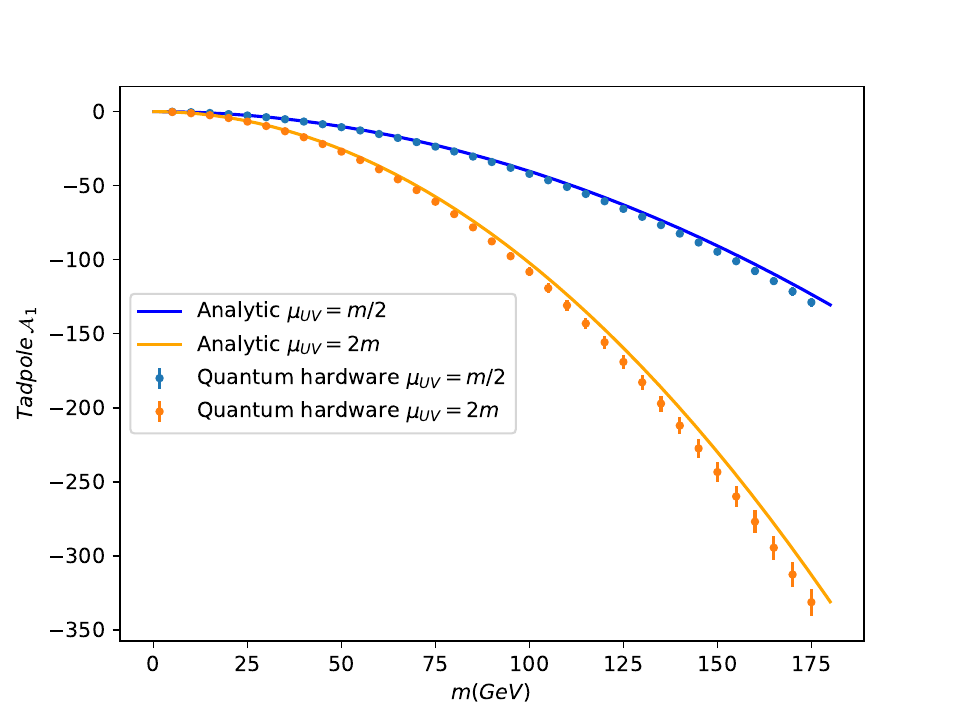}
  \end{subfigure}
  \begin{subfigure}{0.49\textwidth} 
    \centering
    \begin{tabular}{ |p{1.65cm}|p{1.85cm}|p{1.85cm} | }
 \hline
 \centering  & \centering \footnotesize{$\varepsilon$ (\%) \\ $\mu_\uv=m/2$} &  \centering \footnotesize{$\varepsilon$ (\%) \\ $\mu_\uv=2m$}  \cr
 \hline
  \centering \footnotesize{CF \& CI} & \centering 0.0 &  \centering 0.1  \cr 
 \hline
    \centering \footnotesize{CF \& QI} & \centering 5.2 &  \centering 2.8  \cr
\hline
\centering \footnotesize{QF \& CI} & \centering 9.4 &  \centering 3.3  \cr  
\hline
\hline 
\centering \footnotesize{\bf{QF \& QI}} & \centering \bf{4.4} &  \centering \bf{5.9}  \cr
\hline 

\end{tabular}
\vspace{0.35cm}
\label{table:hw_tadpole}
  \end{subfigure}
  \caption{(Left) Quantum integration on \texttt{Qibo} and IBM Quantum hardware of the renormalized tadpole integral ${\cal A}^{(1,\r)}_{1}(p,m;\mu_\uv)$ as a function of the mass $m$ and the renormalization scale $\mu_{UV}$.(Right) Deviations from the analytical value of the integral when the Fourier (F) decomposition or the IQAE integration (I) are performed in a classical (C) or a quantum (Q) computer.}
  \label{fig:hw_tadpole}
\end{figure}

To this aim, we have implemented the QFIAE on hardware. In particular, for the QNN part we have trained the model with considerably lighter hyperparameters than for simulation $n_{data}=15$ points and we have used $layers=3$, $step\_size=0.07$, $max\_steps=60$. However the results obtained show that with these values, the QNN is sufficiently expressive to fit the function in the integration domain [0,1]. In particular, since the access to the hardware is limited, the way to proceed here has been to fit the tadpole integrand and extract its Fourier series for an arbitrary value of the mass, $m_1=5$ and for $\mu_{UV}=m_1/2,2m_1$ and then to obtain the Fourier coefficients for other values as a simple multiplication by the quotient $m^2/m_1^2$. Since if we set $\mu_{UV}=km_1$, where $k$ is a real number, we obtain the following relation,  ${\cal A}^{(1,\r)}_1(m)/{\cal A}^{(1,\r)}_1(m_1)=(m/m_1)^2$, that allows us to extrapolate the results for the entire range of mass [5,175] GeV. The results in Fig. \ref{fig:hw_tadpole} show a relatively small deviation from the analytical value for both $\mu_{UV}=m/2$ and $\mu_{UV}=2m$. This represents a noteworthy achievement as it is the first application of an end-to-end quantum algorithm executed on a quantum computer for estimating loop integrals and the results obtained are satisfactory. Also, it further demonstrates the effectiveness of the error mitigation techniques applied, in particular: pulse transpilation, which involves decomposing circuits into hardware-native gates (as detailed in Sec.\ref{subsubsec:hardware_imp}). While this technique has shown promise in the context of Variational Quantum Circuits~(VQC)~\cite{Melo2023pulseefficient}, its potential for fault-tolerant applications remains largely unexplored. The positive results in Fig.~\ref{fig:hw_tadpole} suggest that this work could serve as a valuable starting point for extending pulse transpilation to fault-tolerant quantum algorithms.

\subsection{Bubble loop integral}\label{subsec:bubble}

\begin{figure}[h]
       \centering
  \includegraphics[width=.491\linewidth]{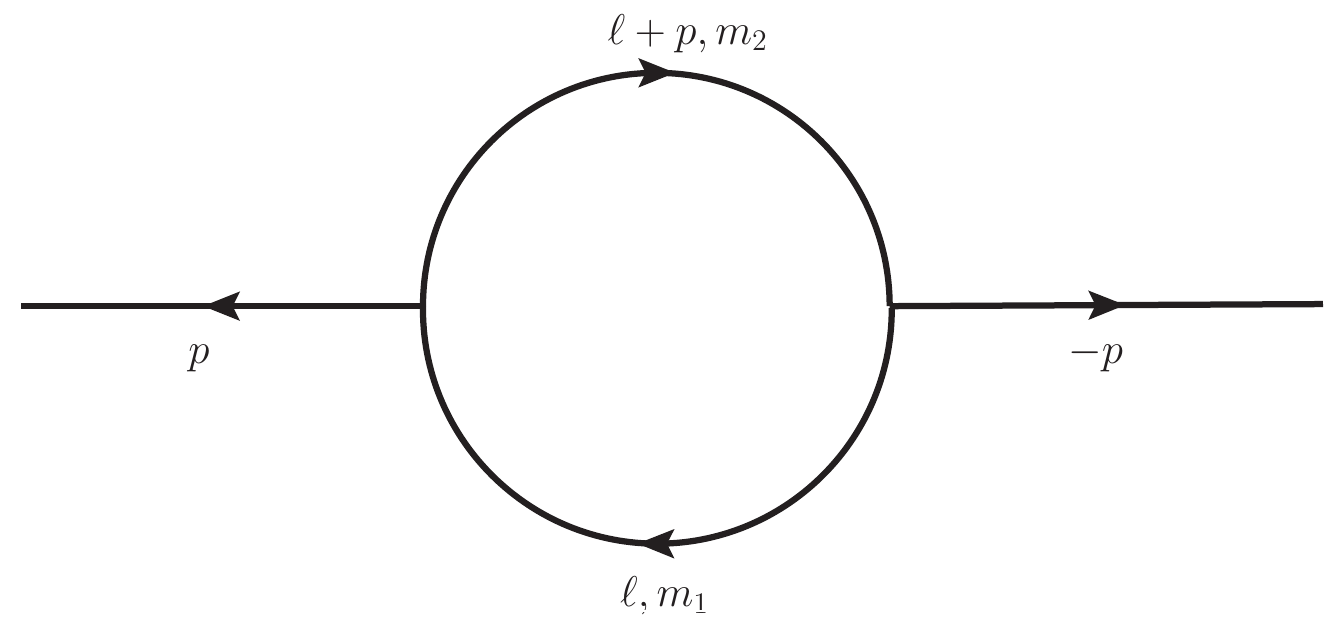}  

        \caption{Bubble Feynman diagram with external momentum $p$ and internal loop momenta $q_1=\ell$ and $q_2=\ell+p$ with masses $m_1$ and $m_2$, respectively. }
        \label{fig:bubble_draw}
\end{figure}

Let us now consider the renormalized one-loop integral corresponding to a bubble topology, see Fig.~\ref{fig:bubble_draw},

\beq
{\cal A}^{(1,\r)}(p,m_1,m_2;\mu_\uv) = {\cal A}^{(1)}(p,m_1,m_2) - {\cal A}^{(1)}_{\uv,2}(\mu_\uv)~,
\label{bubbleR}
\eeq
where
\beq
{\cal A}^{(1)}_2(p,m_1,m_2) = \int_\ell \prod_{i=1}^2 G_F(q_i) = 
\int_\ell \frac{1}{\left(\ell^2-m_1^2+\ii \right)\left((\ell+p)^2-m_2^2+\ii \right)}~.
\label{bubble}
\eeq

The UV counterterm is the same as for the tadpole integral with $a=2$ in~\Eq{eq:A1UV2}. The analytic expression with which we will compare the result of the numerical integration, for equal masses, is
\beq
{\cal A}^{(1,\r)}_2(p,m,m;\mu_\uv) = 
\frac{1}{16\pi^2} 
\left( - \ln{\frac{m^2}{\mu_\uv^2}} + \beta \, \ln{\frac{\beta-1}{\beta+1}} + 2 \right)~, \quad
\beta = \sqrt{1-\frac{4m^2}{p^2+\ii}}~.
\eeq

The LTD representation of \Eq{bubble} is~\cite{Aguilera-Verdugo:2020set}
\beq
{\cal A}^{(1)}_2(p,m_1,m_2) =  \int_{\lb} \frac{1}{x_2} \left( \frac{1}{\lambda^+} 
+ \frac{1}{\lambda^-} \right)~,
\eeq
where 
\beq
x_2 = \prod_{i=1,2} 2 \qon{i}~, \qquad \lambda^\pm = \sum_{i=1,2} \qon{i} \pm p_0~.
\eeq
The on-shell energies are given by 
\beq
\qon{i} = \sqrt{\lb^2+m_i^2-\ii}~, \qquad i=\{1,2\}~,
\label{eq:qon}
\eeq
assuming the external momentum has vanishing spatial components, $p=(p_0,{\bf 0})$.

If $p_0^2< (m_1+m_2)^2$ the integral is purely real. Otherwise, it gets an 
imaginary contribution from the unitary threshold singularity at $\lambda^-\to 0$, assuming $p_0>0$. To deal with this threshold singularity, it is convenient to introduce the following contour deformation in the complex plane, which will smooth the behavior of the function in the vicinity of the pole without altering the result of the integral,
\beq
\lb = \lb' \left( 1 - \frac{\imath \, \delta}{\sqrt{\lb'^2}} \right)~,
\label{eq:contour}
\eeq
where $\delta$ is a parameter with dimensions of mass that controls the size of the deformation. 
With this transformation, the modulus of the loop momentum and the Jacobian of the transformation read
\beq
\lb^2 = \left( \sqrt{\lb'^2}- \imath \, \delta \right)^2~,
\qquad \left| \frac{d^3\lb}{d^3\lb'}\right| = 
\frac{\lb^2}{\lb'^2}~,
\eeq
which is consistent with the $\ii$ prescription 
of the on-shell energies. 

\begin{figure}[h!]
       \centering
  \includegraphics[width=.49\linewidth]{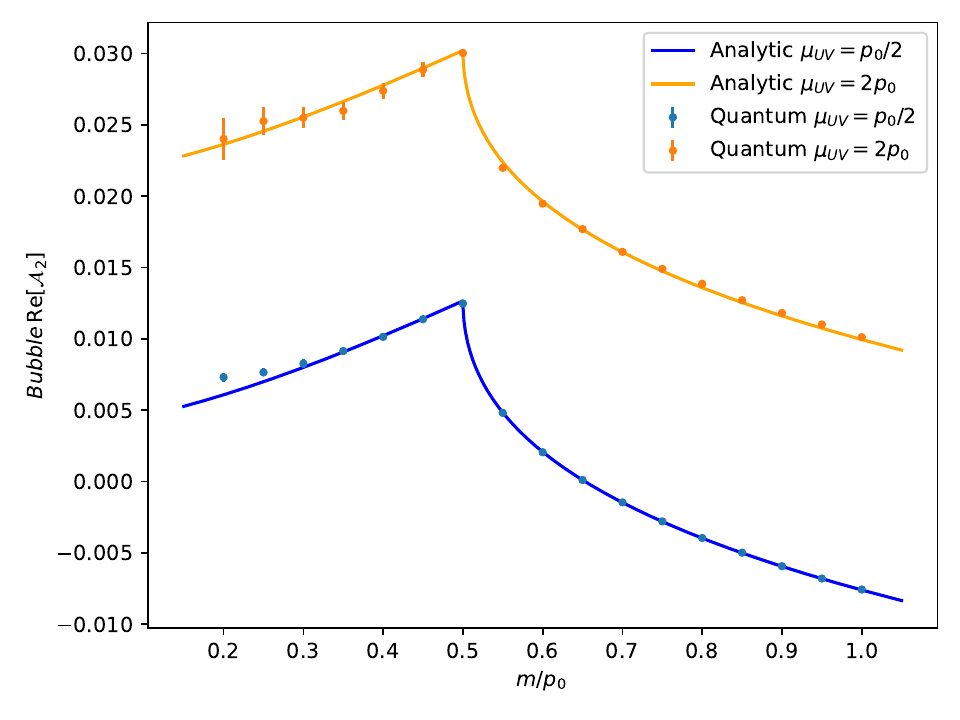}  
  \includegraphics[width=.49\linewidth]{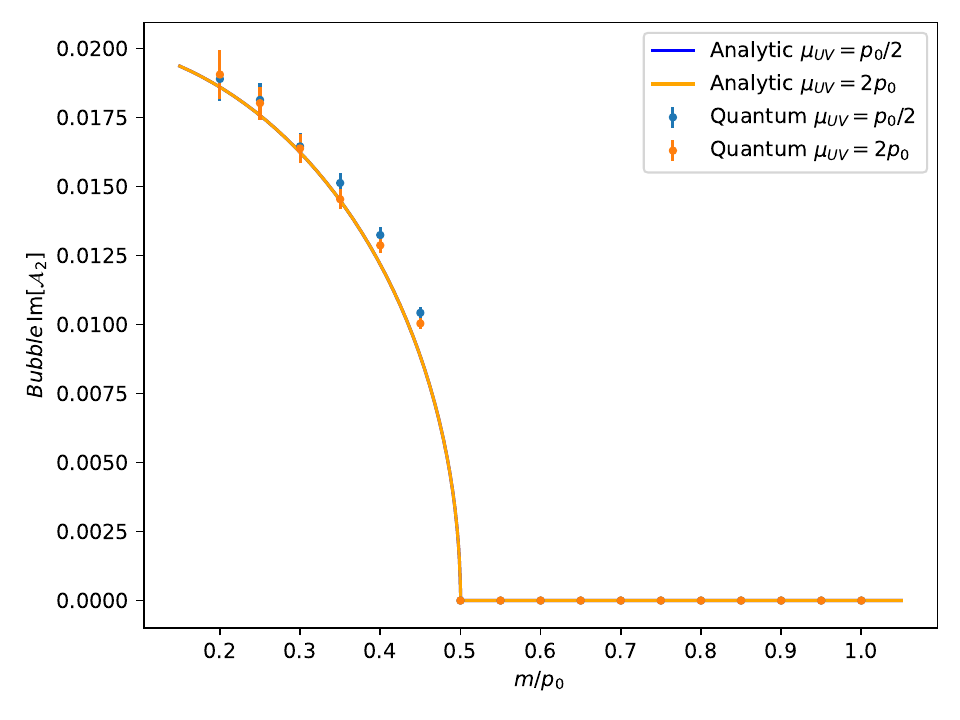}  

        \caption{Quantum integration of the real (left) and imaginary (right) part of the renormalized bubble integral ${\cal A}^{(1,\r)}_{2}(p,m,m;\mu_\uv)$ as a function of the ratio of the mass $m$ to the energy component of the external momentum set at $p_0 = 100$~GeV, and the renormalization scale $\mu_{\uv}$. For the values $m/p_0\geq 0.3$, the parameters used in the quantum implementation are: $max\_steps=200$, $step\_size=0.02$, whereas for $m/p_0< 0.3$ more precision was needed, because of the threshold singularities, so the parameters used are: $max\_steps=450$, $step\_size=0.095$. For both cases, some parameters are equal: $layers=n_{Fourier}=10$, $n_{qubits}=1$ for the quantum Fourier part and $n_{qubits}=5$, $n_{shots}=10^3$,$\epsilon=0.001$, $\alpha=0.05$ for the IQAE part.}
        \label{fig:a1bubble}
\end{figure}

To compute this bubble integral, we 
use the change of variable introduced in~\Eq{eq:changevar_5}. 
Also, we consider $m_i=m$, for $i=1,2$. 
With these considerations, the way to proceed to estimate the integrals is using two different QNN to fit separately the real and the imaginary parts of the bubble integrand and then integrating each Fourier series using IQAE. The results obtained are depicted in Fig.~\ref{fig:a1bubble}, where the real and imaginary parts are displayed on the left and right plots, respectively.
It is important to mention that the parameter $\delta$ has been set to $\delta=0.21 \sqrt{s} $, to enhance the smoothness of the function around the threshold singularity. This choice helps the QNN in its task of fitting the function more effectively. The uncertainties are calculated as in the previous section. It is worth mentioning that uncertainties in the region $m/p_0 < 0.3$, appear to be larger compared to the rest. This is explained by the Fourier coefficients for low frequencies being larger than in the region $m/p_0 \geq 0.3$. Hence the statistical uncertainties of the integrals of the trigonometric functions for low frequencies are intrinsically larger since the integrals we are estimating are also larger, i.e have a larger weight, which is the Fourier coefficient. Another interesting point about these results is that for the mentioned region the QNN fits the target function with a slight drop in performance, hence there seems to be a correlation between the QNN struggling to fit a function and the coefficients of the lower frequencies terms being larger.

Despite this, it is remarkable that the quantum integration values are in agreement with the analytical values within uncertainties. This constitutes another significant achievement since we have successfully circumvented the threshold singularity while applying a quantum algorithm to obtain an estimation of the bubble integral.

\subsubsection{Bubble tensor loop integral}
We now consider the following tensor bubble integral:
\beq
{\cal A}^{(1)}_{2,T}(p,m_1,m_2) = \int_\ell (\ell \cdot p)^2 \prod_{i=1}^2 G_F(q_i) = 
\int_\ell \frac{(\ell \cdot p)^2}{\left(\ell^2-m_1^2+\ii \right)\left((\ell+p)^2-m_2^2+\ii \right)}~.
\label{bubbletensor}
\eeq

\begin{figure}[h]
       \centering
  \includegraphics[width=.49\linewidth]{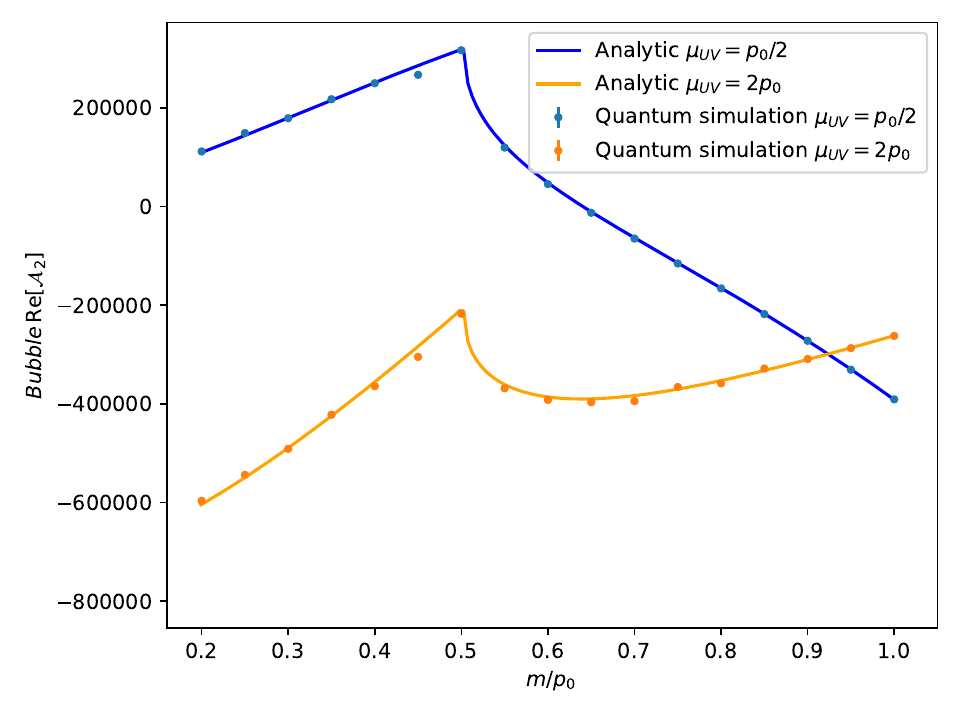}  
  \includegraphics[width=.49\linewidth]{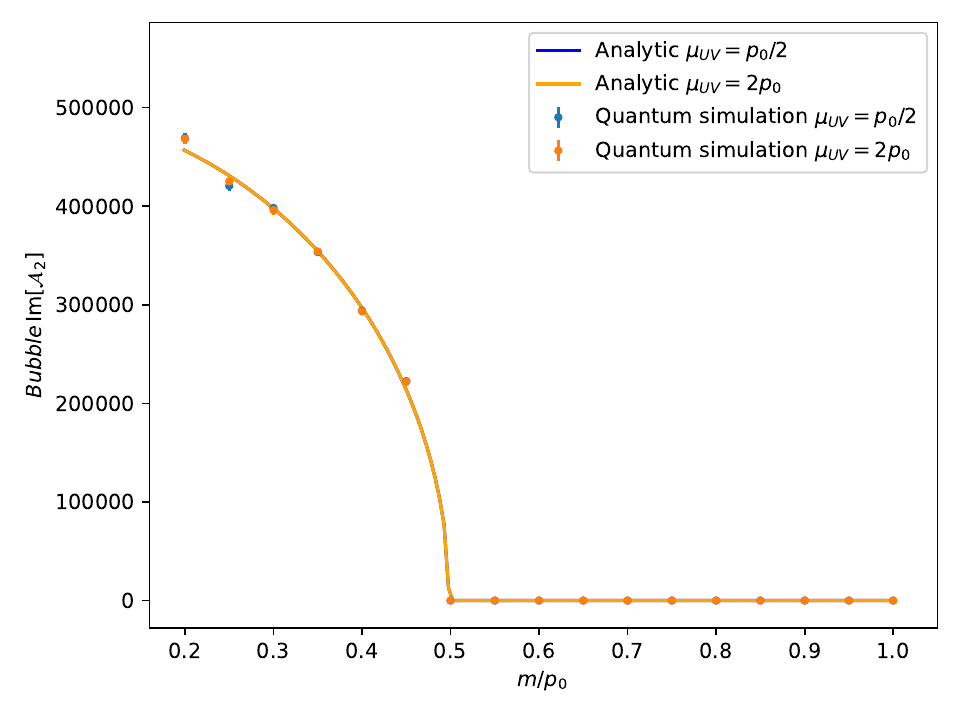}  

        \caption{Quantum integration of the real (left) and imaginary (right) part of the renormalized tensor bubble integral ${\cal A}^{(1,\r)}_{2,T}(p,m,m;\mu_\uv)$ as a function of the ratio of the mass $m$ to the energy component of the external momentum set at $p_0 = 100$~GeV, and the renormalization scale $\mu_{\uv}$.  The parameters used in the quantum implementation are: $max\_steps=5000$, $step\_size=0.0001$, $layers=n_{Fourier}=8$, $n_{qubits}=2$ for the quantum Fourier part and $n_{qubits}=5$, $n_{shots}=10^3$,$\epsilon=0.001$, $\alpha=0.05$ for the IQAE part.}
        \label{fig:tbubble}
\end{figure}

Proceeding in a similar way that with the scalar bubble integral we obtain the results presented in Fig. \ref{fig:tbubble}

\subsection{Triangle loop integral}\label{subsec:triangle}

\begin{figure}[h]
       \centering
  \includegraphics[width=.491\linewidth]{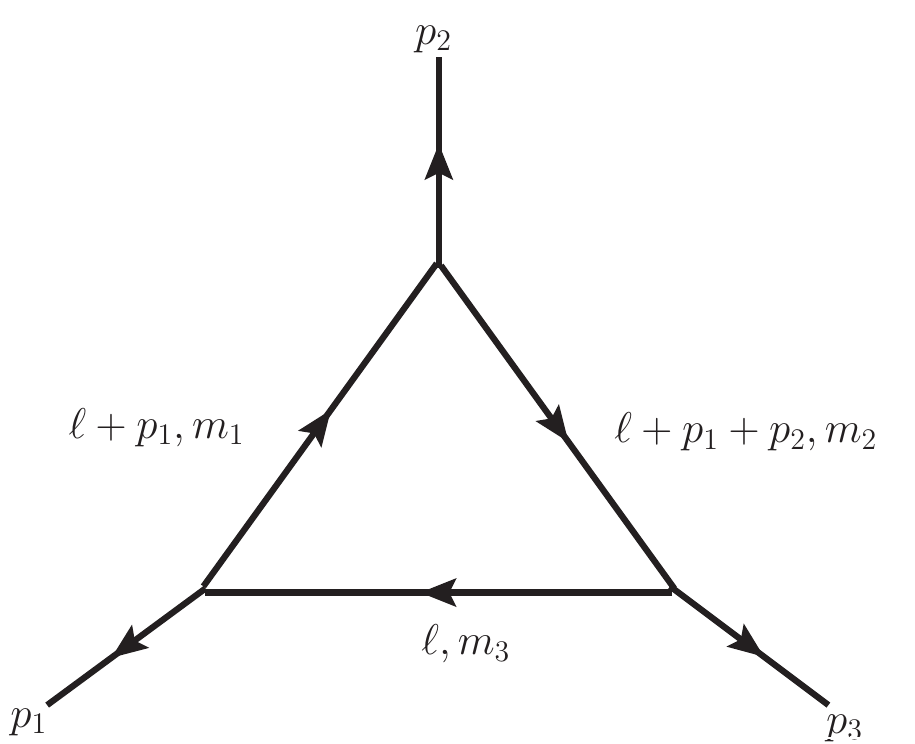}  
        \caption{Triangle Feynman loop diagram. 
        The external particles, with momentum $\{p_i\}_{i=1,2,3}$, are considered massless. The internal loop momenta are  $q_1=\ell+p_1$, $q_2=\ell+p_1+p_2$, and $q_3=\ell$ with masses $m_1$, $m_2$, $m_3$ respectively.
        }
        \label{fig:triangle_draw}
\end{figure}

We now consider the one-loop three-point function, which corresponds to a triangle loop topology, see Fig.~\ref{fig:triangle_draw}, 
\beq
{\cal A}^{(1)}_3(p_1,p_2,m_1,m_2,m_3) = \int_\ell \prod_{i=1}^3 G_F(q_i)~,
\label{triangle}
\eeq
where $q_1=\ell+p_1$, $q_2=\ell+p_{12}$ and $q_3=\ell$, with $p_{12}=p_1+p_2$, and $p_{12}^2=s$. 
This integral is UV finite and does not require any renormalization. The analytic expression for the integrated result reads 
\beq
{\cal A}^{(1)}_3(p_1,p_2,m,m,m) = 
- \frac{1}{16\pi^2 \, s}  
\left[ 
{\rm Li}_2 \left( {\frac{2}{1-\beta}} \right) +
{\rm Li}_2 \left( {\frac{2}{1+\beta}} \right)
\right]~, \qquad
\beta = \sqrt{1-\frac{4m^2}{s+\ii}}~,
\eeq

The causal LTD representation is given by~\cite{Aguilera-Verdugo:2020kzc}
\beq
{\cal A}^{(1)}_3(p_1,p_2,m_1,m_2,m_3) = - \int_{\lb} \frac{1}{x_3} \left( 
\frac{1}{\lambda_{12}^- \lambda_{23}^+} 
+ \frac{1}{\lambda_{23}^+ \lambda_{31}^-} 
+ \frac{1}{\lambda_{31}^- \lambda_{12}^+} 
+ (\lambda_{ij}^+ \leftrightarrow \lambda_{ij}^-) \right)~,
\label{eq:triangle}
\eeq
with $x_3 = \prod_{i=1}^3 2 \qon{i}$, where now $\qon{1}=\sqrt{(\lb+\pb_1)^2+m_1^2-\ii}$, and $ \qon{2}$ and $\qon{3}$ are still described by~\Eq{eq:qon} if we work in the center of mass (cms) frame where $\pb_{12}=0$ and $p_1$ and $p_2$ are back-to-back along the $z$ axis, $\pb_1 = p_{1,0}(0,0,1)$ and $\pb_2 = p_{2,0}(0,0,-1)$. The causal denominators are
\beq
\lambda_{12}^\pm = \qon{1} + \qon{2} \pm p_{2,0}~, \quad
\lambda_{23}^\pm = \qon{2} + \qon{3} \mp p_{12,0}~, \quad
\lambda_{31}^\pm = \qon{3} + \qon{1} \pm p_{1,0}~. 
\eeq

The integration variables in~\Eq{eq:triangle} are the modulus of the loop three-momentum and its polar angle with respect to $p_1$, considering that the azimuthal integration is trivial. This means that the Fourier decomposition is a function of two variables and we have to integrate each of them separately. We apply the change of variable in~\Eq{eq:changevar_5} on the modulus of the loop three-momentum, and the contour deformation in~\Eq{eq:contour} to deal with the unitary threshold singularity at $\lambda_{23}^+ \to 0$, when $m<\sqrt{s}/2$. The contour deformation smoothes the behavior of the integrand over the threshold singularity and therefore significantly improves the quality of the Fourier decomposition. In this case we set $\delta=0.1 \sqrt{s}$.

\begin{figure}[h!]
       \centering
  \includegraphics[width=.49\linewidth]{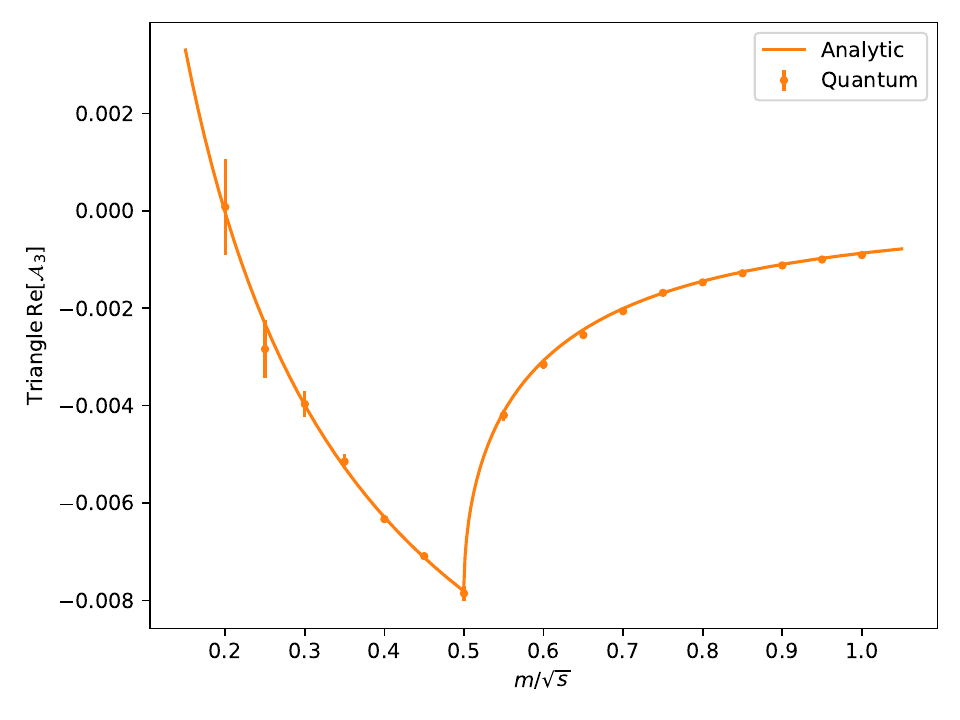}  
  \includegraphics[width=.49\linewidth]{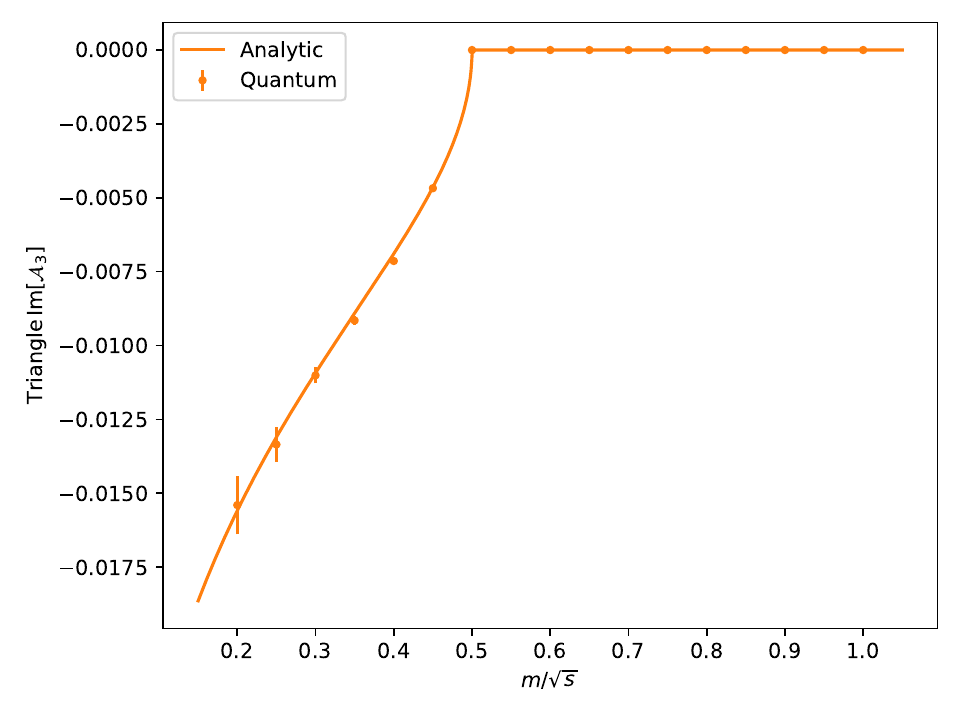}  
        \caption{Quantum integration of the real (left) and imaginary (right) part of the triangle integral ${\cal A}^{(1)}_{3}(p_1,p_2,m,m,m)$ as a function of the ratio of the mass $m$ to the cms energy set at $\sqrt{s}= 2$~GeV and comparison with the analytical result. For the values $ m/\sqrt{s}\geq 0.5$, the parameters used in the quantum implementation are: $max\_steps=400$, $step\_size=0.05$, whereas for $ m/\sqrt{s}\leq 0.5$ more precision was needed, because of the threshold singularity, so the parameters used are:  $max\_steps=450$, $step\_size=0.065$. For both cases, some parameters are equal: $layers=n_{Fourier}=10$, $n_{qubits}=1$ for the QNN and $n_{qubits}=5$, $n_{shots}=10^4$, $\epsilon=0.001$, $\alpha=0.05$ for the IQAE part. 
        }
        \label{fig:a1triangle}
\end{figure}

The obtained results are shown in Figure \ref{fig:a1triangle}~(left) and~(right), illustrating the real and imaginary components, respectively. The uncertainties are also calculated as in the previous sections but are expected to be higher since we are performing a double integral by the IQAE method, and each integration introduces an error. However, the estimated uncertainties align with the deviations observed in the real and imaginary components of the integral. All in all, this represents another noteworthy accomplishment as we have successfully extended for the first time the QFIAE method to a two-dimensional function with a threshold singularity, and utilized a quantum algorithm to approximate the triangle integral with sufficient accuracy.

\subsubsection{Triangle tensor loop integral}
We now consider the following tensor triangle integral:
\beq
{\cal A}^{(1)}_{3,T}(p_1,p_2,m_1,m_2,m_3) = \int_\ell (\ell \cdot p) \prod_{i=1}^3 G_F(q_i)~,
\label{ttriangle}
\eeq

\begin{figure}[h]
       \centering
  \includegraphics[width=.49\linewidth]{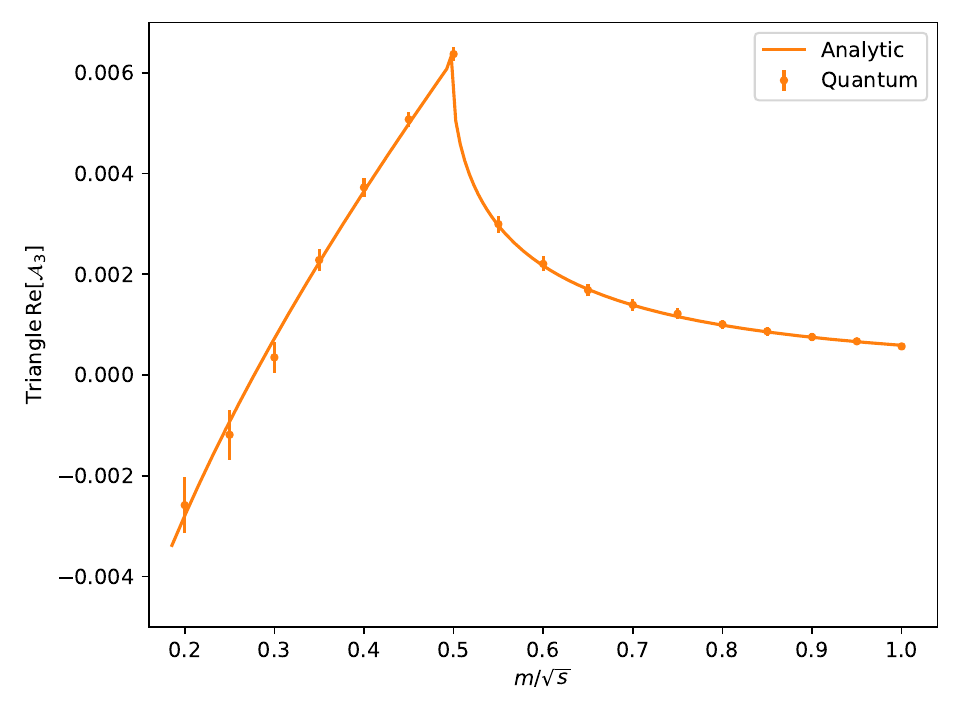}  
  \includegraphics[width=.49\linewidth]{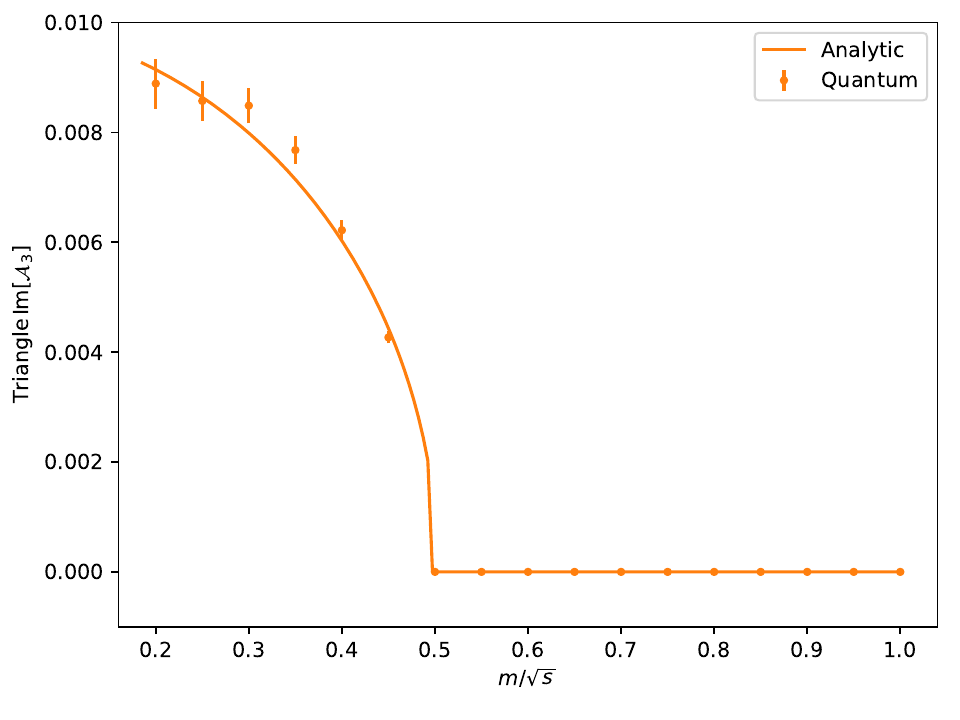}  

        \caption{Quantum integration of the real (left) and imaginary (right) part of the renormalized tensor triangle integral ${\cal A}^{(1,\r)}_{3,T}(p,m,m;\mu_\uv)$ as a function of the ratio of the mass $m$ to the energy component of the external momentum set at $p_0 = 2$~GeV. The parameters used in the quantum implementation are: $max\_steps=15000$, $step\_size=0.001$, $layers=n_{Fourier}=20$, $n_{qubits}=6$ for the quantum Fourier part and $n_{qubits}=5$, $n_{shots}=10^3$,$\epsilon=0.01$, $\alpha=0.05$ for the IQAE part.}
        \label{fig:ttriangle}
\end{figure}

Proceeding in a similar way that with the scalar bubble integral we obtain the results presented in Fig. \ref{fig:ttriangle}

\subsection{Pentagon loop integral}\label{subsec:pentagon}

Five-point integrals are important because they represent a special case that is not trivial to solve (even in the fully massless case). Scalar pentagon integrals have been studied intensively in recent years \cite{Bern:1993kr,Tramontano:2002xn,Denner:2002ii,DelDuca:2009ac,Papadopoulos:2014lla}. Also, cases where the pentagon is free of divergences have been studied many years ago \cite{Melrose:1965kb,tHooft:1978jhc,vanNeerven:1983vr,vanOldenborgh:1989wn}. For some applications to next to leading order (NLO) quantum chromodynamics (QCD) phenomenology, it is possible to reduce the five-point scalar integral to a set of boxes. But beyond NLO, the reduction of pentagon integrals also requires five-point integrals in higher dimensions.

\begin{figure}[ht!]
       \centering

  \includegraphics[width=.491\linewidth]{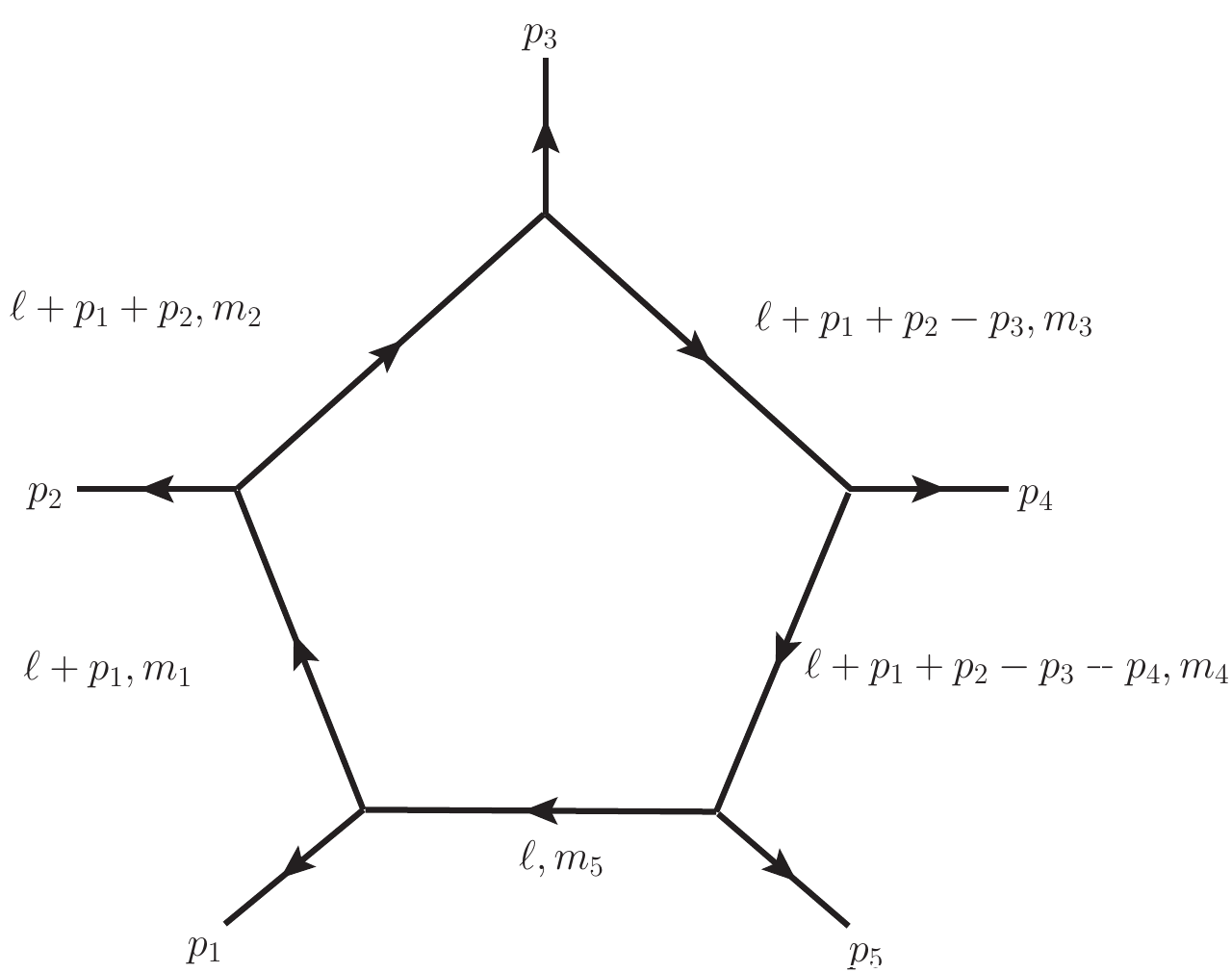}  

        \caption{One-loop pentagon Feynman diagram with massless external momenta $\{p_i\}_{i=1}^5$ particle with momentum $p_1$, $p_2$, $p_3$,$p_4$,$p_5$ and massive internal loop momenta with masses $\{m_i\}_{i=1}^5$. $q_i = \{\ell+p_1, \ell+p_1+p_2, \ell+p_1+p_2-p_3, \ell+p_1+p_2-p_3-p_4, \ell\}$ }
        \label{fig:pentagon_draw}
\end{figure}

If the number of scales (internal and external masses) is not too large, the resulting integral can be obtained in terms of hypergeometric functions~\cite{Bern:1993kr}. However, in cases with several scales, this could not be solved trivially in terms of known functions. For this reason, in the most general case, the use of universal numerical approaches is of interest. In particular, we consider the following one-loop scalar pentagon integral,
\beq
{\cal A}^{(1)}_5(\{p_i,m_i\}_{i=1,\ldots, 5}) = 
\int_\ell 
\prod_{i=1}^5 G_F(q_i)~,
\label{tensorpentagon}
\eeq
which is depicted in Fig. \ref{fig:pentagon_draw}, where
\beq
q_j = \ell + \sum_{i=1}^j p_i~,
\eeq
with $q_5 = \ell$ by momentum conservation. The LTD representation is obtained with {\tt Lotty}~\cite{Bobadilla:2021pvr},
\beq
{\cal A}^{(1)}_5(\{p_i,m_i\}_{i=1,\ldots, 5}) = 
\int_{\lb} 
\frac{1}{x_5}\, \sum \left( L_{ij}^+ L_{kl}^- + L_{ij}^- L_{kl}^+\right)~,
\label{tensorpentagonltd}
\eeq
where $x_5 = \prod_{i=1}^5 2 \qon{i}$ with $\qon{i}=\sqrt{\qb_i^2+m_i^2-\ii}$, and 
\beq
L_{ij}^\pm = \left( \frac{1}{\lambda^\pm_{i}} +
\frac{1}{\lambda^\pm_{j}} \right) \frac{1}{\lambda^\pm_{ij}}~,
\eeq
where the causal denominators are defined as 
\beq
\begin{split}
\lambda^{\pm}_{i} &=\qon{i}+\qon{i+1} \pm p_{i,0}~,  
\\
\lambda^{\pm}_{ij} &= \qon{i}+\qon{j+1} \pm \left(p_{i} + p_{j} \right)_0~, \qquad j = i+1~, \\ 
\lambda^{\pm}_{ij} &= \lambda^{\pm}_{i} + \lambda^{\pm}_{j}~,  \qquad \qquad \qquad \qquad \quad j = i+2~.
\end{split}
\eeq
It is understood that the indices of the on-shell energies are defined cyclically, namely $i=n+1$ with $n=5$ is equivalent to $i=1$. The loop momentum is parametrized in terms of the polar and azimuthal angles, $\lb = |\lb| (\sin \theta \cos \phi, \sin \theta \sin \phi, \cos \theta)$. The integrand in~\Eq{tensorpentagonltd} depends on three independent integration variables.

We consider the specific kinematic configuration P11 defined in \cite{Buchta:2015wna} and compare our results with the numerical values obtained therein. This specific configuration corresponds to:
\beq
\begin{split}
p_1 &= (33.74515, 45.72730, 31.15254, -7.47943)~,\\
p_2 &= (31.36435, -41.50734, 46.47897, 2.04203)~,\\
p_3 &= (4.59005, 17.07010, 32.65403, 41.93628)~,\\
p_4 &= (29.51054, -28.25963, 46.17333, -35.08918)~,\\
m_1 &= m_2 = m_3 = m_4 = m_5 = 5.01213~.
\label{eq:p11buchta}
\end{split}
\eeq

The results presented in Table~\ref{table:pentagon} demonstrate the successful acquisition of the integral result using the quantum method. Although the quantum uncertainty may seem much bigger than the uncertainty estimated by classical methods it could be reduced by adjusting the $\epsilon$ parameter in the IQAE algorithm. This accomplishment is significant as it involves a three-dimensional function integrated through a quantum Monte Carlo approach, potentially paving the way for tackling two-loop or higher-loop integrals in a broader sense, which involve functions with four or more variables.

\begin{table}[h!]
\begin{center}
\begin{tabular}{|p{2cm}|p{3cm}p{4cm}| }
 \hline
 & \centering Method & \centering{Result}  \cr
 \hline
  \centering P11 & \centering FQMCI & $-1.24 (4)\cdot 10^{-13}$  \\
      & \centering LTD$_{\rm class}$ & $-1.24027 (16)\cdot 10^{-13}$  \\
 \hline 
\end{tabular}

\caption{
Quantum integration of a pentagon configuration using the FQMCI method, and comparison with a classical numerical integration \cite{Buchta:2015wna}. With $n_{Fourier}=10$, $n_{qubits}=5$, $n_{shots}=10^4$, $\epsilon=0.001$, $\alpha=0.05$. The uncertainty for the quantum integration has been calculated as in previous sections. }
\label{table:pentagon}
\end{center}
\end{table}
Athough the three-dimensional pentagon integral could also be computed using QFIAE, the linear Ans\"atzes considered in Sec.~\ref{subsec:qfiae_implementation} were not expressive enough to accurately capture the behavior of the pentagon functions. Exploring more expressive Ans\"atzes that ensure efficient trainability for multivariate functions remains an avenue for future work.

\section{Quantum decay rates at second order in perturbation theory}\label{app:qdecaysint}

In this section, we make use of a recently proposed novel approach based on LTD to efficiently recast perturbative theoretical predictions at high-energy colliders, the LTD causal unitary~\cite{Ramirez-Uribe:2024rjg,LTD:2024yrb}, where differential observables, cross sections and decay rates are assembled from the LTD representation of vacuum amplitudes, i.e. scattering amplitudes without external particles. We then combine this new approach of rewriting physical observables in terms of vacuum amplitudes with the QFIAE to perform numerical integration. Therefore, the aim of this section is going one step further and test the performance of QFIAE with physical decay rates at second order in perturbation theory or next-to-leading order (NLO). This involves combining one-loop and tree-level contributions, each individually singular and numerically challenging, though their sum remains finite. The LTD causal unitary approach provides a unified treatment of loop and tree-level terms, leading to relatively flat integrands that are well-suited for numerical integration, particularly through Fourier decomposition.
\subsection{Causal unitary and decay rates at NLO}\label{subsec:causal_unitary}

A vacuum amplitude in the LTD framework, $\ad{\Lambda}$, with $\Lambda$ denoting the number of independent loop four-momenta, is obtained from its Feynman representation by applying the Cauchy residue theorem to integrate out one component of each loop momentum~\cite{Catani:2008xa,Bierenbaum:2010cy}. This is typically done for the energy components, leading to a reformulation where Feynman propagators are replaced by causal propagators of the form~\cite{Aguilera-Verdugo:2020set}
\beq
\frac{1}{\lambda_{i_1 \cdots i_m}} = \left(\sum_{s=1}^m \qon{i_s} \right)^{-1}~,
\label{eq:causalvacuum}
\eeq
where $\qon{i_s} =\sqrt{\qb_{i_s}^2+m_{i_s}^2-\ii}$ represents the on-shell energy of each internal propagator, with $\qb_{i_s}$ being the spatial part of the four-momentum and $m_{i_s}$ its corresponding mass. The numerator of the vacuum amplitude in LTD depends on these on-shell energies as well as the internal masses. The infinitesimal imaginary term in the on-shell energy originates from the prescription used in Feynman propagators. While loop vacuum amplitudes in conventional Feynman representations are defined over a Minkowski space in terms of four-momenta, their LTD counterparts are formulated in an Euclidean space, involving loop three-momenta.

Each causal propagator in \Eq{eq:causalvacuum} corresponds to a subset of internal particles that partitions the vacuum amplitude into two subamplitudes, ensuring that the momenta of all particles within the subset flow in the same direction. Consequently, each term in the vacuum amplitude is a product of causal propagators, maintaining this alignment of momentum flow for shared particles~\cite{Aguilera-Verdugo:2020kzc,Ramirez-Uribe:2020hes,JesusAguilera-Verdugo:2020fsn,Ramirez-Uribe:2022sja,Sborlini:2021owe,TorresBobadilla:2021ivx}. This structure is analogous to the selection of acyclic configurations in a directed graph, as encountered in graph theory~\cite{Ramirez-Uribe:2021ubp,Clemente:2022nll,Ramirez-Uribe:2024wua,Ochoa-Oregon:2025opz}. When a causal propagator becomes singular, all associated particles go on shell. This observation provides a systematic way to generate interference terms in scattering amplitudes with different numbers of final-state particles and loops by taking residues on causal propagators. This is the core idea of LTD causal unitary~\cite{Ramirez-Uribe:2024rjg,LTD:2024yrb}. The vacuum amplitude in LTD acts as a kernel amplitude, which generates all the final states that contribute to a scattering or decay process from all possible residues on causal propagators.

\begin{figure}[ht!]
\begin{center}
\includegraphics[scale=0.38]{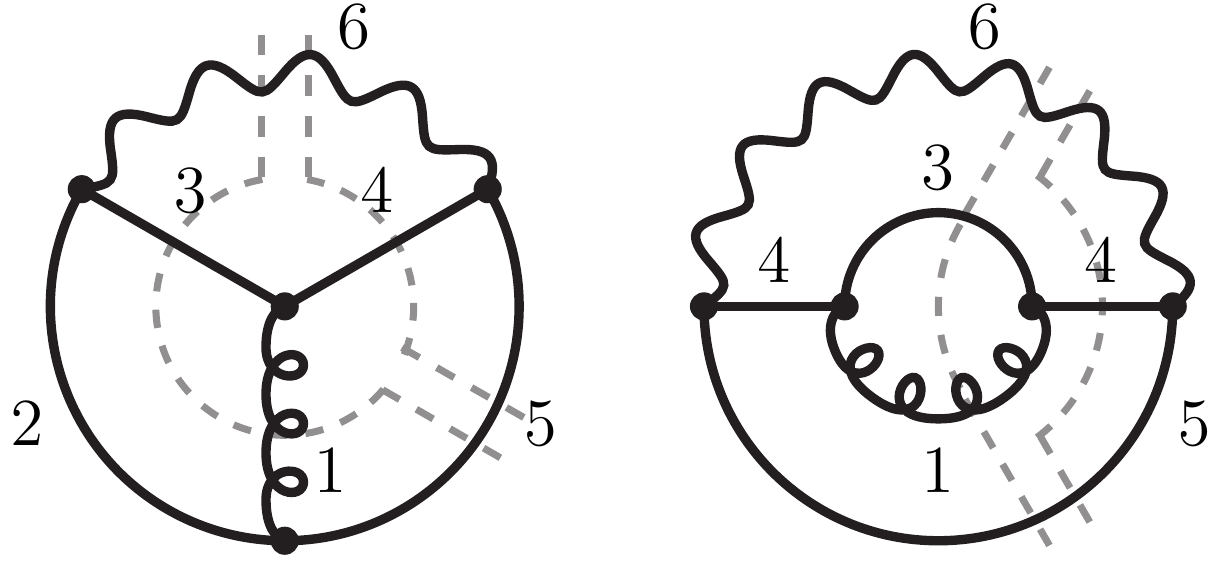}
\caption{Three-loop vacuum diagrams contributing to the decay $\gamma^*\to q \bar q (g)$ at NLO. The gray dashed lines indicate phase-space residues, corresponding to different final states. Similar diagrams contribute to the decays $H\to q\bar q (g)$ and $\Phi\to \phi\phi (\phi)$, where the photon labeled~$6$ is replaced by a Higgs boson or a heavy scalar $\Phi$. In the case of heavy scalar decay, particles~$1$ to $5$ are substituted by light scalars.
\label{fig:qqbar}}
\end{center}
\end{figure}

As benchmark decay rates at NLO, in this section we analyze the decay of a heavy scalar into lighter scalars, as well as the decay of a Higgs boson or an off-shell photon into a pair of massive quarks and antiquarks. These processes have been implemented as a proof of concept for LTD causal unitarity in Ref.~\cite{LTD:2024yrb}, where classical integration techniques were used to compute the total decay rates. A detailed discussion of the expressions utilized in the numerical implementation can be found in Ref.~\cite{LTD:2024yrb}. The vacuum diagrams relevant to the decay process $\gamma^* \to q\bar q (g)$ are depicted in Fig.~\ref{fig:qqbar}. Analogous vacuum diagrams describe the other two decay channels under consideration.

The vacuum diagrams in Fig.~\ref{fig:qqbar} contribute to a vacuum amplitude that, in LTD, depends on the three loop three-momenta $\{\lb_1, \lb_2, \lb_3\}$. The momenta of the internal propagators are given by  
\begin{align*}  
\qb_1 &= \lb_1+\lb_2~, & \qb_2 &= \lb_1+\lb_3~, & \qb_3 &= \lb_1~,  \\  
\qb_4 &= \lb_2~, & \qb_5 &= \lb_2-\lb_3~, & \qb_6 &= \lb_3~,  
\end{align*}  
with on-shell energies $\qon{i} = \sqrt{\qb_i^2+m_i^2-\ii}$. Working in the rest frame of the decaying particle, we set $\lb_3=\boldsymbol{0}$, making the unintegrated decay rate dependent only on $\lb_1$ and $\lb_2$ through the on-shell energies.  

At NLO, the differential decay rate of a particle $a$ is expressed as  
\bea
d\Gamma^{(1)}_{a} &=& \frac{d\Phi_{\lb_1\lb_2}}{2\sqrt{s}} \, \bigg[\Big( \ad{3,a}(456) \, \ps{45\bar 6}  +  \ad{3,a}(1356) \, \ps{135\bar 6}\Big) \nn \\ &+& (5\leftrightarrow 2, 4\leftrightarrow 3)
 \bigg]~,
\label{eq:decayratescalarNLO_5}
\eea
where $\ad{3,a}(456)$ and $\ad{3,a}(1356)$ are phase-space residues of the vacuum amplitude in LTD. These are obtained from the residues of the corresponding causal propagators at $\lambda_{456} \to 0$ and $\lambda_{1356}\to 0$, representing quantum fluctuations at one-loop with two final-state particles and at tree-level with three final-state particles, respectively.  

For the decay of a heavy scalar $\Phi$ into lighter scalars, these phase-space residues take the form  
\begin{align*}  
\ad{3,\Phi}(456) &=  \frac{g_\Phi^{(1)} m_\Phi^2}{x_{12345}}  
\left(L^{13\bar 4}_{23\bar 4 \bar 5, 125} +  
L^{12\bar 5}_{23\bar 4\bar 5, 134} +  
L^{2345}_{134,125} \right)~, \\  
\ad{3,\Phi}(1356) &=  \frac{g_\Phi^{(1)} m_\Phi^2}{x_{135}}  
\left(\frac{1}{\lambda_{13 \bar 4} \lambda_{134} \lambda_{1\bar 2 5} \lambda_{125}} \right)~,  
\end{align*}  
where $g_\Phi^{(1)}$ collects the interaction couplings, $x_{i_1\cdots i_n} = \prod_{s=1}^n 2\qon{i_s}$ is the product of on-shell energies, and $L^i_{j,k} = \lambda_i^{-1} \left(\lambda_j^{-1} + \lambda_k^{-1} \right)$. The quantity  
\begin{equation}  
\lambda_{i_1\cdots i_r \bar i_{r+1} \cdots \bar i_n} = \lambda_{i_1\cdots i_r}- \lambda_{i_{r+1} \cdots i_n}  
\end{equation}  
encodes the structure of the propagator residues. 

The integration measure is expressed in terms of two loop three-momenta:
\begin{equation}
    d\Phi_{\lb_1 \lb_2} = \prod_{j=1}^{2} \frac{d^3 \lb_j}{(2\pi)^3}~,
    \label{eq:integrationmeasure}
\end{equation}
involving six integration variables. However, each term in \Eq{eq:decayratescalarNLO_5} must satisfy energy conservation, encoded as
\begin{equation}
    \ps{i_1\cdots i_n \bar a} = 2\pi \delta(\lambda_{i_1\cdots i_n \bar a})~,
    \label{eq:dirac_5}
\end{equation}
and the decay is isotropic in the rest frame of the decaying particle. As a consequence, the decay rate depends on two independent integration variables, constrained by the Dirac delta functions in \Eq{eq:dirac_5}. These are a polar angle, representing the angle between the two loop three-momenta, typically parametrized as $\cos{\theta} = 1 - 2v$, with $v \in [0,1]$, and the modulus of one of the loop three-momenta, mapped from $[0,\infty)$ to the finite interval $[0,1)$ in numerical implementations. Explicitly,
\begin{equation}
    d\Phi_{\lb_1 \lb_2} \to \frac{1}{4\pi^4}\int_0^\infty \lb_1^2  d|\lb_1| \int_0^\infty \lb_2^2 d|\lb_2| \int_0^1 dv~,
    \label{eq:measure_5}
\end{equation}
where
\begin{equation}
    \ps{45 \bar 6} =  2\pi \,  \delta\left( \sqrt{\lb_2^2+m^2} -\sqrt{s} \right)~,
    \label{eq:delta1_5}
\end{equation}
and
\begin{align}
    \ps{135 \bar 6} &= 2\pi \,  \delta\bigg( |
\lb_1+\lb_2| 
    + \sqrt{\lb_1^2+m^2} + \sqrt{\lb_2^2+m^2} - \sqrt{s} \bigg)~,
    \label{eq:delta2_5}
\end{align}
with $|\lb_1+\lb_2| = \sqrt{\lb_1^2 + \lb_2^2 + 2(1-2v) |\lb_1||\lb_2|}$ and $|\lb_i| = \sqrt{\lb_i^2}$. The two Dirac delta functions in \Eq{eq:delta1_5} and \Eq{eq:delta2_5} allow one of the integration variables in \Eq{eq:measure_5} to be expressed in terms of the remaining two, which is solved analytically in numerical implementations.

A key feature of \Eq{eq:decayratescalarNLO_5} is that loop and tree-level contributions—corresponding to different numbers of final-state particles—are treated under the same integration measure. This ensures local cancellation of singularities encountered in state-of-the-art methods, avoiding the need for intermediate calculations in arbitrary spacetime dimensions~\cite{Bollini:1972ui,tHooft:1972tcz}. Additionally, the resulting integrand is flatter than in other approaches,  allowing for a much faster and more efficient numerical implementation,  making it particularly suitable for QFIAE integration. 

\subsection{Quantum integration of NLO decay rates}\label{subsec:qintdecayrates}

\begin{figure}[h]
\begin{center}
\includegraphics[scale=0.6]{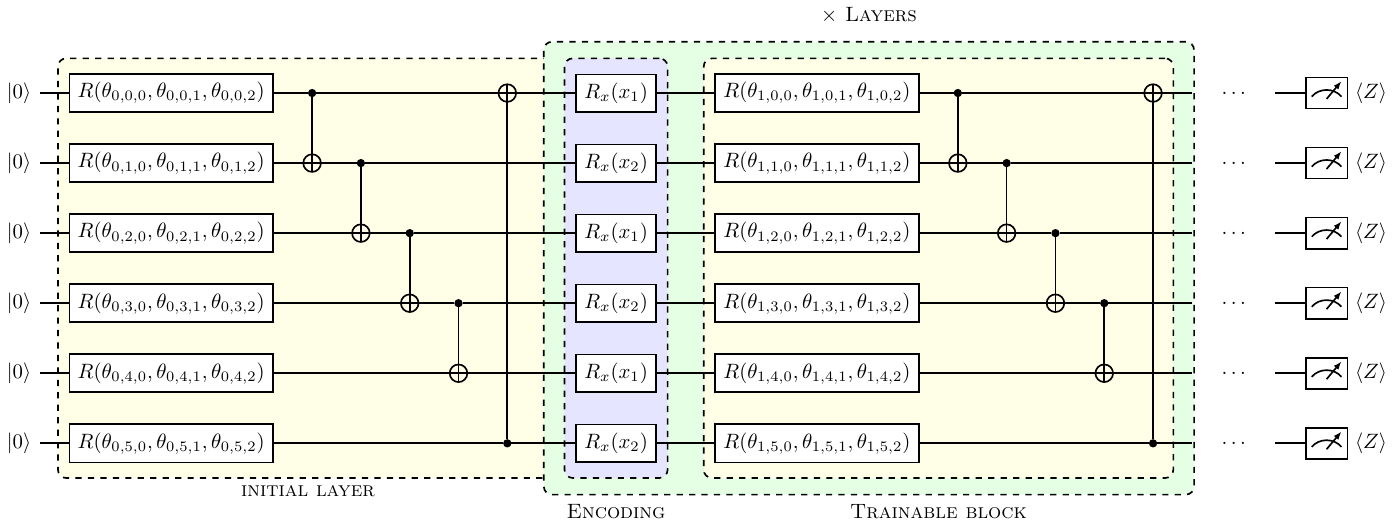}
\caption{Architecture of the QNN employed to fit a 2-dimensional function. 
\label{fig:qnn_6q}}
\end{center}
\end{figure}

In this section, we apply the quantum integration algorithm QFIAE to estimate the total decay rate at NLO for the decay processes discussed in Sec.~\ref{subsec:causal_unitary}. The main challenge in this quantum implementation lies in designing a Quantum Neural Network (QNN) that can effectively model the differential decay rate function. To address this, we introduce a QNN with a general-purpose Ansatz, as shown in Fig. \ref{fig:qnn_6q}. This Ansatz is designed with sufficient entanglement and free parameters, offering high expressibility while ensuring efficient trainability, thus enabling the accurate solution of the regression problem.

We use \texttt{PennyLane}~\cite{pennylane} and \texttt{JAX}~\cite{jax2018github} to build and train the QNN. The architecture, shown in Fig.~\ref{fig:qnn_6q}, consists of a 6-qubit quantum circuit where a specific Ansatz is repeated $n_{layers}$ times. This Ansatz includes two main components: a variational layer with trainable parameters, implemented using \texttt{qml.StronglyEntanglingLayers}, and an encoding block for two input variables. Each variable is encoded three times in parallel across the 6 qubits using \texttt{qml.AngleEmbedding}.

\begin{figure}[h]
\begin{center}
\includegraphics[scale=0.49]{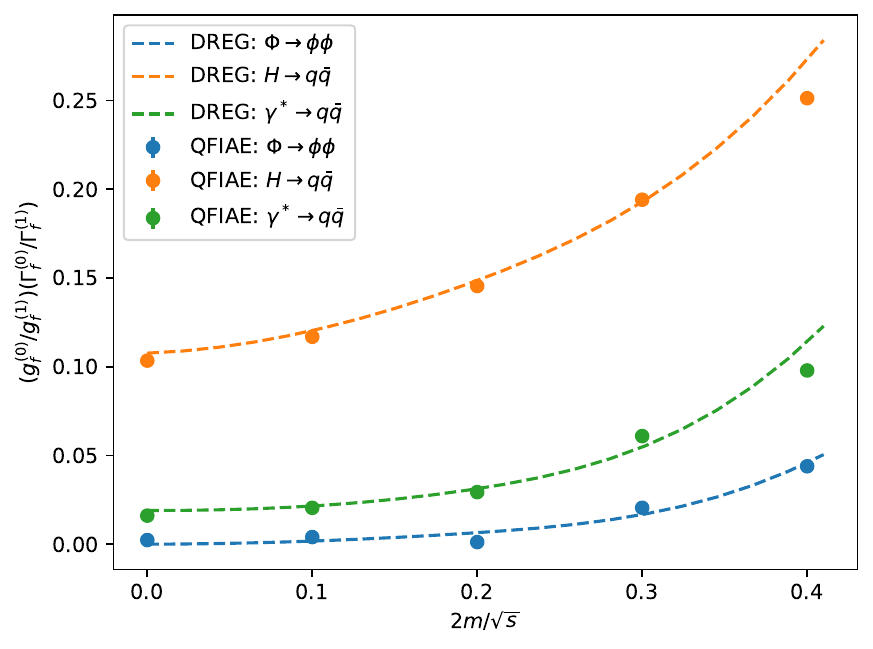}
\includegraphics[scale=0.49]{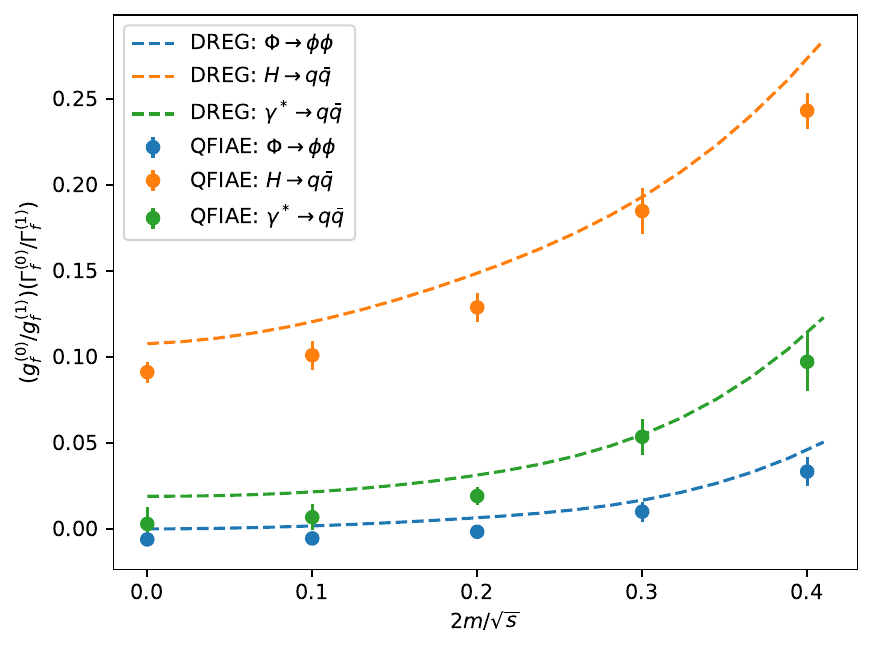}
\caption{Quantum-integrated decay rates for the three processes $H\to q\bar q (g)$, $\gamma^*\to q\bar q (g)$, and $\Phi\to \phi\phi(\phi)$ at NLO as a function of the final state mass. The left  panel shows results from a quantum simulator, while the right panel presents results where the IQAE has been executed on quantum hardware. The calculations use QFIAE with the LTD causal unitary approach, and the dashed lines represent theoretical predictions in dimensional regularization. Parameters for the quantum implementation: $max\_steps=15000$,  $step\_size=0.001$, $layers=n_{Fourier}=20$, $n_{qubits}=6$ for the QNN, and $n_{qubits}=5$, $n_{shots}=10^3$, $\epsilon=0.01$, $\alpha=0.05$ for the IQAE module.}
\label{fig:qint_combined}
\end{center}
\end{figure}

Following the results from Sections \ref{app:qfiae} and \ref{app:lfintqc} on regression with variational quantum circuits, encoding is performed using $R_x$ rotations, and measurements are taken in the Pauli-$Z$ basis, as illustrated in Fig.~\ref{fig:qnn_6q}. In terms of complexity, each layer has a quantum depth of 7, consisting of one encoding step, one variational gate step, and five entangling two-qubit gate steps. For the integrated decay rates presented in Fig~\ref{fig:qint_combined}, we employ 20 layers, resulting in a total quantum depth of 140. To assess the feasibility of such variational circuit to be executed in current devices, we refer to two recent IonQ studies~\cite{ionqarticle,ionqnews}, mentioned in Sec.~\ref{app:qfiae}, where various quantum algorithms were evaluated using QED-C benchmarks. The findings show that our algorithm, which requires a quantum depth of 140 and a low qubit count ($\leq 6$), would achieve a high success probability on IonQ and Quantinuum devices.

Once the QNN successfully approximates the target function, we extract its Fourier series and use it as input for the IQAE subroutine. The IQAE module is designed to operate with low quantum depth and a minimal number of qubits, making it feasible for execution on current quantum hardware. It is implemented with \texttt{Qibo}\cite{qibo_paper} for quantum simulations (Fig.~\ref{fig:qint_combined} left) and with \texttt{Qiskit}\cite{qiskit2024} on real hardware (Fig.~\ref{fig:qint_combined} right). Specifically, we run the IQAE module on the 27-qubit IBM Quantum superconducting processor \textit{ibmq\_mumbai}, using only 5 qubits to sequentially integrate the Fourier terms.

To mitigate quantum noise, we apply the same techniques that in Sec.~\ref{app:lfintqc}. First, pulse-efficient transpilation~\cite{Earnest_2021}, which reduces the number of two-qubit gates by leveraging hardware-native cross-resonance interactions. Additionally, we use Dynamical Decoupling (DD) during circuit execution and apply Zero Noise Extrapolation (ZNE) to the output via the \texttt{Qiskit} Runtime Estimator primitive~\cite{estimator}.

Regarding the hyperparameters employed for this application, for training the QNN we use $max\_steps$, which sets the number of iterations for the ADAM optimizer~\cite{adam}, and $step\_size$, which represents the learning rate. The parameter $layers$ specifies the number of circuit layers, while $n_{Fourier}$ determines the number of Fourier coefficients used in the truncated representation of the circuit. The number of qubits in the variational circuit is defined by $n_{qubits}$. For the IQAE module, $n_{qubits}$ specifies the number of qubits used, while $n_{shots}$ sets the number of measurement samples per circuit run. The error tolerance for each integral is controlled by $\epsilon$, and $\alpha$ defines the confidence interval for the integral results.

The results in Fig.~\ref{fig:qint_combined} show a relatively small deviation from their corresponding analytical values in standard dimensional regularization (DREG). Notably, the left panel presents a systematic deviation compared to the right panel, attributed to hardware noise that remains despite the applied error mitigation techniques. Table~\ref{table:errors} presents the explicit numerical results and uncertainties corresponding to these figures.

\begin{table}[h!]
\begin{center}
\begin{tabular}{ccccc}  \hline
Decay & $2m/\sqrt{s}$ & Hardware & Simulator &  DREG \\ \hline
 $\Phi\to \phi\phi(\phi)$         & $0.0$ & $-0.0061(28)$ &  $0.0023(5)$ &  $0.0000$ \\ 
 & $0.1$ & $-0.0055(31)$ &  $0.0040(6)$ &  $0.0018$ \\
 & $0.2$ & $-0.0016(30)$ &  $0.0011(6)$ &  $0.0065$ \\
 & $0.3$ & $ 0.0101(56)$ &  $0.0205(11)$ &  $0.0167$ \\
 & $0.4$ & $ 0.0333(85)$ &  $0.0439(15)$ &  $0.0459$ \\ \hline
 $H\to q\bar q(g)$            & $0.0$ & $ 0.0911(61)$ &  $0.1034(13)$ &  $0.1077$ \\
 & $0.1$ & $ 0.1009(83)$ &  $0.1169(14)$ &  $0.1204$ \\
 & $0.2$ & $ 0.1288(85)$ &  $0.1455(14)$ &  $0.1486$ \\
 & $0.3$ & $ 0.1847(135)$ &  $0.1941(20)$ &  $0.1928$ \\
 & $0.4$ & $ 0.2431(104)$ &  $0.2513(30)$ &  $0.2730$ \\ \hline
 $\gamma^*\to q\bar q (g)$     & $0.0$ & $ 0.0029(96)$ &  $0.0161(14)$ &  $0.0190$ \\
 & $0.1$ & $ 0.0068(74)$ &  $0.0205(13)$ &  $0.0215$ \\
 & $0.2$ & $ 0.0191(50)$ &  $0.0293(13)$ &  $0.0313$ \\
 & $0.3$ & $ 0.0535(103)$ &  $0.0609(20)$ &  $0.0547$ \\
 & $0.4$ & $ 0.0971(171)$ &  $0.0979(30)$ &  $0.1140$ \\   \hline
\end{tabular}
  \caption{Quantum-integrated decay rates at NLO for the processes $H\to q\bar q (g)$, $\gamma^*\to q\bar q(g)$, and $\Phi\to \phi\phi(\phi)$ as a function of the final state mass, computed using QFIAE and LTD causal unitary. The `Hardware' column reports results from the QNN on a quantum simulator and the IQAE on quantum hardware, while the `Simulator' column shows results where both the QNN and IQAE run on quantum simulators. The `DREG' column provides the exact analytic results at NLO accuracy. 
  }

  \label{table:errors}
\end{center}
\end{table}

Table \ref{table:errors} shows that IQAE executions on quantum hardware present uncertainties about an order of magnitude larger than those on a quantum simulator. This difference is expected, as the inherent quantum noise on real hardware adds to the statistical uncertainty of the IQAE method. Despite this, most of the results are in agreement within the uncertainty bands with the expected values, demonstrating satisfactory agreement given the current limitations of quantum hardware.

\section{Conclusions}\label{app:qintconclusions}

In this Chapter we have discussed a new application of QML and QAE to build a novel Quantum Monte Carlo integrator, called QFIAE~\cite{deLejarza:2024scm,deLejarza:2024pgk,deLejarza:2023IEEE}. This method goes beyond previously defined quantum integrators such as FQMCI, since it exploits QNN to approximate the Fourier series. Furthermore, QFIAE eliminates the need for numerical integration to compute Fourier coefficients for a general multivariate function $f(\vec{x})$, potentially preserving the quadratic quantum speedup achieved through amplitude amplification (a generalization of Grover's algorithm), where other methods fall short. This constitutes, to the best of our knowledge, the very first end-to-end Quantum Monte Carlo integrator with the potential to maintain this speedup.

As an initial testbed, we have successfully applied the quantum integrator to a one-dimensional function that corresponds to the elementary scattering process $e^+e^- \rightarrow q \bar{q}$. We obtained a very precise estimation of the integral with a relative error around~1\%, while maintaining a relatively low quantum depth. In particular, a study of the performance and the quantum depth of both circuits of the algorithm has been conducted and have been compared with the depth and width of the algorithms that can be effectively implemented on NISQ devices. The results showed that our method lies in the region where a quantum algorithm can be executed with a high probability of success. 

Then, we considered more complicated integrals and we have applied the QFIAE to estimate loop Feynman integrals for the first time. Our primary objective was to explore the efficiency of quantum algorithms in this context and evaluate their potential for a quantum speedup.

Specifically, we employed the QFIAE method to integrate the one-dimensional tadpole, and bubble Feynman loop integrals, within the LTD formalism, across various kinematical regions, circumventing different threshold singularities using a contour deformation. For the one-dimensional tadpole integral we succesfully executed the full quantum pipeline on quantum hardwares of \texttt{Qibo} and IBM Quantum obtaining satisfactory results with deviations of around 5\% of the analytical value.

Furthermore, we extended QFIAE to multiple dimensions and particularly utilized it to integrate different configurations of the two-dimensional scalar and tensor triangle Feynman loop integrals. The results obtained demonstrate the remarkable accuracy of this quantum algorithm in estimating these integrals. 

Moreover, we also used the FQMCI method, which relies on numerical integration to calculate the Fourier coefficients, as opposed to the QFIAE method that employs a QNN for that endeavor. With the FQMCI method, we successfully integrated a particular kinematical configuration of a pentagon integral (three-dimensional function). This integral involves a highly intricate three-dimensional integrand, making its integration a significant challenge. Therefore, the successful application of a quantum algorithm to handle such complex functions is a notable achievement. Nevertheless, designing an efficient Ansatz that enables a QNN to accurately learn the function and allow its integration via QFIAE remains an open challenge for future work.\\

In a follow-up study, we went one step further to perform the first quantum computation of a total decay rate at second order in perturbative quantum field theory. Leveraging the LTD framework, we have successfully combined loop and tree-level Feynman diagrams with the quantum algorithm QFIAE on a quantum computer. This methodological advancement is significant from the high-energy physics perspective, as it allows us to integrate a real process with potential for quantum speedup. While we do not claim to have achieved quantum advantage in this work, our results lay the groundwork for future explorations in this direction.
From the perspective of quantum computing, our study marks a noteworthy achievement. By solving a relatively complicated regression problem using a QNN on a realistic dataset, we found a good compromise between trainability and expressibility, a common challenge in QNNs. Most of the results presented are in agreement with the expected values within the uncertainty bands, even when the IQAE part of the algorithm is executed on hardware. This demonstrates the potential of quantum computing to address complex, real-world problems, corresponding to real physical processes, and highlights the importance of continuing to push the boundaries of what quantum technology can achieve.
Our findings underscore the critical importance of integrating quantum computing methodologies with high-energy physics applications. By tackling practical challenges and pushing the limits of current quantum computing capabilities, we can better understand the potential and limitations of quantum technologies and design better algorithms for future applications. 
\chapter{Quantum generative models for Fragmentation Functions}\label{chap:qchebff}

\section{Introduction}\label{app:intro_qcheb}

Fragmentation Functions (FFs) are an essential component in particle physics, specifically in the context of Quantum Chromodynamics (QCD), the Quantum Field Theory that describes the interactions between quarks and gluons~\cite{deFlorian:2007ekg,Aidala:2010bn,deFlorian:2007aj}. FFs describe the hadronization process of partons after a hard-scattering event. Specifically, they quantify the probability that a given parton produces a particular hadron.

In leading-twist collinear QCD~\cite{Collins:1989gx}, collider processes are described using hard scattering partonic cross-sections, initial-state parton distribution functions (PDFs) for the initial hadronic structure, and FFs for the hadronization process~\cite{Collins:1981uw}. These functions are the core components of the theoretical description of hard interactions and are crucial for their success. Future experiments, like those at the Electron-Ion Collider (EIC), aim to improve FF parametrizations and facilitate combined analyses of PDFs and FFs with higher-precision data~\cite{Aschenauer2019}. 

In a typical collider process, partonic cross-sections are calculated perturbatively. However, unlike partonic cross-sections, FFs cannot be calculated directly and must be obtained from experimental data via global analysis across diverse processes~\cite{Rojo:2015acz,Albino:2008aa}. Nevertheless, their energy scale evolution is predicted by QCD.

Several analyses have been proposed to extract FFs from experimental data using statistical tools and heuristic approaches. Initial studies introduced a polynomial-based Ansatz in the parton's momentum fraction~\cite{Gluck:1998xa,Kretzer:2000yf,Kniehl:2000fe,Hirai:2007cx}. Later approaches used Euler Beta function distributions, along with additional parameters to refine the fit~\cite{deFlorian:2007aj,deFlorian:2007ekg,Aidala:2010bn}.
Despite the progress achieved so far, the determinations of FFs might still be affected by sources of procedural bias, including limited functional form flexibility and uncertainty estimation challenges. Machine Learning techniques, particularly neural networks, have emerged to overcome these limitations~\cite{Nocera:2017qgb, Bertone:2017xsf, Bertone:2017tyb, Soleymaninia:2022qjf}.

Although the supervised neural network approach has proven effective in determining FFs, and parton densities (PDF) in hadronic collisions \cite{NNPDF:2021uiq}, some challenges remain. A primary issue lies in the energy-scale evolution. The models are typically trained for a specific range of momentum fraction $z$ and a fix energy scale $Q$. However, evolving a given FF to different energy scales requires solving the DGLAP evolution equations~\cite{Gribov:1972ri,Lipatov:1974qm,Altarelli:1977zs,Dokshitzer:1977sg}, which might be computationally expensive. Additionally, once the FFs are determined, their true value are only known for specific points, necessitating interpolation to estimate the function at other regions, increasing the complexity of the method. 

In the same way that the previous advancements in Machine Learning introduced powerful tools to tackle complex challenges, we are now at the dawn of a new era powered by Quantum Computing (QC) and Quantum Machine Learning (QML). These new techniques, leveraging the properties of quantum mechanics: superposition, entanglement and interference, offer a novel approach to analyzing high-energy-physics (HEP) data \cite{Delgado:2022tpc, DiMeglio:2023nsa, Rodrigo:2024say}. Notable applications include jet clustering~\cite{thaler,delgado_jets,lejarza,deLejarza:2022vhe}, elementary particle process integration~\cite{deLejarza:2023IEEE,AGLIARDI2022137228}, anomaly detection~\cite{Belis:2023atb,Schuhmacher:2023pro}, data classification~\cite{Belis:2024guf, Belis:2021zqi}, studying the causal structure~\cite{Ramirez-Uribe:2021ubp,Clemente:2022nll,Ramirez-Uribe:2024wua,Ochoa-Oregon:2025opz} and integration of multi-loop Feynman diagrams~\cite{deLejarza:2024scm,deLejarza:2024pgk,Pyretzidis:2025stx}. In the context of QML~\cite{carrazza,Cruz-Martinez:2023vgs} and Tensor Networks~\cite{Kang:2025xpz} different methods are being explored for determination and integration of PDFs.

These applications pave the way for analyzing HEP data from collider experiments through the lens of QC. In this context, quantum generative models (QGMs) have shown promising applications in HEP~\cite{Delgado:2022tpc,Delgado:2023ofr,Delgado:2024vne,Bermot:2023kvh, Tuysuz:2024hyl, Rudolph:2023iqf}.

QGMs usually refer to a class of QML models that leverage Parameterized Quantum Circuits (PQCs) and classical optimization subroutines to generate new data samples resembling a given dataset. QGMs may offer potential advantages over classical generative models in terms of computational power, efficiency, and expressiveness. Several QGMs adapt the classical generative frameworks, such as Quantum Boltzmann Machines (QBMs)~\cite{Amin2018,Zoufal2021,Coopmans2024} and Quantum Generative Adversarial Networks (QGANs)~\cite{Lloyd2018,Dallaire-Demers2018,Zoufal2019,Huang2021PRAppl}. In QBMs, a PQC as a Hamiltonian operator is trained such that the resulting Gibbs state matches the target data distribution. Likewise, QGANs are iteratively trained with the generator and discriminator competing against each other to create quantum states and distinguish between generated and real data. Distinct from other QGMs, Quantum Circuit Born Machines (QCBMs) are quantum-inspired generative models that leverage the inherent probabilistic nature of quantum measurement to implicitly treat data represented by output bitstrings as an observable~\cite{Liu2018,Benedetti2019npj,Benedetti_2020}. 

In contrast to implicit QCBMs based on measurement results, the Quantum Chebyshev Probabilistic Model (QCPM) is a type of generative model that belongs to explicit models that treat data as input variable~\cite{Williams:2023cuz,Kyriienko2024,Wu2024QHT}. More specifically, QCPM allows for separation of the training/learning and sampling/generating stages to make QC differentiable in the latent Chebyshev space and to enable dense sampling in the bit-basis via incorporation of extended registers. Furthermore, QCPM allows us to investigate the association between variables and to scale up systemically for modeling of high-dimensional complex distributions owing to its modular circuit architecture. 

In this Chapter, we propose a generalized QCPM framework to study and predict the behavior of FFs~\cite{deLejarza:2025upd}. We also show how to study the correlation between the variables $z$ and $Q$ using this model, and its potential impact on QML.

The code to reproduce the results of this chapter is available in \cite{githubGitHubCERNITINNOVATIONQChPM}.

\section{Fragmentation Functions}\label{app:ffs}

Fragmentation functions (FFs) are an essential tool for producing HEP predictions.  In particle colliders, such as the Large Hadron Collider (LHC) at CERN, over 1 billion collisions occur every second, generating vast amounts of data that need to be processed and analyzed efficiently \cite{cern-no-date}. This highlights the importance of fully understanding each component in the process of producing predictions to streamline calculations and manage the data efficiently.

FFs provide the probability that a parton (quark or gluon) will fragment into a particular hadron. This information is key for computing integrals that define observables, such as the cross-section, which are essential for interpreting the results of particle collisions and understanding the underlying physics.
To understand the role of FFs in computing exclusive observables, it is convenient to define the differential cross-section for the single-inclusive production of an hadron $h$ in electron-positron annihilation:
\begin{equation}
    \frac{d\sigma^h}{dz}(z, Q^2) = \frac{4\pi \alpha^2(Q)}{Q^2} F^h(z, Q^2),
    \label{eq:sigma}
\end{equation}
where $F^h$ is the fragmentation structure function, and $\alpha(Q)$ represents the Quantum Electrodynamics (QED) running coupling. While in the literature $F^h$ is often called fragmentation function, we will denote it as fragmentation structure function to avoid any confusion with the partonic FFs. Now, following the standard collinear factorization~\cite{Ellis:1996mzs} the QCD cross-section in a hadronic collider is expressed as a convolution of perturbatively calculable partonic cross-sections and non-perturbative distribution functions. 

The structure function is defined as a convolution between coefficient functions and FFs:
\begin{equation}
\begin{split}
    \hspace{-0.25cm} F^h(z, Q^2)  &= \frac{1}{n_f} \sum_{q} \hat{e}_q^2  
     \bigg[  D^h_S(z, Q^2) \otimes C_{S2, q}(z, \alpha_s(Q)) \\
    & + D^h_g(z, Q^2) \otimes C_{S2, g}(z, \alpha_s(Q)) \\
    & + D^h_{NS}(z, Q^2) \otimes C_{NS2, q}(z, \alpha_s(Q)) \bigg],
\end{split}
\end{equation}
where $ \hat{e}_q $ are scale-dependent quark electroweak charge factors, defined in \cite{deFlorian:1997zj} and $\alpha_s(Q)$ is the QCD running coupling. 
The sum is performed over the $n_f$ active flavours at the scale $Q$, and $ C_{S2, q}, C_{S2, g}, C_{NS2, q} $
are the coefficient functions corresponding respectively to the singlet and
nonsinglet combinations of FFs, 
\begin{align}
    D^h_S(z, Q^2) &= \sum_q D^h_{q+}(z, Q^2),  \nonumber \\
    \hspace{-0.15cm} D^h_{NS}(z, Q^2) &= \sum_q \frac{\hat{e}_q^2}{\langle e^2 \rangle} \left[ D^h_{q+}(z, Q^2) - D^h_S(z, Q^2) \right],
    \label{eq:ffns}
\end{align}
and to the gluon FF, $D^h_g(z, Q^2)$. Note that in Eq.~\ref{eq:ffns} the notation $D^h_{q+} \equiv D^h_q + D^h_{\bar{q}}$ and $\langle e^2 \rangle \equiv \frac{1}{n_f} \sum_q \hat{e}_q^2$ has been used. The usual convolution integral with respect to $z$ is denoted by $\otimes$ and reads
\begin{equation}
    f(z)\otimes g(z) = \int_z^1 \frac{dy}{y} f(y) g\left(\frac{z}{y}\right).
\end{equation}

Then, to solve Eq.~\ref{eq:sigma} it is needed to perform the evolution of the FFs with the energy scale $ Q $, which follows the DGLAP evolution equations~\cite{Gribov:1972ri, Lipatov:1974qm, Altarelli:1977zs, Dokshitzer:1977sg}. The singlet component $ D^h_S(z, Q^2) $ mixes with the gluon FF, and is given by:
\begin{equation}
\begin{split}
    \frac{\partial}{\partial \ln Q^2} \begin{pmatrix} D^h_S \\ D^h_g \end{pmatrix} (z, Q^2) & = \begin{pmatrix} P_{qq} & 2n_f P_{gq} \\ P_{qg} & P_{gg} \end{pmatrix} (z, \alpha_s)  \\ & \otimes \begin{pmatrix} D^h_S \\ D^h_g \end{pmatrix} (z, Q^2),
\end{split}
\end{equation}
while the non-singlet component $ D^h_{\rm NS}(z, Q^2) $ evolves as:
\begin{equation}
    \frac{\partial}{\partial \ln{Q^2}} D^h_{\rm NS}(z, Q^2) = P_+ (z, \alpha_s) \otimes D^h_{\rm NS}(z, Q^2),
    \label{eq:dglap2}
\end{equation}
with $P_+=P_{qq}+P_{q\bar q}$ and the splitting functions $P_{ji}$ having a perturbative expansion in the strong coupling $ \alpha_s $:
\begin{equation}
    P_{ji}(z, \alpha_s) = \sum_{l=0} a_s^{l+1} P^{(l)}_{ji}(z),
\end{equation}
where $ j, i = g, q $, and $ a_s = \alpha_s/4\pi $. The time-like splitting functions are known up to $\mathcal{O}(a^3)$ in the modified minimal subtraction, or $\overline{\textrm{MS}}$, scheme~\cite{Almasy:2011eq,Moch:2007tx,Mitov:2006ic}.

Now that we understand the contribution that FFs have in computing observables, such as the cross-section in Eq.~\ref{eq:sigma}, it becomes evident the necessity of an effective way to determine them. One of the most effective methods to date is the NNFF1.0 neural network-based analysis developed by the NNPDF collaboration \cite{Bertone:2017tyb}. Nevertheless, it also presents some limitations, such as determining FFs at fixed energy scale $Q$, requiring the DGLAP evolution to obtain values at other scales. Furthermore, to explore a wider range of momentum fractions $z$ and energy scales $Q$, interpolation is required. In the next section, we introduce a quantum generative model that may provide a solution to these challenges, offering a flexible approach for FF scaling and determination.

\section{Quantum Chebyshev Probabilistic Model}\label{app:qcheb}

The capability to sample from probability distributions is central in scientific data. In this regard, quantum generative models have the potential to be useful in this context. They combine the potential of classical generative models with one of the main strengths of quantum computing, sampling.

In this section, we introduce a quantum generative model based on Chebyshev polynomials. The main goal of most quantum generative models is to enable sampling in the computational basis, which consists of the orthonormal states $\{\ket{x_j}\}_{j=0}^{2^N-1}$ satisfying the Kronecker delta condition $\langle x_j | x_{j'} \rangle = \delta_{jj'}$. However, training a model directly in this space is often difficult, as it requires a very fine control over the different probabilities. To address this, one can build and train the model in a different orthonormal basis and then map it to the computational basis for sampling. With this in mind, we employ the orthonormal Chebyshev basis and its corresponding transform to build our generative model \cite{Williams:2023cuz}.

First, we need to build the adequate feature map to encode the data into the Chebyshev space. Hence, we build an (unnormalized) quantum state $\ket{\tau(x)}$ whose amplitudes are determined by Chebyshev polynomials of the first kind, defined as $T_k(x) = \cos(k \arccos(x))$, where $k$ denotes the degree. This state takes the form
\begin{equation}
\ket{\tau(x)} = \frac{1}{2^{N/2}} T_0(x) \ket{\varnothing} + \frac{1}{2^{(N-1)/2}} \sum_{k=1}^{2^N - 1} T_k(x) \ket{k}, \label{eq:chebyshev-state}
\end{equation}
where the amplitude of the zero state is weighted by $T_0(x) = 1$. These polynomials satisfy a discrete orthogonality relation at specific sampling points, known as Chebyshev nodes:
\begin{equation}
\sum_{j=0}^{2^N - 1} T_k(x_j^{\text{Ch}}) T_\ell(x_j^{\text{Ch}}) =
\begin{cases}
0 & \text{if } k \ne \ell, \\
2^N & \text{if } k = \ell = 0, \\
2^{N-1} & \text{if } k = \ell \ne 0.
\end{cases}
\end{equation}

Here, the Chebyshev nodes are given by $x_j^{\text{Ch}} = \cos\left(\pi(2j + 1)/{2^{N+1}}\right)$, corresponding to the roots of Chebyshev polynomials. The resulting states $\{ \ket{\tau(x_j^{\text{Ch}})} \}_{j=0}^{2^N - 1}$ form an orthonormal set, with $\langle \tau(x_j^{\text{Ch}})| \tau(x_{j'}^{\text{Ch}})\rangle = \delta_{jj'}$.

It is worth noting that, unlike the standard computational or Fourier bases, which rely on uniform grids, the Chebyshev nodes form a non-uniform mesh within the interval $(-1, 1)$. Outside these node points, the states $\ket{\tau(x)}$ are no longer orthogonal. An analytical expression for their squared overlap can be derived when one of the arguments is fixed to a Chebyshev node:
\begin{equation}
\left| \braket{\tau(x')}{\tau(x)} \right|^2 =
\frac{\left[T_{2^{N+1} - 1}(x') T_{2^N}(x) - T_{2^N}(x') T_{2^{N+1} - 1}(x)\right]^2}{2^{2N}(x' - x)^2}.
\end{equation}

This identity follows from the Christoffel-Darboux formula for Chebyshev polynomials~\cite{Iten_2016}. Since these states are not normalized in general, generating them with a quantum circuit involves using an ancillary qubit and normalizing the state post-preparation. We denote the normalized version as 
\begin{equation}
    \ket{\tilde{\tau}(x)} = \frac{\ket{\tau(x)}}{\sqrt{\braket{\tau(x)}}},
\end{equation}
which equals the unnormalized version exactly at the Chebyshev nodes and approaches it as $N$ grows large. 

The next step is to design a quantum circuit that prepares the normalized state $\ket{\tilde{\tau}(x)}$, which we will call the feature map. We begin by noting that Chebyshev polynomials can be expressed in terms of cosine functions evaluated on a specific non-uniform grid. Since $\cos(x) = \frac{1}{2}\left(e^{ix} + e^{-ix}\right)$, the amplitudes can be embedded through a combination of exponential functions, each of which can be implemented using a phase feature map~\cite{Kyriienko:2022zyd}. These components can be combined using the Linear Combination of Unitaries (LCU) framework~\cite{lcu}, where the unitary terms are controlled by the state of an ancillary qubit. The desired interference is achieved by post-selecting on the ancilla being measured in the $\ket{0}$ state. Notably, the identity $e^{-ix} = e^{ix} \cdot e^{-i2x}$ allows us to control only one of the exponential components, simplifying the implementation.

\begin{figure}[h]
    \centering
    \includegraphics[width=0.99\linewidth]{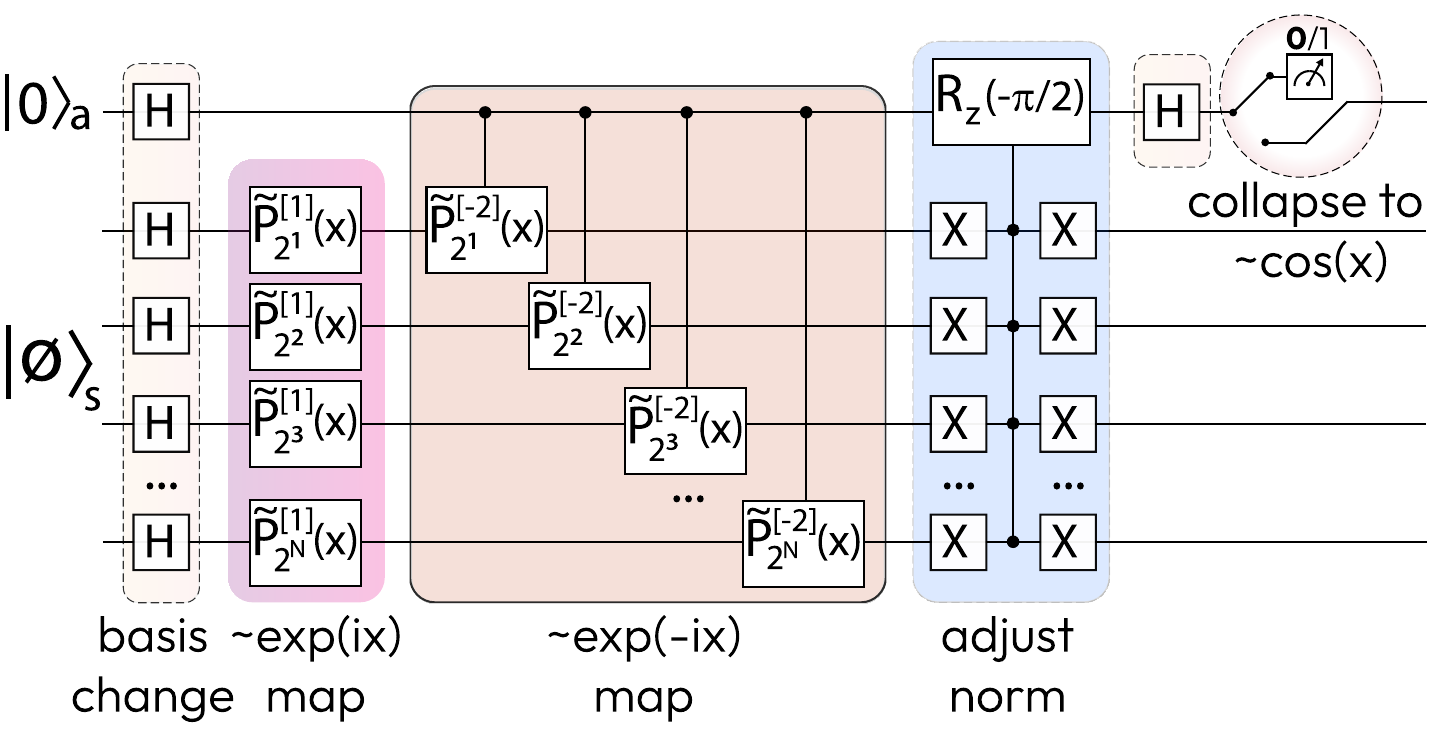}
    \caption{Quantum Chebyshev feature map that prepares a Chebyshev state via a sequence of phase feature maps that embed complex exponentials, a controlled rotation to adjust the normalization of the zero-frequency term, and post-selection on the ancilla measured in the $\ket{0}$ state. The scaled single-qubit phase shift gates in the feature map circuit are defined as $ \tilde{P}^{[s]}_l(x) = \text{diag}\{1, \exp( i s {2^N \arccos(x)/l} ) \},$ where $l$ increases exponentially as $2^j$, with $j$ indexing the qubit number, and $s \in \{1, -2\}$. Here, $H$ and $X$ denote the Hadamard and Pauli $\hat{X}$ gates, respectively. Picture adapted from~\cite{Williams:2023cuz}.
}
    \label{fig:chebfm}
\end{figure}

This approach enables the construction of equal-weighted combinations of exponentials, scaled appropriately to reproduce the amplitudes associated with Chebyshev polynomials $T_k(x)$. To correctly set the amplitude of the constant term $T_0(x)$, an additional adjustment is required. This is achieved by applying a single iteration of a Grover-like rotation circuit~\cite{Grover:1996rk}, which rotates the state around $\ket{\varnothing}$ by a fixed angle. Since the ancilla measurement commutes with this rotation, we put the measurement to the end of the circuit, completing the preparation of the desired feature map. The quantum circuit of this feature map is shown in Fig.~\ref{fig:chebfm}.

\begin{figure}[h]
    \centering
    \includegraphics[width=0.99\linewidth]{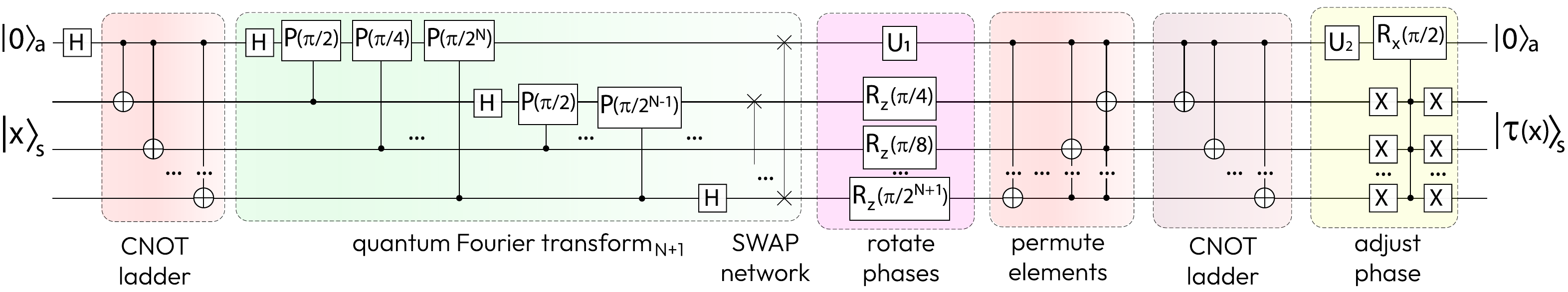}
    \caption{Quantum Chebyshev transform ($\hat{U}_{\text{QChT}}$) circuit, which maps computational basis states $\{\ket{x_j}\}_{j=0}^{2^N-1}$ into Chebyshev states $\{\ket{\tau(x_j^{\text{Ch}})}\}_{j=0}^{2^N-1}$. The transform consists of a quantum Fourier transform (QFT) applied to $N + 1$ qubits (an $N$-qubit system plus one ancillary qubit), followed by a sequence of phase-adjusting and permutation operations. We employ a standard phase shift gate defined as $P(\phi) = \text{diag}\{1, \exp(i\phi)\}$. Additionally, we use local phase rotations acting on the ancilla, given by $U_1 = P(-\pi/2^{N+1}) R_Z(-\pi(2^N - 1)/2^{N+1})$ and $U_2 = P(-\pi/2) R_Y(-\pi/2)$, which can be combined into a single gate for implementation. Picture adapted from~\cite{Williams:2023cuz}.
}
    \label{fig:qchebt}
\end{figure}

Once we have established how to encode data in the Chebyshev basis we need a transformation between that and the computational basis, and vice versa. Once an orthonormal basis is prepared, there exists a one-to-one correspondence (bijection) and a unitary transformation that maps it to any other orthonormal basis. In this context, we define the Chebyshev transform as 
\begin{equation}
    \hat{U}_{\text{QChT}} = \sum_{j=0}^{2^N-1} \ket{\tau(x_j^{\text{Ch}})}\bra{x_j},
\end{equation}
which maps computational basis states $\ket{x_j}$ to Chebyshev states $\ket{\tau(x_j^{\text{Ch}})}$. The corresponding circuit to produce this map is presented in Fig.~\ref{fig:qchebt}.

The Chebyshev transform can be viewed as a variant of the cosine transform~\cite{Klappenecker:2001xto}. Specifically, the amplitude vector of the state $\ket{\tau(x_j^{\text{Ch}})}$ corresponds to the $(j+1)$th column of the type-II discrete cosine transform matrix, $\text{DCT}^{\text{II}}_N$, defined as
\begin{equation}
    \text{DCT}^{\text{II}}_N \equiv 2^{-(N-1)/2} \left\{ c_k \cos\left[\frac{k(j + 1/2)\pi}{2^N}\right] \right\}_{k,j=0}^{2^N - 1},
\end{equation}
where $c_0 = 1/\sqrt{2}$ and $c_k = 1$ for $k \neq 0$. This matrix is closely related to the Fourier transform, but requires mixing and interference of its elements, motivating the design of the extended QFT-based circuit of Fig.~\ref{fig:qchebt}.

The circuit begins with a Hadamard gate applied to the ancilla qubit (the most significant bit), followed by a ladder of CNOT gates, which collectively prepare a cat-like entangled state. A quantum Fourier transform is then applied to all $N + 1$ qubits. This is followed by a sequence of unitary operations that separate and align the real and imaginary parts of the state. In particular, local gates $U_1$ and $R_Z$ are used to adjust relative phases of the components $\ket{0}_a\ket{\Phi}$ and $\ket{1}_a\ket{\Phi}$, where $\ket{\Phi}$ represents an arbitrary $N$-qubit intermediate state.

A permutation subcircuit is then applied to reorder the amplitudes appropriately, followed by an additional CNOT ladder. The circuit concludes with a global phase adjustment via $U_2$ and a set of multi-controlled $R_X$ gates to scale the output amplitudes such that $\ket{0}_a\ket{\Phi}$ and $\ket{1}_a\ket{\Phi}$ are purely real and imaginary, respectively. We note that the ancilla starts and ends in the $\ket{0}$ state, ensuring a "clean" execution. Finally, the complete feature map is obtained by composing the embedding with the transform as $\hat{U}_f(x) = \hat{U}_{\text{QChT}}^\dagger \hat{U}_\tau(x)$.

\begin{figure}[h]
    \centering
    \includegraphics[width=0.7\linewidth]{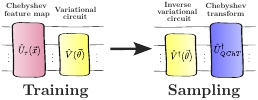}
    \caption{Workflow of the QCPM. \textit{Left:} Training circuit encodes $x$ using a Chebyshev feature map, and includes a variational circuit whose parameters will be tuned in order to reproduce the behavior of the desired distribution. \textit{Right:}  Sampling circuit performs the inverse operations of the trained variational circuit, and the quantum Chebyshev transform, producing data that follows the distribution employed for training.}
    \label{fig:qcpm_sketch_6}
\end{figure}

After we have defined a protocol to encode data into Chebyshev basis states and a map between this basis and the computational one, we can build a probabilistic model that aims to learn in Cheyshev basis and sample in the computational basis. The workflow of this Quantum Chebyshev Probabilistic Model (QCPM) is depicted in Fig.~\ref{fig:qcpm_sketch_6}.

One of the most interesting features of this QCPM of Fig.~\ref{fig:qcpm_sketch_6} is that it allows you to train the data over the target distribution with a limited amount of qubits, which is usually associated to a limited sampling density, and then perform an extended sampling using a QChT with additional qubits. This enables a natural ``quantum interpolation'' that increases the resolution up to the desired level. This improvement comes at the cost of reducing the maximum probability per point, as the output is distributed over a larger number of datapoints. In Section \ref{app:qchebff} we will see a practical example of how this feature can be leveraged.

\newpage
\section{Quantum Chebyshev Probabilistic Models for Fragmentation Functions}\label{app:qchebff}
\subsection{Model}\label{sec:model}
\begin{figure*}[h]
    \includegraphics[width=\textwidth]{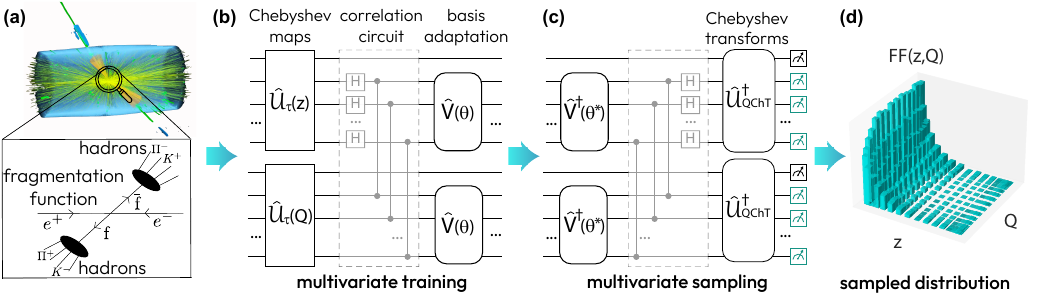} 
    \caption{Workflow for describing fragmentation functions with quantum probabilistic models based on Chebyshev polynomials. (a) The input data for $D_i^h(z,Q)$ is produced for a grid of $z$ and $Q$. (b) Quantum probabilistic model is composed of two Chebyshev feature maps for encoding $z$ and $Q$, a correlation circuit that entangles both registers, and basis adaptation circuits to be trained on $D_i^h(z,Q)$. (c) For sampling we perform the inverse of basis adaptation, the correlation circuit, followed by parallel inverse quantum Chebyshev transforms for mapping the model into the bit basis. (d) Sampling results assembled in a 2-dimensional plot that represents $D_i^h(z,Q)$. }
    \label{fig:workflow}
\end{figure*}

We introduce a quantum generative model that provides a flexible approach for FF scaling and determination. The workflow of the QCPM algorithm is shown in Fig.~\ref{fig:workflow}.

The training circuit is composed of two Chebyshev feature map circuits $\hat{\mathcal{U}}_{\tau}(u)$ and $\hat{\mathcal{U}}_{\tau}(v)$ for encoding two independent variables, $u$ and $v$ in parallel registers, followed by a correlation circuit $\hat{\mathcal{C}}$ and two separate variational  Anz\"atze $\hat{\mathcal{V}}_{\theta}$ and $\hat{\mathcal{V}}_{\vartheta}$, as illustrated in Fig.~\ref{fig:workflow}(b). The correlation circuit generates Bell-like entangled states to make two otherwise independent latent variables correlated for efficient training. The quantum model $p_Q(u,v)= \alpha \, p_{\theta,\vartheta}(u,v)+\beta$ is trained to search for optimal angles ($\theta_\text{opt}$ and $\vartheta _\text{opt}$) and classical weighting parameters ($\alpha_\text{opt}$ and $\beta_\text{opt}$), so that a mean squared error loss function is minimized. On the other hand, the sampling circuit consists of the inverse versions of the trained variational Anz\"atze, followed by two identical sets of extended inverse quantum Chebyshev transform $\hat{\mathcal{U}}_{\mathrm{QChT}}^\dagger$ circuits, a procedure accounting for the basis transformation from the Chebyshev to bitbasis spaces, as illustrated in Fig.~\ref{fig:workflow}(c).

\begin{figure}[h]
\begin{center}
\includegraphics[width=0.8\linewidth]{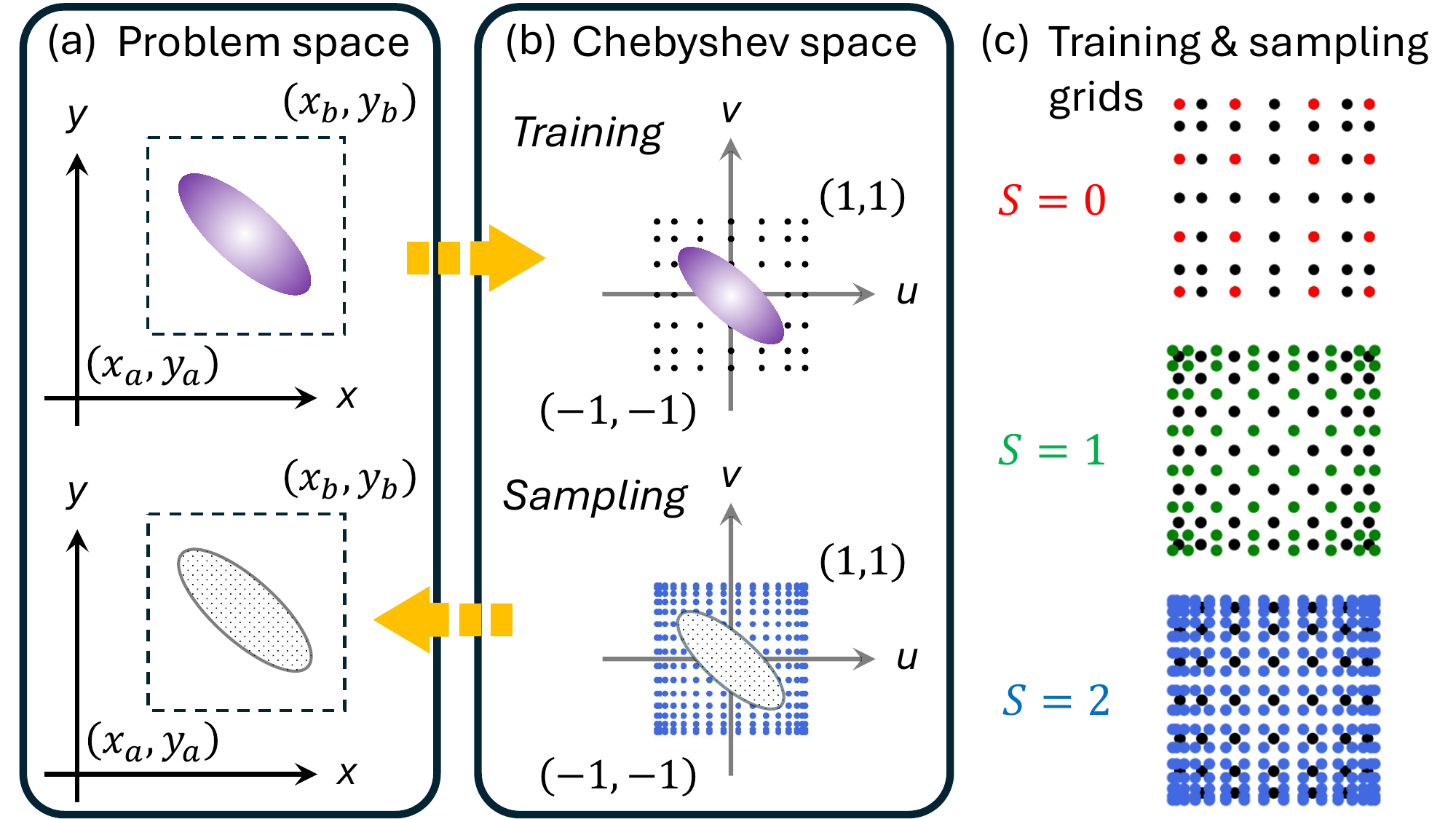}
\end{center} 
\caption{Generalized Quantum Chebyshev Probabilistic modeling. Application of shifted Chebyshev polynomials to the problem (a) and Chebyshev (b) spaces. (c) Comparisons of training and sampling grids for different extended registers.}
\label{fig:SCP}
\end{figure}

Fig.~\ref{fig:SCP} depicts the proposed scheme to make QCPMs more general and applicable to most of probability distributions. Given a probability density function $p(x,y)$ distributed within a bounding box defined by two points, $a$ and $b$ (Fig.~\ref{fig:SCP}(a), top), we build a quantum model $p_Q(x,y)$ with finite $2N$ qubits to approximately represent $p(x,y)$ and express the quantum model as a two-dimensional Chebyshev expansion, $p_Q(x,y) = \sum_{k,l=0}^{2^N-1}c_{k,l}T_k(u)T_l(v)$, where $u=u(x)=\frac{2x-(x_a+x_b)}{x_b-x_a}$ and $v=v(y)=\frac{2y-(y_a+y_b)}{y_b-y_a}$, which is a linear transformation that maps the problem domain 
$\Omega_P = [x_a,x_b] \times [y_a,y_b]$ with coordinates $(x,y)$ to the Chebyshev domain $\Omega_C = [-1,1] \times [-1,1]$ with coordinates $(u,v)$, as shown in the top panels of Fig.~\ref{fig:SCP}(a,b). The reason why this shifted transformation is required is that $T_{k}(x)$ $(T_{l}(y))$, the Chebyshev polynomials of the first kind, only satisfies a discrete orthogonality condition and forms a complete orthogonal basis over the range $x \in [-1,1]$ ($y \in [-1,1]$). In $\Omega_C$, the problem becomes a matter of determining the coefficients $c_{k,l}$, which can be found after a successful training of the parameterized quantum circuits. The model $p_Q(u,v)$ is trained on a \textit{training grid} consisting of 2D Chebyshev nodes $\{u^{\text{Ch}}_i\}_{i=0}^{2^N-1} \times \{v^{\text{Ch}}_j\}_{j=0}^{2^N-1} $ plus additional half-index points $\{u^{\text{Ch}}_{i+1/2}\}_{i=0}^{2^N-2} \times \{v^{\text{Ch}}_{j+1/2}\}_{j=0}^{2^N-2} $. The \textit{training grid} is schematically illustrated as a black-dotted grid within $\Omega_C$ with the total number of training points denoted by $G_T = (2^{N+1}-1)^2$. Once the model got trained, sampling is carried out through projective measurements in the same domain (Fig.~\ref{fig:SCP}(b), bottom). 

Finally, the sampled $p_Q(u,v)$ is mapped back to the problem domain to obtain the sampled histogram $p_Q(x,y)$ (Fig.~\ref{fig:SCP}(a), bottom) by $x=x(u)=\frac{(x_b-x_a)u+(x_a+x_b)}{2}$ and $y=y(v)=\frac{(y_b-y_a)v+(y_a+y_b)}{2}$. Specifically, we perform classical post-processing tasks on a batch of measured binary datasets after projective measurements in the computational basis. This involves periodically dropping out those bits with zero probability, concatenating rest of the bits in a sequential way, and then reshaping the resulting bitstring into a 2D array with the size compatible to the Hilbert space of 2$(N+S)$-qubit states, where $S$ is the number of extended registers. Therefore, the sampled $p_Q(u,v)$ is represented by the processed samples spatially assigned to a \textit{sampling grid} composed of 2D Chebyshev nodes $\{u^{\text{Ch}}_i\}_{i=0}^{2^{(N+S)}-1} \times \{v^{\text{Ch}}_j\}_{j=0}^{2^{(N+S)}-1} $, schematically illustrated as a color-dotted grid within $\Omega_C$ with the total number of sampling points denoted by $G_S=(2^{N+S})^2$. For the purpose of visualization, we set $N=2$ and superimpose both the training and sampling grids for different values of $S$, as shown in Fig.~\ref{fig:SCP}(c). The ratios of $G_S$ to $G_T$ are 0.33, 1.31 and 5.23, respectively, and they converge to a constant value of 0.25, 1 and 4 when $N$ gets larger. Considering that the overall probability sum for all possible outcomes stays at one, this fact implies that the maximum probability will be decreased by four folds as the number of the extended registers increases.

Most importantly, the quantum model is trained on a sparse \textit{training grid} given insufficient training dataset, whereas during sampling stage for $S>0$, the trained model provides predictions about unseen data at a controllable dense \textit{sampling grid} not spatially overlapped with the \textit{training grid}, i.e., the learned distribution of the training data has enough information to mimic the actual distribution of the data, which enables robust representations of generated images through so-called quantum generalization/interpolation. In the following, we assume that the problem domain of interest $\Omega_P = [10^{-2},1] \times [1, 10000]$ is expressed in terms of the momentum fraction and energy scale $(z,Q)$, as shown in Fig.~\ref{fig:workflow}(d).

\begin{figure}[h]
\begin{center}
\includegraphics[width=\linewidth]{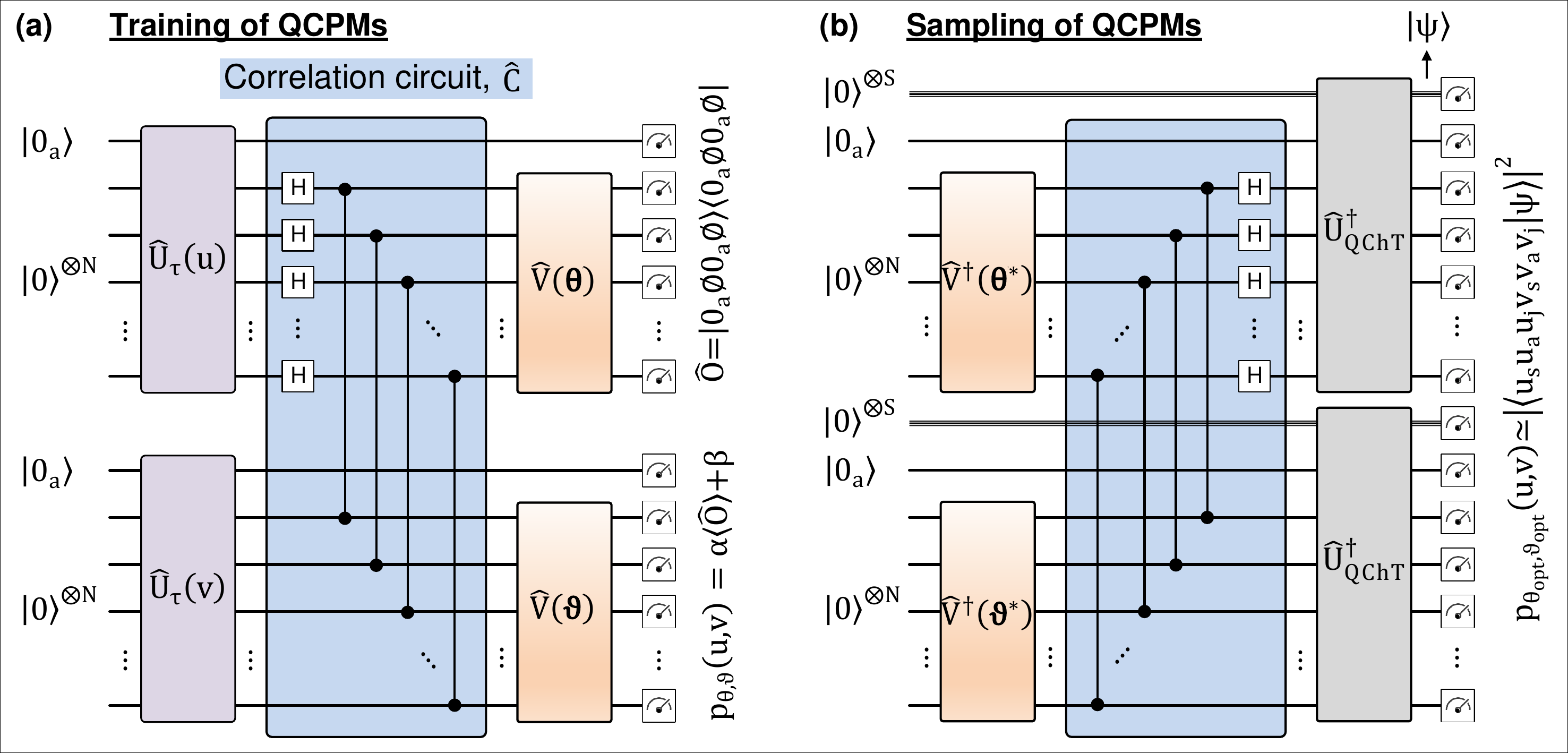}
\end{center} 
    \caption{
    (a) Quantum circuit used to train the multivariate distribution in the QCPM latent space, where a correlation circuit $\hat{\mathcal{C}}$ is sandwiched between two identical sets of feature map circuits and variational Ans\"atze. Measured observable is defined as $\hat{\mathcal{O}} = |0_a \mathrm{\o} 0_a \mathrm{\o} \rangle \langle 0_a \mathrm{\o} 0_a \mathrm{\o}|$, where $|\mathrm{\o}\rangle \equiv |0\rangle^{\otimes N}$. Here, $\alpha$ and $\beta$ are trainable scaling and bias parameters. (b) Quantum circuit used to sample the multivariate distribution from the trained model, where the inverse versions of the same parameterized circuits are applied with $\theta^*$ and $\vartheta^*$ being retrieved after the optimization procedure, followed by the inverse versions of the same correlation and two identical sets of quantum Chebyshev transform  circuits associated with extended registers of $S$ qubits ($\ket{0}^{\otimes S}$) in parallel, 
    for fine sampling in the computational basis $|u_s u_a u_j v_s v_a v_j\rangle$. The quantum state prior to measurement is denoted as $|\psi\rangle$.}
\label{fig:2DQCs}
\end{figure}

The training and sampling quantum circuits that form the QCPM employed here are presented in Fig.~\ref{fig:2DQCs}.

The training circuit (Fig.~\ref{fig:2DQCs}(a)) is composed of two Chebyshev feature map circuits $\hat{\mathcal{U}}_{\tau}(u)$ and $\hat{\mathcal{U}}_{\tau}(v)$ for encoding two independent variables, $u$ and $v$ $\in \Omega_C$, in parallel registers, followed by a correlation circuit $\hat{\mathcal{C}}$ and two separate variational Anz\"atze $\hat{\mathcal{V}}(\theta)$ and $\hat{\mathcal{V}}(\vartheta)$. The correlation circuit generates Bell-like entangled states to make two otherwise independent latent variables correlated for efficient training of fragmentation functions. The quantum model is defined as:
\begin{equation}
    p_{\mathrm{Q}}(u,v) = p_{\theta,\vartheta}(u,v) = \alpha |\langle 0_a \mathrm{\o} 0_a \mathrm{\o} | \bigl( \hat{\text{I}} \otimes \hat{\mathcal{V}}(\theta) \otimes \hat{\text{I}} \otimes \hat{\mathcal{V}}(\vartheta) \bigl) \hat{\mathcal{C}} \bigl( \hat{\mathcal{U}}_{\tau}(u) \otimes \hat{\mathcal{U}}_{\tau}(v) \bigl) |0_a \mathrm{\o} 0_a \mathrm{\o} \rangle|^2 + \beta,
\end{equation}
and it is trained to minimize a mean squared error loss by optimizing the angles $\theta^*$, $\vartheta^*$ and classical weighting parameters $\alpha_\text{opt}$ and $\beta_\text{opt}$. Since  
\begin{align}
    p_{\theta^*,\vartheta^*}(u,v) \simeq\ &\big| \langle u_s u_a u_j v_s v_a v_j| 
    \left( \hat{\mathcal{U}}_{\mathrm{QChT}}^\dagger \otimes \hat{\mathcal{U}}_{\mathrm{QChT}}^\dagger \right)
    \hat{\mathcal{C}}_{\text{ext}}^\dagger \nonumber \\
    &\times \left( \hat{\text{I}}^{\otimes (S+a)} \otimes \hat{\mathcal{V}}^{\dagger}(\theta^*) 
    \otimes \hat{\text{I}}^{\otimes (S+a)} \otimes \hat{\mathcal{V}}^{\dagger}(\vartheta^*) \right)
    |0_s 0_a \varnothing\, 0_s 0_a \varnothing \rangle \big|^2
\end{align}

the sampling circuit (Fig.~\ref{fig:2DQCs}(b)) consists of the inverse operations of the trained variational Anz\"atze, followed by an extended inverse correlation circuit $\hat{\mathcal{C}}^{\dagger}_{\text{ext}}$ and two identical sets of extended inverse quantum Chebyshev transform $\hat{\mathcal{U}}_{\mathrm{QChT}}^\dagger$ circuits. This procedure allows the basis transformation from the Chebyshev to computational basis spaces. In this context, we assume that the problem domain of interest is expressed in terms of the
momentum fraction and energy scale $(z, Q)$ $\in \Omega_P$.

\begin{figure}[h]
\begin{center}
\includegraphics[width=\linewidth]{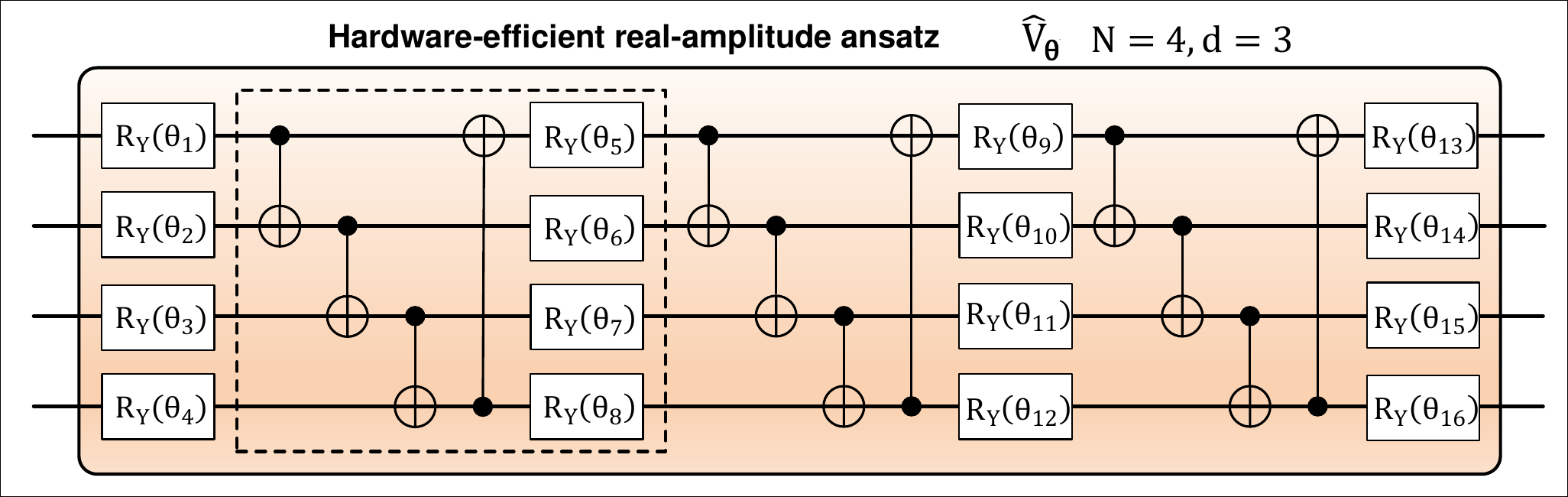}
\end{center}
    \caption{Hardware-Efficient Real-Amplitude (HERA) quantum circuit for \textit{N} = 4 qubits with a circuit quantum depth of \textit{d} = 3. The first block $(d=1)$ is framed by a dashed box. The HERA is composed of $N(d+1)$ tunable single-qubit $\mathrm{R}_\mathrm{Y}$ gates and $N d$ entangling (CNOT) gates. The training parameters are $\{\theta_i\}_{i=1}^{N(d+1)}$.
    }
\label{fig:HERA}
\end{figure}

The training setup for the QCPM is defined as follows. We fix the epoch count of the ADAM~\cite{adam} optimizer to $10^4$, the number of Ansatz layers to 3, and the number of qubits per variable to 4. Hence, both circuits have the same number of free parameters (16 for each variable) and require the same training time.  To optimize performance, we perform a sweep over the learning rate of the ADAM optimizer in the range $[0.1, 1.0]$ and select the model that achieves the highest accuracy in learning the functions $D_i^h(z,Q)$. All quantum simulations are performed using \texttt{Pennylane} \cite{Bergholm:2018cyq}, and the training process is accelerated with \texttt{JAX} \cite{jax2018github}. The structure of the variational Ansatz used for the variational quantum circuit is depicted in Fig.~\ref{fig:HERA}.

In contrast to GAN architectures, which involve the simultaneous training of two neural networks, the QCPM framework separates the training and sampling stages. Because the probability distribution is encoded in the orthogonal Chebyshev basis states provided by the quantum Chebyshev feature map circuit, the model effectively represents a wide range of functions with sufficient training data and learns patterns with a decent number of variational parameters. The number of variational parameters used for training in this work is 32 $(= 2 N (d+1))$ with $N$ = 4 and $d$ = 3. The trained model generates new data that fall within the trained distribution.
Beyond its generative capabilities, the QCPM efficiently encodes probability distributions into quantum states, making it a valuable tool for a broader range of quantum algorithms and applications.

\subsection{Results}\label{sec:results}

In this section, we apply the quantum generative Chebyshev model to Fragmentation Functions (FFs), using data, accessible via the LHAPDF6 interface \cite{Buckley:2014ana,lhapdfLHAPDFMain}. These FF sets are derived from data on hadron production in electron-positron Single-Inclusive Annihilation (SIA), one of the cleanest processes for studying hadron production, as it does not require simultaneous knowledge of PDFs. The process of obtaining FFs for different hadrons incorporates symmetry assumptions and empirical fragmentation preferences. Isospin symmetry is employed to relate FFs for oppositely charged pions and kaons, with the assumption that certain quarks preferentially fragment into specific hadrons, i.e. they are favoured, unfavoured towards a particular hadron. Charge conjugation symmetry is also utilized to simplify the analysis by linking FFs for positive and negative hadrons, allowing the functions for opposite charges to be expressed in terms of each other. 

We use the NNFF10\_PIsum\_nnlo set for pions ($\pi^\pm = \pi^+ + \pi^-$) and the NNFF10\_KAsum\_nnlo set for kaons ($K^\pm = K^+ + K^-$), both at next-to-next-to-leading order (NNLO) in perturbation theory.
Regarding the specific FFs, $D_i^h$ with $i$ being the parton and $h$ the hadron, that will be analyzed, the inclusive SIA data involves a total of five independent combinations of FF:
\begin{equation}
    \{ D_g^h, D_{b^+}^h, D_{c^+}^h, D_{d^++s^+}^h,  D_{u^+}^h    \},
    \label{eq:ffs}
\end{equation}
where $h$ represents the hadron the parton fragments into, $ D^h_{q+} \equiv D^h_q + D^h_{\bar{q}} $, and $ D_{d^++s^+}^h\equiv D^h_{d^+} + D^h_{s^+} $.

\begin{figure}[h]
\begin{center}
\includegraphics[width=0.8\linewidth]{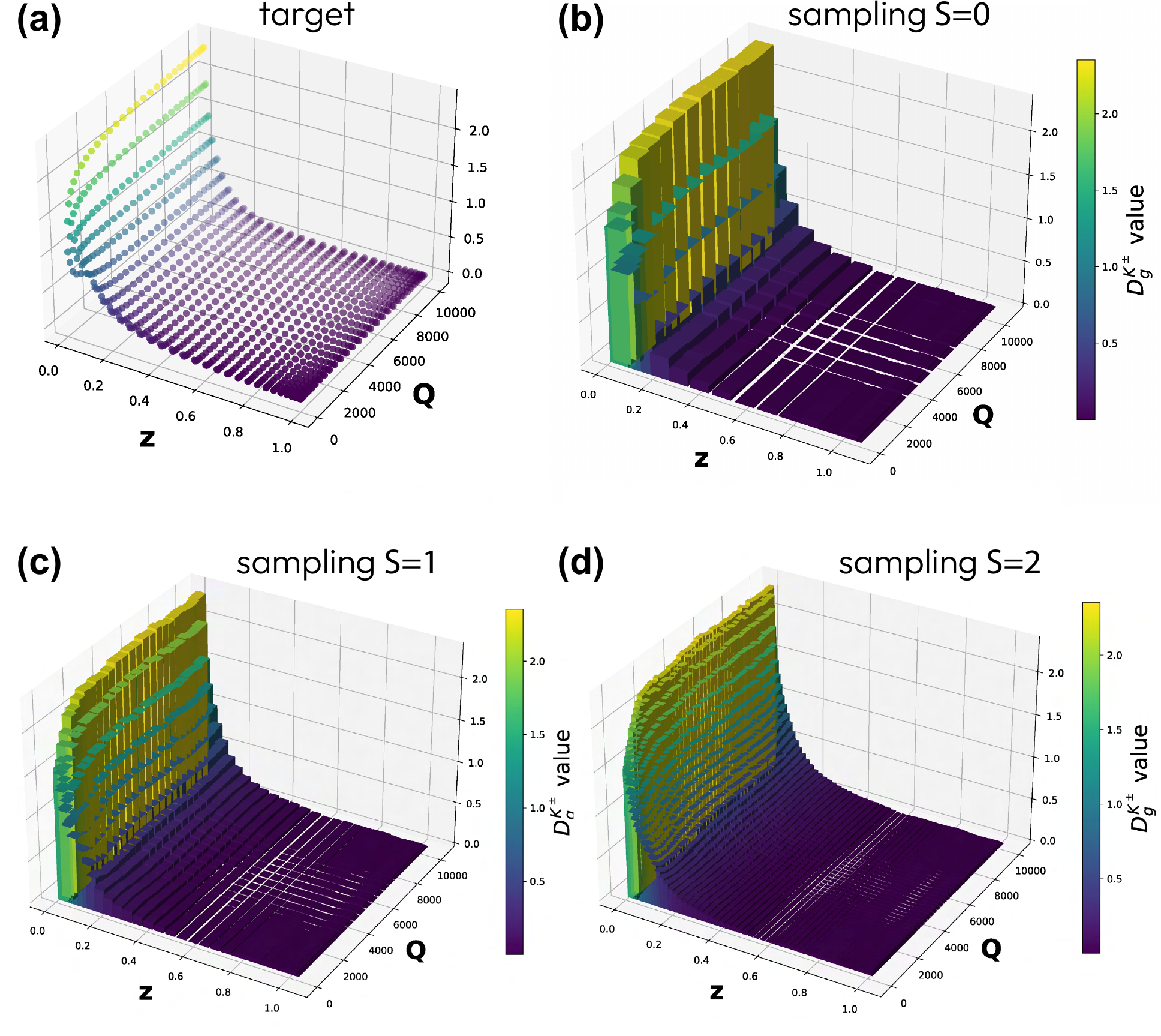}
\end{center}
\caption{Sampling of $D_{g}^{K^\pm}(z,Q)$, the FF of a gluon fragmenting into kaons, with $S=0,1,2$ additional qubits for each variable to interpolate in untrained regions. (a) Target distribution $D_{g}^{K^\pm}(z,Q)$. (b) Samples from trained QCPM of $D_{g}^{K^\pm}(z,Q)$ with the same number of qubits ($S=0$). (c, d)  Samples from $D_{g}^{K^\pm}(z,Q)$ with extended register ($S=1,2$).}
    \label{fig:FFresults}
\end{figure}

In Fig. \ref{fig:FFresults} we present the results of one particular FF as an example of the validity of the method, $D_{g}^{K^\pm}$, which corresponds to the sum of FFs of the gluon $g$ fragmenting into the kaon $K^+$ and its antiparticle $K^-$. Note that the visualization of the sampling for all the other FFs in Eq. \ref{eq:ffs} for $h=K^\pm, \pi^\pm$ are available in the Appendix~\ref{app:appendix_FF}.

The results of Fig.~\ref{fig:FFresults} show that the model correctly captures the behavior of the two-dimensional function $D_{c^+}^{K^\pm}(z,Q)$ in the region of interest. Fig.~\ref{fig:FFresults} (a) depicts the target function that the model is trained to learn, represented by the datapoints in the \textit{training grid}. Fig.~\ref{fig:FFresults} (b) shows the ability of the model to generate samples using the same quantum registers employed during training. Fig.~\ref{fig:FFresults} (c) and~(d) display the sampling performance when the model utilizes one ($S=1$) or two ($S=2$) additional qubits per variable. In the case of $S=2$ the model generalizes the behavior of the target function and makes predictions of the values of $D_{c^+}^{K^\pm}(z,Q)$ in untrained regions. Since one can increase the number of extra qubits for sampling, this approach shows significant potential as an effective method for achieving natural quantum interpolation with the desired accuracy by adding more qubits.

At this point, it is of particular interest to analyze how correlations between the variables affect the training process of the QCPMs. In this study, we explicitly introduce correlations between the variables $z$ and $Q$ in the training circuit and analyze their impact on the performance of the models. The introduction of correlations between variables $z$ and $Q$ in the training circuit is motivated by their analytical relationship through the DGLAP evolution equations. 
In this work, we use a heuristic-based approach, where we explicitly introduce a correlation circuit $\hat{C}$, as illustrated in Fig.~\ref{fig:2DQCs}, to entangle the registers that load $z$ and $Q$, which we refer to as $\mathcal{Z}$ and $\mathcal{Q}$. This circuit combines Hadamard gates ($H$) applied to the first variable with controlled-$Z$ gates ($CZ$). Our aim is to evaluate the impact of these correlations on QCPM performance and infer correlations.  

Specifically, we compare the accuracy of the models with (w/ CC) and without (w/o CC) correlations between the variables and under identical conditions (same number of qubits, trainable parameters and optimizer iterations). To do so, we consider the same setup as in Fig. \ref{fig:FFresults}. Since we are solving a regression problem, we use the coefficient of determination $R^2$ as the accuracy to quantify the goodness of the fit.

\begin{figure}[h]
\begin{center}
 \includegraphics[width=0.7\linewidth]{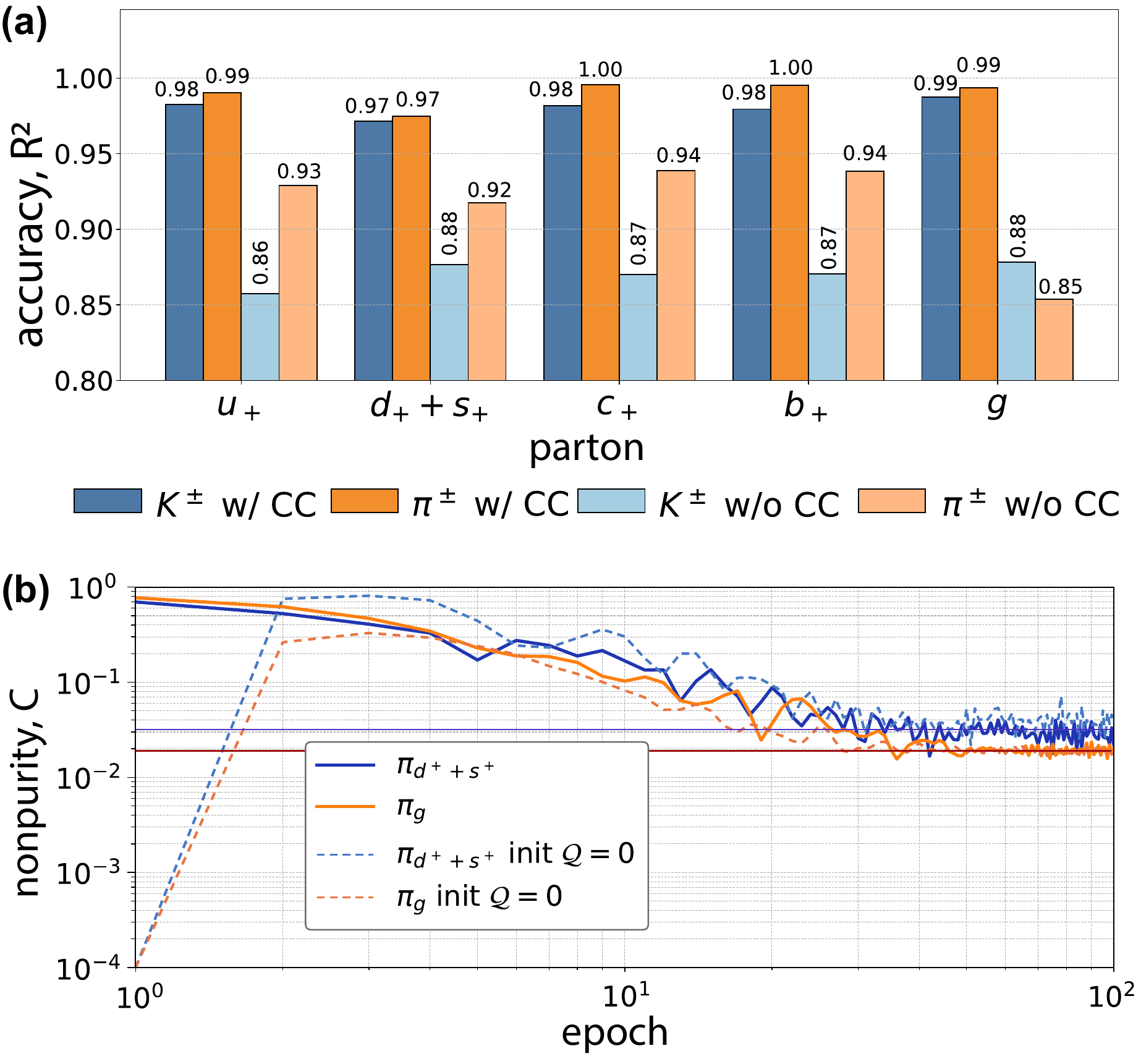}   
\end{center}
    \caption{(a) Accuracy ($R^2$) comparison of QCPMs for learning FFs $D_i^h(z, Q)$ with (w/ CC) and without (w/o CC) correlations between $z$ and $Q$. (b) Nonpuriry coefficient $C = |1-\gamma_{\mathcal{Z}}|$ of system $\mathcal{Z}$ for 100 training epochs with the fixed correlation circuit (solid curves, log-log curves). Dashed curves represent $C$ when the system starts in a product state without entanglement between registers (no correlations). Solid curves highlight the values of nonpurity after training.}
    \label{fig:results_correlations}
\end{figure}

The results presented in Fig. \ref{fig:results_correlations}(a) demonstrate that incorporating the correlation circuit $\hat{C}$ in the QCPM, leads to an improvement in performance across all the FFs studied. While models without correlations could reach similar accuracy with more optimizer iterations or additional trainable parameters, under the same conditions, the models with correlations consistently outperform those without. These findings highlight an important insight offered by the developed quantum probabilistic models: it is possible to infer, indirectly, the degree of correlation between physical variables by analyzing the performance of the quantum model and the role of entanglement between $\mathcal{Z}$ and $\mathcal{Q}$. While correlations are typically estimated from sample statistics and require large datasets, we propose assessing them through the entanglement structure within QCPMs. This can be done using various measures, such as entanglement entropy, purity, or mutual information, this is further explored in Appendix~\ref{app:appendix_corr}. In Fig.  \ref{fig:results_correlations}(b)  we focus on purity ~\cite{Scali2024}, as it can be efficiently evaluated via the SWAP test~\cite{Paine:2022uim}.

Specifically, to quantify the cross-variable correlations we define the nonpurity as $C=|1-\gamma_{\mathcal{Z}}|$, where $\gamma_\mathcal{Z}(\rho)\equiv\textrm{Tr}_{\mathcal{Q}}(\rho_\mathcal{Z}^2)$ is the purity of the subsystem $\mathcal{Z}$, with $\mathcal{Q}$ being traced out. We study the evolution of $C$ during the sampling stage of the QCPM, which includes a fixed correlation circuit, and present the behavior of $C$ over 100 training epochs in Fig.\ref{fig:results_correlations}(b) (solid lines for two representative processes). Initially, the nonpurity is relatively high, around $C_0 \approx 0.7$, but gradually decreases to a small yet non-zero value, $C_{100} \approx 2 \times 10^{-2}$ (see red and blue curves in Fig. \ref{fig:results_correlations}(b)). Note that similar final values of $C$ are observed even when starting from uncorrelated register initializations (dashed curves in Fig. \ref{fig:results_correlations}(b)). Although modest, this residual nonpurity indicates the presence of non-negligible $\mathcal{Z}$–$\mathcal{Q}$ correlations necessary to achieve the model's final accuracy, as shown in Fig.~\ref{fig:results_correlations}(a). Overall, the analysis of QCPM trained on multidimensional datasets leads to quantifiable measures of cross-correlations to high accuracy.

\section{Conclusions}\label{app:concl_qcheb}

In this Chapter we have studied how Quantum Chebyshev Probabilistic Models (QCPMs) learns multivariate HEP distributions, inferring relevant properties, and performing generative modeling by sampling of quantum circuits. In particular, we applied QCPMs to describe fragmentation functions (FFs) of charged pions and kaons~\cite{deLejarza:2025upd}, which are essential for understanding hadron production in high-energy collisions.
This approach aligns well with previous works, which have established that FFs present a polynomial dependence on the momentum fraction $z$. Hence, the Chebyshev basis proves to be particularly suitable for capturing this behavior.

Moreover, the Quantum Chebyshev Probabilistic Model presents some interesting features.
First, we have demonstrated that the generalized QCPM enables the extension of the Chebyshev space for modeling the largest class of probability distributions. Moreover, once the model is trained, it allows for a $2^{2(S-1)}$-fold increase in sampling grids by incorporating extended registers, facilitating a natural ``quantum interpolation'' in regions where the model lacks direct training data. This stands in contrast to previous classical models, which often require the computationally expensive DGLAP evolution for interpolation in untrained regions. On the other hand, the Chebyshev quantum model's structure enables us to study the correlation between the momentum fraction $z$ and the energy scale $Q$. We have performed an analysis to examine how correlations between quantum registers encoding $z$ and $Q$ affect the model's training performance. We found that introducing entanglement between these registers leads to improved model accuracy compared to non-entangled models under identical conditions. This finding extends what was found in previous studies suggesting that entanglement (``quantumness'') can enhance model training. These results underscore the potential of QML in analyzing quantum physical processes, such as HEP ones, and motivate the need for further research into how quantum computing techniques can be leveraged to study quantum systems.

\chapter{Outlook and final remarks}\label{chapter:outlook}

The Standard Model (SM) of particle physics is one of the most successful theories to date, with the discovery of the Higgs boson in 2012, 13 years ago, at the CERN's Large Hadron Collider (LHC) as a major breakthrough in high-energy physics (HEP). Further sophisticated HEP experiments are being planned to dig deeper into the origin of the universe and the fundamental content of matter. In this regard, these more complex experiments, demand advanced computational tools, both to analyze the vast amounts of data produced in high-energy collisions and to generate more accurate and precise theoretical predictions.

In this context, quantum computing has emerged over the past few decades as a promising technology with the potential to surpass the capabilities of classical devices in data analysis. Nevertheless, significant progress is still required. Both in hardware, where better quantum processors must be built, and in software, where more efficient and resilient algorithms are needed to achieve quantum advantage. One particularly interesting application lies in HEP. Inspired by Richard Feynman's pioneering idea of using quantum computers to simulate quantum systems, it seems natural to apply quantum computing to analyze HEP data, which is essentially governed by Quantum Field Theory. In particular in this Thesis we have explored different avenues of research where quantum algorithms can be used for HEP usecases. 

In Chapter~\ref{chap:qjets} we introduced two quantum circuits to speed up clustering algorithms:  a quantum subroutine for computing Minkowski distances and a quantum routine for tracking the maximum value in unsorted data. Employing these quantum subroutines, we designed quantum versions of three widely used clustering algorithms: \texttt{K-means}, Affinity Propagation, and $k_T$-jet clustering. For \texttt{K-means}, we implemented a quantum version that follows the classical structure but replaces Minkowski distance computations and minimum distance calculations with quantum circuits. This approach maintains clustering accuracy while offering exponential speedups in vector dimensionality $d$ and number of clusters $K$ when executed on quantum hardware with qRAM. The quantum Affinity Propagation algorithm follows a similar design, using quantum Minkowski computations to evaluate similarities, again achieving exponential speedup obtaining the same performance. We also introduced a quantum approach to $k_T$-jet clustering, where the minimum distance is computed with our quantum algorithm, that improves the runtime from $\mathcal{O}(N^3)$ classically to $\mathcal{O}(N^2\log N)$ quantumly, with further potential to reach $\mathcal{O}(N\log N)$ by incorporating geometric nearest-neighbor optimizations, analogous to those used in the classical \texttt{FastJet} package. Across all three cases, our quantum simulations demonstrated clustering efficiencies comparable to classical methods, while offering theoretical advantages in scalability and runtime when executed on quantum computers.

Then, in Chapter~\ref{chap:qint} we introduced a novel Quantum Monte Carlo integrator, dubbed Quantum Fourier Iterative Amplitude Estimation (QFIAE), which leverages Quantum Machine Learning~(QML) and Quantum Amplitude Estimation (QAE) to overcome limitations of earlier quantum integration methods like Fourier Quantum Monte Carlo Integration (FQMCI). The key feature of QFIAE is that it uses a Quantum Neural Network (QNN) to learn the target function that we want to integrate and extracts the Fourier series that this circuit represents. Then, the Fourier trigonometric components are easily integrated using QAE surpassing the problem of encoding arbitraryly complicated target functions. Unlike FQCMI, it provides with a full quantum pipeline and eliminates the need for classical numerical integration when estimating Fourier coefficients. This makes QFIAE, to the best of our knowledge, the first end-to-end quantum integrator capable of preserving the full quadratic speedup from amplitude amplification. We began by validating the method on a one-dimensional scattering process, $e^+e^- \rightarrow q \bar{q}$, where QFIAE achieved a relative error of about 1\% with circuits that remain within the depth and width limits of current NISQ devices. Encouraged by these results, we applied the algorithm to more complex tasks: evaluating loop Feynman integrals using the Loop-Tree Duality~(LTD). In one dimension, we successfully ran the full algorithm in the quantum computers of  \texttt{Qibo} and IBM Quantum for tadpole integrals, reaching errors around 5\%. We then extended QFIAE to higher dimensions, applying it to scalar and tensor triangle loop integrals in two dimensions, and demonstrating accurate and reliable estimates. Our most advanced application combined tree and loop-level Feynman diagrams to compute a physical observable for the first time on a quantum compute.  The decay rate at next-to leading order (NLO) in perturbative Quantum Field Theory. This milestone, achieved within the LTD framework using QFIAE, demonstrates the feasibility of applying quantum algorithms to realistic HEP processes. While quantum advantage is not claimed, our results represent a significant step forward, as the results are in agreement with analytical values. From a quantum computing point of view, we found that the QNNs' Ansatz we chose can handle realistic regression problems effectively, with a good balance between trainability and expressibility.

Finally, in Chapter~\ref{chap:qchebff} we 
employed Quantum Chebyshev Probabilistic Models (QCPMs) as a new approach for learning and generating multivariate distributions in HEP. Specifically, we applied these models to fragmentation functions (FFs), which describe how quarks and gluons transform into hadrons during high-energy collisions.
One of the strengths of this method lies in its flexibility, since QCPMs can be pretrained to match known marginal distributions and then fine-tuned to capture inter-variable correlations more accurately. Our results showed that introducing entanglement between the quantum registers representing the variables $z$ and $Q$ led to a noticeable improvement in training performance, supporting the growing evidence that quantum correlations can help in the learning process in quantum models. On the generative side, QCPMs demonstrated strong generalization capabilities, accurately interpolating between training points. Thanks to the Chebyshev basis, our models could sample over finely spaced grids, and the use of additional $S$ qubits in the quantum registers exponentially increased the resolution of the sampling space by a factor of $2^{2(S-1)}$. These results highlight the potential of QCPMs as powerful QML tools for modeling complex, structured distributions in particle physics. By enabling both precise inference and high-resolution generative modeling, this approach opens new directions for using quantum computing to gain deeper insight into fundamental processes in high-energy collisions.

Overall, the work presented in this thesis paves the way for developing more specialized and powerful quantum methods that make full use of core quantum features like superposition and entanglement to improve computational methods to analyze HEP data and Quantum Field Theories. A key direction for future research is to scale these techniques to larger, fault-tolerant quantum systems, where real quantum advatange over classical machines can be fulfilled.

\chapter{Resumen de la tesis}\label{chapter:resumen}

A continuación, se presenta una descripción resumida del contenido de esta tesis. En ella se incluye una introducción a los temas principales abordados, un resumen de los objetivos planteados, una explicación de la metodología empleada, una síntesis de los principales resultados obtenidos y, finalmente, se exponen las conclusiones generales del trabajo.

\section*{Introducción}\label{sec:resumen_intro}

El Modelo Estándar es la teoría más exitosa en el ámbito de la física de partículas. Una de sus mayores fortalezas es la capacidad que tiene de producir una descripción altamente precisa de tres de las cuatro interacciones fundamentales de la naturaleza: la fuerza electromagnética, la interacción nuclear débil y la interacción nuclear fuerte. En el Capítulo~\ref{chap:sm} se presenta una descripción detallada de sus fundamentos y de sus características más relevantes.

No obstante, a pesar de sus numerosos éxitos, existe un consenso en la comunidad científica de que el Modelo Estándar no representa una teoría completa de la naturaleza a nivel fundamental. 
Estas limitaciones continúan siendo un campo activo de investigación y no se ha encontrado una solución hasta la fecha. La falta de contradicciones internas o refutaciones directas deja sin una guía clara para extender el Modelo Estándar, lo que obliga a explorar nuevas posibilidades sin una dirección definida. En esta línea, se plantean nuevos experimentos en colisionadores de alta energía como el Large Hadron Collider (LHC) para llegar a regímenes de energía no explorados que puedan arrojar algo de luz sobre cuales son las piezas que faltan en el rompecabezas que conforma el Modelo Estándar de la física de partículas.

Paralelamente, se está produciendo lo que ha sido bautizado como la ``segunda revolución cuántica''. Esta hace referencia a la capacidad de controlar sistemas cuánticos de forma coherente aprovechando sus propiedades inherentes para almanecenar, manipular y leer información. Este campo recibe el nombre de información cuántica. Dentro del mismo, una de las aplicaciones más immediatas es la capacidad de realizar cálculos y operaciones de interés, lo cual recibe el nombre de computación cuántica. La computación cuántica y los ordenadores cuánticos, que son los dispositivos capaces de realizar estas operaciones, han emergido en los últimos años como una tecnología incipiente con mucho potencial para revolucionar la forma en que resolvemos ciertos problemas y analizamos los datos. Sin embargo, la computación cuántica es un campo de investigación que, con apenas un par de décadas de historia, aún se encuentra en sus incicios. Hacen falta muchos avances tanto a nivel de desarrollo de algoritmos más eficientes y con capacidad de aprovechar las diferentes propiedades cuánticas, como a nivel de desarrollo de dispositivos capaces de ejecutarlos. Actualmente se dice que nos encontramos en lo que se llama la era NISQ (Noise Intermediate Scale Quantum), en la que existen ordenadores cuánticos de un número limitado de bits cuánticos (qubits) que además son propensos a errores. En este paradigma es dificil pensar que podamos obtener una ventaja cuántica en el futuro imediato. No obstante, el campo está evolucionando a una velocidad vertiginosa y no es para nada descabellado pensar en obtener resultados útiles de aplicación industrial y científica en las próximas décadas.

Con todo esto en mente e inspirados por la célebre frase de Richard Feynman: ``La naturaleza no es clásica, maldita sea, y si quieres hacer una simulación de la naturaleza, más te vale que sea cuántica'', resulta natural considerar la computación cuántica como una herramienta prometedora para analizar datos de física de altas energías, ya que estos están descritos por la Teoría Cuántica de Campos. En esta tesis, exploramos distintas líneas de investigación donde los algoritmos cuánticos pueden aplicarse a problemas relevantes en la física de partículas.

\section*{Objetivos}\label{sec:resumen_objetivos}

El objetivo principal de esta tesis es identificar problemas en física de partículas con un coste computacional significativo y explorar su resolución mediante algoritmos cuánticos que puedan ofrecer una ventaja, ya sea teórica o heurística. Asimismo, se aborda el diseño y la aplicación de dichos algoritmos a estos problemas concretos. 

En particular, podemos diferenciar los algoritmos estudiados en dos tipos. Por un lado se estudian algoritmos cuánticos que ofrecen una ventaja teórica, los cuales garantizan una mejora al ser ejecutados en ordenadores cuánticos suficiente grandes. Por otro lado, también se diseñan algortimos cuánticos variacionales, los cuales funcionan de forma parecida a métodos ampliamente empleados en el aprendizaje automático (Machine Learning), como las redes neuronales, y por este motivo se asocian habitualmente a lo que se conoce como el aprendizaje automático cuántico (Quantum Machine Learning). Estos algoritmos variacionales, al igual que sus análogos clásicos, no garantizan una ventaja en el rendimiento al ejecutarse en ordenadores cuánticos de tamaño suficientemente grande. En su lugar, ofrecen soluciones alternativas de carácter ``heurístico'', que, al escalar el problema, suelen mostrar ventajas computacionales significativas frente a otros métodos. Aunque la ausencia de una prueba matemática que asegure una mejora en la escalabilidad podría parecer una debilidad, lo cierto es que esta situación también se da en el aprendizaje automático clásico, donde, a pesar de la falta de garantías teóricas, las soluciones que producen han demostrado funcionar en una amplia gama de problemas computacionales.

Respecto a los casos de uso que se estudian en esta tesis, nos centraremos en los siguientes tres. Clusterización de jets de partículas, en el Capítulo~\ref{chap:qjets}, integración de diagramas de Feynman y tasas de desintegración a next-to-leading order (NLO), en el Capítulo~\ref{chap:qint}, y generación de funciones de fragmentación, en el Capítulo~\ref{chap:qchebff}.

La clusterización de jets es un proceso que consiste en agrupar partículas que comparten ciertas propiedades para analizarlas como un solo objeto, conocido como jet. La mayoria de algoritmos empleados para resolver este problema se basan en calcular las distancias entre las partículas y agrupar aquellas que se encuentran más próximas. En este contexto, el objetivo es desarrollar y aplicar dos algoritmos cuánticos: uno para calcular eficientemente las distancias entre partículas, y otro que identifique el mínimo de esas distancias para determinar qué partículas deben formar parte del mismo jet.

En cuanto al Capítulo~\ref{chap:qint}, el objetivo es el diseño de algoritmo cuántico de integración tipo Monte Carlo para poderlo aplicar a integrales de Feynmann de loop y al cálculo de observables físicos como las tasas de desintegración. Este algoritmo cuántico combina una red neuronal cuántica (algoritmo variacional) con una variante del algoritmo de Grover (algoritmo de ventaja teórica).

Finalmente, el objetivo del Capítulo~\ref{chap:qchebff} es desarrollar un algoritmo variacional capaz de aprender las funciones de fragmentación asociadas a distintos partones al fragmentarse en hadrones, y de reproducir estos resultados con una resolución superior a la utilizada durante el entrenamiento. Esto permitiría realizar una interpolación ``cuántica'' de los valores que toman dichas funciones en las regiones de interés.

\section*{Metodología}\label{sec:resumen_metodologia}

Para obtener los resultados presentados en esta tesis, se ha empleado una metodología precisa y específica que consta de algoritmos deterministas, de aprendizaje automático y de algoritmos cuánticos. Podemos dividir los métodos empleados por secciones.

\subsection*{Algoritmos para jet clustering}\label{subsec:algos_jet}

Los algoritmos para clusterización de jets empleados en el capítulo \ref{chap:qjets} de esta Tesis son modificationes de: \texttt{K-means}, Affinity Propagation y $k_T$-jet. En particular las modificaciones de estos algoritmos consiste en sustituir las rutinas clásicas que calculan distancias y mínimos por algoritmos cuánticos integrados en la ``pipeline'' de cada método.

En primer lugar, en esta tesis presentamos un algoritmo cuántico que permite el cálculo de una distancia de Minkowski entre dos puntos. La motivación detrás de usar este tipo de distancia es que en física de altas energías (HEP), los vectores suelen definirse en un espacio-tiempo de cuatro dimensiones con métrica de Minkowski. Estos vectores se expresan como $x_i = (x_{i,0}, {\bf x}_i)$, donde $x_{i,0}$ es la componente temporal y ${\bf x}_i$ representa las tres componentes espaciales. En general, supondremos $d$ dimensiones espacio-temporales, con $d-1$ indicando el número de componentes espaciales. El análogo de la distancia Euclídea clásica en el espacio de Minkowski es la suma invariante al cuadrado $s_{ij}^{\rm (C)}$, también conocida como masa invariante al cuadrado cuando se trabaja con cuadrimomentos:
\begin{equation}
s_{ij}^{\rm (C)} = (x_{0,i}+x_{0,j})^2 
- |{\bf x}_i + {\bf x}_j|^2~.
\end{equation}

Esta cantidad invariante de Lorentz se utiliza como métrica de similitud entre momentos de partículas. Además, corresponde a la distancia empleada en ciertos algoritmos tradicionales de clusterización de jets en colisionadores $e^+e^-$~\cite{JADE:1982ttq,Bethke:1991wk,Rodrigo:1999qg}. Calcular esta distancia tipo Minkowski mediante un algoritmo cuántico requiere dos aplicaciones del circuito cuántico \textit{SwapTest} (detallado en el Apéndice A), una para las componentes espaciales y otra para las temporales. La distancia espacial se obtiene a partir de los estados:
\begin{equation}
\ket{\psi_1} =\frac{1}{\sqrt{2}} \left( \ket{0, x_i} + \ket{1, x_j} \right), \qquad
\ket{\psi_2} =\frac{1}{\sqrt{Z_{ij}}} \left( |{\bf x}_i| \ket{0} -|{\bf x}_j| \ket{1} \right),
\label{eq:varphi}
\end{equation}
donde $Z_{ij}=|{\bf x}_i|^2+|{\bf x}_j|^2$. Mientras que la distancia temporal se obtiene a partir de:
\begin{equation}
\ket{\varphi_1} = H \ket{0} = \frac{1}{\sqrt{2}} \left( \ket{0} + \ket{1} \right)~, \qquad 
\ket{\varphi_2} =\frac{1}{\sqrt{Z_0}} \left(x_{0,i} \ket{0} +x_{0,j} \ket{1} \right)~,
\label{eq:minkstates}
\end{equation}
donde $Z_{0}=x_{0,i}^2+x_{0,j}^2$. Aplicando el \textit{SwapTest} a estos estados se obtiene:
\begin{equation}
\begin{split}
P(|0\rangle|_{\text{time}})&=\frac{1}{2}+\frac{1}{2} |\langle \varphi_1| \varphi_2 \rangle|^2 \ , \\
P(|0\rangle|_{\text{spatial}})&=\frac{1}{2}+\frac{1}{2} |\langle \psi_1| \psi_2 \rangle|^2 \ ,
\end{split}
\end{equation}
donde los overlap están dados por
\begin{equation}
\
|\langle \varphi_1| \varphi_2 \rangle|^2 = \frac{1}{2Z_0}(x_{0,i} +x_{0,j})^2. \quad
|\langle\psi_1|\psi_2\rangle |^2 
= \frac{1}{2Z_{ij}}|{\bf x}_i-{\bf x}_j|^2.
\end{equation}

Por lo tanto la versión cuántica de la suma invariante al cuadrado resulta:
\begin{equation}
s_{ij}^{\rm (Q)} =2\big( 
Z_0(2P_{\Psi_3}(|0\rangle|_{\text{time}})-1)-Z_{ij}(2P_{\Psi_3}(|0\rangle|_{\text{spatial}})-1)\big). 
\label{eq:qdistancemink}
\end{equation}
Para los $\log_2 (d-1)$ primeros qubits, se aplica el \textit{SwapTest} a las componentes espaciales, asumiendo que los estados $\psi_1$ y $\psi_2$ se han cargado desde una qRAM en un tiempo de $\mathcal{O}(\log (d-1))$, ya que $\psi_1$ se codifica en $\log_2 (d-1)$ qubits. Para el último qubit, el \textit{SwapTest} se usa para las componentes temporales, y en este caso el coste es $\mathcal{O}(1)$, ya que solo se usan estados de una dimensión.

En segundo lugar, en esta tesis presentamos un algoritmo cuántico sencillo para identificar el valor máximo en una lista no ordenada $L[0, \ldots, N-1]$ de $N$ elementos. El procedimiento consta de dos pasos: 
\begin{enumerate}
    \item Los datos se codifican en un estado de $\log(N)$ qubits mediante codificación por amplitudes:
\begin{equation}
\ket{\Psi} = \frac{1}{\sqrt{L_{\text{sum}}}}\sum_{j=0}^{N-1}L[j] \, \ket{j}~,
\label{eq:amplitude encoding}
\end{equation}
donde $L_{\text{sum}} = \sum_{j=0}^{N-1}L[j]^2$ es una constante de normalización
    \item  El estado resultante se mide múltiples veces, y el índice observado con mayor frecuencia se asocia con el valor máximo.
\end{enumerate}

El cuello de botella del procedimiento radica en la codificación de los datos en el estado cuántico. Asumiendo que los datos están almacenados en una qRAM, como sería el caso en un ordenador cuántico universal real, la codificación requiere $\mathcal{O}(\log(N))$ pasos~\cite{lloyd2013quantum,2008QRAM,2008Arch,demartini2009experimental,2010Mag,PhysRevLett.105.140501,PhysRevLett.105.140502,PhysRevLett.105.140503}. Dado que el mejor algoritmo clásico requiere $\mathcal{O}(N)$ pasos, la mejora cuántica es exponencial bajo estas suposiciones.

Este método es probabilísticos y puede fallar en identificar correctamente el valor máximo absoluto cuando la desviación estándar de los datos es baja o los extremos son similares. Sin embargo, como se muestra en la Sección~\ref{app:qclusteringalgos}, estas limitaciones no afectan significativamente la aplicación en clusterización de jets. Más allá de HEP, este algoritmo puede ser útil en áreas como la Teoría de Valores Extremos (EVT)~\cite{smith1990extreme,Gumbel1958}, relevante en predicción de eventos raros a partir de grandes volúmenes de datos. Esto incluye aplicaciones en finanzas, meteorología, ingeniería, geología, tráfico, entre otras~\cite{Reiss2007,Coles2001,castillo2004}.

\subsection*{Algoritmos para  integración de funciones multidimensionales}\label{subsec:algos_int}

En el Capítulo~\ref{chap:qint} se describe el método de integración cuántica QFIAE introducido en~\cite{deLejarza:2023IEEE}, y que se utiliza para calcular integrales relevantes en física de partículas. Este algoritmo cuántico combina de forma coherente el poder de los circuitos cuánticos parametrizados ~PQC), el análisis de Fourier y la estimación de amplitud cuántica para aproximar integrales con alta precisión. En particular, se aplica a integrales Monte Carlo de la forma:
\begin{equation}
     I=\int_{x_{min}}^{x_{max}} p(x)f(x) dx.
\end{equation}
La estimación de esta integral se realiza evaluando el valor esperado de una función $0\leq f(x) \leq 1$ sobre entradas de $n$ bits $x\in \{ 0,1\}^n$, con probabilidad $p(x)$:
\begin{equation}
    \mathbb{E}[f(x)]=\sum_{x=0}^{2^n-1} p(x)f(x).
\end{equation}

El método ofrece una aceleración cuadrática en el número de muestras necesarias, gracias al uso de la estimación de amplitud cuántica basada en el algoritmo de Grover~\cite{Grover:1997fa}. Para ello, se definen los operadores:
\begin{equation}
    \mathcal{P}|0\rangle_n=\sum_{x=0}^{2^n-1} \sqrt{p(x)}|x\rangle_n ,
\end{equation}
\begin{equation}
    \mathcal{R}|x\rangle_n|0\rangle=|x\rangle_n \left(  \sqrt{f(x)}|1\rangle +\sqrt{1-f(x)}|0\rangle \right),
\end{equation}
que permiten construir el operador $\mathcal{A}$, el cual actúa como:
\beq
\begin{split}
|\psi\rangle &= \mathcal{A}|0\rangle_{n+1}=\mathcal{R}(\mathcal{P}\otimes\mathbb{I}^1)|0\rangle_{n+1}\\
&=\sum_{x=0}^{2^n-1} \sqrt{p(x)}|x\rangle_n \left( \sqrt{f(x)}|1\rangle +\sqrt{1-f(x)}|0\rangle \right).
\end{split}
\eeq

Definimos:
\begin{equation}
a=\sum_{x=0}^{2^n-1} p(x)f(x)= \mathbb{E}[f(x)]~,
\end{equation}
\begin{equation}
    |\tilde{\psi}_1\rangle=\frac{1}{\sqrt{a}}\sum_{x=0}^{2^n-1} \sqrt{p(x)}\sqrt{f(x)}|x\rangle_n~,
\end{equation}
\begin{equation}
 |\tilde{\psi}_0\rangle=\frac{1}{\sqrt{1-a}}\sum_{x=0}^{2^n-1} \sqrt{p(x)}\sqrt{1-f(x)}|x\rangle_n~,
\end{equation}
y el estado:
\begin{equation}
 |\psi\rangle=\sqrt{a}|\tilde{\psi}_1\rangle|1\rangle+\sqrt{1-a}|\tilde{\psi}_0\rangle|0\rangle~.
\end{equation}

Ahora se define el operador de Grover:
\begin{equation}
\mathcal{Q}=U_{\psi}U_{\psi_0}~,  
\end{equation}
donde:
\begin{equation}
 U_{\psi_0}=\mathbb{I}_{n+1}-  2\mathbb{I}_n|0\rangle \langle 0|~, \qquad U_{\psi}=\mathbb{I}_{n+1}-2|\psi\rangle  \langle\psi|~.
\end{equation}

Usando $a=\sin^2{\theta_a}$, se aplica el algoritmo IQAE. Este consta de tres pasos: determinar el número óptimo de aplicaciones de $\mathcal{Q}$, aplicar $\mathcal{A}$ y $\mathcal{Q}^k$, y finalmente extraer el valor de $a$:
\begin{equation}
    \mathcal{Q}^k\mathcal{A}|0\rangle_{n+1}=\sin{((2k+1)\theta_a)}|\tilde{\psi}_1\rangle|1\rangle +\cos{((2k+1)\theta_a)}|\tilde{\psi}_0\rangle|0\rangle~.
\end{equation}

Una estrategia eficiente para codificar funciones consiste en descomponer $f(x)$ como serie de Fourier:
\begin{equation}
f(x)=c+ \sum_{i=1}^\infty a_n\cos(n\omega x)+ b_n\sin(n\omega x)~.
\end{equation}
Esto permite usar QAE para estimar los componentes trigonométricos y combinarlos.

Para evitar el costo de calcular los coeficientes de Fourier por métodos clásicos, en~\cite{deLejarza:2023IEEE} se propone usar una red neuronal cuántica (QNN) para obtener la serie de Fourier. Se entrena un circuito siguiendo la arquitectura de ``data-reuploading''~\cite{P_rez_Salinas_2020}, que alterna bloques de codificación $\mathcal{S}(\vec{x})$ y bloques entrenables $\mathcal{A}(\vec{\theta_l})$:
\begin{equation}
    U_0 \equiv \mathcal{A}(\vec{\theta_0})~, \qquad U_l \equiv \mathcal{A}(\vec{\theta_l})\mathcal{S}(\vec{x})~.
\end{equation}

El circuito consta de dos componentes principales: un operador inicial, $U_0$, que prepara un estado en superposición, y un operador $U_l$, aplicado de forma repetida en capas con distintos parámetros entrenables $\vec{\theta_l}$. Este operador se encarga de mapear los datos de entrada y se optimiza según una métrica definida. La función de coste es el error cuadrático entre el valor esperado del circuito y el valor real de la función. Como muestra~\cite{Schuld_2021}, este valor esperado se puede expresar como:
\begin{equation}
    \langle \textit{M} \rangle (\vec{x}, {\vec{\theta}}) = \sum_{\vec{w} \in \Omega} c_{\vec{w}} e^{i\vec{x}\vec{w}}~,
\end{equation}
donde los elementos de la serie de Fourier se obtienen directamente del circuito cuántico. 

\begin{figure}[h]
       \centering
  \includegraphics[width=0.79\linewidth]{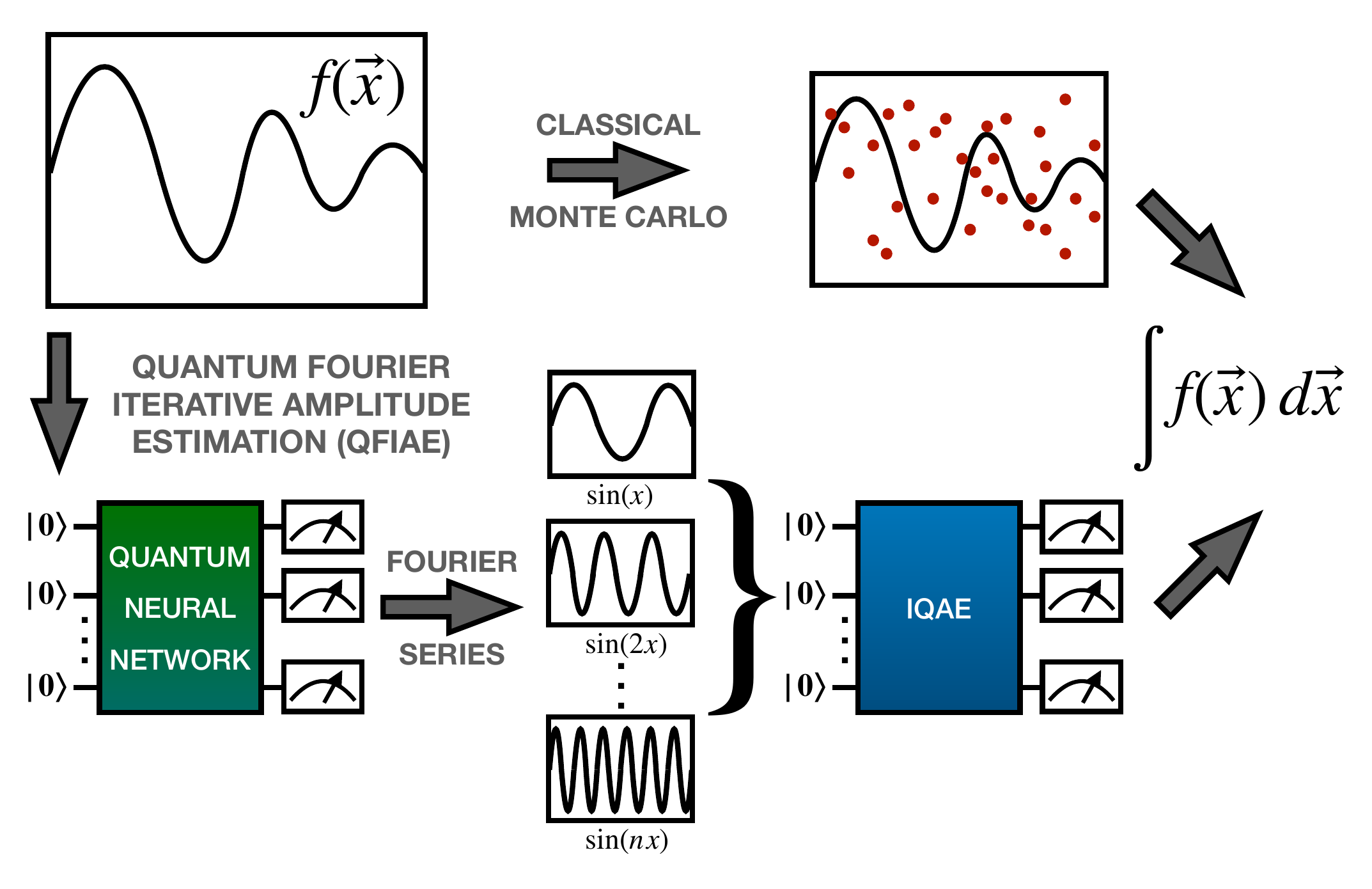}  
        \caption{Comparación entre los flujos de trabajo de la integración Monte Carlo clásica y el algoritmo cuántico QFIAE. En QFIAE, se introducen como entrada la función objetivo, la distribución de probabilidad y el intervalo de integración. A continuación, una QNN ajusta la función y permite extraer su serie de Fourier. Posteriormente, se utilizan IQAE para calcular los integrales de cada componente trigonométrico en el intervalo especificado. Finalmente, se combinan estos resultados ponderados por sus coeficientes para obtener el valor final de la integral.} 
        \label{fig:sketch_qfiae}
\end{figure}

En resumen, el método QFIAE consiste en descomponer la función objetivo en su serie de Fourier mediante el uso de una QNN. Posteriormente, cada componente trigonométrico se integra utilizando IQAE. El funcionamiento de este algoritmo se ilustra en la Fig.~\ref{fig:sketch_qfiae}, que ofrece una representación visual de su operativa.

\subsection*{Algoritmos para generación de funciones de fragmentación}

En el Capítulo~\ref{chap:qchebff} introducimos un modelo generativo cuántico basado en polinomios de Chebyshev. El objetivo principal de la mayoría de los modelos generativos cuánticos es permitir el muestreo en la base computacional, que consiste en los estados ortonormales $\{\ket{x_j}\}_{j=0}^{2^N-1}$ que satisfacen la condición delta de Kronecker $\langle x_j | x_{j'} \rangle = \delta_{jj'}$. Sin embargo, entrenar un modelo directamente en este espacio suele ser difícil, ya que requiere un control muy fino sobre las distintas probabilidades. Para abordar esto, se puede construir y entrenar el modelo en una base ortonormal diferente y luego mapearlo a la base computacional para realizar el muestreo. Con esto en mente, empleamos la base ortonormal de Chebyshev y su correspondiente transformada para construir nuestro modelo generativo~\cite{Williams:2023cuz}.

Primero, construimos el mapa de características adecuado para codificar los datos en el espacio de Chebyshev. Así, codificamos un estado cuántico (no normalizado) $\ket{\tau(x)}$ cuyas amplitudes estén determinadas por polinomios de Chebyshev de primer tipo, definidos como $T_k(x) = \cos(k \arccos(x))$, donde $k$ denota el grado. Este estado toma la forma
\begin{equation}
\ket{\tau(x)} = \frac{1}{2^{N/2}} T_0(x) \ket{\varnothing} + \frac{1}{2^{(N-1)/2}} \sum_{k=1}^{2^N - 1} T_k(x) \ket{k}, \label{eq:chebyshev-state}
\end{equation}
donde la amplitud del estado cero está ponderada por $T_0(x) = 1$. Estos polinomios satisfacen una relación de ortogonalidad discreta en puntos de muestreo específicos, conocidos como nodos de Chebyshev:
\begin{equation}
\sum_{j=0}^{2^N - 1} T_k(x_j^{\text{Ch}}) T_\ell(x_j^{\text{Ch}}) =
\begin{cases}
0 & \text{si } k \ne \ell, \\
2^N & \text{si } k = \ell = 0, \\
2^{N-1} & \text{si } k = \ell \ne 0.
\end{cases}
\end{equation}

Aquí, los nodos de Chebyshev están dados por $x_j^{\text{Ch}} = \cos\left(\pi(2j + 1)/{2^{N+1}}\right)$, correspondientes a las raíces de los polinomios de Chebyshev. Los estados resultantes $\{ \ket{\tau(x_j^{\text{Ch}})} \}_{j=0}^{2^N - 1}$ forman un conjunto ortonormal, con $\langle \tau(x_j^{\text{Ch}})| \tau(x_{j'}^{\text{Ch}})\rangle = \delta_{jj'}$.

Cabe señalar que, a diferencia de las bases computacional o de Fourier, que se basan en mallas uniformes, los nodos de Chebyshev forman una malla no uniforme dentro del intervalo $(-1, 1)$. Fuera de estos puntos nodales, los estados $\ket{\tau(x)}$ ya no son ortogonales. Se puede derivar una expresión analítica para su superposición al cuadrado cuando uno de los argumentos se fija en un nodo de Chebyshev:
\begin{equation}
\left| \braket{\tau(x')}{\tau(x)} \right|^2 =
\frac{\left[T_{2^{N+1} - 1}(x') T_{2^N}(x) - T_{2^N}(x') T_{2^{N+1} - 1}(x)\right]^2}{2^{2N}(x' - x)^2}.
\end{equation}

Esta identidad se deduce de la fórmula de Christoffel-Darboux para los polinomios de Chebyshev~\cite{Iten_2016}. Dado que estos estados no están normalizados en general, generarlos con un circuito cuántico implica usar un qubit auxiliar y normalizar el estado después de su preparación. Denotamos la versión normalizada como 
\begin{equation}
    \ket{\tilde{\tau}(x)} = \frac{\ket{\tau(x)}}{\sqrt{\braket{\tau(x)}}},
\end{equation}
que coincide con la versión no normalizada exactamente en los nodos de Chebyshev y se le aproxima a medida que $N$ crece.

El siguiente paso es diseñar un circuito cuántico que prepare el estado normalizado $\ket{\tilde{\tau}(x)}$, al que llamaremos mapa de características. Las amplitudes pueden ser incorporadas mediante una combinación de funciones exponenciales, cada una de las cuales puede implementarse usando un mapa de fase~\cite{Kyriienko:2022zyd}. Estos componentes pueden combinarse utilizando el marco de Combinación Lineal de Unitarios (LCU)~\cite{lcu}, donde los términos unitarios son controlados por el estado de un qubit auxiliar. La interferencia deseada se logra seleccionando la medición del auxiliar en el estado $\ket{0}$. En particular, la identidad $e^{-ix} = e^{ix} \cdot e^{-i2x}$ permite controlar solo uno de los componentes exponenciales, simplificando la implementación. Este enfoque permite construir combinaciones de exponenciales con pesos iguales, escaladas apropiadamente para reproducir las amplitudes asociadas con los polinomios de Chebyshev $T_k(x)$. Para ajustar correctamente la amplitud del término constante $T_0(x)$, se requiere una corrección adicional. Esto se logra aplicando una única iteración de un circuito de rotación tipo Grover~\cite{Grover:1996rk}, que rota el estado alrededor de $\ket{\varnothing}$ por un ángulo fijo. Dado que la medición del qubit auxiliar conmuta con esta rotación, colocamos la medición al final del circuito, completando la preparación del mapa de características deseado.

Una vez que hemos establecido cómo codificar los datos en la base de Chebyshev, necesitamos una transformación entre esta y la base computacional, y viceversa. Una vez que una base ortonormal está preparada, existe una correspondencia biunívoca (biyección) y una transformación unitaria que la mapea a cualquier otra base ortonormal. En este contexto, definimos la transformada de Chebyshev como
\begin{equation}
    \hat{U}_{\text{QChT}} = \sum_{j=0}^{2^N-1} \ket{\tau(x_j^{\text{Ch}})}\bra{x_j},
\end{equation}
que mapea los estados de la base computacional $\ket{x_j}$ a los estados de Chebyshev $\ket{\tau(x_j^{\text{Ch}})}$. La transformada de Chebyshev puede verse como una variante de la transformada del coseno~\cite{Klappenecker:2001xto}. Específicamente, el vector de amplitudes del estado $\ket{\tau(x_j^{\text{Ch}})}$ corresponde a la columna $(j+1)$-ésima de la matriz de la transformada discreta del coseno de tipo II, $\text{DCT}^{\text{II}}_N$, definida como
\begin{equation}
    \text{DCT}^{\text{II}}_N \equiv 2^{-(N-1)/2} \left\{ c_k \cos\left[\frac{k(j + 1/2)\pi}{2^N}\right] \right\}_{k,j=0}^{2^N - 1},
\end{equation}
donde $c_0 = 1/\sqrt{2}$ y $c_k = 1$ para $k \ne 0$. Esta matriz está estrechamente relacionada con la transformada de Fourier, pero requiere mezclar e interferir sus elementos, lo que motiva el diseño del circuito extendido basado en la QFT.

El circuito comienza con una compuerta de Hadamard aplicada al qubit auxiliar (el bit más significativo), seguida por una escalera de compuertas CNOT, que en conjunto preparan un estado entrelazado tipo gato. A continuación, se aplica una transformada cuántica de Fourier a todos los $N + 1$ qubits. Esto es seguido por una secuencia de operaciones unitarias que separan y alinean las partes real e imaginaria del estado. En particular, compuertas locales $U_1$ y $R_Z$ se utilizan para ajustar las fases relativas de los componentes $\ket{0}_a\ket{\Phi}$ y $\ket{1}_a\ket{\Phi}$, donde $\ket{\Phi}$ representa un estado intermedio arbitrario de $N$ qubits.

A continuación, se aplica un subcircuito de permutación para reordenar las amplitudes de manera apropiada, seguido por otra escalera de CNOTs. El circuito concluye con un ajuste global de fase mediante $U_2$ y un conjunto de compuertas $R_X$ controladas por múltiples qubits, con el fin de escalar las amplitudes de salida de modo que $\ket{0}_a\ket{\Phi}$ y $\ket{1}_a\ket{\Phi}$ sean puramente reales e imaginarias, respectivamente. Observamos que el qubit auxiliar comienza y termina en el estado $\ket{0}$, asegurando una ejecución ``limpia''. Finalmente, el mapa de características completo se obtiene componiendo el embedding con la transformada como $\hat{U}_f(x) = \hat{U}_{\text{QChT}}^\dagger \hat{U}_\tau(x)$.

\begin{figure}[h]
    \centering
    \includegraphics[width=0.7\linewidth]{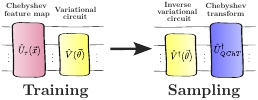}
    \caption{Funcionamiento del QCPM. \textit{Izquierda:} El circuito de entrenamiento codifica $x$ utilizando un mapa de características de Chebyshev, e incluye un circuito variacional cuyos parámetros se ajustarán para reproducir el comportamiento de la distribución objetivo. \textit{Derecha:} El circuito de muestreo realiza las operaciones inversas del circuito variacional entrenado y de la transformada cuántica de Chebyshev, generando datos que siguen la distribución utilizada durante el entrenamiento.}
    \label{fig:qcpm_sketch}
\end{figure}

Una vez que hemos definido un protocolo para codificar datos en los estados base de Chebyshev y un mapeo entre esta base y la base computacional, podemos construir un modelo probabilístico que busque aprender en la base de Chebyshev y muestrear en la base computacional. El funcionamiento de este Modelo Probabilístico Cuántico de Chebyshev (QCPM, por sus siglas en inglés) se muestra en la Fig.~\ref{fig:qcpm_sketch}.

Una de las características más interesantes de este QCPM es que permite entrenar los datos sobre la distribución objetivo usando una cantidad limitada de qubits, lo cual normalmente implica una densidad de muestreo limitada, y luego realizar un muestreo extendido utilizando una QChT con qubits adicionales. Esto permite una ``interpolación cuántica'' natural que incrementa la resolución hasta el nivel deseado.

\section*{Resultados}\label{sec:resumen_resultados}

Los resultados más relevantes de esta tesis pueden dividirse en tres secciones que se corresponden con la aplicación de las tres metodologías introducidas anteriormente.

\subsection*{Clusterización de jets}

En el Capítulo~\ref{chap:qjets} aplicamos los algoritmos cuánticos descritos en la metodología de cálculo de distancia y cálculo de mínimos a problemas reales de clusterización de jets con datos de simulaciones de eventos que ocurren en las colisiones de partículas del Gran Colisionador de Hadrones (LHC) del CERN.

En particular, simulamos un evento de $n$ partículas en el LHC utilizando un generador de eventos de espacio de fases personalizado, implementado en \texttt{C++} y basado en \texttt{ROOT}~\cite{Brun:1997pa}. Este generador puede manejar eventos con hasta decenas de miles de partículas, permitiendo una elección flexible de configuraciones en el estado final, incluyendo partones de QCD sin masa, partones de QCD con fotones, bosones vectoriales masivos o quarks top. Concretamente, estudiamos la producción de estados finales sin masa de $n$ partículas en colisiones protón-protón, aunque nuestro estudio también es aplicable a colisionadores $e^{+}e^{-}$, a una energía en el centro de masas de $\sqrt{s} = 14$ TeV. La selección del estado final sigue los siguientes criterios: los jets se identifican utilizando el algoritmo de jets $k_T$ con un momento transversal mínimo de $p_{T, \text{min}} \geq 10$ GeV y un radio de $R = 1$. Nos centramos en eventos con $n = 128$ partículas sin masa en el estado final.

\begin{figure}[h]
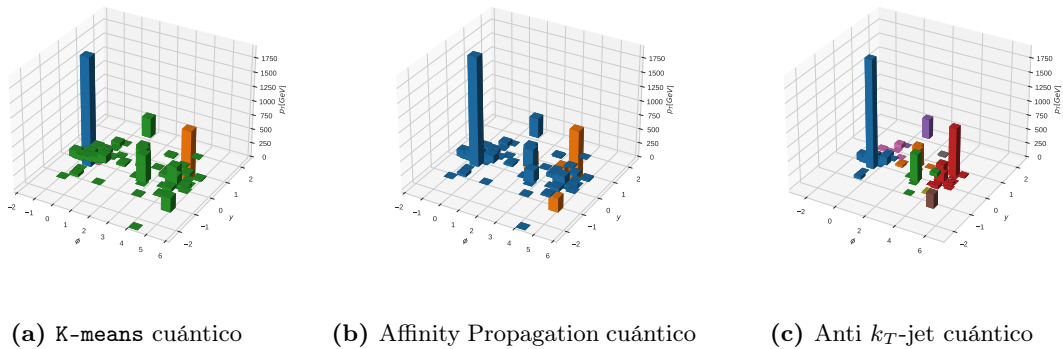

    \centering
    \begin{subfigure}[b]{0.32\textwidth}
        \centering
        \includegraphics[width=\linewidth]{Figures/jets/qks_realdata_660_50000_++.pdf}
        \caption{\texttt{K-means} cuántico}
    \end{subfigure}
    \hfill
    \begin{subfigure}[b]{0.32\textwidth}
        \centering
        \includegraphics[width=\linewidth]{Figures/jets/qap_realdata_quantum.pdf}
        \caption{Affinity Propagation cuántico}
    \end{subfigure}
    \hfill
    \begin{subfigure}[b]{0.32\textwidth}
        \centering
        \includegraphics[width=\linewidth]{Figures/jets/quantum_anti-kt_nt2.pdf}
        \caption{Anti $k_T$-jet cuántico}
    \end{subfigure}
    \caption{Proceso de clusterización de jets de distintos algoritmos cuánticos.}
    \label{fig:resumen_jets}
\end{figure}

Los resultados de aplicar los algoritmos \texttt{K-means}, Affinity Propagation y anti $k_T$-jet a estos datos aparecen reflejados en la Fig~\ref{fig:resumen_jets}.

En Fig. \ref{fig:resumen_jets} podemos observar como cada algoritmo clasifica los datos de forma distinta y en un número de jets diferente. En concreto podría parecer que la Fig. \ref{fig:resumen_jets} (c) realiza una agrupación más adecuada con un mayor número de jets. Esto es un efecto visual, ya que la agrupación de jets se representa gráficamente en tres dimensiones, lo cual coincide con la dimensionalidad de la métrica $k_T$, mientras que \texttt{K-means} y Affinity Propagation utilizan una distancia de Minkowski en cuatro dimensiones, lo que dificulta una representación visualmente atractiva en tres dimensiones. 

Por otra parte, la comparación más adecuada es la de estos algoritmos cuánticos con sus análogos clásicos, ya que se trata de métodos probabilísticos cuyo mejor resultado posible es replicar el comportamiento de los algoritmos clásicos, que son deterministas. Estos resultados pueden verse en la Tabla~\ref{table:resumen_jets_efs}.

\begin{longtable}{| p{0.45cm} | p{2cm} | p{2cm} | p{2cm} | p{2cm} | p{2cm} | }
\hline
& \centering{\texttt{K-means} cuántico} & \centering {Affinity Propagation cuántico}  & \centering {$k_T$ jet cuántico} & \centering{ Anti-$k_T$ jet cuántico}  & \centering { Cam/Aachen jet cuántico}   \cr   \hline 
 \centering $\varepsilon_c$ &\centering 0.94 &\centering 1.00 &\centering 0.98 &\centering 0.99 &\centering 0.98 \cr   \hline
 \omit
    \\  
\caption{Eficiencias de los algoritmos de clusterización de jet al compararse con los resultados que obtienen sus análogos clásicos.}
\label{table:resumen_jets_efs}
\end{longtable}

En la Tabla~\ref{table:resumen_jets_efs} se recogen las eficiencias de los algoritmos clásicos, incluyendo las versiones de $k_T$-jet que no se habían incluido gráficamente en la Fig.~\ref{fig:resumen_jets}. Los resultados muestran que en todos los casos las eficiencias son muy altas, cercanas a 1, por lo que podemos concluir que los métodos cuánticos, pese a ser probabilísticos no conducen a una clusterización significativamente diferente de los métodos clásicos. Finalmente, la complejidad temporal de estos algoritmos es comparada con la de sus análogos clásicos en función de que subrutinas cuánticas se utilizan como se muestra en la Tabla~\ref{table:resumen_jets_complex}.

\begin{longtable}{|  p{3.5cm} | p{2.51cm}| p{2.51cm} | p{4.0cm} | }
\hline
\centering{Algoritmo de clusterización de jets}  & \centering{Subrutina cuántica} & \centering{ Versión clásica}& \centering {Versión \\ cuántica}     \cr   \hline 
 \centering \texttt{K-means}  & \centering{Ambos}&\centering $\mathcal{O}(NKd)  $&\centering $\mathcal{O}(N\log K\log(d-1)) $  \cr   \hline
 \centering \texttt{AP}  & \centering{ Distancia} &\centering $\mathcal{O}(N^2Td) $&\centering $ \mathcal{O}(N^2T\log(d-1)) $  \cr   \hline
 \centering Anti-$k_T$ \texttt{Jet} & \centering{ Máximo}&\centering$\mathcal{O}(N^2) $&\centering $\mathcal{O}(N\log N) $  \cr   \hline
 \centering Anti-$k_T$ \texttt{FastJet} & \centering{ Máximo}&\centering $\mathcal{O}(N\log N) $&\centering $\mathcal{O}(N\log N) $  \cr   \hline 
 \omit
    \\  
   \caption{Complejidad temporal de los algoritmos clásicos y cuánticos de clusterización de jets cuando se aplica una o ambas subrutinas cuánticas.}
\label{table:resumen_jets_complex}
\end{longtable}

\subsection*{Integración de diagramas de Feynmann y tasas de desintegración}

En el Capítulo~\ref{chap:qint} empleamos el algoritmo de integración cuántica QFIAE para integrar secciones eficaces de procesos elementales de física de partículas, integrales de Feynman y tasas de desintegración de partículas elementales a segundo orden en teoría de perturbaciones. 

La primers integral de Feynman que abordamos es el integral de un lazo (loop) del tipo renacuajo (tadpole):
\beq
{\cal A}^{(1)}_1(m) = \int_{\ell} \frac{1}{\ell^2 - m^2 + \ii}~,
\eeq
con masa interna $m$. El factor $\ii$ corresponde a la prescripción compleja habitual de Feynman para la continuación analítica en distintas regiones cinemáticas. El momento de lazo a integrar es $\ell$. En LTD (Loop-Tree Duality), la expresión matemática correspondiente tiene soporte sobre el momento de lazo tridimensional, ya que la componente energética se integra~\cite{Catani:2008xa}. Si se introduce un contratérmino ultravioleta local, entonces su representación en LTD es
\begin{align}
{\cal A}^{(1,\r)}_1(m;\mu_\uv) 
= -\frac{1}{2} \int_{\lb} \left[
\frac{1}{\qon{1}} - \frac{1}{\qon{\uv}} \left(1 + \frac{\mu_\uv^2 - m^2}{2(\qon{\uv})^2} \right) \right]~,  
\label{eq:A1}
\end{align}
la cual está bien definida en cuatro dimensiones físicas, donde la medida de integración es $\int_{\lb} = \int d^3\lb/(2\pi)^3$. En la Ecuación~\eqref{eq:A1}, definimos las energías en la capa de masa como $\qon{1}=\sqrt{\lb^2+m^2 - \ii}$ y $\qon{\uv}=\sqrt{\lb^2 + \mu_\uv^2 - \ii}$, donde $\mu_\uv$ es la escala de renormalización. La integral renacuajo es unidimensional porque el integrando es independiente del ángulo sólido, y solo depende del módulo del momento tridimensional. El cambio de variable
\beq
|\lb| = \frac{m \, z}{1-z}~, \qquad z \in [0,1)~,
\label{eq:changevar}
\eeq
remapea la variable de integración a un rango finito donde se define la serie de Fourier de la función objetivo.

Para implementar QFIAE en una computadora cuántica real, debemos considerar que los dispositivos de la era NISQ se enfrentan a diversas fuentes de ruido, desde efectos cuánticos como la decoherencia hasta errores específicos del hardware, incluyendo errores en las compuertas, lectura y calibración. Trabajando con la misma tecnología (qubits superconductores), abogamos por una estrategia de implementación independiente del hardware. Específicamente, proponemos una aproximación donde los dos módulos de QFIAE se implementan en dos computadores cuánticos distintos suministrados por diferentes proveedores. Esto permite minimizar el impacto del ruido específico del hardware en el desempeño algorítmico total.

Una QNN de un qubit es entrenada usando una versión actualizada del método de descenso por gradiente Adam. El algoritmo se ha escrito usando \texttt{Qibo}, mientras que \texttt{Qibolab}~\cite{qibolab} y \texttt{Qibocal}~\cite{qibocal} se usan respectivamente para controlar y calibrar el dispositivo cuántico superconductor de 5 qubits alojado en el Quantum Research Centre (QRC) del Technology Innovation Institute (TII). Por otro lado, un algoritmo IQAE de 5 qubits ha sido ejecutado usando \texttt{Qiskit}~\cite{Qiskit} en el dispositivo superconductivo IBM Quantum de 27 qubits \textit{ibmq\_mumbai}. 

\begin{table}[ht]
\centering
\begin{tabular}{cccc}  
\toprule
Fourier & IQAE & \parbox{3cm}{${\cal A}^{(1,\r)}_1(m;\mu_\uv)$\\
$m=5, \mu_\uv=m/2$} &  
\parbox{3cm}{${\cal A}^{(1,\r)}_1(m;\mu_\uv)$\\
$m=5, \mu_\uv=2m$} \\ 
\midrule
C  & C & $-0.106$ &  $-0.258$  \\ 
S  & S & $-0.101(3)$ &  $-0.254(9)$  \\ 
S  & Q & $-0.108(4)$ &  $-0.270(12)$  \\ 
Q  & S & $-0.105(2)$ &  $-0.252(6)$  \\
Q  & Q & $-0.106(3)$ &  $-0.270(9)$  \\ 
\midrule 
\multicolumn{2}{c}{Analítico} & $-0.1007$ &  $-0.2554$  \\ 
\bottomrule
\end{tabular}
\caption{Integral renormalizada del tipo renacuajo ${\cal A}^{(1,\r)}_1(m;\mu_\uv)$ evaluada en \texttt{Qibo} e IBM Quantum, como función del cociente $\mu_\uv/m$, donde se fija $m=5$~GeV. La descomposición de Fourier y la integración mediante IQAE se realizan ya sea con un simulador cuántico (S) o en un dispositivo cuántico real (Q). En la primera fila ambas componentes se realizan con métodos clásicos (C).}
\label{table:resumen_hw_tadpole}
\end{table}

Los resultados presentados en la Tabla~\ref{table:resumen_hw_tadpole} muestran una desviación relativamente pequeña con respecto al valor analítico tanto para $\mu_\uv = m/2$ como para $\mu_\uv = 2m$. En particular, el acuerdo con los valores analíticos cuando ambos módulos del algoritmo se ejecutan sobre un computador cuántico (penúltima fila) es mejor que $1.7$ desviaciones estándar en todos los casos. Este resultado representa un logro significativo dado el estado actual de la tecnología de computación cuántica. Además, hay que tener en cuenta que se trata de la primera aplicación de un algoritmo cuántico extremo a extremo ejecutado en un computador cuántico para la estimación de integrales de lazos de Feynman. Parte de la desviación se debe también a la propia aproximación por serie de Fourier, como se observa en la primera fila de la Tabla~\ref{table:resumen_hw_tadpole}, lo cual constituye una prueba de la robustez del algoritmo cuántico propuesto.

\begin{figure}[h!]
       \centering
  \includegraphics[width=.49\linewidth]{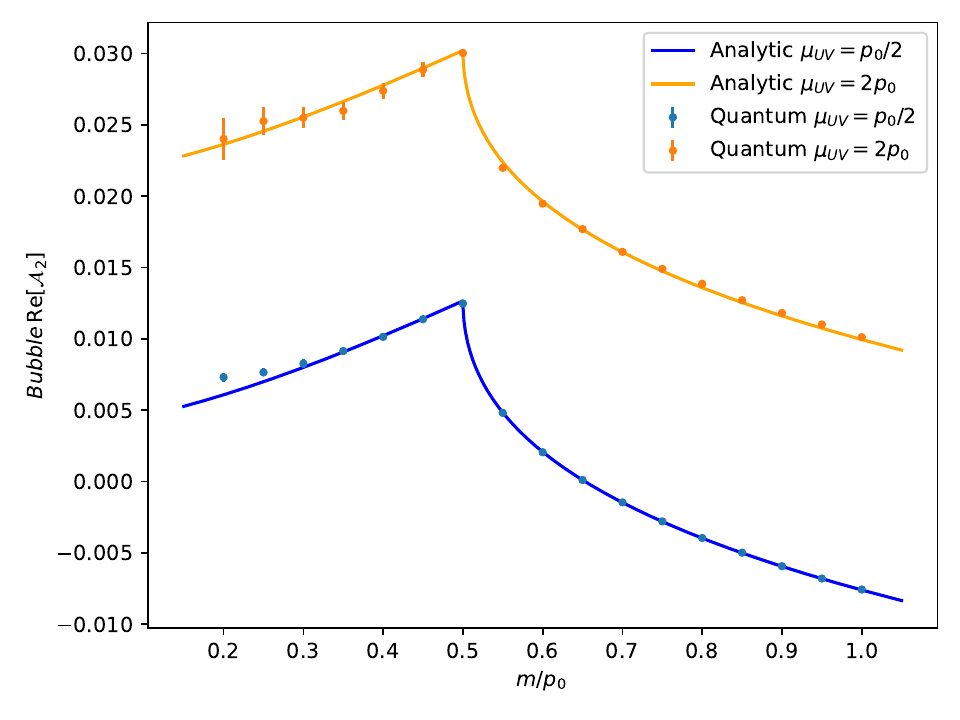}  
  \includegraphics[width=.49\linewidth]{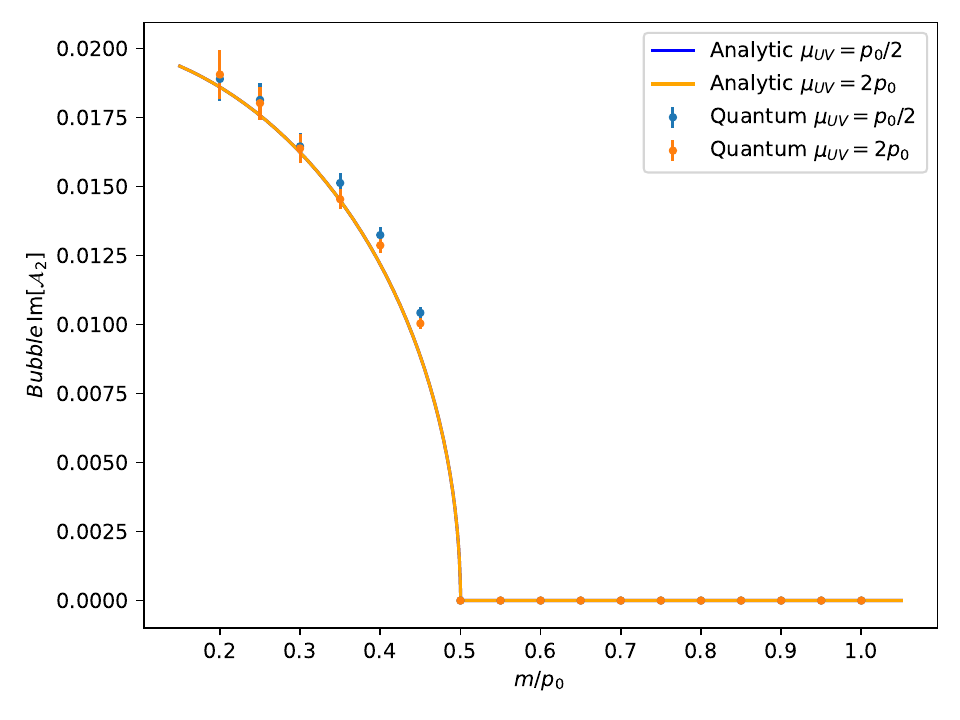}  

        \caption{Integración cuántica de la parte real (izquierda) e imaginaria (derecha) de la integral de burbuja renormalizada ${\cal A}^{(1,\r)}{2}(p,m,m;\mu_\uv)$ como función del cociente entre la masa $m$ y la componente de energía del momento externo, fijado en $p_0 = 100$~GeV, y de la escala de renormalización $\mu_{\uv}$.}
        \label{fig:resumen_a1bubble}
\end{figure}

A continuación procedemos a integrar una integral de Feynmann más compleja, tipo burbuja (bubble) con singularidades de umbral (threshold). Su representación en LTD es~\cite{Aguilera-Verdugo:2020set}
\vspace{-0.1cm}
\beq
{\cal A}^{(1,\r)}_2(p,m_1,m_2) =  \int_{\lb} \left[\frac{1}{x_2} \left( \frac{1}{\lambda^+} 
+ \frac{1}{\lambda^-} \right)-\frac{1}{4(\qon{\uv})^{3}} \right]~,
\eeq
donde  $x_2 = \prod_{i=1,2} 2 \qon{i}$, $\lambda^\pm = \sum_{i=1,2} \qon{i} \pm p_0$, y las energías on-shell están dadas por $\qon{i} = \sqrt{\lb^2 + m_i^2 - \ii}$ con $i \in \{1,2\}$,  asumiendo que el momento externo tiene componentes espaciales nulas, $p=(p_0, {\bf 0})$.  
Si $p_0^2 < (m_1 + m_2)^2$, la integral es puramente real. En caso contrario, aparece una contribución imaginaria debida a la singularidad en el umbral unitario en $\lambda^- \to 0$, suponiendo $p_0 > 0$.  Para tratar esta singularidad de umbral, es conveniente introducir una deformación de contorno $\delta$ en el plano complejo, con el fin de suavizar el comportamiento de la función en las cercanías del umbral sin alterar el resultado de la integral. Consideramos el caso $m_1 = m_2 = m$ y procedemos a estimar la integral usando dos redes neuronales cuánticas (QNNs) diferentes para ajustar por separado las partes real e imaginaria del integrando, e integrar cada serie de Fourier utilizando IQAE.  Hemos implementado QFIAE en dos simuladores cuánticos diferentes. Para las QNN usamos \texttt{Pennylane}~\cite{pennylane}, mientras que para IQAE empleamos \texttt{Qibo}. Los resultados se muestran en la Fig.~\ref{fig:resumen_a1bubble}, donde se representan las partes real e imaginaria en los gráficos izquierdo y derecho, respectivamente. Estos resultados muestran un acuerdo entre los valores integrados por QFIAE y los resultados analíticos.

En tercer lugar, consideramos una integral bidimensional de Feynmann de tipo triángulo (triangle). La representación LTD de esta integral está dada por~\cite{Aguilera-Verdugo:2020kzc}
\beq
\begin{split}
{\cal A}^{(1)}_3&(p_1,p_2,m_1, m_2,m_3) = - \int_{\lb} \frac{1}{x_3}
\left( 
\frac{1}{\lambda_{12}^- \lambda_{23}^+} \right. \\
&\left.+ \frac{1}{\lambda_{23}^+ \lambda_{31}^-}  
+  \frac{1}{\lambda_{31}^- \lambda_{12}^+} 
+ (\lambda_{ij}^+ \leftrightarrow \lambda_{ij}^-) \right)~,
\end{split}
\label{eq:triangle}
\eeq
con $x_3 = \prod_{i=1}^3 2 \qon{i}$, donde ahora las energías son $\qon{1}=\sqrt{(\lb+\pb_1)^2+m_1^2-\ii}$ y $\qon{i} = \sqrt{\lb^2+m_i^2-\ii}$ para $i\in\{2,3\}$. Trabajamos en el sistema del centro de masas, donde $\pb_{12} = \pb_1+\pb_2 = 0$, $p_{12,0} = p_{1,0} + p_{2,0} = \sqrt{s}$, y los momentos externos $p_1$ y $p_2$ están en direcciones opuestas a lo largo del eje $z$. Los denominadores causales son
\beq
\begin{split}
\lambda_{31}^\pm &= \qon{3} + \qon{1} \pm p_{1,0},~ 
\lambda_{12}^\pm = \qon{1} + \qon{2} \pm p_{2,0},\\ 
  \lambda_{23}^\pm &= \qon{2} + \qon{3} \mp p_{12,0}.
\end{split}
\eeq

\begin{figure}[h]
       \centering
  \includegraphics[width=.49\linewidth]{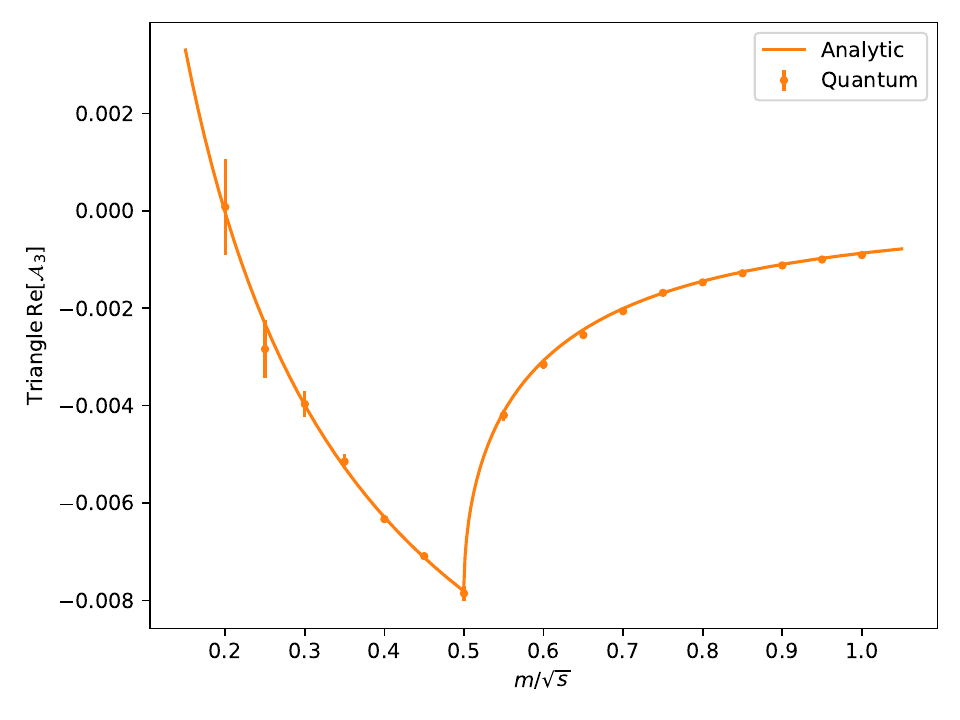}
  \includegraphics[width=.49\linewidth]{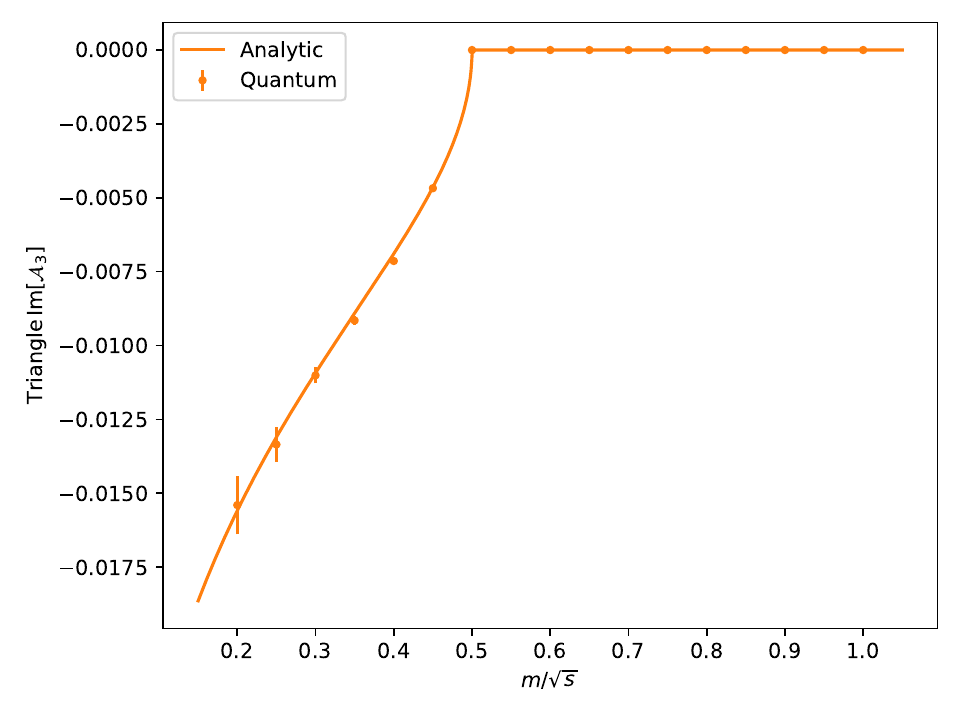}
        \caption{Integración cuántica de la parte real (izquierda) e imaginaria (derecha) de la integral del triángulo ${\cal A}^{(1)}_{3}(p_1,p_2,m,m,m)$ como función del cociente entre la masa $m$ y la energía del centro de masas fijada en $\sqrt{s}= 2$~GeV.}
        \label{fig:resumen_a1triangle}
\end{figure}

Las variables de integración en~\Eq{eq:triangle} son el módulo del momento de lazo tridimensional y su ángulo polar respecto a $p_1$, considerando que la integración azimutal es trivial. Esto implica que la descomposición en series de Fourier depende de dos variables y debemos integrar cada una por separado. También aplicamos una deformación del contorno para tratar la singularidad de umbral unitaria en $\lambda_{23}^+ \to 0$, cuando $s > (m_2 + m_3)^2$ que suaviza el comportamiento del integrando cerca de la singularidad de umbral y mejora significativamente la calidad de la descomposición en Fourier. Las estimaciones de la integral del triángulo, obtenidas en los simuladores \texttt{Pennylane} y \texttt{Qibo}, se muestran en las Figs.~\ref{fig:resumen_a1triangle}(izquierda) y~\ref{fig:resumen_a1triangle}(derecha), ilustrando las partes real e imaginaria, respectivamente. 

A partir de aquí, podemos ir un paso más allá y estudiar aplicaciones del algoritmo QFIAE de integrales que se correspondan con observables físicos en las desintegraciones de partículas elementales. Para ello hacemos uso del método causal unitario de LTD que emplea amplitudes del vacío cuántico para expresar integrales de Feymann. Con este formalismo podemos escribir la tasa de desintegración a segundo orden en teoría de perturbaciones de una partícula elemental $a$:
\bea
&& d\Gamma^{(1)}_{a} = \frac{d\Phi_{\lb_1\lb_2}}{2\sqrt{s}} \, \bigg[
\Big( \ad{3,a}(456) \, \ps{45\bar 6}  + \ad{3,a}(1356) \, \ps{135\bar 6} \Big)  
+ (5\leftrightarrow 2, 4\leftrightarrow 3) \bigg]~, \nn \\
\label{eq:decayratescalarNLO} 
\eea
donde $\ad{3,a}(456)$ y $\ad{3,a}(1356)$ son los residuos en el espacio de fases de la amplitud de vacío en LTD, cuando $\lambda_{456} \to 0$ y $\lambda_{1356} \to 0$. Estas contribuciones representan, respectivamente, las fluctuaciones cuánticas perturbativas a un lazo con dos partículas en el estado final y a nivel árbol con tres partículas en el estado final. En una teoría escalar:
\bea
&& \ad{3,\Phi}(456)  =  \frac{g_\Phi^{(1)} m_\Phi^2}{x_{12345}}
\left(L^{13\bar 4}_{23\bar 4 \bar 5, 125} +
L^{12\bar 5}_{23\bar 4\bar 5, 134} + 
L^{2345}_{134,125} \right)~, \nn \\ 
&& \ad{3,\Phi}(1356)  =  \frac{g_\Phi^{(1)} m_\Phi^2}{x_{135}}
\left(\frac{1}{\lambda_{13 \bar 4} \lambda_{134} \lambda_{1\bar 2 5} \lambda_{125}} \right)~,
\eea
donde $x_{i_1\cdots i_n} = \prod_{s=1}^n 2\qon{i_s}$ y $L^i_{j,k} = \lambda_i^{-1} \left(\lambda_j^{-1} + \lambda_k^{-1} \right)$, con
\beq
\lambda_{i_1\cdots i_r \bar i_{r+1} \cdots \bar i_n} = \lambda_{i_1\cdots i_r}- \lambda_{i_{r+1} \cdots i_n}~.
\eeq

La medida de integración es
\beq
d\Phi_{\lb_1 \lb_2} = \prod_{j=1}^{2} \frac{d^3 \lb_j}{(2\pi)^3}~, 
\label{eq:integrationmeasure}
\eeq
y usando simetría rotacional y la delta de Dirac:
\beq
\ps{i_1\cdots i_n \bar a} =  2\pi \,  \delta(\lambda_{i_1\cdots i_n \bar a})~, 
\label{eq:dirac}
\eeq
lo que reduce la integral a dos variables independientes:
\beq
d\Phi_{\lb_1 \lb_2} \to \frac{1}{4\pi^4}\int_0^\infty \lb_1^2  d|\lb_1| \int_0^\infty \lb_2^2 d|\lb_2| \int_0^1 dv~,
\label{eq:measure}
\eeq
con
\beq
\ps{45 \bar 6} =  2\pi \,  \delta\left( \sqrt{\lb_2^2+m^2} -\sqrt{s} \right)~,
\label{eq:delta1}
\eeq
y
\bea
\ps{135 \bar 6} &=&  2\pi \,  \delta\bigg( |\lb_1+\lb_2| + \sqrt{\lb_1^2+m^2}+\sqrt{\lb_2^2+m^2} -\sqrt{s} \bigg)~,
\label{eq:delta2}
\eea
donde $|\lb_1+\lb_2| = \sqrt{\lb_1^2 + \lb_2^2 + 2(1-2v)|\lb_1||\lb_2|}$. Las deltas se usan para resolver una variable en términos de las otras dos.

La ventaja clave de esta formulación es que permite tratar simultáneamente contribuciones a lazo y árbol, lo que garantiza cancelaciones locales de singularidades sin necesidad de usar regularización dimensional~\cite{Bollini:1972ui,tHooft:1972tcz}, y además produce integrandos más suaves y eficientes numéricamente.

\begin{figure}[h] 
\begin{center} 
\includegraphics[scale=0.49]{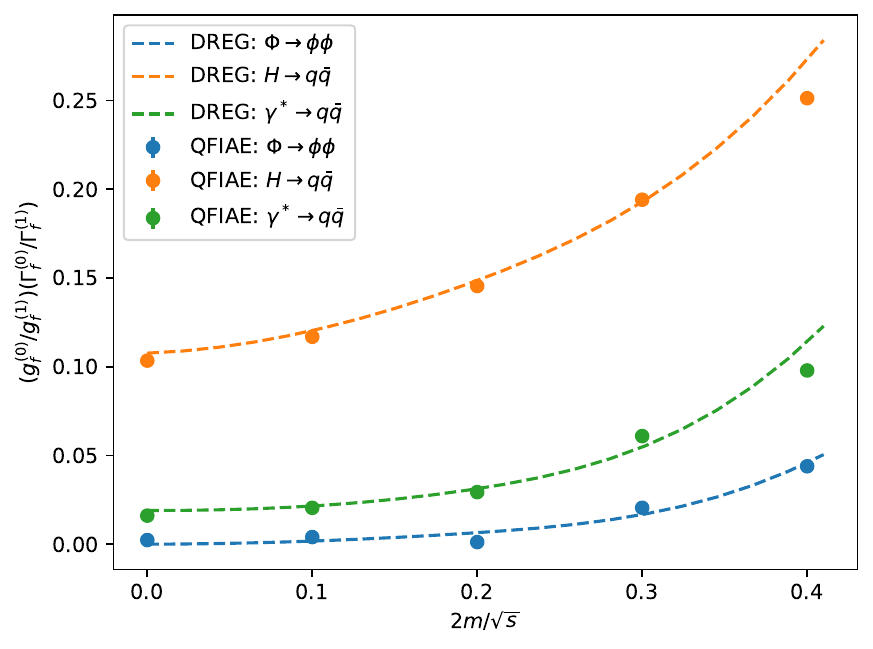} \includegraphics[scale=0.49]{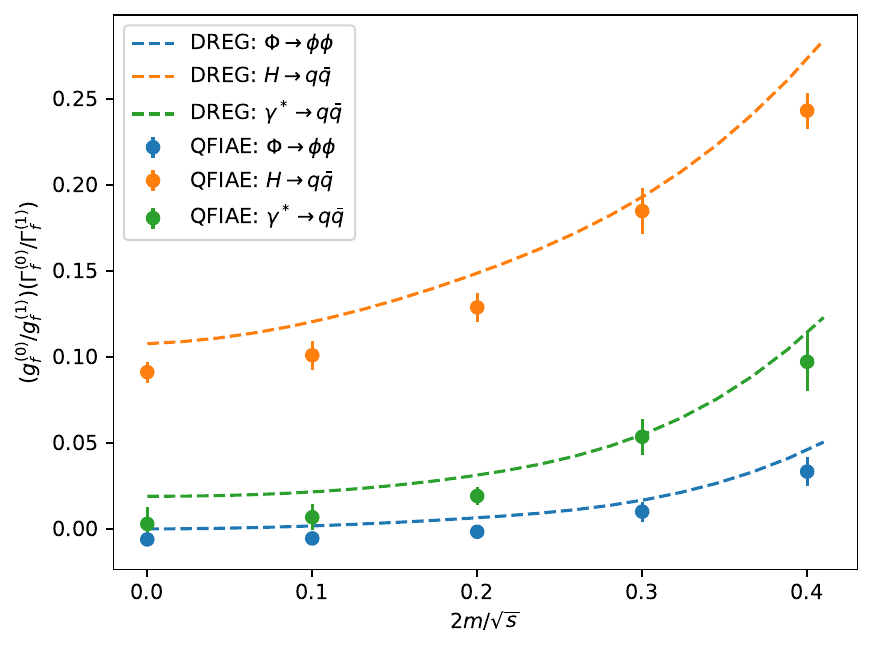} \caption{Tasas de desintegración integradas cuánticamente para los tres procesos $H\to q\bar q (g)$, $\gamma^*\to q\bar q (g)$ y $\Phi\to \phi\phi(\phi)$ a segundo orden en teoría de perturbaciones, en función de la masa del estado final. El panel izquierdo muestra resultados obtenidos con un simulador cuántico, mientras que el derecho presenta resultados en los que el algoritmo IQAE se ejecutó en hardware cuántico. Los cálculos emplean QFIAE con el enfoque causal unitario de LTD, y las líneas discontinuas representan las predicciones teóricas en regularización dimensional.} 
\label{fig:resumen_qint_combined} 
\end{center} 
\end{figure}

Hemos integrado estas tasas de desintegración para diferentes procesos. Los resultados que se presentan en la Fig.~\ref{fig:resumen_qint_combined} muestran una desviación relativamente pequeña respecto a sus valores analíticos correspondientes en Regularización Dimensional estándar (DREG). Notablemente, el panel izquierdo presenta una desviación sistemática en comparación con el panel derecho, atribuida al ruido del hardware que persiste a pesar de las técnicas de mitigación de errores aplicadas. 

\subsection*{Generación de funciones de fragmentación}

En el Capítulo~\ref{chap:qchebff} empleamos el método QCPM para aprender y generar funciones de fragmentación de diferentes partones que hadronizan en piones y kaones. Concretamente, utilizamos el conjunto de datos NNFF10\_PIsum\_nnlo para piones ($\pi^\pm = \pi^+ + \pi^-$) y el conjunto NNFF10\_KAsum\_nnlo para kaones ($K^\pm = K^+ + K^-$), ambos a tercer orden (NNLO) en teoría de perturbaciones. Las funciones de fragmentación (FFs) específicas que se analizan, extraidas de datos de producción de hadrones en aniquilación inclusiva simple electrón-positrón (SIA), involucran un total de cinco combinaciones independientes de FFs:
\begin{equation} 
{ D_g^h, D_{b^+}^h, D_{c^+}^h, D_{d^++s^+}^h, D_{u^+}^h }, 
\label{eq:ffs} 
\end{equation}
donde $h$ representa el hadrón en el que fragmenta el partón, $D^h_{q+} \equiv D^h_q + D^h_{\bar{q}}$, y $D_{d^++s^+}^h \equiv D^h_{d^+} + D^h_{s^+}$. 

\begin{figure}[h!]
        \includegraphics[width=1.0\linewidth]{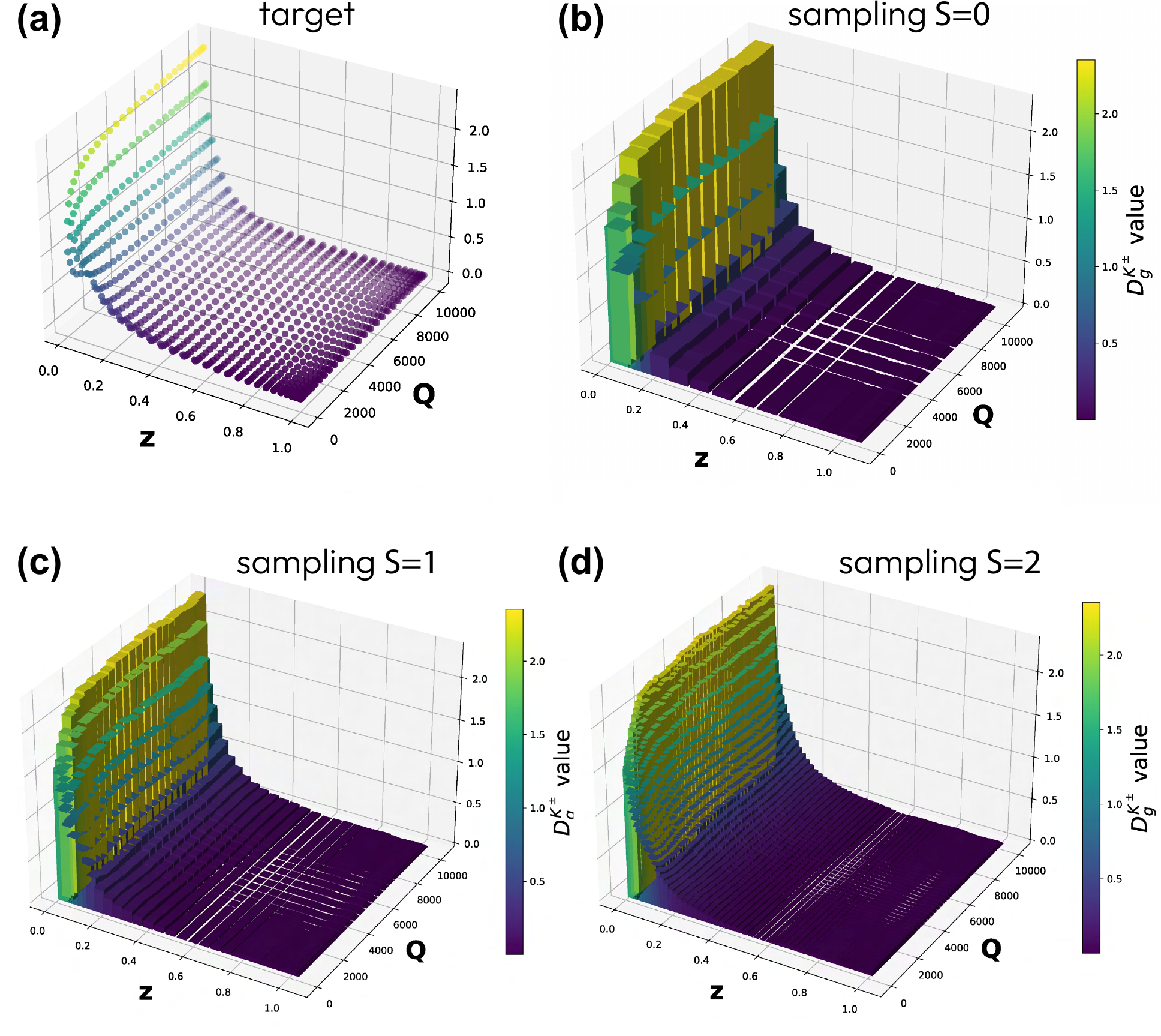}
        \caption{Muestreo de $D_{g}^{K^\pm}(z,Q)$, la FF de un gluón fragmentando en kaones, con $S=0,1,2$ qubits adicionales por variable para interpolar en regiones no entrenadas. (a) Distribución objetivo $D_{g}^{K^\pm}(z,Q)$. (b) Muestreo del modelo QCPM entrenado de $D_{g}^{K^\pm}(z,Q)$ con el mismo número de qubits ($S=0$). (c,d) Muestreo de $D_{g}^{K^\pm}(z,Q)$ con registro extendido ($S=1,2$).}
    \label{fig:resumen_FFresults}
\end{figure}

En la Fig. \ref{fig:resumen_FFresults} presentamos los resultados de una función de fragmentación (FF) en particular como ejemplo de la validez del método, $D_{g}^{K^\pm}$, que corresponde a la suma de las FF del gluón $g$ fragmentando en el kaón $K^+$ y su antipartícula $K^-$.

Los resultados en la Fig.~\ref{fig:resumen_FFresults} muestran que QCPM captura correctamente el comportamiento de la función bidimensional $D_{g}^{K^\pm}(z,Q)$ en la región de interés. La Fig.~\ref{fig:resumen_FFresults}(a) muestra la distribución objetivo sobre los datos de entrenamiento. La Fig.~\ref{fig:resumen_FFresults}(b) demuestra la capacidad del modelo para generar muestras utilizando los mismos registros cuánticos empleados durante el entrenamiento. Las Fig.~\ref{fig:resumen_FFresults}(c) y \ref{fig:resumen_FFresults}(d) presentan el rendimiento del muestreo cuando el modelo utiliza uno ($S=1$) y dos ($S=2$) qubits adicionales por variable, respectivamente, donde el modelo realiza predicciones en regiones no entrenadas y, en general, muestra una excelente correspondencia con los datos de referencia.

\section*{Conclusiones}\label{sec:resumen_conclusiones}

El Modelo Estándar (SM) de la física de partículas es una teoría muy exitosa, con el descubrimiento del bosón de Higgs en 2012, hace 13 años, en el Gran Colisionador de Hadrones (LHC) del CERN, como el mayor avance reciente en la física de altas energías (HEP). Se están planificando experimentos HEP aún más sofisticados para profundizar en el origen del universo y el contenido fundamental de la materia. En este sentido, estos experimentos más complejos exigen herramientas computacionales avanzadas, tanto para analizar las enormes cantidades de datos producidos en las colisiones de alta energía como para generar predicciones teóricas más precisas y exactas. En este contexto, la computación cuántica surge como una tecnología prometedora para complementar los métodos clásicos en el análisis de datos y la generación de simulaciones.

Esta tesis explora diversas aplicaciones de algoritmos cuánticos en física de altas energías. Primero, se presentan algoritmos cuánticos para clusterización de jets como \texttt{K-means}, Affinity Propagation y $k_T$-jet clustering, mostrando ventajas teóricas en escalabilidad y tiempos de ejecución. Posteriormente, se desarrolló un nuevo integrador cuántico basado en estimación de amplitud y redes neuronales cuánticas (QFIAE), que permite integrar funciones multidimensionales, que ha sido aplicado en ordenadores cuánticos actuales a cálculos de integrales de Feynman así como al cálculo de observables físicos como tasas de desintegración a segundo orden perturbativo en Teoría Cuántica de Campos. Finalmente, se emplearon Modelos Probabilísticos Cuánticos de Chebyshev (QCPMs) para aprender y generar funciones de fragmentación, obteniendo unos resultados satisfactorios en precisión y capacidad generativa. Además, el uso de qubits adicionales permitió aumentar de forma exponencial la resolución del muestreo.

En definitiva, este trabajo demuestra el potencial de la computación cuántica para abordar problemas complejos en física de partículas y abre nuevas direcciones hacia aplicaciones más sofisticadas, especialmente en un futuro a medio plazo cuando se disponga de dispositivos cuánticos tolerantes a fallos.



\counterwithin{equation}{chapter}
\counterwithin{figure}{chapter}
\appendix 

\chapter{Controlled \textit{SwapTest}}
\label{app:swaptest}
\begin{figure}[h]
    \centering
    \includegraphics[width=0.45\textwidth]{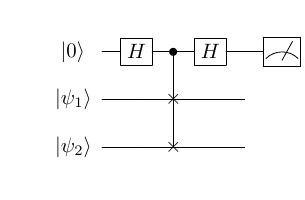}
    \caption{Quantum circuit of the \textit{SwapTest}.}  
    \label{fig:swaptest}
\end{figure}

A well-known procedure for determining the entanglement between two quantum states is the controlled \textit{SwapTest}~\cite{Buhrman:2001} method. The corresponding quantum circuit associated to the \textit{SwapTest} method is shown in Fig.~\ref{fig:swaptest}.

This method allows us to quantify the overlap between $\ket{\psi_1}$ and $\ket{\psi_2}$, which are two input general quantum states of $n$ and $m$ qubits respectively such that $n\geq m$ (otherwise we exchange the labels $1$ and $2$), by measuring an ancillary qubit.
The controlled \textit{SwapTest} proceeds in three steeps starting from the initial state
\beq
\ket{\Psi_0} = \ket{0, \psi_1, \psi_2}~,
\eeq
where the ancillary qubit has been initialized to $\ket{0}$. 
In the first step, a Hadamard ($H$) gate is applied to the ancillary qubit,
while the states to be probed are left unchanged, resulting 
in the new state
\beq
\ket{\Psi_1} = \left( H \otimes \id^{\otimes n+m}\right) \ket{\Psi_0}
=\frac{1}{\sqrt{2}} \left( \ket{0 ,\psi_1, \psi_2}
+\ket{1 ,\psi_1, \psi_2}\right)\,,
\eeq
where the identity $\id^{\otimes n+m}$ acts over the $\ket{\psi_1}$ and $\ket{\psi_2}$ states and the tensor product $\otimes$ is omitted in the composed states (e.g $\ket{0} \otimes \ket{\psi_1} \otimes \ket{ \psi_2} = \ket{0 ,\psi_1, \psi_2}$).
A controlled swap gate (CSWAP) is then applied to $\ket{\Psi_1}$ where all the $m$ qubits of $\ket{\psi_2}$ are swapped with the $m$ first qubits of $\ket{\psi_1}$, leading to 
\beq
\ket{\Psi_2} = {\rm CSWAP} \ket{\Psi_1}
=\frac{1}{\sqrt{2}} \left( \ket{0, \psi_1, \psi_2}
+\ket{1 ,\psi_2 ,\psi_1'}\right)~,
\eeq
where $\psi_i'$, is the swapped state of $\psi_i$, i.e., a state where the $m$ first qubits of $\psi_1$ have been swapped with the rest $n-m$ qubits. The final step consist of 
applying again a Hadamard gate to the ancillary qubit
\beq
\ket{\Psi_3} = \left( H \otimes \id^{\otimes n+m}\right) \ket{\Psi_2}
=\frac{1}{2} \left( \ket{0} \otimes \left(\ket{\psi_1 ,\psi_2} + \ket{\psi_2, \psi_1'}\right)
+\ket{1} \otimes \left( \ket{\psi_1, \psi_2} - \ket{\psi_2, \psi_1'} \right)
\right)~.
\eeq
The resulting probability of measuring the ancillary qubit in the 
state $\ket{0}$ is given by 
\beq
\begin{split}
P_{\Psi_3}(\ket{0}) &= \left|\langle 0\ket{\Psi_3} \right|^2 = 
\frac{1}{4} \left| \ket{\psi_1, \psi_2} + \ket{\psi_2, \psi_1'} \right|^2 
=\frac{1}{2}+\frac{1}{2} {\rm Re} 
\left[ \la \psi_2, \psi_1' \ket{\psi_1, \psi_2} \right]\\
&=\frac{1}{2}+\frac{1}{2} \la \psi_1'| \psi_2 \ra \la \psi_2| \psi_1 \ra ~,
\end{split}
\label{eq:p0swaptest}
\eeq
which turns out to be as follows if $m=n$, thus $|\psi_1' \ra$=$|\psi_1 \ra$
\beq
P_{\Psi_3}(\ket{0}) =
\frac{1}{2}+\frac{1}{2} \left|\la \psi_1| \psi_2 \ra \right|^2~,
\label{eq:p0swaptestsimple}
\eeq
and this provides the squared inner product between 
the two states with an uncertainty of ${\cal O}(\epsilon)$
after ${\cal O}(\epsilon^{-2})$ shots.

\chapter{Training and sampling of fragmentation functions of kaon and pion}
\label{app:appendix_FF}

In this Appendix, we present a visualization of the training (implicitly) and sampling (explicitly) processes for all fragmentation functions (FFs) considered in Section~\ref{app:qchebff}, corresponding to partons fragmenting into kaons and pions. Specifically, the ten FF distributions under study are: \begin{equation}
\begin{split} 
{ D_g^{h}, D_{b^+}^{h}, D_{c^+}^{h}, D_{d^++s^+}^{h}, D_{u^+}^{h} }, \qquad h = K^\pm, \pi^\pm~, 
\end{split} 
\end{equation} where $D^h_{q+} \equiv D^h_q + D^h_{\bar{q}}$ and $D_{d^++s^+}^h \equiv D^h_{d^+} + D^h_{s^+}$. Additionally, we define $D_i^{h^\pm} = D_i^{h^+} + D_i^{h^-}$, with $i$ denoting the fragmenting parton.

\begin{figure}[h]
\includegraphics[width=.99\textwidth]{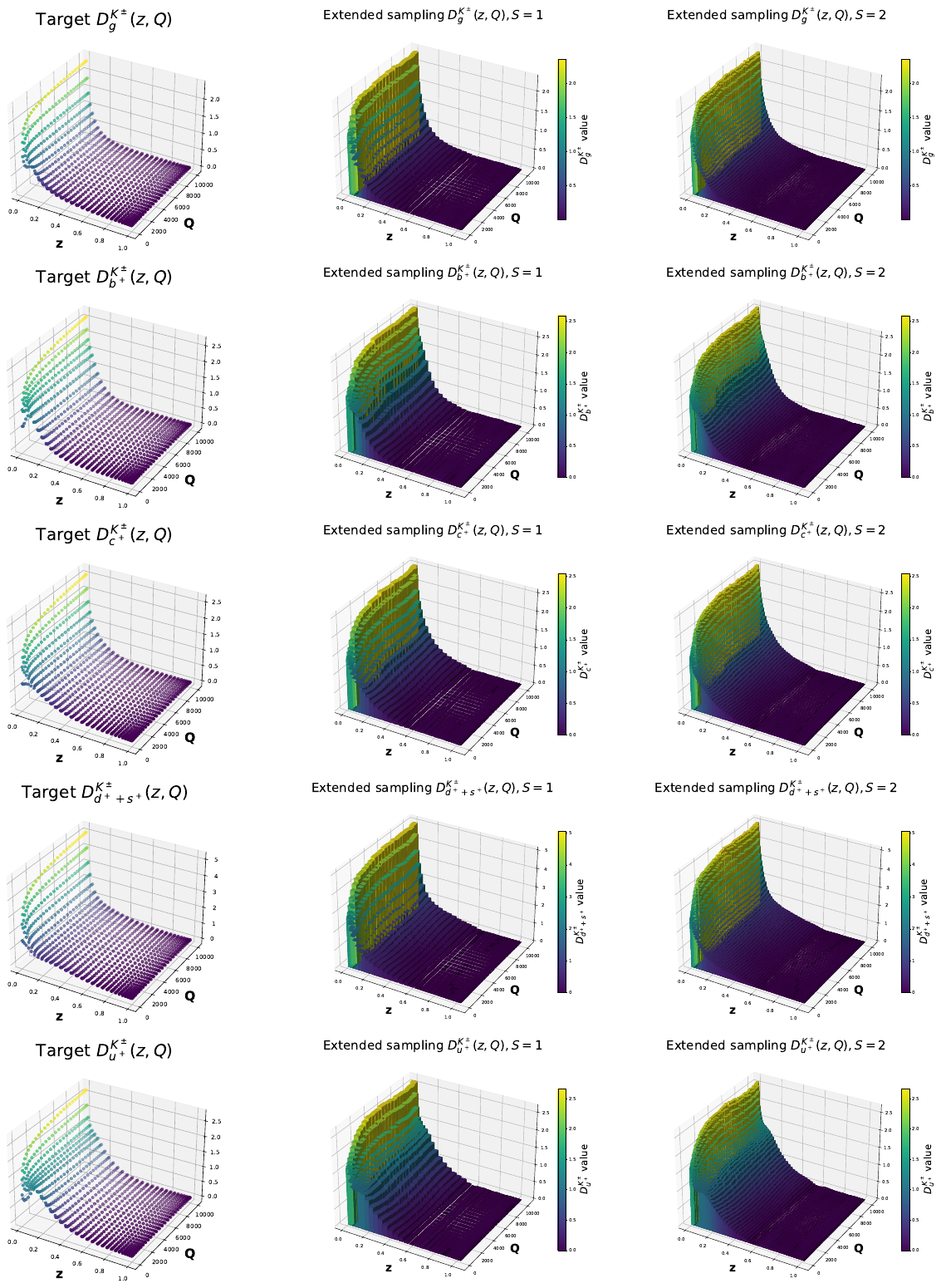}
\caption{Sampling of fragmentation functions of partons into kaons $K^+$ and $K^-$.}
\label{fig:FF_KA_results}
\end{figure}

\begin{figure}[h]
\includegraphics[width=.99\textwidth]{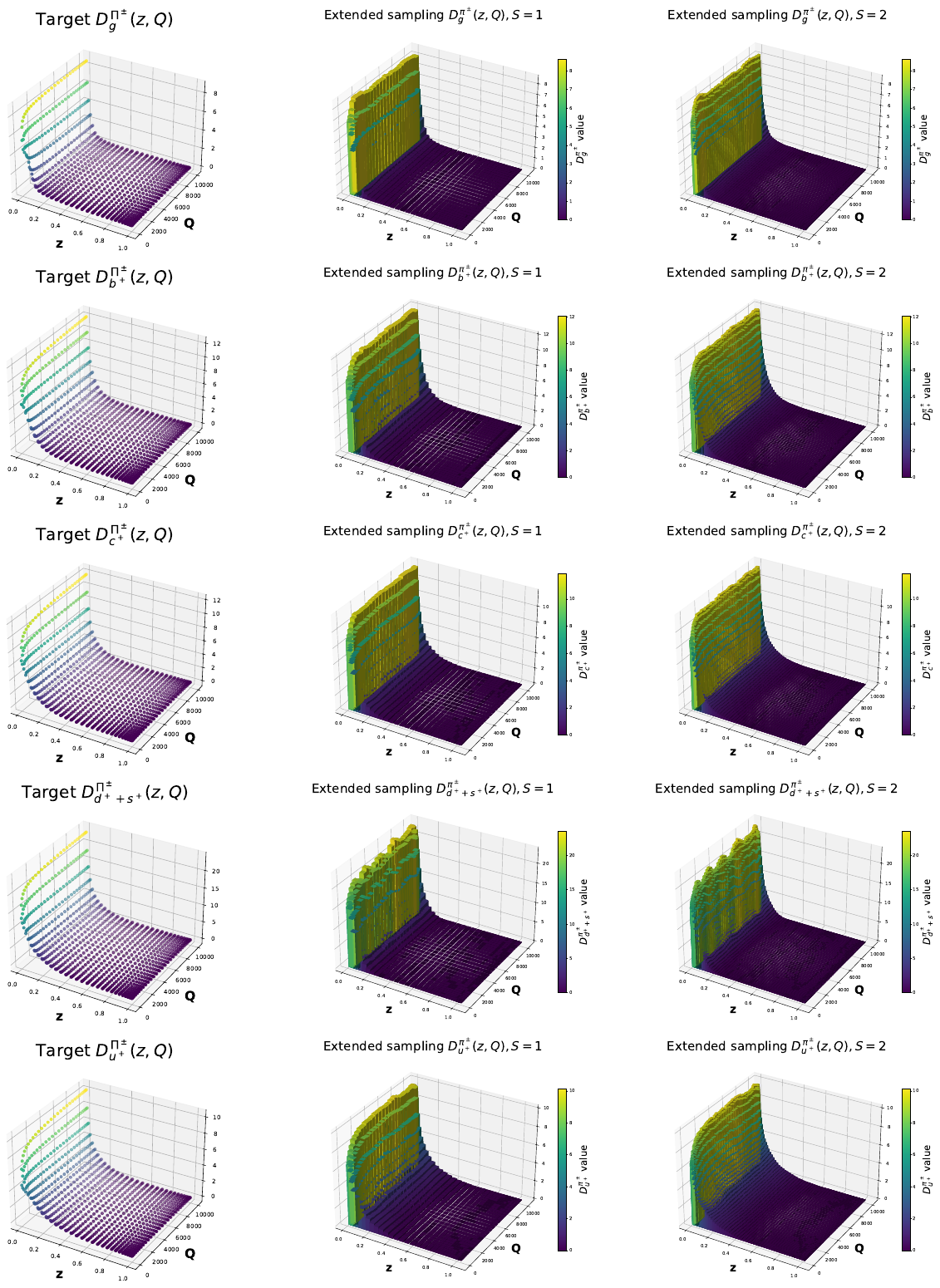}
\caption{Sampling of fragmentation functions of partons into pions $\pi^+$ and $\pi^-$.}
\label{fig:FF_PI_results}
\end{figure}

Figures~\ref{fig:FF_KA_results} and~\ref{fig:FF_PI_results} demonstrate that the quantum model effectively learns the full set of FF distributions and successfully generates samples from them.

Regarding the resources used for sampling, we have employed $10^6$ shots to obtain enough data points to provide a smooth description of the target function that correctly mimics the FF behavior. When performing extended sampling with additional qubits for each variable the number of qubits is increased $2S$ times, expanding the Hilbert space. Specifically, for $S=1$, the Hilbert space increases by a factor of~$4$, and for $S=2$, it expands by a factor of~$16$. This increase in the Hilbert space translates to a need for 4 times more shots when using $S=1$ and 16 times more shots when using $S=2$ to maintain the same level of accuracy. Additionally, this enlargement of the Hilbert space results in smaller measurement bins, with the area of each bin becoming 4 times smaller for $S=1$ and 16 times smaller for $S=2$.
\chapter{Correlations in Quantum Chebyshev Probabilistic Models}
\label{app:appendix_corr}

In this Appendix we study the correlations between the variables $z$ and $Q$ corresponding to the momentum fraction and the energy scale of the Fragmentation Functions studied in Section~\ref{app:qchebff}.
\section{Von Neumann entropy in quantum registers}

In this section, we quantify the entanglement within each of the registers encoding the variables $z$ and $Q$ during the training phase shown in Fig.\ref{fig:2DQCs}(a), using the von Neumann entropy. This quantity is a fundamental measure of entanglement for pure bipartite states in quantum information theory\cite{Neumann1927, nielsen}. Given a quantum state described by a density matrix $\rho$, the von Neumann entropy is defined as
\begin{equation} 
S(\rho) = -\text{Tr}(\rho \log_2 \rho) = -\sum_i \lambda_i \log_2 \lambda_i~, \end{equation} 
where $\lambda_i$ are the eigenvalues of $\rho$. For a pure state $\ket{\psi}$, the entropy of either subsystem provides a quantitative measure of entanglement between them~\cite{Bennett1996}. Specifically, in a bipartite setting, the entanglement entropy is given by
\begin{equation} 
S(\rho_A) = S(\rho_B) = -\text{Tr}(\rho_A \log_2 \rho_A) = -\text{Tr}(\rho_B \log_2 \rho_B), 
\end{equation} 
where $\rho_A$ and $\rho_B$ are the reduced density matrices of the respective subsystems $A$ and $B$\cite{Hill1997}.

In our setup, we consider two quantum systems, $\mathcal{Z}$ and $\mathcal{Q}$, each represented by $N = 4$ qubits and encoding the variables $z$ and $Q$, respectively. To analyze entanglement within these systems, we partition each register into two subsystems of $N/2 = 2$ qubits, allowing us to assess internal correlations independently in $\mathcal{Z}$ and $\mathcal{Q}$.

It is important to clarify that we do not compute the entanglement between $\mathcal{Z}$ and $\mathcal{Q}$ as a whole. Since these systems are entangled via controlled-$Z$ ($CZ$) gates, the combined system remains in a pure state, and its entropy would be zero. Instead, our analysis targets the internal entanglement within each register.

\begin{figure}[h]
\begin{center}
\includegraphics[width=.8\textwidth]{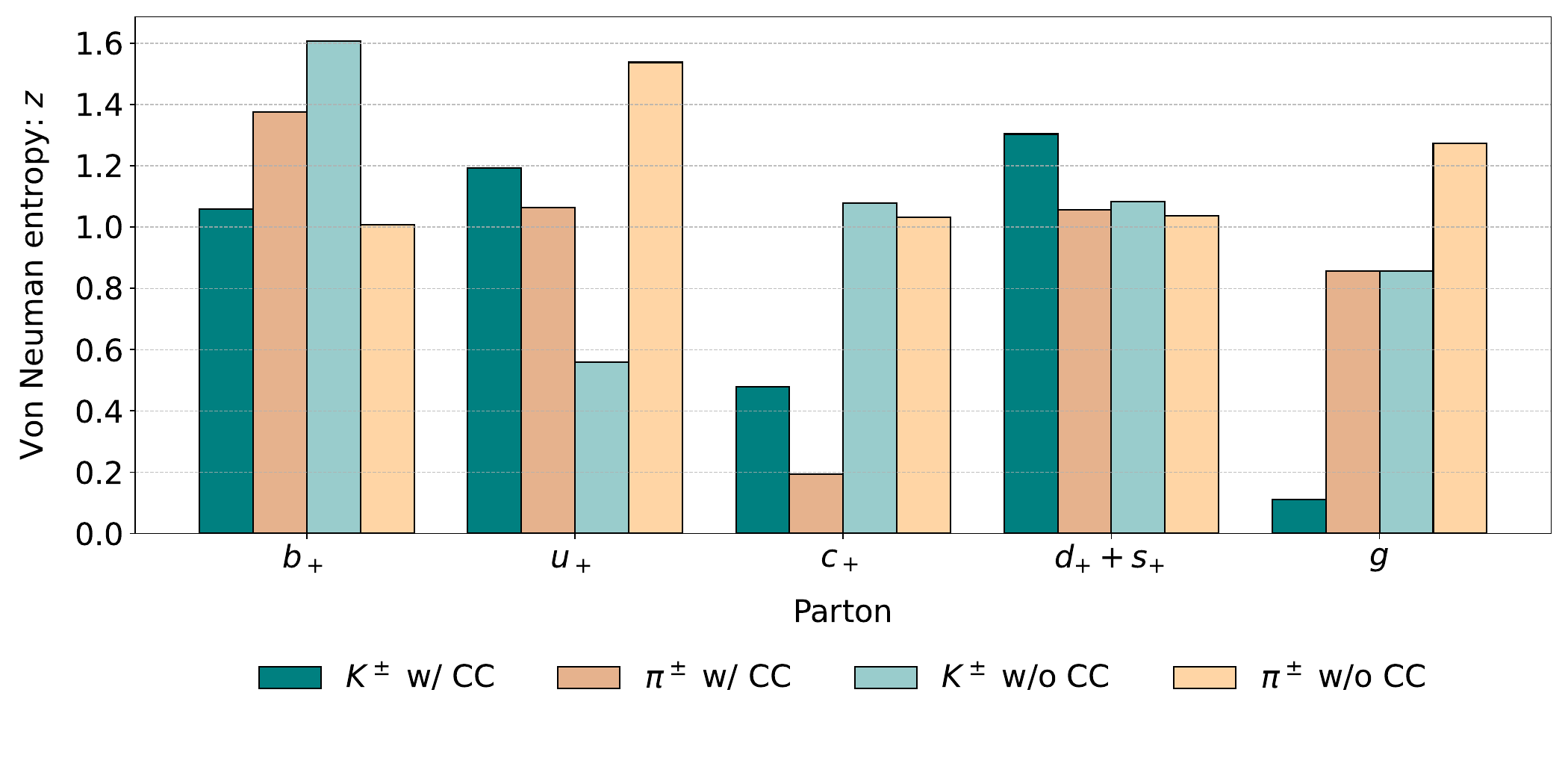}
 \includegraphics[width=.8\textwidth]{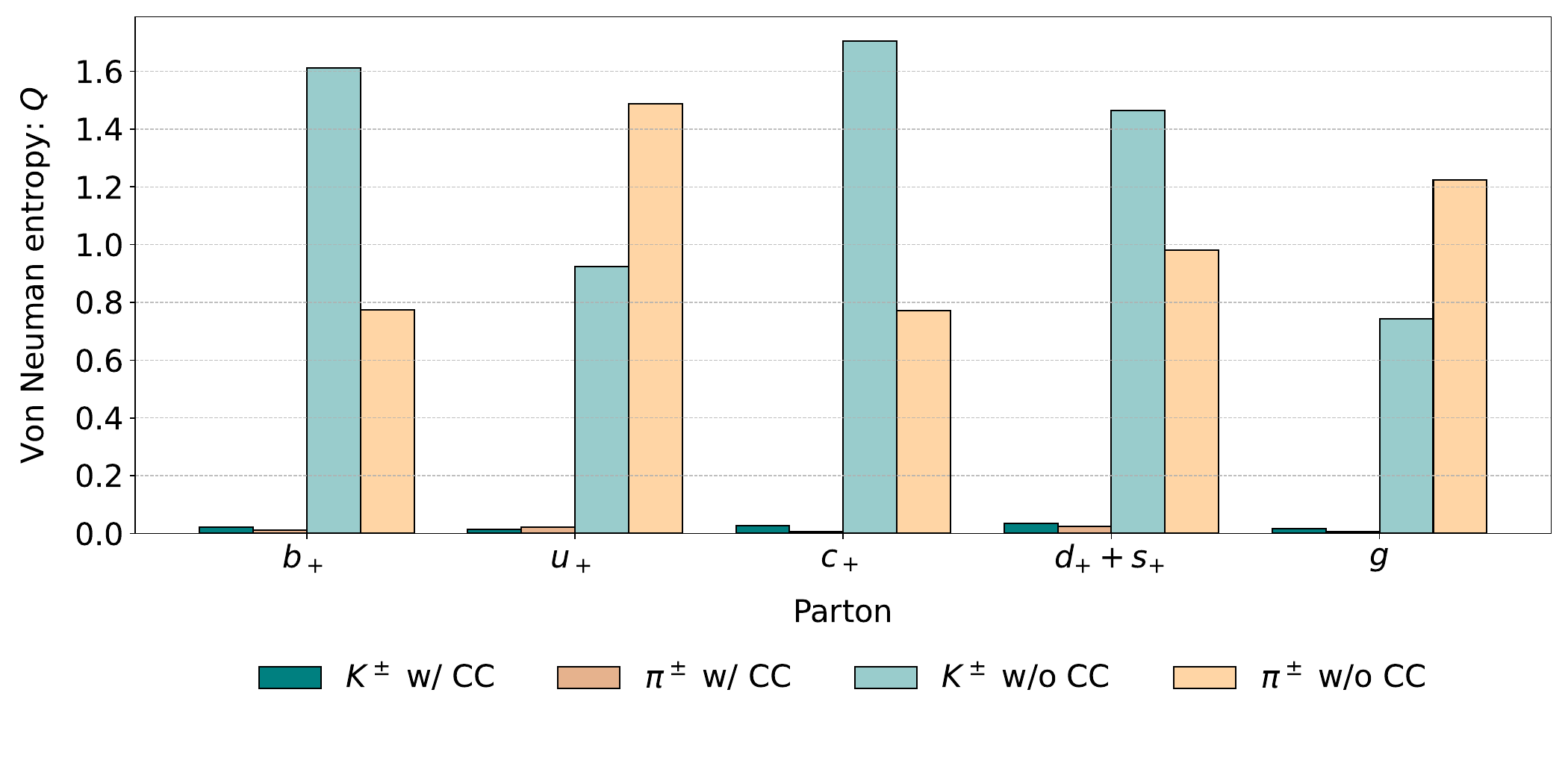}
\end{center}
\caption{Von Neumann entropies of the circuits in $z$ registers (top) and $Q$ registers (bottom) with (w/ CC) and without (w/o CC) correlations between $z$ and $Q$.}
\label{fig:vonneuman}
\end{figure}

To perform these calculations, we use the \texttt{qml.math.vn\_entropy} function provided by \texttt{PennyLane}~\cite{vn_entropy}. The corresponding results are presented in Fig.~\ref{fig:vonneuman}.

The results presented in Fig. \ref{fig:vonneuman}(a) show significant variations in the von Neumann entropy within the $\mathcal{Z}$ system across the different circuits that have learned the FFs. However, these variations do not appear to correlate with the inclusion or exclusion of the correlations circuit between systems $\mathcal{Z}$ and $\mathcal{Q}$. This suggests that while the internal entanglement structure of the $\mathcal{Z}$ system is dynamic, it is not driven by the inter-system correlations.

On the other hand, the results of Fig. \ref{fig:vonneuman}(b) reveal a strong dependence of the von Neumann entropy in the $\mathcal{Q}$ system on the inclusion of the correlations circuit. When the correlations circuit is not present, we observe a high von Neumann entropy, indicating a substantial level of internal entanglement within the $\mathcal{Q}$ system.  Nonetheless, when the correlations circuit is included, the von Neumann entropy drops drastically, approaching almost zero. This significant reduction in entropy suggests a transfer or ``absorption'' of entanglement from the $\mathcal{Q}$ system to the $\mathcal{Z}$ system. Such behavior is consistent with the principle of entanglement monogamy, wherein the entanglement shared between two systems constrains the entanglement each system can maintain internally~\cite{Zong_2022}.\\

\section{Purity in quantum registers}

In quantum information theory, purity serves as a measure of how mixed or coherent a quantum state is. For a system described by a density matrix $\rho$, purity is defined as
\begin{equation} \gamma(\rho) = \text{Tr}(\rho^2), \end{equation}
where $\rho$ is the state’s density operator. A purity value of $\gamma(\rho) = 1$ indicates a pure state, meaning the system resides in a well-defined quantum state. Conversely, if $\gamma(\rho) < 1$, the state is mixed, reflecting a probabilistic combination of different quantum states. In the extreme case of a maximally mixed state, purity reaches its minimum value of $\gamma(\rho) = 1/d$, where $d$ denotes the dimension of the system’s Hilbert space\cite{Jaeger2007-JAEQIA}. In the context of quantum circuits, purity offers useful insight into the system’s internal coherence and entanglement structure. Greater mixing generally implies higher entanglement, while lower mixing corresponds to a more separable (and hence less entangled) quantum state.

In this section, we quantify the entanglement between the registers encoding $z$ and $Q$, during the training phase of Fig. \ref{fig:2DQCs}(a), using their purity. In particular, we calculate the purity of the system $\mathcal{Z}$, of $N=5$ qubits, that encodes the variable $z$ which by construction is equivalent to the system $\mathcal{Q}$ that encodes the variable $Q$. To evaluate the purity of $\mathcal{Z}$, we trace out the degrees of freedom associated with $\mathcal{Q}$, effectively isolating the subsystem of interest.

\begin{equation}
    \gamma_\mathcal{Z}(\rho)\equiv\textrm{Tr}(\rho_\mathcal{Z}^2)=\textrm{Tr}(\rho_\mathcal{Q}^2)\equiv\gamma_\mathcal{Q}(\rho).
    \label{eq:purities}
\end{equation}

\begin{figure}[h]
    \centering
    \includegraphics[width=0.5\linewidth]{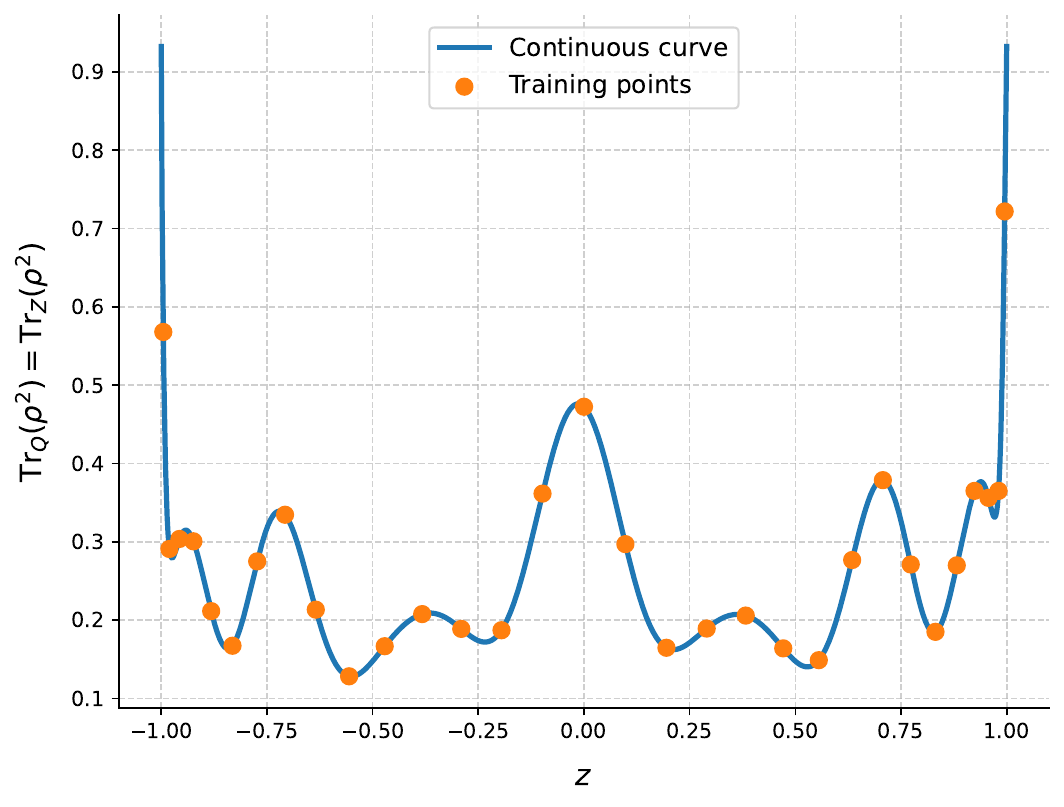}
    \caption{Purity of the quantum register $\mathcal{Z}$ for different $z$ values. Orange points represent the training points in Chebyshev nodes and halfpoints. Blue line represents a continuous curve in the range $[-1,1]$.}
    \label{fig:purity}
\end{figure}

We now examine how the purity of the register $\mathcal{Z}$ evolves as a function of the momentum fraction $z$. As appears in Fig.~\ref{fig:purity}, the purity evaluated at the Chebyshev nodes (every second orange point) presents an alternating pattern of local maxima and minima. In particular, purity reaches its highest values at the boundaries $z = -1$ and $z = 1$, as well as at $z = 0$, indicating minimal entanglement. In contrast, the lowest purity values (signaling maximal entanglement) are observed around $z \approx 0.5$ and $z \approx 0.8$.

\section{Mutual information in quantum registers}

In quantum information theory, quantum mutual information measures the total correlations, both classical and quantum, between two subsystems of a quantum state \cite{nielsen}. For a bipartite system represented by a density matrix $\rho_{\mathcal{ZQ}}$, the quantum mutual information between subsystems $\mathcal{Z}$ and $\mathcal{Q}$ is expressed as:
\begin{equation} 
I(\mathcal{Z}:\mathcal{Q}) = S(\rho_\mathcal{Z}) + S(\rho_\mathcal{Q}) - S(\rho_{\mathcal{ZQ}}) 
\label{eq:qmi}
\end{equation}
where $S(\rho) = -\text{Tr}(\rho \log \rho)$ is the von Neumann entropy. A large mutual information indicates strong correlations between the two subsystems, suggesting significant entanglement. On the other hand, when $I(\mathcal{Z}:\mathcal{Q}) = 0$, it implies that the subsystems are entirely independent.

A notable simplification of Eq. \ref{eq:qmi} occurs when the matrix $\rho_{\mathcal{ZQ}}$ represents a pure state, as is the case here. In this scenario, $S(\rho_{\mathcal{ZQ}}) = 0$ and $S(\rho_{\mathcal{Z}}) = S(\rho_{\mathcal{Q}})$. Therefore, Eq. \ref{eq:qmi} is as follows:
\begin{equation} 
I(\mathcal{Z}:\mathcal{Q}) = S(\rho_\mathcal{Z}) + S(\rho_\mathcal{Q}) = 2 S(\rho_\mathcal{Z})= -2 \text{Tr}_\mathcal{Z}(\rho\log_2 \rho).
\label{eq:qmi2}
\end{equation}

\begin{figure}[h] 
\centering 
\includegraphics[width=0.5\linewidth]{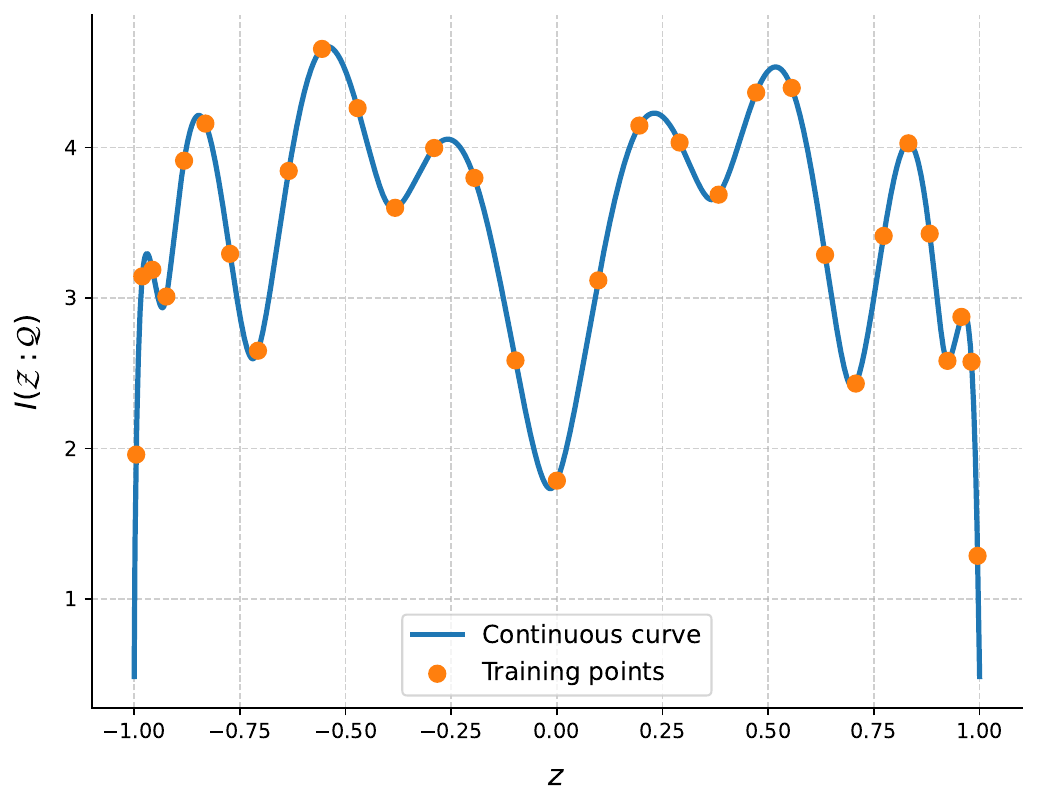} 
\caption{Quantum mutual information between registers $\mathcal{Z}$ and $\mathcal{Q}$ for different $z$  values. Orange points are training points in Chebyshev nodes and halfpoints. Blue line represents a continuous curve over the range $[-1,1]$.} 
\label{fig:mutual} 
\end{figure}

In this section, we quantify the entanglement between the registers encoding $z$ and $Q$ during the training stage of Fig. \ref{fig:2DQCs}(a), using their mutual information. Specifically, we calculate the quantum mutual information for the system $\mathcal{Z}$, consisting of $N=5$ qubits, which depends on the variable $z$. As illustrated in Fig. \ref{fig:mutual}, the mutual information presents an alternating pattern of local maxima and minima at the Chebyshev nodes. The strongest correlations are observed at intermediate values of $z$, where the entanglement is most pronounced, while minimal correlations occur at the boundaries, $-1$ and $1$, as well as at $z = 0$. 

\listoffigures 

\listoftables 

\printbibliography[heading=bibintoc]

@article{deLejarza:2022bwc,
    author = "de Lejarza, Jorge J. Mart{\'\i}nez and Cieri, Leandro and Rodrigo, Germ{\'a}n",
    title = "{Quantum clustering and jet reconstruction at the LHC}",
    eprint = "2204.06496",
    archivePrefix = "arXiv",
    primaryClass = "hep-ph",
    reportNumber = "IFIC/22-14 FTUV-22-0413.2034",
    doi = "10.1103/PhysRevD.106.036021",
    journal = "Phys. Rev. D",
    volume = "106",
    number = "3",
    pages = "036021",
    year = "2022"
}

@article{Ochoa-Oregon:2025opz,
    author = "Ochoa-Oregon, Salvador A. and Uribe-Ram{\'\i}rez, Juan P. and Hern{\'a}ndez-Pinto, Roger J. and Ram{\'\i}rez-Uribe, Selomit and Rodrigo, Germ{\'a}n",
    title = "{Graph theory-based automated quantum algorithm for efficient querying of acyclic and multiloop causal configurations}",
    eprint = "2508.04019",
    archivePrefix = "arXiv",
    primaryClass = "quant-ph",
    month = "8",
    year = "2025"
}

@article{maxplanck,
author = {Nauenberg, Michael},
year = {2016},
month = {09},
pages = {709-720},
title = {Max Planck and the birth of the quantum hypothesis},
volume = {84},
journal = {American Journal of Physics},
doi = {10.1119/1.4955146}
}

@article{aHeisenberg:1927zz,
    author = "a Heisenberg, W.",
    title = "{Uber den anschaulichen Inhalt der quantentheoretischen Kinematik und Mechanik}",
    doi = "10.1007/BF01397280",
    journal = "Z. Phys.",
    volume = "43",
    pages = "172--198",
    year = "1927"
}

@article{Pyretzidis:2025stx,
    author = "Pyretzidis, Konstantinos and de Lejarza, Jorge J. Mart{\'\i}nez and Rodrigo, Germ{\'a}n",
    title = "{Unlocking Multi-Dimensional Integration with Quantum Adaptive Importance Sampling}",
    eprint = "2506.19965",
    archivePrefix = "arXiv",
    primaryClass = "quant-ph",
    month = "6",
    year = "2025"
}

@article{DiMeglio:2023nsa,
    author = "Di Meglio, Alberto and others",
    title = "{Quantum Computing for High-Energy Physics: State of the Art and Challenges}",
    eprint = "2307.03236",
    archivePrefix = "arXiv",
    primaryClass = "quant-ph",
    reportNumber = "FERMILAB-PUB-23-468-ETD",
    doi = "10.1103/PRXQuantum.5.037001",
    journal = "PRX Quantum",
    volume = "5",
    number = "3",
    pages = "037001",
    year = "2024"
}

@misc{pennylaneQuantumCircuit,
	author = {},
	title = {{Q}uantum {C}ircuit {B}orn {M}achines | {P}enny{L}ane {D}emos --- pennylane.ai},
	howpublished = {\url{https://pennylane.ai/qml/demos/tutorial_qcbm}},
	year = {},
	note = {[Accessed 25-03-2025]},
}

@article{Arute:2019zxq,
    author = "Arute, Frank and others",
    title = "{Quantum supremacy using a programmable superconducting processor}",
    eprint = "1910.11333",
    archivePrefix = "arXiv",
    primaryClass = "quant-ph",
    doi = "10.1038/s41586-019-1666-5",
    journal = "Nature",
    volume = "574",
    number = "7779",
    pages = "505--510",
    year = "2019"
}

@article{Bond_Taylor_2022,
   title={Deep Generative Modelling: A Comparative Review of VAEs, GANs, Normalizing Flows, Energy-Based and Autoregressive Models},
   volume={44},
   ISSN={1939-3539},
   url={http://dx.doi.org/10.1109/TPAMI.2021.3116668},
   DOI={10.1109/tpami.2021.3116668},
   number={11},
   journal={IEEE Transactions on Pattern Analysis and Machine Intelligence},
   publisher={Institute of Electrical and Electronics Engineers (IEEE)},
   author={Bond-Taylor, Sam and Leach, Adam and Long, Yang and Willcocks, Chris G.},
   year={2022},
   month=nov, pages={7327–7347} }

@misc{sengar2024generativeartificialintelligencesystematic,
      title={Generative Artificial Intelligence: A Systematic Review and Applications}, 
      author={Sandeep Singh Sengar and Affan Bin Hasan and Sanjay Kumar and Fiona Carroll},
      year={2024},
      eprint={2405.11029},
      archivePrefix={arXiv},
      primaryClass={cs.LG},
      url={https://arxiv.org/abs/2405.11029}, 
}

@article{Einstein:1905vqn,
    author = "Einstein, A.",
    title = {{Zur Elektrodynamik bewegter K\"orper}},
    doi = "10.1002/andp.19053221004",
    journal = "Annalen Phys.",
    volume = "322",
    number = "10",
    pages = "891--921",
    year = "1905"
}

@InCollection{church-turing,
	author       =	{Copeland, B. Jack},
	title        =	{{The Church-Turing Thesis}},
	booktitle    =	{The {Stanford} Encyclopedia of Philosophy},
	editor       =	{Edward N. Zalta and Uri Nodelman},
	howpublished =	{\url{https://plato.stanford.edu/archives/win2024/entries/church-turing/}},
	year         =	{2024},
	edition      =	{{W}inter 2024},
	publisher    =	{Metaphysics Research Lab, Stanford University}
}

@InCollection{turing-machine,
	author       =	{De Mol, Liesbeth},
	title        =	{{Turing Machines}},
	booktitle    =	{The {Stanford} Encyclopedia of Philosophy},
	editor       =	{Edward N. Zalta and Uri Nodelman},
	howpublished =	{\url{https://plato.stanford.edu/archives/win2024/entries/turing-machine/}},
	year         =	{2024},
	edition      =	{{W}inter 2024},
	publisher    =	{Metaphysics Research Lab, Stanford University}
}

@book{nielsen,
  Author = {Michael A. Nielsen and Isaac L. Chuang},
  Title = {Quantum Computation and Quantum Information: 10th Anniversary Edition},
  Publisher = {Cambridge University Press},
  Year = {2011},
  ISBN = {9781107002173},
  URL = {https://www.amazon.com/Quantum-Computation-Information-10th-Anniversary/dp/1107002176?SubscriptionId=AKIAIOBINVZYXZQZ2U3A&tag=chimbori05-20&linkCode=xm2&camp=2025&creative=165953&creativeASIN=1107002176}
}

@article{Peruzzo:2013bzg,
    author = "Peruzzo, Alberto and McClean, Jarrod and Shadbolt, Peter and Yung, Man-Hong and Zhou, Xiao-Qi and Love, Peter J. and Aspuru-Guzik, Al{\'a}n and O'Brien, Jeremy L.",
    title = "{A variational eigenvalue solver on a photonic quantum processor}",
    eprint = "1304.3061",
    archivePrefix = "arXiv",
    primaryClass = "quant-ph",
    doi = "10.1038/ncomms5213",
    journal = "Nature Commun.",
    volume = "5",
    number = "1",
    pages = "4213",
    year = "2014"
}

@article{deLejarza:2024scm,
    author = "de Lejarza, Jorge J. Mart\'\i{}nez and Renter\'\i{}a-Estrada, David F. and Grossi, Michele and Rodrigo, Germ\'an",
    title = "{Quantum integration of decay rates at second order in perturbation theory}",
    eprint = "2409.12236",
    archivePrefix = "arXiv",
    primaryClass = "quant-ph",
    doi = "10.1088/2058-9565/ada9c5",
    journal = "Quantum Sci. Technol.",
    volume = "10",
    number = "2",
    pages = "025026",
    year = "2025"
}

@article{LTD:2024yrb,
    author = "Ram\'\i{}rez-Uribe, Selomit and Renter\'\i{}a-Olivo, Andr\'es E. and Renter\'\i{}a-Estrada, David F. and de Lejarza, Jorge J. Mart\'\i{}nez and Dhani, Prasanna K. and Cieri, Leandro and Hern\'andez-Pinto, Roger J. and Sborlini, German F. R. and Torres Bobadilla, William J. and Rodrigo, Germ\'an",
    collaboration = "LTD",
    title = "{Vacuum amplitudes and time-like causal unitary in the loop-tree duality}",
    eprint = "2404.05492",
    archivePrefix = "arXiv",
    primaryClass = "hep-ph",
    doi = "10.1007/JHEP01(2025)103",
    journal = "JHEP",
    volume = "01",
    pages = "103",
    year = "2025"
}

@inproceedings{Pich:2012sx,
    author = "Pich, Antonio",
    title = "{The Standard Model of Electroweak Interactions}",
    booktitle = "{2010 European School of High Energy Physics}",
    eprint = "1201.0537",
    archivePrefix = "arXiv",
    primaryClass = "hep-ph",
    reportNumber = "IFIC-11-73, FTUV-12-0102",
    pages = "1--50",
    month = "1",
    year = "2012"
}

@misc{apuntesew,
author = {Claudia Hagedorn},
title = {``\emph{Chapter 11: Physics beyond the Standard Model". Electroweak course. Masters degree in Advanced Physics, Universitat de València (2021)}},
}

@article{higgsdiscovery,
    author = "Aad, Georges and others",
    collaboration = "ATLAS",
    title = "{Observation of a new particle in the search for the Standard Model Higgs boson with the ATLAS detector at the LHC}",
    eprint = "1207.7214",
    archivePrefix = "arXiv",
    primaryClass = "hep-ex",
    reportNumber = "CERN-PH-EP-2012-218",
    doi = "10.1016/j.physletb.2012.08.020",
    journal = "Phys. Lett. B",
    volume = "716",
    pages = "1--29",
    year = "2012"
}

@article{CMS:2012qbp,
    author = "Chatrchyan, Serguei and others",
    collaboration = "CMS",
    title = "{Observation of a New Boson at a Mass of 125 GeV with the CMS Experiment at the LHC}",
    eprint = "1207.7235",
    archivePrefix = "arXiv",
    primaryClass = "hep-ex",
    reportNumber = "CMS-HIG-12-028, CERN-PH-EP-2012-220",
    doi = "10.1016/j.physletb.2012.08.021",
    journal = "Phys. Lett. B",
    volume = "716",
    pages = "30--61",
    year = "2012"
}

@article{Bjorken:1964gz,
    author = "Bjorken, J. D. and Glashow, S. L.",
    title = "{Elementary Particles and SU(4)}",
    doi = "10.1016/0031-9163(64)90433-0",
    journal = "Phys. Lett.",
    volume = "11",
    pages = "255--257",
    year = "1964"
}

@inproceedings{Novaes:1999yn,
    author = "Novaes, S. F.",
    title = "{Standard model: An Introduction}",
    booktitle = "{10th Jorge Andre Swieca Summer School: Particle and Fields}",
    eprint = "hep-ph/0001283",
    archivePrefix = "arXiv",
    reportNumber = "IFT-P-010-2000",
    pages = "5--102",
    month = "1",
    year = "1999"
}

@article{Preskill:2018jim,
    author = "Preskill, John",
    title = "{Quantum Computing in the NISQ era and beyond}",
    eprint = "1801.00862",
    archivePrefix = "arXiv",
    primaryClass = "quant-ph",
    doi = "10.22331/q-2018-08-06-79",
    journal = "Quantum",
    volume = "2",
    pages = "79",
    year = "2018"
}

@article{PhysRevLett.104.063603,
  title = {Machine Learning for Precise Quantum Measurement},
  author = {Hentschel, Alexander and Sanders, Barry C.},
  journal = {Phys. Rev. Lett.},
  volume = {104},
  issue = {6},
  pages = {063603},
  numpages = {4},
  year = {2010},
  month = {Feb},
  publisher = {American Physical Society},
  doi = {10.1103/PhysRevLett.104.063603},
  url = {https://link.aps.org/doi/10.1103/PhysRevLett.104.063603}
}

@inproceedings{Schuld2017QuantumML,
  title={Quantum machine learning for supervised pattern recognition.},
  author={Maria Schuld},
  year={2017},
  url={https://api.semanticscholar.org/CorpusID:69768800}
}

@article{panella,
  author       = {Massimo Panella and
                  Giuseppe Martinelli},
  title        = {Neural networks with quantum architecture and quantum learning},
  journal      = {Int. J. Circuit Theory Appl.},
  volume       = {39},
  number       = {1},
  pages        = {61--77},
  year         = {2011},
  url          = {https://doi.org/10.1002/cta.619},
  doi          = {10.1002/CTA.619},
}

@article{Farhi2018ClassificationWQ,
  title={Classification with Quantum Neural Networks on Near Term Processors},
  author={Edward Farhi and Hartmut Neven},
  journal={arXiv: Quantum Physics},
  year={2018},
  url={https://api.semanticscholar.org/CorpusID:119037649}
}

@inproceedings{Rojas1996TheBA,
  title={The Backpropagation Algorithm},
  author={Ra{\'u}l Rojas},
  year={1996},
  url={https://api.semanticscholar.org/CorpusID:62185690}
}

@phdthesis{Perez-Salinas:2021blv,
    author = "P\'erez-Salinas, Adri\'an",
    title = "{Algorithmic strategies for seizing quantum computing}",
    eprint = "2112.15175",
    archivePrefix = "arXiv",
    primaryClass = "quant-ph",
    school = "Barcelona, Autonoma U., Barcelona U.",
    year = "2021"
}

@article{Schuld:2020enb,
    author = "Schuld, Maria and Sweke, Ryan and Meyer, Johannes Jakob",
    title = "{Effect of data encoding on the expressive power of variational quantum-machine-learning models}",
    eprint = "2008.08605",
    archivePrefix = "arXiv",
    primaryClass = "quant-ph",
    doi = "10.1103/PhysRevA.103.032430",
    journal = "Phys. Rev. A",
    volume = "103",
    number = "3",
    pages = "032430",
    year = "2021"
}

@article{Riofrio:2023ncy,
    author = "Riofr\'\i{}o, Carlos A. and Mitevski, Oliver and Jones, Caitlin and Krellner, Florian and Vu\v{c}kovi\'c, Aleksandar and Doetsch, Joseph and Klepsch, Johannes and Ehmer, Thomas and Luckow, Andre",
    title = "{A Characterization of Quantum Generative Models}",
    eprint = "2301.09363",
    archivePrefix = "arXiv",
    primaryClass = "quant-ph",
    reportNumber = "Article No.: 12, Pages 1 - 34",
    doi = "10.1145/3655027",
    journal = "ACM Trans. Quant. Comput.",
    volume = "5",
    number = "2",
    pages = "12",
    year = "2024"
}

@phdthesis{Zoufal:2021pbi,
    author = "Zoufal, Christa",
    title = "{Generative Quantum Machine Learning}",
    eprint = "2111.12738",
    archivePrefix = "arXiv",
    primaryClass = "quant-ph",
    doi = "10.3929/ethz-b-000514692",
    school = "Zurich, ETH",
    year = "2021"
}

@article{Ostaszewski:2019vnn,
    author = "Ostaszewski, Mateusz and Grant, Edward and Benedetti, Marcello",
    title = "{Structure optimization for parameterized quantum circuits}",
    eprint = "1905.09692",
    archivePrefix = "arXiv",
    primaryClass = "quant-ph",
    doi = "10.22331/q-2021-01-28-391",
    journal = "Quantum",
    volume = "5",
    pages = "391",
    year = "2021"
}

@article{Biamonte_2017,
   title={Quantum machine learning},
   volume={549},
   ISSN={1476-4687},
   url={http://dx.doi.org/10.1038/nature23474},
   DOI={10.1038/nature23474},
   number={7671},
   journal={Nature},
   publisher={Springer Science and Business Media LLC},
   author={Biamonte, Jacob and Wittek, Peter and Pancotti, Nicola and Rebentrost, Patrick and Wiebe, Nathan and Lloyd, Seth},
   year={2017},
   month=sep, pages={195–202} }

@article{P_rez_Salinas_2020,
   title={Data re-uploading for a universal quantum classifier},
   volume={4},
   ISSN={2521-327X},
   url={http://dx.doi.org/10.22331/q-2020-02-06-226},
   DOI={10.22331/q-2020-02-06-226},
   journal={Quantum},
   publisher={Verein zur Forderung des Open Access Publizierens in den Quantenwissenschaften},
   author={Pérez-Salinas, Adrián and Cervera-Lierta, Alba and Gil-Fuster, Elies and Latorre, José I.},
   year={2020},
   month=feb, pages={226} }

@misc{kerasKerasDocumentation,
	author = {Keras},
	title = {{K}eras documentation: {T}he {S}equential model },
	howpublished = {\url{https://keras.io/guides/sequential_model/}},
	year = {2023},
	note = {[Accessed 05-02-2025]},
}

@article{SCHMIDHUBER201585,
title = {Deep learning in neural networks: An overview},
journal = {Neural Networks},
volume = {61},
pages = {85-117},
year = {2015},
issn = {0893-6080},
doi = {https://doi.org/10.1016/j.neunet.2014.09.003},
url = {https://www.sciencedirect.com/science/article/pii/S0893608014002135},
author = {Jürgen Schmidhuber},
}

@inproceedings{GmezRamos2013ARO,
  title={A Review of Artificial Neural Networks: How Well Do They Perform in Forecasting Time Series?},
  author={Elsy G{\'o}mez-Ramos and Francisco Venegas-Mart{\'i}nez},
  year={2013},
  url={https://api.semanticscholar.org/CorpusID:58795942}
}

@article{Tacchino_2020,
doi = {10.1088/2058-9565/abb8e4},
url = {https://dx.doi.org/10.1088/2058-9565/abb8e4},
year = {2020},
month = {oct},
publisher = {IOP Publishing},
volume = {5},
number = {4},
pages = {044010},
author = {Tacchino, Francesco and Barkoutsos, Panagiotis and Macchiavello, Chiara and Tavernelli, Ivano and Gerace, Dario and Bajoni, Daniele},
title = {Quantum implementation of an artificial feed-forward neural network},
journal = {Quantum Science and Technology},
}

@article{Cao:2017tnw,
    author = "Cao, Yudong and Guerreschi, Gian Giacomo and Aspuru-Guzik, Al\'an",
    title = "{Quantum Neuron: an elementary building block for machine learning on quantum computers}",
    eprint = "1711.11240",
    archivePrefix = "arXiv",
    primaryClass = "quant-ph",
    month = "11",
    year = "2017"
}

@article{andrecut,
author = {ANDRECUT, M. and ALI, M. K.},
title = {A QUANTUM NEURAL NETWORK MODEL},
journal = {International Journal of Modern Physics C},
volume = {13},
number = {01},
pages = {75-88},
year = {2002},
doi = {10.1142/S0129183102002948},

URL = { 
    
        https://doi.org/10.1142/S0129183102002948
    
    

},
eprint = { 
    
        https://doi.org/10.1142/S0129183102002948
    
    

}
}

@article{Rebentrost:2013bin,
    author = "Rebentrost, Patrick and Mohseni, Masoud and Lloyd, Seth",
    title = "{Quantum Support Vector Machine for Big Data Classification}",
    eprint = "1307.0471",
    archivePrefix = "arXiv",
    primaryClass = "quant-ph",
    doi = "10.1103/physrevlett.113.130503",
    journal = "Phys. Rev. Lett.",
    volume = "113",
    number = "13",
    pages = "130503",
    year = "2014"
}

@article{Cortes1995SupportVectorN,
  title={Support-Vector Networks},
  author={Corinna Cortes and Vladimir Naumovich Vapnik},
  journal={Machine Learning},
  year={1995},
  volume={20},
  pages={273-297},
  url={https://api.semanticscholar.org/CorpusID:52874011}
}

@book{Alba_2019, title={La máquina que cambió el mundo: Génesis, desarrollo y evolución del ordenador}, author={Alba, Salvador Lucas}, year={2019}, language={en} }

@article{Hubregtsen:2021lqn,
    author = "Hubregtsen, Thomas and Wierichs, David and Gil-Fuster, Elies and Derks, Peter-Jan H. S. and Faehrmann, Paul K. and Meyer, Johannes Jakob",
    title = "{Training quantum embedding kernels on near-term quantum computers}",
    eprint = "2105.02276",
    archivePrefix = "arXiv",
    primaryClass = "quant-ph",
    doi = "10.1103/PhysRevA.106.042431",
    journal = "Phys. Rev. A",
    volume = "106",
    number = "4",
    pages = "042431",
    year = "2022"
}

@misc{stoudenmire2017supervisedlearningquantuminspiredtensor,
      title={Supervised Learning with Quantum-Inspired Tensor Networks}, 
      author={E. Miles Stoudenmire and David J. Schwab},
      year={2017},
      eprint={1605.05775},
      archivePrefix={arXiv},
      primaryClass={stat.ML},
      url={https://arxiv.org/abs/1605.05775}, 
}

@article{Carrasquilla_2017,
   title={Machine learning phases of matter},
   volume={13},
   ISSN={1745-2481},
   url={http://dx.doi.org/10.1038/nphys4035},
   DOI={10.1038/nphys4035},
   number={5},
   journal={Nature Physics},
   publisher={Springer Science and Business Media LLC},
   author={Carrasquilla, Juan and Melko, Roger G.},
   year={2017},
   month=feb, pages={431–434} }

@article{Schuld:2018aiz,
    author = "Schuld, Maria and Bergholm, Ville and Gogolin, Christian and Izaac, Josh and Killoran, Nathan",
    title = "{Evaluating analytic gradients on quantum hardware}",
    eprint = "1811.11184",
    archivePrefix = "arXiv",
    primaryClass = "quant-ph",
    doi = "10.1103/PhysRevA.99.032331",
    journal = "Phys. Rev. A",
    volume = "99",
    number = "3",
    pages = "032331",
    year = "2019"
}

@article{Schuld_2014,
   title={An introduction to quantum machine learning},
   volume={56},
   ISSN={1366-5812},
   url={http://dx.doi.org/10.1080/00107514.2014.964942},
   DOI={10.1080/00107514.2014.964942},
   number={2},
   journal={Contemporary Physics},
   publisher={Informa UK Limited},
   author={Schuld, Maria and Sinayskiy, Ilya and Petruccione, Francesco},
   year={2014},
   month=oct, pages={172–185} }

@article{Cerezo:2020jpv,
    author = "Cerezo, M. and others",
    title = "{Variational quantum algorithms}",
    eprint = "2012.09265",
    archivePrefix = "arXiv",
    primaryClass = "quant-ph",
    reportNumber = "LA-UR-20-30142",
    doi = "10.1038/s42254-021-00348-9",
    journal = "Nature Rev. Phys.",
    volume = "3",
    number = "9",
    pages = "625--644",
    year = "2021"
}

@misc{pennylaneFromNISQ,
	author = {Juan Miguel Arrazola},
	title = {{F}rom {N}{I}{S}{Q} to {I}{S}{Q} | {P}enny{L}ane {B}log --- pennylane.ai},
	howpublished = {\url{https://pennylane.ai/blog/2023/06/from-nisq-to-isq}},
	year = {2023},
	note = {[Accessed 14-01-2025]},
}

@article{Brassard:2000xvp,
    author = "Brassard, Gilles and Hoyer, Peter and Mosca, Michele and Tapp, Alain",
    title = "{Quantum amplitude amplification and estimation}",
    eprint = "quant-ph/0005055",
    archivePrefix = "arXiv",
    doi = "10.1090/conm/305/05215",
    month = "5",
    year = "2000"
}

@article{Montanaro_2015,
   title={Quantum speedup of Monte Carlo methods},
   volume={471},
   ISSN={1471-2946},
   url={http://dx.doi.org/10.1098/rspa.2015.0301},
   DOI={10.1098/rspa.2015.0301},
   number={2181},
   journal={Proceedings of the Royal Society A: Mathematical, Physical and Engineering Sciences},
   publisher={The Royal Society},
   author={Montanaro, Ashley},
   year={2015},
   month=sep, pages={20150301} }

@misc{pennylaneBasicArithmetic,
	author = {Guillermo Alonso-Linaje},
	title = {{B}asic arithmetic with the quantum {F}ourier transform ({Q}{F}{T}) | {P}enny{L}ane {D}emos --- pennylane.ai},
	howpublished = {\url{https://pennylane.ai/qml/demos/tutorial_qft_arithmetics}},
	year = {2022},
	note = {[Accessed 14-01-2025]},
}

@article{Buhrman:2001rma,
    author = "Buhrman, Harry and Cleve, Richard and Watrous, John and de Wolf, Ronald",
    title = "{Quantum Fingerprinting}",
    eprint = "quant-ph/0102001",
    archivePrefix = "arXiv",
    doi = "10.1103/PhysRevLett.87.167902",
    journal = "Phys. Rev. Lett.",
    volume = "87",
    number = "16",
    pages = "167902",
    year = "2001"
}

@article{Nielsen:1996pv,
    author = "Nielsen, M. A. and Caves, Carlton M. and Schumacher, Benjamin and Barnum, Howard",
    title = "{Information theoretic approach to quantum error correction and reversible measurement}",
    eprint = "quant-ph/9706064",
    archivePrefix = "arXiv",
    doi = "10.1098/rspa.1998.0160",
    journal = "Proc. Roy. Soc. Lond. A",
    volume = "454",
    pages = "277",
    year = "1998"
}

@article{Perl:1975bf,
    author = "Perl, Martin L. and others",
    title = "{Evidence for Anomalous Lepton Production in e+ - e- Annihilation}",
    reportNumber = "SLAC-PUB-1626, LBL-4228",
    doi = "10.1103/PhysRevLett.35.1489",
    journal = "Phys. Rev. Lett.",
    volume = "35",
    pages = "1489--1492",
    year = "1975"
}

@article{UA1:1983crd,
    author = "Arnison, G. and others",
    collaboration = "UA1",
    title = "{Experimental Observation of Isolated Large Transverse Energy Electrons with Associated Missing Energy at $\sqrt{s}= 540$ GeV}",
    reportNumber = "CERN-EP-83-13",
    doi = "10.1016/0370-2693(83)91177-2",
    journal = "Phys. Lett. B",
    volume = "122",
    pages = "103--116",
    year = "1983"
}

@article{UA1:1983mne,
    author = "Arnison, G. and others",
    collaboration = "UA1",
    title = "{Experimental Observation of Lepton Pairs of Invariant Mass Around 95-GeV/c**2 at the CERN SPS Collider}",
    reportNumber = "CERN-EP-83-73",
    doi = "10.1016/0370-2693(83)90188-0",
    journal = "Phys. Lett. B",
    volume = "126",
    pages = "398--410",
    year = "1983"
}

@article{CDF:1995wbb,
    author = "Abe, F. and others",
    collaboration = "CDF",
    title = "{Observation of top quark production in $\bar{p}p$ collisions}",
    eprint = "hep-ex/9503002",
    archivePrefix = "arXiv",
    reportNumber = "FERMILAB-PUB-95-022-E, CDF-PUB-TOP-PUBLIC-3040, ANL-HEP-PR-95-44",
    doi = "10.1103/PhysRevLett.74.2626",
    journal = "Phys. Rev. Lett.",
    volume = "74",
    pages = "2626--2631",
    year = "1995"
}

@article{PhysRevLett.13.508,
  title = {Broken Symmetries and the Masses of Gauge Bosons},
  author = {Higgs, Peter W.},
  journal = {Phys. Rev. Lett.},
  volume = {13},
  issue = {16},
  pages = {508--509},
  numpages = {0},
  year = {1964},
  month = {Oct},
  publisher = {American Physical Society},
  doi = {10.1103/PhysRevLett.13.508},
  url = {https://link.aps.org/doi/10.1103/PhysRevLett.13.508}
}

@article{PhysRevLett.13.321,
  title = {Broken Symmetry and the Mass of Gauge Vector Mesons},
  author = {Englert, F. and Brout, R.},
  journal = {Phys. Rev. Lett.},
  volume = {13},
  issue = {9},
  pages = {321--323},
  numpages = {0},
  year = {1964},
  month = {Aug},
  publisher = {American Physical Society},
  doi = {10.1103/PhysRevLett.13.321},
  url = {https://link.aps.org/doi/10.1103/PhysRevLett.13.321}
}

@article{Goldstone:1961eq,
    author = "Goldstone, J.",
    title = "{Field Theories with Superconductor Solutions}",
    doi = "10.1007/BF02812722",
    journal = "Nuovo Cim.",
    volume = "19",
    pages = "154--164",
    year = "1961"
}

@article{ml4hep,
    author = "Albertsson, Kim and others",
    title = "{Machine Learning in High Energy Physics Community White Paper}",
    eprint = "1807.02876",
    archivePrefix = "arXiv",
    primaryClass = "physics.comp-ph",
    reportNumber = "FERMILAB-PUB-18-318-CD-DI-PPD",
    doi = "10.1088/1742-6596/1085/2/022008",
    journal = "J. Phys. Conf. Ser.",
    volume = "1085",
    number = "2",
    pages = "022008",
    year = "2018"
}

@misc{cern-no-date,
	author = {CERN},
	title = {{Facts and figures about the LHC}},
	url = {https://home.cern/resources/faqs/facts-and-figures-about-lhc},
        note = {Accessed: 2024-11-11}
        
}

@misc{geeksforgeeksLogicGate,
	author = {},
	title = {{L}ogic {G}ate ({A}{N}{D}, {O}{R}, {X}{O}{R}, {N}{O}{T}, {N}{A}{N}{D}, {N}{O}{R} and {X}{N}{O}{R}) - {G}eeksfor{G}eeks --- geeksforgeeks.org},
	howpublished = {\url{https://www.geeksforgeeks.org/logic-gates/}},
	year = {},
	note = {[Accessed 03-01-2025]},
}

@misc{ibmWhatQuantum,
	author = {IBM},
	title = {{W}hat {I}s {Q}uantum {C}omputing? | {I}{B}{M} --- ibm.com},
	howpublished = {\url{https://www.ibm.com/think/topics/quantum-computing}},
	year = {2024},
	note = {[Accessed 02-01-2025]},
}

@article{FCC:2018evy,
    author = "Abada, A. and others",
    collaboration = "FCC",
    title = "{FCC-ee: The Lepton Collider}: {Future Circular Collider Conceptual Design Report Volume 2}",
    reportNumber = "CERN-ACC-2018-0057",
    doi = "10.1140/epjst/e2019-900045-4",
    journal = "Eur. Phys. J. ST",
    volume = "228",
    number = "2",
    pages = "261--623",
    year = "2019"
}

@article{FCC:2018vvp,
    author = "Abada, A. and others",
    collaboration = "FCC",
    title = "{FCC-hh: The Hadron Collider}: {Future Circular Collider Conceptual Design Report Volume 3}",
    reportNumber = "CERN-ACC-2018-0058",
    doi = "10.1140/epjst/e2019-900087-0",
    journal = "Eur. Phys. J. ST",
    volume = "228",
    number = "4",
    pages = "755--1107",
    year = "2019"
}

@article{FCC:2018bvk,
    author = "Abada, A. and others",
    collaboration = "FCC",
    title = "{HE-LHC: The High-Energy Large Hadron Collider}: {Future Circular Collider Conceptual Design Report Volume 4}",
    reportNumber = "CERN-ACC-2018-0059",
    doi = "10.1140/epjst/e2019-900088-6",
    journal = "Eur. Phys. J. ST",
    volume = "228",
    number = "5",
    pages = "1109--1382",
    year = "2019"
}

@article{Bambade:2019fyw,
    author = "Bambade, Philip and others",
    title = "{The International Linear Collider: A Global Project}",
    eprint = "1903.01629",
    archivePrefix = "arXiv",
    primaryClass = "hep-ex",
    reportNumber = "DESY 19-037, DESY-19-037, FERMILAB-FN-1067-PPD, IFIC/19-10, IRFU-19-10,
  JLAB-PHY-19-2854, KEK Preprint 2018-92, JLAB-PHY-19-2854, KEK
  Preprint 2018-92, LAL/RT 19-001, PNNL-SA-142168,
  SLAC-PUB-17412, SLAC-PUB-17412",
    month = "3",
    year = "2019"
}

@article{Roloff:2018dqu,
    editor = "Roloff, P. and Franceschini, R. and Schnoor, U. and Wulzer, A.",
    collaboration = "CLIC, CLICdp",
    title = "{The Compact Linear e$^+$e$^-$ Collider (CLIC): Physics Potential}",
    eprint = "1812.07986",
    archivePrefix = "arXiv",
    primaryClass = "hep-ex",
    month = "12",
    year = "2018"
}

@article{CEPCStudyGroup:2018rmc,
    collaboration = "CEPC Study Group",
    title = "{CEPC Conceptual Design Report: Volume 1 - Accelerator}",
    eprint = "1809.00285",
    archivePrefix = "arXiv",
    primaryClass = "physics.acc-ph",
    reportNumber = "IHEP-CEPC-DR-2018-01, IHEP-AC-2018-01",
    month = "9",
    year = "2018"
}

@article{Abel:2022lqr,
    author = "Abel, Steve and Criado, Juan C. and Spannowsky, Michael",
    title = "{Completely quantum neural networks}",
    eprint = "2202.11727",
    archivePrefix = "arXiv",
    primaryClass = "quant-ph",
    doi = "10.1103/PhysRevA.106.022601",
    journal = "Phys. Rev. A",
    volume = "106",
    number = "2",
    pages = "022601",
    year = "2022"
}

@article{Araz:2022haf,
    author = "Araz, Jack Y. and Spannowsky, Michael",
    title = "{Classical versus quantum: Comparing tensor-network-based quantum circuits on Large Hadron Collider data}",
    eprint = "2202.10471",
    archivePrefix = "arXiv",
    primaryClass = "quant-ph",
    reportNumber = "IPPP/22/06",
    doi = "10.1103/PhysRevA.106.062423",
    journal = "Phys. Rev. A",
    volume = "106",
    number = "6",
    pages = "062423",
    year = "2022"
}

@article{Ngairangbam:2021yma,
    author = "Ngairangbam, Vishal S. and Spannowsky, Michael and Takeuchi, Michihisa",
    title = "{Anomaly detection in high-energy physics using a quantum autoencoder}",
    eprint = "2112.04958",
    archivePrefix = "arXiv",
    primaryClass = "hep-ph",
    reportNumber = "OU-HET-1125, IPPP/21/54",
    doi = "10.1103/PhysRevD.105.095004",
    journal = "Phys. Rev. D",
    volume = "105",
    number = "9",
    pages = "095004",
    year = "2022"
}

@article{Williams:2021lvr,
    author = "Williams, Simon and Malik, Sarah and Spannowsky, Michael and Bepari, Khadeejah",
    title = "A quantum walk approach to simulating parton showers",
    eprint = "2109.13975",
    archivePrefix = "arXiv",
    primaryClass = "hep-ph",
    month = "9",
    year = "2021"
}

@article{Araz:2021zwu,
    author = "Araz, Jack Y. and Spannowsky, Michael",
    title = "{Quantum-inspired event reconstruction with Tensor Networks: Matrix Product States}",
    eprint = "2106.08334",
    archivePrefix = "arXiv",
    primaryClass = "hep-ph",
    reportNumber = "IPPP/20/114",
    doi = "10.1007/JHEP08(2021)112",
    journal = "JHEP",
    volume = "08",
    pages = "112",
    year = "2021"
}

@article{Blance:2020nhl,
    author = "Blance, Andrew and Spannowsky, Michael",
    title = "{Quantum Machine Learning for Particle Physics using a Variational Quantum Classifier}",
    eprint = "2010.07335",
    archivePrefix = "arXiv",
    primaryClass = "hep-ph",
    reportNumber = "IPPP/20/48",
    doi = "10.1007/JHEP02(2021)212",
    journal = "JHEP",
    volume = "02",
    pages = "212",
    year = "2021"
}

@article{Bepari:2020xqi,
    author = "Bepari, Khadeejah and Malik, Sarah and Spannowsky, Michael and Williams, Simon",
    title = "{Towards a quantum computing algorithm for helicity amplitudes and parton showers}",
    eprint = "2010.00046",
    archivePrefix = "arXiv",
    primaryClass = "hep-ph",
    reportNumber = "IPPP/20/41",
    doi = "10.1103/PhysRevD.103.076020",
    journal = "Phys. Rev. D",
    volume = "103",
    number = "7",
    pages = "076020",
    year = "2021"
}

@article{Blance:2020ktp,
    author = "Blance, Andrew and Spannowsky, Michael",
    title = "{Unsupervised event classification with graphs on classical and photonic quantum computers}",
    eprint = "2103.03897",
    archivePrefix = "arXiv",
    primaryClass = "hep-ph",
    doi = "10.1007/JHEP08(2021)170",
    journal = "JHEP",
    volume = "21",
    pages = "170",
    year = "2020"
}

@inproceedings{Delgado:2022tpc,
    author = "Delgado, Andrea and others",
    title = "{Quantum Computing for Data Analysis in High-Energy Physics}",
    booktitle = "{2022 Snowmass Summer Study}",
    eprint = "2203.08805",
    archivePrefix = "arXiv",
    primaryClass = "physics.data-an",
    month = "3",
    year = "2022"
}

@article{ball1967clustering,
author = {Ball, Geoffrey H. and Hall, David J.},
title = {A clustering technique for summarizing multivariate data},
journal = {Behavioral Science},
volume = {12},
number = {2},
pages = {153-155},
doi = {https://doi.org/10.1002/bs.3830120210},
year = {1967}
}

@ARTICLE{Lloyd:1982ls,
  author={Lloyd, S.},
  journal={IEEE Transactions on Information Theory}, 
  title={Least squares quantization in PCM}, 
  year={1982},
  volume={28},
  number={2},
  pages={129-137},
  doi={10.1109/TIT.1982.1056489}
}

@article{Chekanov:2005cq,
    author = "Chekanov, S.",
    title = "{A New jet algorithm based on the k-means clustering for the reconstruction of heavy states from jets}",
    eprint = "hep-ph/0512027",
    archivePrefix = "arXiv",
    reportNumber = "ANL-HEP-PR-05-118",
    doi = "10.1140/epjc/s2006-02618-3",
    journal = "Eur. Phys. J. C",
    volume = "47",
    pages = "611--616",
    year = "2006"
}

@article{Thaler:2011gf,
    author = "Thaler, Jesse and Van Tilburg, Ken",
    title = "{Maximizing Boosted Top Identification by Minimizing N-subjettiness}",
    eprint = "1108.2701",
    archivePrefix = "arXiv",
    primaryClass = "hep-ph",
    reportNumber = "MIT-CTP-4287",
    doi = "10.1007/JHEP02(2012)093",
    journal = "JHEP",
    volume = "02",
    pages = "093",
    year = "2012"
}

@inproceedings{macqueen1967some,
  title={Some methods for classification and analysis of multivariate observations},
  author={MacQueen, James and others},
  booktitle={Proceedings of the fifth Berkeley symposium on mathematical statistics and probability},
  year={1967},
  organization={Oakland, CA, USA}
}

@article{Drineas:2004fa,
author = {Drineas, Petros and Frieze, Alan and Kannan, Ravindran and Vempala, S. and Vinay, V.},
year = {2004},
month = {01},
pages = {},
title = {Clustering Large Graphs via the Singular Value Decomposition: Theoretical Advances in Data Clustering (Guest Editors: Nina Mishra and Rajeev Motwani)},
volume = {56},
journal = {Machine Learning},
doi = {10.1023/B:MACH.0000033113.59016.96}
}

@article{Catani:1991hj,
    author = "Catani, S. and Dokshitzer, Yuri L. and Olsson, M. and Turnock, G. and Webber, B. R.",
    title = "{New clustering algorithm for multi - jet cross-sections in e+ e- annihilation}",
    reportNumber = "CAVENDISH-HEP-91-5",
    doi = "10.1016/0370-2693(91)90196-W",
    journal = "Phys. Lett. B",
    volume = "269",
    pages = "432--438",
    year = "1991"
}

@article{Cacciari:2008gp,
    author = "Cacciari, Matteo and Salam, Gavin P. and Soyez, Gregory",
    title = "{The anti-$k_t$ jet clustering algorithm}",
    eprint = "0802.1189",
    archivePrefix = "arXiv",
    primaryClass = "hep-ph",
    reportNumber = "LPTHE-07-03",
    doi = "10.1088/1126-6708/2008/04/063",
    journal = "JHEP",
    volume = "04",
    pages = "063",
    year = "2008"
}

@article{Dokshitzer:1997in,
    author = "Dokshitzer, Yuri L. and Leder, G. D. and Moretti, S. and Webber, B. R.",
    title = "{Better jet clustering algorithms}",
    eprint = "hep-ph/9707323",
    archivePrefix = "arXiv",
    reportNumber = "CAVENDISH-HEP-97-06",
    doi = "10.1088/1126-6708/1997/08/001",
    journal = "JHEP",
    volume = "08",
    pages = "001",
    year = "1997"
}

@article{Jordan:2012xnu,
    author = "Jordan, Stephen P. and Lee, Keith S. M. and Preskill, John",
    title = "{Quantum Algorithms for Quantum Field Theories}",
    eprint = "1111.3633",
    archivePrefix = "arXiv",
    primaryClass = "quant-ph",
    doi = "10.1126/science.1217069",
    journal = "Science",
    volume = "336",
    pages = "1130--1133",
    year = "2012"
}

@article{Bauer:2019qxa,
    author = "Bauer, Christian W. and de Jong, Wibe A. and Nachman, Benjamin and Provasoli, Davide",
    title = "{Quantum Algorithm for High Energy Physics Simulations}",
    eprint = "1904.03196",
    archivePrefix = "arXiv",
    primaryClass = "hep-ph",
    doi = "10.1103/PhysRevLett.126.062001",
    journal = "Phys. Rev. Lett.",
    volume = "126",
    number = "6",
    pages = "062001",
    year = "2021"
}

@article{Ramirez-Uribe:2021ubp,
    author = "Ram{\'\i}rez-Uribe, Selomit and Renter{\'\i}a-Olivo, Andr{\'e}s E. and Rodrigo, Germ{\'a}n and Sborlini, German F. R. and Vale Silva, Luiz",
    title = "{Quantum algorithm for Feynman loop integrals}",
    eprint = "2105.08703",
    archivePrefix = "arXiv",
    primaryClass = "hep-ph",
    reportNumber = "IFIC/21-15, DESY 21-067",
    doi = "10.1007/JHEP05(2022)100",
    journal = "JHEP",
    volume = "05",
    pages = "100",
    year = "2022"
}

@article{Clemente:2022nll,
    author = "Clemente, Giuseppe and Crippa, Arianna and Jansen, Karl and Ram{\'\i}rez-Uribe, Selomit and Renter{\'\i}a-Olivo, Andr{\'e}s E. and Rodrigo, Germ{\'a}n and Sborlini, German F. R. and Vale Silva, Luiz",
    title = "{Variational quantum eigensolver for causal loop Feynman diagrams and directed acyclic graphs}",
    eprint = "2210.13240",
    archivePrefix = "arXiv",
    primaryClass = "hep-ph",
    reportNumber = "IFIC/22-28",
    doi = "10.1103/PhysRevD.108.096035",
    journal = "Phys. Rev. D",
    volume = "108",
    number = "9",
    pages = "096035",
    year = "2023"
}

@article{Byrnes:2005qx,
    author = "Byrnes, Tim and Yamamoto, Yoshihisa",
    title = "{Simulating lattice gauge theories on a quantum computer}",
    eprint = "quant-ph/0510027",
    archivePrefix = "arXiv",
    doi = "10.1103/PhysRevA.73.022328",
    journal = "Phys. Rev. A",
    volume = "73",
    pages = "022328",
    year = "2006"
}

@article{Zohar:2015hwa,
    author = "Zohar, Erez and Cirac, J. Ignacio and Reznik, Benni",
    title = "{Quantum Simulations of Lattice Gauge Theories using Ultracold Atoms in Optical Lattices}",
    eprint = "1503.02312",
    archivePrefix = "arXiv",
    primaryClass = "quant-ph",
    doi = "10.1088/0034-4885/79/1/014401",
    journal = "Rept. Prog. Phys.",
    volume = "79",
    number = "1",
    pages = "014401",
    year = "2016"
}

@article{Banuls:2019bmf,
    author = "Ba\~nuls, M. C. and others",
    title = "{Simulating Lattice Gauge Theories within Quantum Technologies}",
    eprint = "1911.00003",
    archivePrefix = "arXiv",
    primaryClass = "quant-ph",
    doi = "10.1140/epjd/e2020-100571-8",
    journal = "Eur. Phys. J. D",
    volume = "74",
    number = "8",
    pages = "165",
    year = "2020"
}

@article{Felser:2020mka,
    author = "Felser, Timo and Trenti, Marco and Sestini, Lorenzo and Gianelle, Alessio and Zuliani, Davide and Lucchesi, Donatella and Montangero, Simone",
    title = "{Quantum-inspired machine learning on high-energy physics data}",
    eprint = "2004.13747",
    archivePrefix = "arXiv",
    primaryClass = "stat.ML",
    doi = "10.1038/s41534-021-00443-w",
    journal = "npj Quantum Inf.",
    volume = "7",
    number = "1",
    pages = "111",
    year = "2021"
}

@article{Guan:2020bdl,
    author = "Guan, Wen and Perdue, Gabriel and Pesah, Arthur and Schuld, Maria and Terashi, Koji and Vallecorsa, Sofia and Vlimant, Jean-Roch",
    title = "{Quantum Machine Learning in High Energy Physics}",
    eprint = "2005.08582",
    archivePrefix = "arXiv",
    primaryClass = "quant-ph",
    reportNumber = "FERMILAB-PUB-20-184-QIS",
    doi = "10.1088/2632-2153/abc17d",
    journal = "Mach. Learn. Sci. Tech.",
    volume = "2",
    pages = "011003",
    year = "2021"
}

@article{Wu:2020cye,
    author = "Wu, Sau Lan and others",
    title = "{Application of quantum machine learning using the quantum variational classifier method to high energy physics analysis at the LHC on IBM quantum computer simulator and hardware with 10 qubits}",
    eprint = "2012.11560",
    archivePrefix = "arXiv",
    primaryClass = "quant-ph",
    reportNumber = "FERMILAB-PUB-20-675-DI-QIS",
    doi = "10.1088/1361-6471/ac1391",
    journal = "J. Phys. G",
    volume = "48",
    number = "12",
    pages = "125003",
    year = "2021"
}

@article{DeJong:2020riy,
    author = "De Jong, Wibe A. and Metcalf, Mekena and Mulligan, James and P\l{}osko\'n, Mateusz and Ringer, Felix and Yao, Xiaojun",
    title = "{Quantum simulation of open quantum systems in heavy-ion collisions}",
    eprint = "2010.03571",
    archivePrefix = "arXiv",
    primaryClass = "hep-ph",
    reportNumber = "MIT-CTP/5247",
    doi = "10.1103/PhysRevD.104.L051501",
    journal = "Phys. Rev. D",
    volume = "104",
    number = "5",
    pages = "051501",
    year = "2021"
}

@article{Pires:2021fka,
    author = "Pires, Diogo and Bargassa, Pedrame and Seixas, Jo\~ao and Omar, Yasser",
    title = "{A Digital Quantum Algorithm for Jet Clustering in High-Energy Physics}",
    eprint = "2101.05618",
    archivePrefix = "arXiv",
    primaryClass = "physics.data-an",
    month = "1",
    year = "2021"
}

@article{Perez-Salinas:2020nem,
    author = "P\'erez-Salinas, Adri\'an and Cruz-Martinez, Juan and Alhajri, Abdulla A. and Carrazza, Stefano",
    title = "{Determining the proton content with a quantum computer}",
    eprint = "2011.13934",
    archivePrefix = "arXiv",
    primaryClass = "hep-ph",
    reportNumber = "TIF-UNIMI-2020-30",
    doi = "10.1103/PhysRevD.103.034027",
    journal = "Phys. Rev. D",
    volume = "103",
    number = "3",
    pages = "034027",
    year = "2021"
}

@article{Barata:2021yri,
    author = "Barata, Jo\~ao and Salgado, Carlos A.",
    title = "{A quantum strategy to compute the jet quenching parameter $\hat{q}$}",
    eprint = "2104.04661",
    archivePrefix = "arXiv",
    primaryClass = "hep-ph",
    doi = "10.1140/epjc/s10052-021-09674-9",
    journal = "Eur. Phys. J. C",
    volume = "81",
    number = "10",
    pages = "862",
    year = "2021"
}

@article{Pires:2020urc,
    author = "Pires, Diogo and Omar, Yasser and Seixas, Jo\~ao",
    title = "{Adiabatic Quantum Algorithm for Multijet Clustering in High Energy Physics}",
    eprint = "2012.14514",
    archivePrefix = "arXiv",
    primaryClass = "hep-ex",
    month = "12",
    year = "2020"
}

@article{Wei:2019rqy,
    author = "Wei, Annie Y. and Naik, Preksha and Harrow, Aram W. and Thaler, Jesse",
    title = "{Quantum Algorithms for Jet Clustering}",
    eprint = "1908.08949",
    archivePrefix = "arXiv",
    primaryClass = "hep-ph",
    reportNumber = "MIT-CTP 5137",
    doi = "10.1103/PhysRevD.101.094015",
    journal = "Phys. Rev. D",
    volume = "101",
    number = "9",
    pages = "094015",
    year = "2020"
}

@article{FCC:2018byv,
    author = "Abada, A. and others",
    collaboration = "FCC",
    title = "{FCC Physics Opportunities}: {Future Circular Collider Conceptual Design Report Volume 1}",
    reportNumber = "CERN-ACC-2018-0056",
    doi = "10.1140/epjc/s10052-019-6904-3",
    journal = "Eur. Phys. J. C",
    volume = "79",
    number = "6",
    pages = "474",
    year = "2019"
}

@article{CEPCStudyGroup:2018ghi,
    author = "Dong, Mingyi and others",
    editor = "Guimar\~aes da Costa, Jo\~ao Barreiro and others",
    collaboration = "CEPC Study Group",
    title = "{CEPC Conceptual Design Report: Volume 2 - Physics \& Detector}",
    eprint = "1811.10545",
    archivePrefix = "arXiv",
    primaryClass = "hep-ex",
    reportNumber = "IHEP-CEPC-DR-2018-02, IHEP-EP-2018-01, IHEP-TH-2018-01",
    month = "11",
    year = "2018"
}

@article{EPPPG:2019qin,
    author = "Ellis, Richard Keith and others",
    title = "{Physics Briefing Book}: {Input for the European Strategy for Particle Physics Update 2020}",
    eprint = "1910.11775",
    archivePrefix = "arXiv",
    primaryClass = "hep-ex",
    reportNumber = "CERN-ESU-004",
    month = "10",
    year = "2019"
}

@article{Feynman:1981tf,
    author = "Feynman, Richard P.",
    editor = "Brown, L. M.",
    title = "{Simulating physics with computers}",
    doi = "10.1007/BF02650179",
    journal = "Int. J. Theor. Phys.",
    volume = "21",
    pages = "467--488",
    year = "1982"
}

@article{Grover:1997fa,
    author = "Grover, Lov K.",
    title = "{Quantum mechanics helps in searching for a needle in a haystack}",
    eprint = "quant-ph/9706033",
    archivePrefix = "arXiv",
    doi = "10.1103/PhysRevLett.79.325",
    journal = "Phys. Rev. Lett.",
    volume = "79",
    pages = "325--328",
    year = "1997"
}

@article{Shor:1994jg,
    author = "Shor, Peter W.",
    title = "{Polynomial time algorithms for prime factorization and discrete logarithms on a quantum computer}",
    eprint = "quant-ph/9508027",
    archivePrefix = "arXiv",
    doi = "10.1137/S0097539795293172",
    journal = "SIAM J. Sci. Statist. Comput.",
    volume = "26",
    pages = "1484",
    year = "1997"
}

@article{Larose:2019,
	Author = {LaRose, Ryan and Tikku, Arkin and O'Neel-Judy, {\'E}tude and Cincio, Lukasz and Coles, Patrick J.},
	Da = {2019/06/26},
	Doi = {10.1038/s41534-019-0167-6},
	Id = {LaRose2019},
	Isbn = {2056-6387},
	Journal = {npj Quantum Information},
	Number = {1},
	Pages = {57},
	Title = {Variational quantum state diagonalization},
	Ty = {JOUR},
	Url = {https://doi.org/10.1038/s41534-019-0167-6},
	Volume = {5},
	Year = {2019}
	}

@article{PhysRevA.101.062310,
  title = {Quantum singular value decomposer},
  author = {Bravo-Prieto, Carlos  and Garc\'{\i}a-Mart\'{\i}n, Diego and Latorre, Jos\'e I.},
  journal = {Phys. Rev. A},
  volume = {101},
  issue = {6},
  pages = {062310},
  numpages = {6},
  year = {2020},
  month = {Jun},
  publisher = {American Physical Society},
  doi = {10.1103/PhysRevA.101.062310},
  url = {https://link.aps.org/doi/10.1103/PhysRevA.101.062310}
}

@article{Kokail:2018eiw,
    author = "Kokail, Christian and others",
    title = "{Self-verifying variational quantum simulation of lattice models}",
    eprint = "1810.03421",
    archivePrefix = "arXiv",
    primaryClass = "quant-ph",
    doi = "10.1038/s41586-019-1177-4",
    journal = "Nature",
    volume = "569",
    number = "7756",
    pages = "355--360",
    year = "2019"
}

@article{montanaro:2015,
	doi = {10.1098/rspa.2015.0301},
  
	url = {https://doi.org/10.1098%2Frspa.2015.0301},
  
	year = 2015,
	month = {sep},
  
	publisher = {The Royal Society},
  
	volume = {471},
  
	number = {2181},
  
	pages = {20150301},
  
	author = {Ashley Montanaro},
  
	title = {Quantum speedup of Monte Carlo methods},
  
	journal = {Proceedings of the Royal Society A: Mathematical, Physical and Engineering Sciences}
}

@article{Orus:2019xrh,
    author = "Or\'us, Rom\'an and Mugel, Samuel and Lizaso, Enrique",
    title = "{Quantum computing for finance: Overview and prospects}",
    doi = "10.1016/j.revip.2019.100028",
    journal = "Rev. Phys.",
    volume = "4",
    pages = "100028",
    year = "2019"
}

@article{Holland:2019zju,
    author = "Holland, Eric T. and Wendt, Kyle A. and Kravvaris, Konstantinos and Wu, Xian and Erich Ormand, W. and DuBois, Jonathan L and Quaglioni, Sofia and Pederiva, Francesco",
    title = "{Optimal Control for the Quantum Simulation of Nuclear Dynamics}",
    eprint = "1908.08222",
    archivePrefix = "arXiv",
    primaryClass = "quant-ph",
    reportNumber = "LLNL-JRNL-787600",
    doi = "10.1103/PhysRevA.101.062307",
    journal = "Phys. Rev. A",
    volume = "101",
    number = "6",
    pages = "062307",
    year = "2020"
}

@article{Lynn:2019rdt,
    author = "Lynn, J. E. and Tews, I. and Gandolfi, S. and Lovato, A.",
    title = "{Quantum Monte Carlo Methods in Nuclear Physics: Recent Advances}",
    eprint = "1901.04868",
    archivePrefix = "arXiv",
    primaryClass = "nucl-th",
    reportNumber = "LA-UR-19-20209",
    doi = "10.1146/annurev-nucl-101918-023600",
    journal = "Ann. Rev. Nucl. Part. Sci.",
    volume = "69",
    pages = "279--305",
    year = "2019"
}

@article{Liu:2020eoa,
    author = "Liu, Junyu and Xin, Yuan",
    title = "{Quantum simulation of quantum field theories as quantum chemistry}",
    eprint = "2004.13234",
    archivePrefix = "arXiv",
    primaryClass = "hep-th",
    reportNumber = "CALT-TH-2020-009",
    doi = "10.1007/JHEP12(2020)011",
    journal = "JHEP",
    volume = "12",
    pages = "011",
    year = "2020"
}

@article{kmeans,
 ISSN = {00359254, 14679876},
 URL = {http://www.jstor.org/stable/2346830},
 author = {J. A. Hartigan and M. A. Wong},
 journal = {Journal of the Royal Statistical Society. Series C (Applied Statistics)},
 number = {1},
 pages = {100--108},
 publisher = {[Royal Statistical Society, Oxford University Press]},
 title = {Algorithm AS 136: A K-Means Clustering Algorithm},
 urldate = {2025-07-22},
 volume = {28},
 year = {1979}
}

@inproceedings{K-Means++,
author = {Arthur, David and Vassilvitskii, Sergei},
title = {K-Means++: The Advantages of Careful Seeding},
booktitle = {Proceedings of the Eighteenth Annual ACM-SIAM Symposium on Discrete Algorithms},
organization = {New Orleans, Louisiana, USA},
year = {2007}
}

@article{Cacciari:2011ma,
    author = "Cacciari, Matteo and Salam, Gavin P. and Soyez, Gregory",
    title = "{FastJet User Manual}",
    eprint = "1111.6097",
    archivePrefix = "arXiv",
    primaryClass = "hep-ph",
    reportNumber = "CERN-PH-TH-2011-297",
    doi = "10.1140/epjc/s10052-012-1896-2",
    journal = "Eur. Phys. J. C",
    volume = "72",
    pages = "1896",
    year = "2012"
}

@article{Buhrman:2001,
  title = {Quantum Fingerprinting},
  author = {Buhrman, Harry and Cleve, Richard and Watrous, John and de Wolf, Ronald},
  journal = {Phys. Rev. Lett.},
  volume = {87},
  issue = {16},
  pages = {167902},
  numpages = {4},
  year = {2001},
  month = {Sep},
  publisher = {American Physical Society},
  doi = {10.1103/PhysRevLett.87.167902},
  url = {https://link.aps.org/doi/10.1103/PhysRevLett.87.167902}
}

@article{kopczyk2018quantum,
      title={Quantum machine learning for data scientists}, 
      author={Dawid Kopczyk},
      year={2018},
      eprint={1804.10068},
      archivePrefix={arXiv},
      primaryClass={quant-ph}
}

@article{Foulds:2020ajt,
    author = "Foulds, Steph and Kendon, Viv and Spiller, Tim",
    title = "{The controlled SWAP test for determining quantum entanglement}",
    eprint = "2009.07613",
    archivePrefix = "arXiv",
    primaryClass = "quant-ph",
    doi = "10.1088/2058-9565/abe458",
    journal = "Quantum Sci. Technol.",
    volume = "6",
    pages = "035002",
    year = "2021"
}

@article{Abhi:2020,
author = {Sarma, Abhijat and Chatterjee, Rupak and Gili, Kaitlin and Yu, Ting},
year = {2020},
month = {06},
pages = {541-552},
title = {Quantum unsupervised and supervised learning on superconducting processors},
volume = {20},
journal = {Quantum Information and Computation},
doi = {10.26421/QIC20.7-8-1}
}

@article{Durr:1996nx,
    author = "Dürr, Christoph and Høyer, Peter",
    title = "{A Quantum algorithm for finding the minimum}",
    eprint = "quant-ph/9607014",
    archivePrefix = "arXiv",
    month = "7",
    year = "1996"
}

@article{JADE:1982ttq,
    author = "Bartel, W. and others",
    collaboration = "JADE",
    title = "{Experimental Evidence for Differences in $p_T$ Between Quark Jets and Gluon Jets}",
    reportNumber = "DESY-82-086",
    doi = "10.1016/0370-2693(83)90994-2",
    journal = "Phys. Lett. B",
    volume = "123",
    pages = "460--466",
    year = "1983"
}

@article{lloyd2013quantum,
      title={Quantum algorithms for supervised and unsupervised machine learning}, 
      author={Seth Lloyd and Masoud Mohseni and Patrick Rebentrost},
      year={2013},
      eprint={1307.0411},
      archivePrefix={arXiv},
      primaryClass={quant-ph}
}

@article{2006,
   title="{Dispelling the $N^3$ myth for the $k_T$ jet-finder}",
   volume={641},
   ISSN={0370-2693},
   url={http://dx.doi.org/10.1016/j.physletb.2006.08.037},
   DOI={10.1016/j.physletb.2006.08.037},
   number={1},
   journal={Physics Letters B},
   publisher={Elsevier BV},
   author={Cacciari, Matteo and Salam, Gavin P.},
   year={2006},
   month={Sep},
   pages={57–61}
}

@article{Grover:1996rk,
    author = "Grover, Lov K.",
    title = "{A Fast quantum mechanical algorithm for database search}",
    eprint = "quant-ph/9605043",
    archivePrefix = "arXiv",
    month = "5",
    year = "1996"
}

@article{2008QRAM,
   title={Quantum Random Access Memory},
   volume={100},
   ISSN={1079-7114},
   url={http://dx.doi.org/10.1103/PhysRevLett.100.160501},
   DOI={160501},
   number={16},
   journal={Physical Review Letters},
   publisher={American Physical Society (APS)},
   author={Giovannetti, V.  and Lloyd, S.   and Maccone, L.},
   year={2008},
   month={Apr}
}

@article{Catani:1993hr,
    author = "Catani, S. and Dokshitzer, Yuri L. and Seymour, M. H. and Webber, B. R.",
    title = "{Longitudinally invariant $K_t$ clustering algorithms for hadron hadron collisions}",
    reportNumber = "CERN-TH-6775-93, LU-TP-93-2",
    doi = "10.1016/0550-3213(93)90166-M",
    journal = "Nucl. Phys. B",
    volume = "406",
    pages = "187--224",
    year = "1993"
}

@article{Ellis:1993tq,
    author = "Ellis, Stephen D. and Soper, Davison E.",
    title = "{Successive combination jet algorithm for hadron collisions}",
    eprint = "hep-ph/9305266",
    archivePrefix = "arXiv",
    reportNumber = "CERN-TH-6860-93",
    doi = "10.1103/PhysRevD.48.3160",
    journal = "Phys. Rev. D",
    volume = "48",
    pages = "3160--3166",
    year = "1993"
}

@article{2008Arch,
   title={Architectures for a quantum random access memory},
   volume={78},
   ISSN={1094-1622},
   url={http://dx.doi.org/10.1103/PhysRevA.78.052310},
   DOI={10.1103/physreva.78.052310},
   number={5},
   journal={Physical Review A},
   publisher={American Physical Society (APS)},
   author={Vittorio Giovannetti  and Seth Lloyd  and Lorenzo Maccone },
   year={2008},
   month={Nov}
}

@article{demartini2009experimental,
  title = {Experimental quantum private queries with linear optics},
  author = {Francesco DeMartini and Vittorio Giovannetti and Seth Lloyd and Lorenzo Maccone and Eleonora Nagali  and Linda Sansoni and Fabio Sciarrino },
  journal = {Phys. Rev. A},
  volume = {80},
  issue = {1},
  pages = {010302},
  numpages = {4},
  year = {2009},
  month = {Jul},
  publisher = {American Physical Society},
  doi = {10.1103/PhysRevA.80.010302},
  url = {https://link.aps.org/doi/10.1103/PhysRevA.80.010302}
}

@article{2010Mag,
   title={Magnetic strong coupling in a spin-photon system and transition to classical regime},
   volume={82},
   ISSN={1550-235X},
   url={http://dx.doi.org/10.1103/PhysRevB.82.024413},
   DOI={10.1103/physrevb.82.024413},
   number={2},
   journal={Physical Review B},
   publisher={American Physical Society (APS)},
   author={I. Chiorescu  and N. Groll and S. Bertaina and T. Mori and  S. Miyashita},
   year={2010},
   month={Jul}
}

@article{PhysRevLett.105.140501,
  title = {High-Cooperativity Coupling of Electron-Spin Ensembles to Superconducting Cavities},
  author = {Schuster, D. I. and Sears, A. P. and Ginossar, E. and DiCarlo, L. and Frunzio, L. and Morton, J. J. L. and Wu, H. and Briggs, G. A. D. and Buckley, B. B. and Awschalom, D. D. and Schoelkopf, R. J.},
  journal = {Phys. Rev. Lett.},
  volume = {105},
  issue = {14},
  pages = {140501},
  numpages = {4},
  year = {2010},
  month = {Sep},
  publisher = {American Physical Society},
  doi = {10.1103/PhysRevLett.105.140501},
  url = {https://link.aps.org/doi/10.1103/PhysRevLett.105.140501}
}

@article{PhysRevLett.105.140502,
  title = {Strong Coupling of a Spin Ensemble to a Superconducting Resonator},
  author = {Kubo, Y. and Ong, F. R. and Bertet, P. and Vion, D. and Jacques, V. and Zheng, D. and Dr\'eau, A. and Roch, J.-F. and Auffeves, A. and Jelezko, F. and Wrachtrup, J. and Barthe, M. F. and Bergonzo, P. and Esteve, D.},
  journal = {Phys. Rev. Lett.},
  volume = {105},
  issue = {14},
  pages = {140502},
  numpages = {4},
  year = {2010},
  month = {Sep},
  publisher = {American Physical Society},
  doi = {10.1103/PhysRevLett.105.140502},
  url = {https://link.aps.org/doi/10.1103/PhysRevLett.105.140502}
}

@article{PhysRevLett.105.140503,
  title = {Storage of Multiple Coherent Microwave Excitations in an Electron Spin Ensemble},
  author = {Wu, Hua and George, Richard E. and Wesenberg, Janus H. and M\o{}lmer, Klaus and Schuster, David I. and Schoelkopf, Robert J. and Itoh, Kohei M. and Ardavan, Arzhang and Morton, John J. L. and Briggs, G. Andrew D.},
  journal = {Phys. Rev. Lett.},
  volume = {105},
  issue = {14},
  pages = {140503},
  numpages = {4},
  year = {2010},
  month = {Sep},
  publisher = {American Physical Society},
  doi = {10.1103/PhysRevLett.105.140503},
  url = {https://link.aps.org/doi/10.1103/PhysRevLett.105.140503}
}

@article{smith1990extreme,
  title={Extreme value theory},
  author={Smith, Richard L},
  journal={Handbook of applicable mathematics},
  volume={7},
  year={1990},
  publisher={Wiley Chichester}
}

@book{Gumbel1958,
author = {E. J. Gumbel},
title = {Statistics of Extremes},
year = {1958},
publisher = {Columbia University Press},
}

@book{castillo2004,
author = {Castillo  E. and Hadi, Ali S and Balakrishnan, Narayanaswamy and Sarabia, Jos{\'e}-Mari{\'a}},
title = {Extreme Value and Related Models with Applications in Engineering and Science},
year = {2004},
publisher = {Wiley},
ISBN = {9780471671725}
}

@book{Coles2001,
  author = {Coles, Stuart},
  isbn = {1-85233-459-2},
  mrclass = {62-01 (60G70 62G32)},
  mrnumber = {1932132 (2003h:62002)},
  publisher = {Springer-Verlag},
  title = {An introduction to statistical modeling of extreme values},
  year = 2001
}

@book{Reiss2007,
    author = {Reiss  R-D. , Thomas  M.},
    title = {Statistical Analysis of Extreme Values With Applications to Insurance, Finance, Hidrology and Other Fields},
    year = {2007},
    publisher = {Birkhauser Verlag},
    ISBN = {9783764372309}
}

@article{Brun:1997pa,
    author = "Brun, R. and Rademakers, F.",
    editor = "Werlen, M. and Perret-Gallix, D.",
    title = "{ROOT: An object oriented data analysis framework}",
    doi = "10.1016/S0168-9002(97)00048-X",
    journal = "Nucl. Instrum. Meth. A",
    volume = "389",
    pages = "81--86",
    year = "1997"
}

@article{DBLP:journals/corr/abs-1804-03719,
  author    = {Patrick J. Coles and
               Stephan J. Eidenbenz and
               Scott Pakin and
               Adetokunbo Adedoyin and
               John Ambrosiano and
               Petr M. Anisimov and
               William Casper and
               Gopinath Chennupati and
               Carleton Coffrin and
               Hristo N. Djidjev and
               David Gunter and
               Satish Karra and
               Nathan Lemons and
               Shizeng Lin and
               Andrey Y. Lokhov and
               Alexander Malyzhenkov and
               David Dennis Lee Mascarenas and
               Susan M. Mniszewski and
               Balu Nadiga and
               Dan O'Malley and
               Diane Oyen and
               Lakshman Prasad and
               Randy Roberts and
               Philip Romero and
               Nandakishore Santhi and
               Nikolai Sinitsyn and
               Pieter Swart and
               Marc Vuffray and
               Jim Wendelberger and
               Boram Yoon and
               Richard J. Zamora and
               Wei Zhu},
  title     = {Quantum Algorithm Implementations for Beginners},
  year      = {2018},
  url       = {http://arxiv.org/abs/1804.03719},
  eprinttype = {arXiv},
  eprint    = {1804.03719},
}

@article{Frey2007ClusteringBP,
author = {Brendan J. Frey  and Delbert Dueck },
title = {Clustering by Passing Messages Between Data Points},
journal = {Science},
volume = {315},
number = {5814},
pages = {972-976},
year = {2007},
doi = {10.1126/science.1136800},
}

@article{Bethke:1991wk,
    author = "Bethke, S. and Kunszt, Z. and Soper, D. E. and Stirling, W. James",
    title = "{New jet cluster algorithms: Next-to-leading order QCD and hadronization corrections}",
    reportNumber = "CERN-TH-6222-91",
    doi = "10.1016/0550-3213(92)90289-N",
    journal = "Nucl. Phys. B",
    volume = "370",
    pages = "310--334",
    year = "1992",
    note = "[Erratum: Nucl.Phys.B 523, 681--681 (1998)]"
}

@article{Leone_2007,
	doi = {10.1093/bioinformatics/btm414},
  
	url = {https://doi.org/10.10932Fbioinformatics2Fbtm414},
  
	year = 2007,
	month = {sep},
  
	publisher = {Oxford University Press ({OUP})},
  
	volume = {23},
  
	number = {20},
  
	pages = {2708--2715},
  
	author = {M. Leone and  Sumedha and M. Weigt},
  
	title = {Clustering by soft-constraint affinity propagation: applications to gene-expression data},
  
	journal = {Bioinformatics}
}

@article{Bailly_Bechet_2009,
	doi = {10.1088/1742-5468/2009/12/p12010},
  
	url = {https://doi.org/10.10882F1742-54682F20092F122Fp12010},
  
	year = 2009,
	month = {dec},
  
	publisher = {{IOP} Publishing},
  
	volume = {2009},
  
	number = {12},
  
	pages = {P12010},
  
	author = {M Bailly-Bechet and S Bradde and A Braunstein and A Flaxman and L Foini and R Zecchina},
  
	title = {Clustering with shallow trees},
  
	journal = {Journal of Statistical Mechanics: Theory and Experiment}
}

@article{Rodrigo:1999qg,
    author = "Rodrigo, German and Bilenky, Mikhail S. and Santamaria, Arcadi",
    title = "{Quark mass effects for jet production in e+ e- collisions at the next-to-leading order: Results and applications}",
    eprint = "hep-ph/9905276",
    archivePrefix = "arXiv",
    reportNumber = "FTUV-99-2, IFIC-99-2",
    doi = "10.1016/S0550-3213(99)00293-X",
    journal = "Nucl. Phys. B",
    volume = "554",
    pages = "257--297",
    year = "1999"
}

@article{Sumedha_2008,
	doi = {10.1140/epjb/e2008-00381-8},
  
	url = {https://doi.org/10.11402Fepjb2Fe2008-00381-8},
  
	year = 2008,
	month = {oct},
  
	publisher = {Springer Science and Business Media {LLC}
},
  
	volume = {66},
  
	number = {1},
  
	pages = {125--135},
  
	author = {M. Leone Sumedha and M. Weigt},
  
	title = {Unsupervised and semi-supervised clustering by message passing: soft-constraint affinity propagation},
  
	journal = {The European Physical Journal B}
}

@article{GonzlezMartn2017,
	doi = {10.3847/1538-4357/aa6f16},
  
	url = {https://doi.org/10.38472F1538-43572Faa6f16},
  
	year = 2017,
	month = {may},
  
	publisher = {American Astronomical Society},
  
	volume = {841},
  
	number = {1},
  
	pages = {37},
  
	author = {Omaira Gonz{\'{a}
}lez-Mart{\'{\i}}n and Josefa Masegosa and Antonio Hern{\'{a}}n-Caballero and Isabel M{\'{a}}rquez and Cristina Ramos Almeida and Almudena Alonso-Herrero and Itziar Aretxaga and Jos{\'{e}} Miguel Rodr{\'{\i}}guez-Espinosa and Jose Antonio Acosta-Pulido and Lorena Hern{\'{a}}ndez-Garc{\'{\i}}a and Donaji Esparza-Arredondo and Mariela Mart{\'{\i}}nez-Paredes and Paolo Bonfini and Alice Pasetto and Deborah Dultzin},
  

	
    title = {Hints on the gradual re-sizing of the torus in AGN by decomposing IRS/Spitzer spectra},
	journal = {The Astrophysical Journal}
}

@inproceedings{Carrazza:2016sgh,
    author = "Carrazza, Stefano and Latorre, Jos\'e I.",
    title = "{Towards the compression of parton densities through machine learning algorithms}",
    booktitle = "{51st Rencontres de Moriond on QCD and High Energy Interactions}",
    eprint = "1605.04345",
    archivePrefix = "arXiv",
    primaryClass = "hep-ph",
    reportNumber = "CERN-TH-2016-115",
    pages = "235--238",
    month = "5",
    year = "2016"
}

@article{HEPSoftwareFoundation:2017ggl,
    author = "Albrecht, Johannes and others",
    collaboration = "HEP Software Foundation",
    title = "{A Roadmap for HEP Software and Computing R\&D for the 2020s}",
    eprint = "1712.06982",
    archivePrefix = "arXiv",
    primaryClass = "physics.comp-ph",
    reportNumber = "HSF-CWP-2017-01, HSF-CWP-2017-001, FERMILAB-PUB-17-607-CD",
    doi = "10.1007/s41781-018-0018-8",
    journal = "Comput. Softw. Big Sci.",
    volume = "3",
    number = "1",
    pages = "7",
    year = "2019"
}

@techreport{Collaboration:2802918,
      author        = "ATLAS Collaboration",
      title         = "{ATLAS Software and Computing HL-LHC Roadmap}",
      institution   = "CERN",
      address       = "Geneva",
      number        ="182",
      reportNumber  = "CERN-LHCC-2022-005, LHCC-G-182",
      month         = "Mar",
      year          = "2022",
      url           = "https://cds.cern.ch/record/2802918"
}

@book{Evans2009TheLH,
  title={The large hadron collider : a marvel of technology},
  author={Lyndon R. Evans},
  publisher={EPFL Press},
  edition={2nd},
  year={2018}
}

@article{Wong:2018frb,
    author = "Wong, Cheuk-Yin and Jiang, Hanpu and Yao, Nanxi and Wen, Liwen and Wang, Gang and Zhong Huang, Huan",
    title = "{Clustering properties of produced particles in high-energy $pp$ collisions}",
    eprint = "1801.00759",
    archivePrefix = "arXiv",
    primaryClass = "hep-ph",
    doi = "10.1103/PhysRevD.102.054007",
    journal = "Phys. Rev. D",
    volume = "102",
    number = "5",
    pages = "054007",
    year = "2020"
}

@misc{githubGitHubGmlejarzaQuantumjetclustering,
	author = {Jorge J. Martinez de Lejarza},
	title = {{G}it{H}ub - gmlejarza/quantum-jet-clustering},
	howpublished = {\url{https://github.com/gmlejarza/Quantum-jet-clustering}},
	year = {},
	note = {[Accessed 31-03-2025]},
}

@misc{githubGitHubCERNITINNOVATIONQChPM,
	author = {Jorge J. Martinez de Lejarza},
	title = {{G}it{H}ub - {C}{E}{R}{N}-{I}{T}-{I}{N}{N}{O}{V}{A}{T}{I}{O}{N}/{Q}{C}h{P}{M}: {Q}uantum {C}hebyshev {P}robabilistic {M}odels for {F}ragmentation {F}unctions --- github.com},
	howpublished = {\url{https://github.com/CERN-IT-INNOVATION/QChPM}},
	year = {2025},
	note = {[Accessed 12-08-2025]},
}

@article{Salam:2010nqg,
    author = "Salam, Gavin P.",
    title = "{Towards Jetography}",
    eprint = "0906.1833",
    archivePrefix = "arXiv",
    primaryClass = "hep-ph",
    doi = "10.1140/epjc/s10052-010-1314-6",
    journal = "Eur. Phys. J. C",
    volume = "67",
    pages = "637--686",
    year = "2010"
}

@article{ATLAS:2013bqs,
    author = "Aad, Georges and others",
    collaboration = "ATLAS",
    title = "{Performance of jet substructure techniques for large-$R$ jets in proton-proton collisions at $\sqrt{s}$ = 7 TeV using the ATLAS detector}",
    eprint = "1306.4945",
    archivePrefix = "arXiv",
    primaryClass = "hep-ex",
    reportNumber = "CERN-PH-EP-2013-069",
    doi = "10.1007/JHEP09(2013)076",
    journal = "JHEP",
    volume = "09",
    pages = "076",
    year = "2013"
}

@article{Gras:2017jty,
    author = {Gras, Philippe and H\"oche, Stefan and Kar, Deepak and Larkoski, Andrew and L\"onnblad, Leif and Pl\"atzer, Simon and Si\'odmok, Andrzej and Skands, Peter and Soyez, Gregory and Thaler, Jesse},
    title = "{Systematics of quark/gluon tagging}",
    eprint = "1704.03878",
    archivePrefix = "arXiv",
    primaryClass = "hep-ph",
    reportNumber = "MIT-CTP-4885, COEPP-MN-17-2, MCNET-17-04",
    doi = "10.1007/JHEP07(2017)091",
    journal = "JHEP",
    volume = "07",
    pages = "091",
    year = "2017"
}

@misc{cernEventDisplayStandAlonelt,
	author = {ATLAS Collaboration},
	title = { {E}vent{D}isplay{S}tand{A}lone \&lt; {A}tlas{P}ublic \&lt; {T}{W}iki --- twiki.cern.ch},
	howpublished = {\url{https://twiki.cern.ch/twiki/bin/view/AtlasPublic/EventDisplayStandAlone}},
	year = {},
	note = {[Accessed 01-04-2025]},
}

@inproceedings{Kar:2015nxu,
    author = "Kar, Deepak",
    title = "{Jet substructure: a discovery tool}",
    booktitle = "{60th Annual Conference of the South African Institute of Physics}",
    pages = "175--179",
    year = "2015"
}

@book{banfi,
author = {Banfi, Andrea},
title = {Hadronic Jets (Second Edition)},
publisher = {IOP Publishing},
year = {2022},
series = {2053-2563},
isbn = {978-0-7503-4737-2},
url = {https://dx.doi.org/10.1088/978-0-7503-4737-2},
doi = {10.1088/978-0-7503-4737-2}
}

@misc{qiskit2024,
      title={Quantum computing with {Q}iskit},
      author={Javadi-Abhari, Ali and Treinish, Matthew and Krsulich, Kevin and Wood, Christopher J. and Lishman, Jake and Gacon, Julien and Martiel, Simon and Nation, Paul D. and Bishop, Lev S. and Cross, Andrew W. and Johnson, Blake R. and Gambetta, Jay M.},
      year={2024},
      doi={10.48550/arXiv.2405.08810},
      eprint={2405.08810},
      archivePrefix={arXiv},
      primaryClass={quant-ph}
}

@article{applot,
author = {Langerak, Thomas and Berendsen, Floris and Heide, Uulke and Kotte, Alexis and Pluim, Josien},
year = {2013},
month = {08},
pages = {091701},
title = {Multiatlas‐based segmentation with preregistration atlas selection},
volume = {40},
journal = {Medical Physics},
doi = {10.1118/1.4816654}
}

@article{Becchetti:tab,
    author = "Becchetti, Matteo and Bonciani, Roberto and Cieri, Leandro and Coro, Federico and Ripani, Federico",
    title = "{Two-loop form factors for diphoton production in quark annihilation channel with heavy quark mass dependence, in preparation}",
    reportNumber = "FTUV-23-0808.6381"
}

@article{Becchetti:2020wof,
    author = "Becchetti, Matteo and Bonciani, Roberto and Del Duca, Vittorio and Hirschi, Valentin and Moriello, Francesco and Schweitzer, Armin",
    title = "{Next-to-leading order corrections to light-quark mixed QCD-EW contributions to Higgs boson production}",
    eprint = "2010.09451",
    archivePrefix = "arXiv",
    primaryClass = "hep-ph",
    doi = "10.1103/PhysRevD.103.054037",
    journal = "Phys. Rev. D",
    volume = "103",
    number = "5",
    pages = "054037",
    year = "2021"
}

@article{Bonciani:2021zzf,
    author = "Bonciani, Roberto and Buonocore, Luca and Grazzini, Massimiliano and Kallweit, Stefan and Rana, Narayan and Tramontano, Francesco and Vicini, Alessandro",
    title = "{Mixed Strong-Electroweak Corrections to the Drell-Yan Process}",
    eprint = "2106.11953",
    archivePrefix = "arXiv",
    primaryClass = "hep-ph",
    doi = "10.1103/PhysRevLett.128.012002",
    journal = "Phys. Rev. Lett.",
    volume = "128",
    number = "1",
    pages = "012002",
    year = "2022"
}

@article{Armadillo:2022bgm,
    author = "Armadillo, Tommaso and Bonciani, Roberto and Devoto, Simone and Rana, Narayan and Vicini, Alessandro",
    title = "{Two-loop mixed QCD-EW corrections to neutral current Drell-Yan}",
    eprint = "2201.01754",
    archivePrefix = "arXiv",
    primaryClass = "hep-ph",
    reportNumber = "TIF-UNIMI-2022-1",
    doi = "10.1007/JHEP05(2022)072",
    journal = "JHEP",
    volume = "05",
    pages = "072",
    year = "2022"
}

@article{Lee:2017qql,
    author = "Lee, Roman N. and Smirnov, Alexander V. and Smirnov, Vladimir A.",
    title = "{Solving differential equations for Feynman integrals by expansions near singular points}",
    eprint = "1709.07525",
    archivePrefix = "arXiv",
    primaryClass = "hep-ph",
    reportNumber = "TTP16-055, MITP-16-144",
    doi = "10.1007/JHEP03(2018)008",
    journal = "JHEP",
    volume = "03",
    pages = "008",
    year = "2018"
}

@article{Mandal:2018cdj,
    author = "Mandal, Manoj K. and Zhao, Xiaoran",
    title = "{Evaluating multi-loop Feynman integrals numerically through differential equations}",
    eprint = "1812.03060",
    archivePrefix = "arXiv",
    primaryClass = "hep-ph",
    reportNumber = "CP3-18-71, MCNET-18-32",
    doi = "10.1007/JHEP03(2019)190",
    journal = "JHEP",
    volume = "03",
    pages = "190",
    year = "2019"
}

@article{Moriello:2019yhu,
    author = "Moriello, Francesco",
    title = "{Generalised power series expansions for the elliptic planar families of Higgs + jet production at two loops}",
    eprint = "1907.13234",
    archivePrefix = "arXiv",
    primaryClass = "hep-ph",
    doi = "10.1007/JHEP01(2020)150",
    journal = "JHEP",
    volume = "01",
    pages = "150",
    year = "2020"
}

@article{Chetyrkin:1981qh,
    author = "Chetyrkin, K. G. and Tkachov, F. V.",
    title = "{Integration by Parts: The Algorithm to Calculate beta Functions in 4 Loops}",
    doi = "10.1016/0550-3213(81)90199-1",
    journal = "Nucl. Phys. B",
    volume = "192",
    pages = "159--204",
    year = "1981"
}

@article{Kotikov:1990kg,
    author = "Kotikov, A. V.",
    title = "{Differential equations method: New technique for massive Feynman diagrams calculation}",
    reportNumber = "ITF-90-31E",
    doi = "10.1016/0370-2693(91)90413-K",
    journal = "Phys. Lett. B",
    volume = "254",
    pages = "158--164",
    year = "1991"
}

@article{Remiddi:1997ny,
    author = "Remiddi, Ettore",
    title = "{Differential equations for Feynman graph amplitudes}",
    eprint = "hep-th/9711188",
    archivePrefix = "arXiv",
    reportNumber = "DFUB-97-15, DFUB 97-15",
    doi = "10.1007/BF03185566",
    journal = "Nuovo Cim. A",
    volume = "110",
    pages = "1435--1452",
    year = "1997"
}

@article{Goncharov:2001iea,
    author = "Goncharov, A. B.",
    title = "{Multiple polylogarithms and mixed Tate motives}",
    eprint = "math/0103059",
    archivePrefix = "arXiv",
    month = "3",
    year = "2001"
}

@article{Adams:2017tga,
    author = "Adams, Luise and Chaubey, Ekta and Weinzierl, Stefan",
    title = "{Simplifying Differential Equations for Multiscale Feynman Integrals beyond Multiple Polylogarithms}",
    eprint = "1702.04279",
    archivePrefix = "arXiv",
    primaryClass = "hep-ph",
    doi = "10.1103/PhysRevLett.118.141602",
    journal = "Phys. Rev. Lett.",
    volume = "118",
    number = "14",
    pages = "141602",
    year = "2017"
}

@article{Broedel:2014vla,
    author = "Broedel, Johannes and Mafra, Carlos R. and Matthes, Nils and Schlotterer, Oliver",
    title = "{Elliptic multiple zeta values and one-loop superstring amplitudes}",
    eprint = "1412.5535",
    archivePrefix = "arXiv",
    primaryClass = "hep-th",
    reportNumber = "AEI-2014-066, DAMTP-2014-95",
    doi = "10.1007/JHEP07(2015)112",
    journal = "JHEP",
    volume = "07",
    pages = "112",
    year = "2015"
}

@article{Broedel:2017kkb,
    author = "Broedel, Johannes and Duhr, Claude and Dulat, Falko and Tancredi, Lorenzo",
    title = "{Elliptic polylogarithms and iterated integrals on elliptic curves. Part I: general formalism}",
    eprint = "1712.07089",
    archivePrefix = "arXiv",
    primaryClass = "hep-th",
    reportNumber = "CERN-TH-2017-273, CP3-17-57, HU-EP-17-29, HU-Mathematik-2017-09, SLAC-PUB-17194",
    doi = "10.1007/JHEP05(2018)093",
    journal = "JHEP",
    volume = "05",
    pages = "093",
    year = "2018"
}

@article{Adams:2014vja,
    author = "Adams, Luise and Bogner, Christian and Weinzierl, Stefan",
    title = "{The two-loop sunrise graph in two space-time dimensions with arbitrary masses in terms of elliptic dilogarithms}",
    eprint = "1405.5640",
    archivePrefix = "arXiv",
    primaryClass = "hep-ph",
    doi = "10.1063/1.4896563",
    journal = "J. Math. Phys.",
    volume = "55",
    number = "10",
    pages = "102301",
    year = "2014"
}

@article{Frellesvig:2023iwr,
    author = "Frellesvig, Hjalte and Weinzierl, Stefan",
    title = "{On $\varepsilon$-factorised bases and pure Feynman integrals}",
    eprint = "2301.02264",
    archivePrefix = "arXiv",
    primaryClass = "hep-th",
    reportNumber = "MITP/23-001",
    doi = "10.21468/SciPostPhys.16.6.150",
    journal = "SciPost Phys.",
    volume = "16",
    number = "6",
    pages = "150",
    year = "2024"
}

@article{Gorges:2023zgv,
    author = {G{\"o}rges, Lennard and Nega, Christoph and Tancredi, Lorenzo and Wagner, Fabian J.},
    title = "{On a procedure to derive {\ensuremath{\epsilon}}-factorised differential equations beyond polylogarithms}",
    eprint = "2305.14090",
    archivePrefix = "arXiv",
    primaryClass = "hep-th",
    doi = "10.1007/JHEP07(2023)206",
    journal = "JHEP",
    volume = "07",
    pages = "206",
    year = "2023"
}

@article{Duhr:2022pch,
    author = "Duhr, Claude and Klemm, Albrecht and Loebbert, Florian and Nega, Christoph and Porkert, Franziska",
    title = "{Yangian-Invariant Fishnet Integrals in Two Dimensions as Volumes of Calabi-Yau Varieties}",
    eprint = "2209.05291",
    archivePrefix = "arXiv",
    primaryClass = "hep-th",
    reportNumber = "BONN-TH-2022-19",
    doi = "10.1103/PhysRevLett.130.041602",
    journal = "Phys. Rev. Lett.",
    volume = "130",
    number = "4",
    pages = "041602",
    year = "2023"
}

@article{Duhr:2022dxb,
    author = "Duhr, Claude and Klemm, Albrecht and Nega, Christoph and Tancredi, Lorenzo",
    title = "{The ice cone family and iterated integrals for Calabi-Yau varieties}",
    eprint = "2212.09550",
    archivePrefix = "arXiv",
    primaryClass = "hep-th",
    reportNumber = "BONN-TH-2022-24, TUM-HEP-1444/22",
    doi = "10.1007/JHEP02(2023)228",
    journal = "JHEP",
    volume = "02",
    pages = "228",
    year = "2023"
}

@inproceedings{Bourjaily:2022bwx,
    author = "Bourjaily, Jacob L. and others",
    title = "{Functions Beyond Multiple Polylogarithms for Precision Collider Physics}",
    booktitle = "{Snowmass 2021}",
    eprint = "2203.07088",
    archivePrefix = "arXiv",
    primaryClass = "hep-ph",
    reportNumber = "BONN-TH-2022-05, UUITP-11/22, CERN-TH-2022-029, TUM-HEP-1391/22,
  HU-EP-22/08, MITP-22-022",
    month = "3",
    year = "2022"
}

@article{Chawdhry:2019bji,
    author = "Chawdhry, Herschel A. and Czakon, Micha L. and Mitov, Alexander and Poncelet, Rene",
    title = "{NNLO QCD corrections to three-photon production at the LHC}",
    eprint = "1911.00479",
    archivePrefix = "arXiv",
    primaryClass = "hep-ph",
    reportNumber = "Cavendish-HEP-19/17, TTK-19-45, P3H-19-041",
    doi = "10.1007/JHEP02(2020)057",
    journal = "JHEP",
    volume = "02",
    pages = "057",
    year = "2020"
}

@article{Czakon:2021mjy,
    author = "Czakon, Michal and Mitov, Alexander and Poncelet, Rene",
    title = "{Next-to-Next-to-Leading Order Study of Three-Jet Production at the LHC}",
    eprint = "2106.05331",
    archivePrefix = "arXiv",
    primaryClass = "hep-ph",
    reportNumber = "Cavendish-HEP-21/09, P3H-21-043, TTK-21-20",
    doi = "10.1103/PhysRevLett.127.152001",
    journal = "Phys. Rev. Lett.",
    volume = "127",
    number = "15",
    pages = "152001",
    year = "2021",
    note = "[Erratum: Phys.Rev.Lett. 129, 119901 (2022), Erratum: Phys.Rev.Lett. 129, 119901 (2022)]"
}

@article{Abreu:2022vei,
    author = "Abreu, Samuel and Becchetti, Matteo and Duhr, Claude and Ozcelik, Melih A.",
    title = "{Two-loop master integrals for pseudo-scalar quarkonium and leptonium production and decay}",
    eprint = "2206.03848",
    archivePrefix = "arXiv",
    primaryClass = "hep-ph",
    reportNumber = "BONN-TH-2022-14, CERN-TH-2022-093, TTP22-037",
    doi = "10.1007/JHEP09(2022)194",
    journal = "JHEP",
    volume = "09",
    pages = "194",
    year = "2022"
}

@article{Abreu:2022cco,
    author = "Abreu, Samuel and Becchetti, Matteo and Duhr, Claude and Ozcelik, Melih A.",
    title = "{Two-loop form factors for pseudo-scalar quarkonium production and decay}",
    eprint = "2211.08838",
    archivePrefix = "arXiv",
    primaryClass = "hep-ph",
    reportNumber = "BONN-TH-2022-23, CERN-TH-2022-187, TTP22-068",
    doi = "10.1007/JHEP02(2023)250",
    journal = "JHEP",
    volume = "02",
    pages = "250",
    year = "2023"
}

@article{Badger:2023mgf,
    author = "Badger, Simon and Czakon, Micha{\l} and Hartanto, Heribertus Bayu and Moodie, Ryan and Peraro, Tiziano and Poncelet, Rene and Zoia, Simone",
    title = "{Isolated photon production in association with a jet pair through next-to-next-to-leading order in QCD}",
    eprint = "2304.06682",
    archivePrefix = "arXiv",
    primaryClass = "hep-ph",
    reportNumber = "Cavendish-HEP-23/02, P3H-23-022, TTK-23-09",
    doi = "10.1007/JHEP10(2023)071",
    journal = "JHEP",
    volume = "10",
    pages = "071",
    year = "2023"
}

@article{Chetyrkin:1979bj,
    author = "Chetyrkin, K. G. and Kataev, A. L. and Tkachov, F. V.",
    title = "{Higher Order Corrections to Sigma-t (e+ e- ---\ensuremath{>} Hadrons) in Quantum Chromodynamics}",
    reportNumber = "IYaI-P-0126",
    doi = "10.1016/0370-2693(79)90596-3",
    journal = "Phys. Lett. B",
    volume = "85",
    pages = "277--279",
    year = "1979"
}

@inproceedings{Brassard,
	doi = {10.1109/istcs.1997.595153},
  
	url = {https://doi.org/10.1109%2Fistcs.1997.595153},
  
	publisher = {{IEEE} Comput. Soc},
  
	author = {G. Brassard and P. Hoyer},
  
	title = {An exact quantum polynomial-time algorithm for Simon{\textquotesingle}s problem},
  
	booktitle = {Proceedings of the Fifth Israeli Symposium on Theory of Computing and Systems}
}

@misc{Brassard_2002,
	doi = {10.1090/conm/305/05215},
  
	url = {https://doi.org/10.1090%2Fconm%2F305%2F05215},
  
	year = 2002,
	publisher = {American Mathematical Society},
  
	pages = {53--74},
  
	author = {Gilles Brassard and Peter H{\o}yer and Michele Mosca and Alain Tapp},
  
	title = {Quantum amplitude amplification and estimation}
}

@article{PhysRevLett.80.4329,
  title = {Quantum Computers Can Search Rapidly by Using Almost Any Transformation},
  author = {Grover, Lov K.},
  journal = {Phys. Rev. Lett.},
  volume = {80},
  issue = {19},
  pages = {4329--4332},
  numpages = {0},
  year = {1998},
  month = {May},
  publisher = {American Physical Society},
  doi = {10.1103/PhysRevLett.80.4329},
  url = {https://link.aps.org/doi/10.1103/PhysRevLett.80.4329}
}

@inproceedings{10.1145/237814.237866,
author = {Grover, Lov K.},
title = {A Fast Quantum Mechanical Algorithm for Database Search},
year = {1996},
isbn = {0897917855},
publisher = {Association for Computing Machinery},
address = {New York, NY, USA},
url = {https://doi.org/10.1145/237814.237866},
doi = {10.1145/237814.237866},
booktitle = {Proceedings of the Twenty-Eighth Annual ACM Symposium on Theory of Computing},
pages = {212–219},
numpages = {8},
location = {Philadelphia, Pennsylvania, USA},
series = {STOC '96}
}

@article{Grinko_2021,
	doi = {10.1038/s41534-021-00379-1},
  
	url = {https://doi.org/10.10382Fs41534-021-00379-1},
  
	year = 2021,
	month = {mar},
  
	publisher = {Springer Science and Business Media {LLC}
},
  
	volume = {7},
  
	number = {1},
  
	author = {Dmitry Grinko and Julien Gacon and Christa Zoufal and Stefan Woerner},
  
	title = {Iterative quantum amplitude estimation},
  
	journal = {npj Quantum Information}
}

@article{Herbert_2022,
	doi = {10.22331/q-2022-09-29-823},
  
	url = {https://doi.org/10.223312Fq-2022-09-29-823},
  
	year = 2022,
	month = {sep},
  
	publisher = {Verein zur Forderung des Open Access Publizierens in den Quantenwissenschaften},
  
	volume = {6},
  
	pages = {823},
  
	author = {Steven Herbert},
  
	title = {Quantum Monte Carlo Integration: The Full Advantage in Minimal Circuit Depth},
  
	journal = {Quantum}
}

@article{Aguilera-Verdugo:2020set,
    author = "Aguilera-Verdugo, J. Jesus and Driencourt-Mangin, Felix and Hern\'andez-Pinto, Roger J. and Plenter, Judith and Ramirez-Uribe, Selomit and Renteria Olivo, Andres E. and Rodrigo, German and Sborlini, German F. R. and Torres Bobadilla, William J. and Tracz, Szymon",
    title = "{Open Loop Amplitudes and Causality to All Orders and Powers from the Loop-Tree Duality}",
    eprint = "2001.03564",
    archivePrefix = "arXiv",
    primaryClass = "hep-ph",
    reportNumber = "IFIC/20-02",
    doi = "10.1103/PhysRevLett.124.211602",
    journal = "Phys. Rev. Lett.",
    volume = "124",
    number = "21",
    pages = "211602",
    year = "2020"
}

@book{9781107002173,
  Author = {Michael A. Nielsen and Isaac L. Chuang},
  Title = {Quantum Computation and Quantum Information: 10th Anniversary Edition},
  Publisher = {Cambridge University Press},
  Year = {2011},
  ISBN = {9781107002173},
  URL = {https://www.amazon.com/Quantum-Computation-Information-10th-Anniversary/dp/1107002176?SubscriptionId=AKIAIOBINVZYXZQZ2U3A&tag=chimbori05-20&linkCode=xm2&camp=2025&creative=165953&creativeASIN=1107002176},
}

@inproceedings{Kitaev2002ClassicalAQ,
  title={Classical and Quantum Computation},
  author={Alexei Y. Kitaev and A. H. Shen and Mikhail N. Vyalyi},
  booktitle={Graduate Studies in Mathematics},
  year={2002}
}

@article{wie2019simpler,
  title={Simpler quantum counting},
  author={Chu-Ryang Wie},
  journal={Quantum Inf. Comput.},
  year={2019},
  volume={19},
  pages={967-983},
  url={https://api.semanticscholar.org/CorpusID:197545059}
}

@article{Suzuki2020,
	doi = {10.1007/s11128-019-2565-2},
  
	url = {https://doi.org/10.10072Fs11128-019-2565-2},
  
	year = 2020,
	month = {jan},
  
	publisher = {Springer Science and Business Media {LLC}
},
  
	volume = {19},
  
	number = {2},
  
	author = {Yohichi Suzuki and Shumpei Uno and Rudy Raymond and Tomoki Tanaka and Tamiya Onodera and Naoki Yamamoto},
  
	title = {Amplitude estimation without phase estimation},
  
	journal = {Quantum Information Processing}
}

@article{AGLIARDI2022137228,
title = {Quantum integration of elementary particle processes},
journal = {Physics Letters B},
volume = {832},
pages = {137228},
year = {2022},
issn = {0370-2693},
doi = {https://doi.org/10.1016/j.physletb.2022.137228},
url = {https://www.sciencedirect.com/science/article/pii/S0370269322003628},
author = {Gabriele Agliardi and Michele Grossi and Mathieu Pellen and Enrico Prati},
}

@inbook{aaronson,
author = {Scott Aaronson and Patrick Rall},
title = {Quantum Approximate Counting, Simplified},
booktitle = {Symposium on Simplicity in Algorithms (SOSA)},
chapter = {},
pages = {24-32},
doi = {10.1137/1.9781611976014.5},
URL = {https://epubs.siam.org/doi/abs/10.1137/1.9781611976014.5},
eprint = {https://epubs.siam.org/doi/pdf/10.1137/1.9781611976014.5},
}

@article{Montanaro2015,
	doi = {10.1098/rspa.2015.0301},
  
	url = {https://doi.org/10.10982Frspa.2015.0301},
  
	year = 2015,
	month = {sep},
  
	publisher = {The Royal Society},
  
	volume = {471},
  
	number = {2181},
  
	author = {Ashley Montanaro},
  
	title = {Quantum speedup of Monte Carlo methods},
  
	journal = {Proceedings of the Royal Society A: Mathematical, Physical and Engineering Sciences}
}

@inproceedings{Pooja,
author = {Pooja Rao and Kwangmin Yu and Hyunkyung Lim and Dasol Jin and Deokkyu Choi},
title = {{Quantum amplitude estimation algorithms on IBM quantum devices}},
volume = {11507},
booktitle = {Quantum Communications and Quantum Imaging XVIII},
editor = {Keith S. Deacon},
organization = {International Society for Optics and Photonics},
publisher = {SPIE},
pages = {115070O},
year = {2020},
doi = {10.1117/12.2568748},
URL = {https://doi.org/10.1117/12.2568748}
}

@article{Aguilera-Verdugo:2020kzc,
    author = "Aguilera-Verdugo, J. Jesus and Hernandez-Pinto, Roger J. and Rodrigo, German and Sborlini, German F. R. and Torres Bobadilla, William J.",
    title = "{Causal representation of multi-loop Feynman integrands within the loop-tree duality}",
    eprint = "2006.11217",
    archivePrefix = "arXiv",
    primaryClass = "hep-ph",
    reportNumber = "IFIC/20-27",
    doi = "10.1007/JHEP01(2021)069",
    journal = "JHEP",
    volume = "01",
    pages = "069",
    year = "2021"
}

@article{Bobadilla:2021pvr,
    author = "Bobadilla, William J. Torres",
    title = "{Lotty \textendash{} The loop-tree duality automation}",
    eprint = "2103.09237",
    archivePrefix = "arXiv",
    primaryClass = "hep-ph",
    reportNumber = "MPP-2021-11",
    doi = "10.1140/epjc/s10052-021-09235-0",
    journal = "Eur. Phys. J. C",
    volume = "81",
    number = "6",
    pages = "514",
    year = "2021"
}

@article{Melo2023pulseefficient,
  doi = {10.22331/q-2023-10-09-1130},
  url = {https://doi.org/10.22331/q-2023-10-09-1130},
  title = {Pulse-efficient quantum machine learning},
  author = {Melo, Andr{\'{e}} and Earnest-Noble, Nathan and Tacchino, Francesco},
  journal = {{Quantum}},
  issn = {2521-327X},
  publisher = {{Verein zur F{\"{o}}rderung des Open Access Publizierens in den Quantenwissenschaften}},
  volume = {7},
  pages = {1130},
  month = oct,
  year = {2023}
}

@article{Earnest_2021,
	doi = {10.1103/physrevresearch.3.043088},
  
	url = {https://doi.org/10.1103%2Fphysrevresearch.3.043088},
  
	year = 2021,
	month = {oct},
  
	publisher = {American Physical Society ({APS})},
  
	volume = {3},
  
	number = {4},
  
	author = {Nathan Earnest and Caroline Tornow and Daniel J. Egger},
  
	title = {Pulse-efficient circuit transpilation for quantum applications on cross-resonance-based hardware},
  
	journal = {Physical Review Research}
}

@article{Casas:2023ure,
	doi = {10.1103/physreva.107.062612},
  
	url = {https://doi.org/10.1103%2Fphysreva.107.062612},
  
	year = 2023,
	month = {jun},
  
	publisher = {American Physical Society ({APS})},
  
	volume = {107},
  
	number = {6},
  
	author = {Berta Casas and Alba Cervera-Lierta},
  
	title = {Multidimensional Fourier series with quantum circuits},
  
	journal = {Physical Review A}
}

@article{Atchade-Adelomou:2023mjf,
    author = "Atchade-Adelomou, Parfait and Larson, Kent",
    title = "{Fourier series weight in quantum machine learning}",
    eprint = "2302.00105",
    archivePrefix = "arXiv",
    primaryClass = "quant-ph",
    month = "1",
    year = "2023"
}

@article{deJesusAguilera-Verdugo:2021mvg,
    author = "de Jes\'us Aguilera-Verdugo, Jos\'e and others",
    title = "{A Stroll through the Loop-Tree Duality}",
    eprint = "2104.14621",
    archivePrefix = "arXiv",
    primaryClass = "hep-ph",
    reportNumber = "IFIC/21-11, DESY-21-056, MPP-2021-65",
    doi = "10.3390/sym13061029",
    journal = "Symmetry",
    volume = "13",
    number = "6",
    pages = "1029",
    year = "2021"
}

@article{Catani:2008xa,
    author = "Catani, Stefano and Gleisberg, Tanju and Krauss, Frank and Rodrigo, German and Winter, Jan-Christopher",
    title = "{From loops to trees by-passing Feynman's theorem}",
    eprint = "0804.3170",
    archivePrefix = "arXiv",
    primaryClass = "hep-ph",
    reportNumber = "FERMILAB-PUB-08-092-T, IFIC-08-21, IPPP-08-22, SLAC-PUB-13218",
    doi = "10.1088/1126-6708/2008/09/065",
    journal = "JHEP",
    volume = "09",
    pages = "065",
    year = "2008"
}

@article{Bierenbaum:2010cy,
    author = "Bierenbaum, Isabella and Catani, Stefano and Draggiotis, Petros and Rodrigo, German",
    title = "{A Tree-Loop Duality Relation at Two Loops and Beyond}",
    eprint = "1007.0194",
    archivePrefix = "arXiv",
    primaryClass = "hep-ph",
    reportNumber = "IFIC-10-17",
    doi = "10.1007/JHEP10(2010)073",
    journal = "JHEP",
    volume = "10",
    pages = "073",
    year = "2010"
}

@article{Bierenbaum:2012th,
    author = "Bierenbaum, Isabella and Buchta, Sebastian and Draggiotis, Petros and Malamos, Ioannis and Rodrigo, German",
    title = "{Tree-Loop Duality Relation beyond simple poles}",
    eprint = "1211.5048",
    archivePrefix = "arXiv",
    primaryClass = "hep-ph",
    reportNumber = "LPN-12-057, IFIC-12-44, DESY-12-208",
    doi = "10.1007/JHEP03(2013)025",
    journal = "JHEP",
    volume = "03",
    pages = "025",
    year = "2013"
}

@article{Buchta:2015wna,
    author = "Buchta, Sebastian and Chachamis, Grigorios and Draggiotis, Petros and Rodrigo, German",
    title = "{Numerical implementation of the loop\textendash{}tree duality method}",
    eprint = "1510.00187",
    archivePrefix = "arXiv",
    primaryClass = "hep-ph",
    reportNumber = "IFIC-15-69",
    doi = "10.1140/epjc/s10052-017-4833-6",
    journal = "Eur. Phys. J. C",
    volume = "77",
    number = "5",
    pages = "274",
    year = "2017"
}

@article{Driencourt-Mangin:2017gop,
    author = "Driencourt-Mangin, F\'elix and Rodrigo, Germ\'an and Sborlini, Germ\'an F. R.",
    title = "{Universal dual amplitudes and asymptotic expansions for $gg\rightarrow H$ and $H\rightarrow \gamma \gamma $ in four dimensions}",
    eprint = "1702.07581",
    archivePrefix = "arXiv",
    primaryClass = "hep-ph",
    reportNumber = "IFIC-17-07, TIF-UNIMI-2017-1",
    doi = "10.1140/epjc/s10052-018-5692-5",
    journal = "Eur. Phys. J. C",
    volume = "78",
    number = "3",
    pages = "231",
    year = "2018"
}

@article{Plenter:2020lop,
    author = "Plenter, Judith and Rodrigo, Germ\'an",
    title = "{Asymptotic expansions through the loop-tree duality}",
    eprint = "2005.02119",
    archivePrefix = "arXiv",
    primaryClass = "hep-ph",
    reportNumber = "IFIC/20-18",
    doi = "10.1140/epjc/s10052-021-09094-9",
    journal = "Eur. Phys. J. C",
    volume = "81",
    number = "4",
    pages = "320",
    year = "2021"
}

@article{Driencourt-Mangin:2019aix,
    author = "Driencourt-Mangin, Felix and Rodrigo, Germ\'an and Sborlini, Germ\'an F. R. and Torres Bobadilla, William J.",
    title = "{Universal four-dimensional representation of $H \to \gamma \gamma$ at two loops through the Loop-Tree Duality}",
    eprint = "1901.09853",
    archivePrefix = "arXiv",
    primaryClass = "hep-ph",
    reportNumber = "IFIC/18-31, TIF-UNIMI-2018-6",
    doi = "10.1007/JHEP02(2019)143",
    journal = "JHEP",
    volume = "02",
    pages = "143",
    year = "2019"
}

@article{Hernandez-Pinto:2015ysa,
    author = "Hernandez-Pinto, Roger J. and Sborlini, German F. R. and Rodrigo, German",
    title = "{Towards gauge theories in four dimensions}",
    eprint = "1506.04617",
    archivePrefix = "arXiv",
    primaryClass = "hep-ph",
    reportNumber = "IFIC-15-35, LPN15-025",
    doi = "10.1007/JHEP02(2016)044",
    journal = "JHEP",
    volume = "02",
    pages = "044",
    year = "2016"
}

@article{Sborlini:2016gbr,
    author = "Sborlini, German F. R. and Driencourt-Mangin, Felix and Hernandez-Pinto, Roger and Rodrigo, German",
    title = "{Four-dimensional unsubtraction from the loop-tree duality}",
    eprint = "1604.06699",
    archivePrefix = "arXiv",
    primaryClass = "hep-ph",
    reportNumber = "IFIC-15-73",
    doi = "10.1007/JHEP08(2016)160",
    journal = "JHEP",
    volume = "08",
    pages = "160",
    year = "2016"
}

@article{Prisco:2020kyb,
    author = "Prisco, Renato Maria and Tramontano, Francesco",
    title = "{Dual subtractions}",
    eprint = "2012.05012",
    archivePrefix = "arXiv",
    primaryClass = "hep-ph",
    doi = "10.1007/JHEP06(2021)089",
    journal = "JHEP",
    volume = "06",
    pages = "089",
    year = "2021"
}

@article{Heinrich:2020ybq,
    author = "Heinrich, Gudrun",
    title = "{Collider Physics at the Precision Frontier}",
    eprint = "2009.00516",
    archivePrefix = "arXiv",
    primaryClass = "hep-ph",
    reportNumber = "KA-TP-13-2020, P3H-20-044",
    doi = "10.1016/j.physrep.2021.03.006",
    journal = "Phys. Rept.",
    volume = "922",
    pages = "1--69",
    year = "2021"
}

@article{shor,
author = {Shor, Peter W.},
title = {Polynomial-Time Algorithms for Prime Factorization and Discrete Logarithms on a Quantum Computer},
journal = {SIAM Journal on Computing},
volume = {26},
number = {5},
pages = {1484-1509},
year = {1997},
doi = {10.1137/S0097539795293172},

URL = { 
    
        https://doi.org/10.1137/S0097539795293172
    
    

},
eprint = { 
    
        https://doi.org/10.1137/S0097539795293172
    
    

}}

@article{Bepari:2021kwv,
    author = "Bepari, Khadeejah and Malik, Sarah and Spannowsky, Michael and Williams, Simon",
    title = "{Quantum walk approach to simulating parton showers}",
    eprint = "2109.13975",
    archivePrefix = "arXiv",
    primaryClass = "hep-ph",
    doi = "10.1103/PhysRevD.106.056002",
    journal = "Phys. Rev. D",
    volume = "106",
    number = "5",
    pages = "056002",
    year = "2022"
}

@article{Schuhmacher:2023pro,
    author = "Schuhmacher, Julian and Boggia, Laura and Belis, Vasilis and Puljak, Ema and Grossi, Michele and Pierini, Maurizio and Vallecorsa, Sofia and Tacchino, Francesco and Barkoutsos, Panagiotis and Tavernelli, Ivano",
    title = "{Unravelling physics beyond the standard model with classical and quantum anomaly detection}",
    eprint = "2301.10787",
    archivePrefix = "arXiv",
    primaryClass = "hep-ex",
    doi = "10.1088/2632-2153/ad07f7",
    journal = "Mach. Learn. Sci. Tech.",
    volume = "4",
    number = "4",
    pages = "045031",
    year = "2023"
}

@article{Intallura:2023yvu,
    author = "Intallura, Philip and Korpas, Georgios and Chakraborty, Sudeepto and Kungurtsev, Vyacheslav and Marecek, Jakub",
    title = "{A Survey of Quantum Alternatives to Randomized Algorithms: Monte Carlo Integration and Beyond}",
    eprint = "2303.04945",
    archivePrefix = "arXiv",
    primaryClass = "quant-ph",
    month = "3",
    year = "2023"
}

@INPROCEEDINGS{deLejarza:2023IEEE,
  author={de Lejarza, Jorge J. Martínez and Grossi, Michele and Cieri, Leandro and Rodrigo, Germán},
  booktitle={2023 IEEE International Conference on Quantum Computing and Engineering (QCE)}, 
  title={Quantum Fourier Iterative Amplitude Estimation}, 
  year={2023},
  volume={01},
  number={},
  pages={571-579},
  doi={10.1109/QCE57702.2023.00071}}

@article{Plekhanov:2021kir,
    author = "Plekhanov, Kirill and Rosenkranz, Matthias and Fiorentini, Mattia and Lubasch, Michael",
    title = "{Variational quantum amplitude estimation}",
    eprint = "2109.03687",
    archivePrefix = "arXiv",
    primaryClass = "quant-ph",
    doi = "10.22331/q-2022-03-17-670",
    journal = "Quantum",
    volume = "6",
    pages = "670",
    year = "2022"
}

@article{Ghosh:2023qze,
    author = "Ghosh, Kumar J. B. and Yogaraj, Kavitha and Agliardi, Gabriele and Sabino, Piergiacomo and Fern{\'a}ndez-Campoamor, Marina and Bernab{\'e}-Moreno, Juan and Cortiana, Giorgio and Shehab, Omar and O'Meara, Corey",
    title = "{Energy Risk Analysis With Dynamic Amplitude Estimation and Piecewise Approximate Quantum Compiling}",
    eprint = "2305.09501",
    archivePrefix = "arXiv",
    primaryClass = "quant-ph",
    doi = "10.1109/TQE.2024.3425969",
    journal = "IEEE Trans. Quantum Eng.",
    volume = "5",
    pages = "1--17",
    year = "2024"
}

@article{carrazza,
  title = {Determining the proton content with a quantum computer},
  author = {P\'erez-Salinas, Adri\'an and Cruz-Martinez, Juan and Alhajri, Abdulla A. and Carrazza, Stefano},
  journal = {Phys. Rev. D},
  volume = {103},
  issue = {3},
  pages = {034027},
  numpages = {14},
  year = {2021},
  month = {Feb},
  publisher = {American Physical Society},
  doi = {10.1103/PhysRevD.103.034027},
  url = {https://link.aps.org/doi/10.1103/PhysRevD.103.034027}
}

@article{lejarza,
  title = {Quantum clustering and jet reconstruction at the LHC},
  author = {Mart\'{\i}nez de Lejarza, Jorge J. and Cieri, Leandro and Rodrigo, Germ\'an},
  journal = {Phys. Rev. D},
  volume = {106},
  issue = {3},
  pages = {036021},
  numpages = {16},
  year = {2022},
  month = {Aug},
  publisher = {American Physical Society},
  doi = {10.1103/PhysRevD.106.036021},
  url = {https://link.aps.org/doi/10.1103/PhysRevD.106.036021}
}

@article{delgado_jets,
  title = {Quantum annealing for jet clustering with thrust},
  author = {Delgado, Andrea and Thaler, Jesse},
  journal = {Phys. Rev. D},
  volume = {106},
  issue = {9},
  pages = {094016},
  numpages = {15},
  year = {2022},
  month = {Nov},
  publisher = {American Physical Society},
  doi = {10.1103/PhysRevD.106.094016},
  url = {https://link.aps.org/doi/10.1103/PhysRevD.106.094016}
}

@article{thaler,
  title = {Quantum algorithms for jet clustering},
  author = {Wei, Annie Y. and Naik, Preksha and Harrow, Aram W. and Thaler, Jesse},
  journal = {Phys. Rev. D},
  volume = {101},
  issue = {9},
  pages = {094015},
  numpages = {20},
  year = {2020},
  month = {May},
  publisher = {American Physical Society},
  doi = {10.1103/PhysRevD.101.094015},
  url = {https://link.aps.org/doi/10.1103/PhysRevD.101.094015}
}

@article{preskill,
author = {Stephen P. Jordan  and Keith S. M. Lee  and John Preskill },
title = {Quantum Algorithms for Quantum Field Theories},
journal = {Science},
volume = {336},
number = {6085},
pages = {1130-1133},
year = {2012},
doi = {10.1126/science.1217069},
URL = {https://www.science.org/doi/abs/10.1126/science.1217069},
eprint = {https://www.science.org/doi/pdf/10.1126/science.1217069}
}

@article{Zohar_2016,
doi = {10.1088/0034-4885/79/1/014401},
url = {https://dx.doi.org/10.1088/0034-4885/79/1/014401},
year = {2015},
month = {dec},
publisher = {IOP Publishing},
volume = {79},
number = {1},
pages = {014401},
author = {Erez Zohar and J Ignacio Cirac and Benni Reznik},
title = {Quantum simulations of lattice gauge theories using ultracold atoms in optical lattices},
journal = {Reports on Progress in Physics}
}

@article{qannealing,
  title = {Quantum annealing in the transverse Ising model},
  author = {Kadowaki, Tadashi and Nishimori, Hidetoshi},
  journal = {Phys. Rev. E},
  volume = {58},
  issue = {5},
  pages = {5355--5363},
  numpages = {0},
  year = {1998},
  month = {Nov},
  publisher = {American Physical Society},
  doi = {10.1103/PhysRevE.58.5355},
  url = {https://link.aps.org/doi/10.1103/PhysRevE.58.5355}
}

@article{Schuld_2021,
	doi = {10.1103/physreva.103.032430},
  
	url = {https://doi.org/10.11032Fphysreva.103.032430},
  
	year = 2021,
	month = {mar},
  
	publisher = {American Physical Society ({APS})},
  
	volume = {103},
  
	number = {3},
  
	author = {Maria Schuld and Ryan Sweke and Johannes Jakob Meyer},
  
	title = {Effect of data encoding on the expressive power of variational quantum-machine-learning models},
  
	journal = {Physical Review A}
}

@article{pennylane,
    author = "Bergholm, Ville and others",
    title = "{PennyLane: Automatic differentiation of hybrid quantum-classical computations}",
    eprint = "1811.04968",
    archivePrefix = "arXiv",
    primaryClass = "quant-ph",
    month = "11",
    year = "2018"
}

@article{qibo_paper,
    doi       = {10.1088/2058-9565/ac39f5},
    url       = {https://doi.org/10.1088/2058-9565/ac39f5},
    year      = 2021,
    month     = {dec},
    publisher = {{IOP} Publishing},
    volume    = {7},
    number    = {1},
    pages     = {015018},
    author    = {Stavros Efthymiou and
                 Sergi Ramos-Calderer and
                 Carlos Bravo-Prieto and
                 Adri{\'{a}}n P{\'{e}}rez-Salinas and
                 Diego Garc{\'{\i}}a-Mart{\'{\i}}n and
                 Artur Garcia-Saez and
                 Jos{\'{e}} Ignacio Latorre and
                 Stefano Carrazza},
    title     = {Qibo: a framework for quantum simulation with hardware acceleration},
    journal   = {Quantum Science and Technology},
}

@misc{tutorial_qfiae,
  author       = {Jorge J. Martínez de Lejarza},
  title        = {{Tutorial: Quantum Fourier Iterative Amplitude 
                   Estimation}},
  month        = jun,
  year         = 2023,
  publisher    = {Zenodo},
  doi          = {10.5281/zenodo.8130980},
  url          = {https://doi.org/10.5281/zenodo.8130980}
}

@misc{estimator,
    title = {Qiskit Runtime Estimator primitive service},  
    author = {IBMQ},  
    year = {2023},
    howpublished={\url{https://qiskit.org/ecosystem/ibm-runtime/stubs/qiskit_ibm_runtime.Estimator.html}},
    note = {Accessed: 2023-11-3}

}

@article{Schenk:2022pgo,
    author = "Schenk, Michael and Combarro, El{\'\i}as F. and Grossi, Michele and Kain, Verena and Li, Kevin Shing Bruce and Popa, Mircea-Marian and Vallecorsa, Sofia",
    title = "{Hybrid actor-critic algorithm for quantum reinforcement learning at CERN beam lines}",
    eprint = "2209.11044",
    archivePrefix = "arXiv",
    primaryClass = "quant-ph",
    doi = "10.1088/2058-9565/ad261b",
    journal = "Quantum Sci. Technol.",
    volume = "9",
    number = "2",
    pages = "025012",
    year = "2024"
}

@inproceedings{Bermot:2023kvh,
    author = "Bermot, Elie and Zoufal, Christa and Grossi, Michele and Schuhmacher, Julian and Tacchino, Francesco and Vallecorsa, Sofia and Tavernelli, Ivano",
    title = "{Quantum Generative Adversarial Networks For Anomaly Detection In High Energy Physics}",
    booktitle = "{2023 International Conference on Quantum Computing and Engineering}",
    eprint = "2304.14439",
    archivePrefix = "arXiv",
    primaryClass = "quant-ph",
    doi = "10.1109/QCE57702.2023.00045",
    month = "4",
    year = "2023"
}

@article{Barata:2023clv,
    author = "Barata, Jo{\~a}o and Du, Xiaojian and Li, Meijian and Qian, Wenyang and Salgado, Carlos A.",
    title = "{Quantum simulation of in-medium QCD jets: Momentum broadening, gluon production, and entropy growth}",
    eprint = "2307.01792",
    archivePrefix = "arXiv",
    primaryClass = "hep-ph",
    doi = "10.1103/PhysRevD.108.056023",
    journal = "Phys. Rev. D",
    volume = "108",
    number = "5",
    pages = "056023",
    year = "2023"
}

@article{Barata:2022wim,
    author = "Barata, Jo\~ao and Du, Xiaojian and Li, Meijian and Qian, Wenyang and Salgado, Carlos A.",
    title = "{Medium induced jet broadening in a quantum computer}",
    eprint = "2208.06750",
    archivePrefix = "arXiv",
    primaryClass = "hep-ph",
    doi = "10.1103/PhysRevD.106.074013",
    journal = "Phys. Rev. D",
    volume = "106",
    number = "7",
    pages = "074013",
    year = "2022"
}

@article{Aguilera-Verdugo:2019kbz,
    author = "Aguilera-Verdugo, J. Jes\'us and Driencourt-Mangin, F\'elix and Plenter, Judith and Ram\'\i{}rez-Uribe, Selomit and Rodrigo, Germ\'an and Sborlini, Germ\'an F. R. and Torres Bobadilla, William J. and Tracz, Szymon",
    title = "{Causality, unitarity thresholds, anomalous thresholds and infrared singularities from the loop-tree duality at higher orders}",
    eprint = "1904.08389",
    archivePrefix = "arXiv",
    primaryClass = "hep-ph",
    reportNumber = "IFIC/19-22",
    doi = "10.1007/JHEP12(2019)163",
    journal = "JHEP",
    volume = "12",
    pages = "163",
    year = "2019"
}

@article{Buchta:2014dfa,
    author = "Buchta, Sebastian and Chachamis, Grigorios and Draggiotis, Petros and Malamos, Ioannis and Rodrigo, Germ\'an",
    title = "{On the singular behaviour of scattering amplitudes in quantum field theory}",
    eprint = "1405.7850",
    archivePrefix = "arXiv",
    primaryClass = "hep-ph",
    reportNumber = "LPN14-078, IFIC-13-85",
    doi = "10.1007/JHEP11(2014)014",
    journal = "JHEP",
    volume = "11",
    pages = "014",
    year = "2014"
}

@article{Ramirez-Uribe:2020hes,
    author = "Ram\'\i{}rez-Uribe, Selomit and Hern\'andez-Pinto, Roger J. and Rodrigo, German and Sborlini, Germ\'an F. R. and Torres Bobadilla, William J.",
    title = "{Universal opening of four-loop scattering amplitudes to trees}",
    eprint = "2006.13818",
    archivePrefix = "arXiv",
    primaryClass = "hep-ph",
    reportNumber = "IFIC/20-29",
    doi = "10.1007/JHEP04(2021)129",
    journal = "JHEP",
    volume = "04",
    pages = "129",
    year = "2021"
}

@article{JesusAguilera-Verdugo:2020fsn,
    author = "Jes\'us Aguilera-Verdugo, J. and Hern\'andez-Pinto, Roger J. and Rodrigo, Germ\'an and Sborlini, German F. R. and Torres Bobadilla, William J.",
    title = "{Mathematical properties of nested residues and their application to multi-loop scattering amplitudes}",
    eprint = "2010.12971",
    archivePrefix = "arXiv",
    primaryClass = "hep-ph",
    reportNumber = "IFIC/20-30; DESY 20-172; MPP-2020-184",
    doi = "10.1007/JHEP02(2021)112",
    journal = "JHEP",
    volume = "02",
    pages = "112",
    year = "2021"
}

@article{Sborlini:2021owe,
    author = "Sborlini, German F. R.",
    title = "{Geometrical approach to causality in multiloop amplitudes}",
    eprint = "2102.05062",
    archivePrefix = "arXiv",
    primaryClass = "hep-ph",
    reportNumber = "DESY-21-017, DESY 21-017",
    doi = "10.1103/PhysRevD.104.036014",
    journal = "Phys. Rev. D",
    volume = "104",
    number = "3",
    pages = "036014",
    year = "2021"
}

@article{Driencourt-Mangin:2019yhu,
    author = "Driencourt-Mangin, Felix and Rodrigo, German and Sborlini, German F. R. and Torres Bobadilla, William J.",
    title = "{Interplay between the loop-tree duality and helicity amplitudes}",
    eprint = "1911.11125",
    archivePrefix = "arXiv",
    primaryClass = "hep-ph",
    reportNumber = "IFIC/19-53",
    doi = "10.1103/PhysRevD.105.016012",
    journal = "Phys. Rev. D",
    volume = "105",
    number = "1",
    pages = "016012",
    year = "2022"
}

@article{TorresBobadilla:2021ivx,
    author = "Torres Bobadilla, William J.",
    title = "{Loop-tree duality from vertices and edges}",
    eprint = "2102.05048",
    archivePrefix = "arXiv",
    primaryClass = "hep-ph",
    reportNumber = "MPP-2021-14",
    doi = "10.1007/JHEP04(2021)183",
    journal = "JHEP",
    volume = "04",
    pages = "183",
    year = "2021"
}

@article{Sborlini:2016hat,
    author = "Sborlini, German F. R. and Driencourt-Mangin, Felix and Rodrigo, German",
    title = "{Four-dimensional unsubtraction with massive particles}",
    eprint = "1608.01584",
    archivePrefix = "arXiv",
    primaryClass = "hep-ph",
    reportNumber = "IFIC-16-45",
    doi = "10.1007/JHEP10(2016)162",
    journal = "JHEP",
    volume = "10",
    pages = "162",
    year = "2016"
}

@article{Bollini:1972ui,
    author = "Bollini, C. G. and Giambiagi, J. J.",
    title = "{Dimensional Renormalization: The Number of Dimensions as a Regularizing Parameter}",
    doi = "10.1007/BF02895558",
    journal = "Nuovo Cim. B",
    volume = "12",
    pages = "20--26",
    year = "1972"
}

@article{tHooft:1972tcz,
    author = "'t Hooft, Gerard and Veltman, M. J. G.",
    title = "{Regularization and Renormalization of Gauge Fields}",
    doi = "10.1016/0550-3213(72)90279-9",
    journal = "Nucl. Phys. B",
    volume = "44",
    pages = "189--213",
    year = "1972"
}

@article{Melrose:1965kb,
    author = "Melrose, D. B.",
    title = "{Reduction of Feynman diagrams}",
    doi = "10.1007/BF02832919",
    journal = "Nuovo Cim.",
    volume = "40",
    pages = "181--213",
    year = "1965"
}

@article{tHooft:1978jhc,
    author = "'t Hooft, Gerard and Veltman, M. J. G.",
    title = "{Scalar One Loop Integrals}",
    reportNumber = "PRINT-79-0134 (UTRECHT)",
    doi = "10.1016/0550-3213(79)90605-9",
    journal = "Nucl. Phys. B",
    volume = "153",
    pages = "365--401",
    year = "1979"
}

@article{vanNeerven:1983vr,
    author = "van Neerven, W. L. and Vermaseren, J. A. M.",
    title = "{Large loop integrals}",
    reportNumber = "NIKHEF-H/83-22",
    doi = "10.1016/0370-2693(84)90237-5",
    journal = "Phys. Lett. B",
    volume = "137",
    pages = "241--244",
    year = "1984"
}

@article{vanOldenborgh:1989wn,
    author = "van Oldenborgh, G. J. and Vermaseren, J. A. M.",
    title = "{New Algorithms for One Loop Integrals}",
    reportNumber = "NIKHEF-H/89-17",
    doi = "10.1007/BF01621031",
    journal = "Z. Phys. C",
    volume = "46",
    pages = "425--438",
    year = "1990"
}

@article{Bern:1993kr,
    author = "Bern, Zvi and Dixon, Lance J. and Kosower, David A.",
    title = "{Dimensionally regulated pentagon integrals}",
    eprint = "hep-ph/9306240",
    archivePrefix = "arXiv",
    reportNumber = "SLAC-PUB-5947, SACLAY-SPH-T-92-048, UCLA-92-043",
    doi = "10.1016/0550-3213(94)90398-0",
    journal = "Nucl. Phys. B",
    volume = "412",
    pages = "751--816",
    year = "1994"
}

@article{Papadopoulos:2014lla,
    author = "Papadopoulos, Costas G.",
    title = "{Simplified differential equations approach for Master Integrals}",
    eprint = "1401.6057",
    archivePrefix = "arXiv",
    primaryClass = "hep-ph",
    doi = "10.1007/JHEP07(2014)088",
    journal = "JHEP",
    volume = "07",
    pages = "088",
    year = "2014"
}

@article{DelDuca:2009ac,
    author = "Del Duca, Vittorio and Duhr, Claude and Nigel Glover, E. W. and Smirnov, Vladimir A.",
    title = "{The One-loop pentagon to higher orders in epsilon}",
    eprint = "0905.0097",
    archivePrefix = "arXiv",
    primaryClass = "hep-th",
    reportNumber = "IPPP-09-25, CP3-09-15",
    doi = "10.1007/JHEP01(2010)042",
    journal = "JHEP",
    volume = "01",
    pages = "042",
    year = "2010"
}

@article{Tramontano:2002xn,
    author = "Tramontano, Francesco",
    title = "{Pentagon integrals for heavy quark physics}",
    eprint = "hep-ph/0211390",
    archivePrefix = "arXiv",
    reportNumber = "DSF-22-2002",
    doi = "10.1103/PhysRevD.67.114005",
    journal = "Phys. Rev. D",
    volume = "67",
    pages = "114005",
    year = "2003"
}

@article{Denner:2002ii,
    author = "Denner, Ansgar and Dittmaier, S.",
    title = "{Reduction of one loop tensor five point integrals}",
    eprint = "hep-ph/0212259",
    archivePrefix = "arXiv",
    reportNumber = "MPI-PHT-2002-63, PSI-PR-02-21",
    doi = "10.1016/S0550-3213(03)00184-6",
    journal = "Nucl. Phys. B",
    volume = "658",
    pages = "175--202",
    year = "2003"
}

@article{rtqem,
    author = "Robbiati, Matteo and Sopena, Alejandro and Papaluca, Andrea and Carrazza, Stefano",
    title = "{Real-time error mitigation for variational optimization on quantum hardware}",
    eprint = "2311.05680",
    archivePrefix = "arXiv",
    primaryClass = "quant-ph",
    reportNumber = "TIF-UNIMI-2023-10, CERN-TH-2023-207",
    month = "11",
    year = "2023"
}

@misc{Qiskit,
    author = {{Qiskit contributors}},
    title = {Qiskit: An Open-source Framework for Quantum Computing},
    year = {2023},
    doi = {10.5281/zenodo.2573505}
}

@inproceedings{qibocal,
    author = "Pasquale, Andrea and Efthymiou, Stavros and Ramos-Calderer, Sergi and Wilkens, Jadwiga and Roth, Ingo and Carrazza, Stefano",
    title = "{Towards an open-source framework to perform quantum calibration and characterization}",
    booktitle = "{21th International Workshop on Advanced Computing and Analysis Techniques in Physics Research}: {AI meets Reality}",
    eprint = "2303.10397",
    archivePrefix = "arXiv",
    primaryClass = "quant-ph",
    reportNumber = "TIF-UNIMI-2023-7",
    month = "3",
    year = "2023"
}

@article{Robbiati:2022dkg,
    author = "Robbiati, Matteo and Efthymiou, Stavros and Pasquale, Andrea and Carrazza, Stefano",
    title = "{A quantum analytical Adam descent through parameter shift rule using Qibo}",
    eprint = "2210.10787",
    archivePrefix = "arXiv",
    primaryClass = "quant-ph",
    reportNumber = "CERN-TH-2022-168, TIF-UNIMI-2022-20",
    doi = "10.22323/1.414.0206",
    journal = "PoS",
    volume = "ICHEP2022",
    pages = "206",
    year = "2022"
}

@article{qibolab,
    author = "Efthymiou, Stavros and others",
    title = "{Qibolab: an open-source hybrid quantum operating system}",
    eprint = "2308.06313",
    archivePrefix = "arXiv",
    primaryClass = "quant-ph",
    reportNumber = "TIF-UNIMI-2023-14, CERN-TH-2023-142",
    doi = "10.22331/q-2024-02-12-1247",
    journal = "Quantum",
    volume = "8",
    pages = "1247",
    year = "2024"
}

@misc{githubGitHubGmlejarzaQuantumFourierIterativeAmplitudeEstimation,
	author = {},
	title = {{G}it{H}ub - gmlejarza/{Q}uantum-{F}ourier-{I}terative-{A}mplitude-{E}stimation},
	howpublished = {\url{https://github.com/gmlejarza/Quantum-Fourier-Iterative-Amplitude-Estimation}},
	year = {},
	note = {[Accessed 04-04-2025]},
}

@article{holmes2022connecting,
  title={Connecting ansatz expressibility to gradient magnitudes and barren plateaus},
  author={Holmes, Zoe and Sharma, Kunal and Cerezo, Marco and Coles, Patrick J},
  journal={PRX Quantum},
  volume={3},
  number={1},
  pages={010313},
  year={2022},
  publisher={APS}
}

@article{ionqarticle,
  author={Lubinski, Thomas and Johri, Sonika and Varosy, Paul and Coleman, Jeremiah and Zhao, Luning and Necaise, Jason and Baldwin, Charles H. and Mayer, Karl and Proctor, Timothy},
  journal={IEEE Transactions on Quantum Engineering}, 
  title={Application-Oriented Performance Benchmarks for Quantum Computing}, 
  year={2023},
  volume={4},
  number={},
  pages={1-32},
  doi={10.1109/TQE.2023.3253761}}

@misc{ionqnews,
    title = {Algorithmic Qubits: A Better Single-Number Metric},  
    author = {IonQ},  
    year = {2023},
    howpublished={\url{https://ionq.com/resources/algorithmic-qubits-a-better-single-number-metric}},
    note = {Accessed: 2023-04-30}

}

@article{Ramirez-Uribe:2024rjg,
    author = "Ram\'\i{}rez-Uribe, Selomit and Dhani, Prasanna K. and Sborlini, German F. R. and Rodrigo, Germ\'an",
    title = "{Rewording Theoretical Predictions at Colliders with Vacuum Amplitudes}",
    eprint = "2404.05491",
    archivePrefix = "arXiv",
    primaryClass = "hep-ph",
    doi = "10.1103/PhysRevLett.133.211901",
    journal = "Phys. Rev. Lett.",
    volume = "133",
    number = "21",
    pages = "211901",
    year = "2024"
}

@article{IBMQV,
  title = {Validating quantum computers using randomized model circuits},
  author = {Cross, Andrew W. and Bishop, Lev S. and Sheldon, Sarah and Nation, Paul D. and Gambetta, Jay M.},
  journal = {Phys. Rev. A},
  volume = {100},
  issue = {3},
  pages = {032328},
  numpages = {11},
  year = {2019},
  month = {Sep},
  publisher = {American Physical Society},
  doi = {10.1103/PhysRevA.100.032328},
  url = {https://link.aps.org/doi/10.1103/PhysRevA.100.032328}
}

@article{Feynman:1963ax,
    author = "Feynman, R. P.",
    editor = "Hsu, Jong-Ping and Fine, D.",
    title = "{Quantum theory of gravitation}",
    journal = "Acta Phys. Polon.",
    volume = "24",
    pages = "697--722",
    year = "1963"
}

@article{Ramirez-Uribe:2022sja,
    author = "Ram\'\i{}rez-Uribe, Selomit and Hern\'andez-Pinto, Roger Jos\'e and Rodrigo, Germ\'an and Sborlini, German F. R.",
    title = "{From Five-Loop Scattering Amplitudes to Open Trees with the Loop-Tree Duality}",
    eprint = "2211.03163",
    archivePrefix = "arXiv",
    primaryClass = "hep-ph",
    reportNumber = "IFIC/20-30",
    doi = "10.3390/sym14122571",
    journal = "Symmetry",
    volume = "14",
    number = "12",
    pages = "2571",
    year = "2022"
}

@article{Ramirez-Uribe:2024wua,
    author = "Ram\'\i{}rez-Uribe, Selomit and Renter\'\i{}a-Olivo, Andr\'es E. and Rodrigo, Germ\'an",
    title = "{Quantum querying based on multicontrolled Toffoli gates for causal Feynman loop configurations and directed acyclic graphs}",
    eprint = "2404.03544",
    archivePrefix = "arXiv",
    primaryClass = "quant-ph",
    month = "4",
    year = "2024"
}

@software{jax2018github,
  author = {James Bradbury and Roy Frostig and Peter Hawkins and Matthew James Johnson and Chris Leary and Dougal Maclaurin and George Necula and Adam Paszke and Jake Vander{P}las and Skye Wanderman-{M}ilne and Qiao Zhang},
  title = {{JAX}: composable transformations of {P}ython+{N}um{P}y programs},
  url = {http://github.com/jax-ml/jax},
  version = {0.3.13},
  year = {2018},
}

@article{deLejarza:2025upd,
    author = "de Lejarza, Jorge J. Mart{\'\i}nez and Wu, Hsin-Yu and Kyriienko, Oleksandr and Rodrigo, Germ{\'a}n and Grossi, Michele",
    title = "{Quantum Chebyshev probabilistic models for fragmentation functions}",
    eprint = "2503.16073",
    archivePrefix = "arXiv",
    primaryClass = "quant-ph",
    doi = "10.1038/s42005-025-02361-1",
    journal = "Commun. Phys.",
    volume = "8",
    number = "1",
    pages = "448",
    year = "2025"
}

@article{Collins:1989gx,
    author = "Collins, John C. and Soper, Davison E. and Sterman, George F.",
    title = "{Factorization of Hard Processes in QCD}",
    eprint = "hep-ph/0409313",
    archivePrefix = "arXiv",
    reportNumber = "ITP-SB-89-31",
    doi = "10.1142/9789814503266_0001",
    journal = "Adv. Ser. Direct. High Energy Phys.",
    volume = "5",
    pages = "1--91",
    year = "1989"
}

@article{deFlorian:2007ekg,
    author = "de Florian, Daniel and Sassot, Rodolfo and Stratmann, Marco",
    title = "{Global analysis of fragmentation functions for protons and charged hadrons}",
    eprint = "0707.1506",
    archivePrefix = "arXiv",
    primaryClass = "hep-ph",
    doi = "10.1103/PhysRevD.76.074033",
    journal = "Phys. Rev. D",
    volume = "76",
    pages = "074033",
    year = "2007"
}

@article{Aidala:2010bn,
    author = "Aidala, Christine A. and Ellinghaus, Frank and Sassot, Rodolfo and Seele, Joseph P. and Stratmann, Marco",
    title = "{Global Analysis of Fragmentation Functions for Eta Mesons}",
    eprint = "1009.6145",
    archivePrefix = "arXiv",
    primaryClass = "hep-ph",
    doi = "10.1103/PhysRevD.83.034002",
    journal = "Phys. Rev. D",
    volume = "83",
    pages = "034002",
    year = "2011"
}

@article{deFlorian:2007aj,
    author = "de Florian, Daniel and Sassot, Rodolfo and Stratmann, Marco",
    title = "{Global analysis of fragmentation functions for pions and kaons and their uncertainties}",
    eprint = "hep-ph/0703242",
    archivePrefix = "arXiv",
    doi = "10.1103/PhysRevD.75.114010",
    journal = "Phys. Rev. D",
    volume = "75",
    pages = "114010",
    year = "2007"
}

@article{Collins:1981uw,
    author = "Collins, John C. and Soper, Davison E.",
    title = "{Parton Distribution and Decay Functions}",
    reportNumber = "OITS-166",
    doi = "10.1016/0550-3213(82)90021-9",
    journal = "Nucl. Phys. B",
    volume = "194",
    pages = "445--492",
    year = "1982"
}

@article{Rojo:2015acz,
    author = "Rojo, Juan and others",
    title = "{The PDF4LHC report on PDFs and LHC data: Results from Run I and preparation for Run II}",
    eprint = "1507.00556",
    archivePrefix = "arXiv",
    primaryClass = "hep-ph",
    reportNumber = "OUTP-15-11P, LCTS-2015-14, GLAS-PPE-2015-01, DESY-15-088, CERN-PH-TH-2015-150, JLAB-THY-15-2064",
    doi = "10.1088/0954-3899/42/10/103103",
    journal = "J. Phys. G",
    volume = "42",
    pages = "103103",
    year = "2015"
}

@article{Albino:2008aa,
    author = "Albino, S. and others",
    title = "{Parton fragmentation in the vacuum and in the medium}",
    eprint = "0804.2021",
    archivePrefix = "arXiv",
    primaryClass = "hep-ph",
    month = "4",
    year = "2008"
}

@article{Aschenauer2019,
	author = {Aschenauer, Elke C. and Borsa, Ignacio and Sassot, Rodolfo and Van Hulse, Charlotte},
	title = {Semi-inclusive deep-inelastic scattering, parton distributions, and fragmentation functions at a future electron-ion collider},
	year = {2019},
	journal = {Physical Review D},
	volume = {99},
	number = {9},
	doi = {10.1103/PhysRevD.99.094004},
	url = {https://www.scopus.com/inward/record.uri?eid=2-s2.0-85066947666&doi=10.1103%2fPhysRevD.99.094004&partnerID=40&md5=2fdbe22a723d31038bd6cf13cfad22de},
	type = {Article},
	publication_stage = {Final},
	source = {Scopus},
	note = {Cited by: 26; All Open Access, Green Open Access, Hybrid Gold Open Access}
}

@article{Gluck:1998xa,
    author = {Gl\"uck, M. and Reya, E. and Vogt, A.},
    title = "{Dynamical parton distributions revisited}",
    eprint = "hep-ph/9806404",
    archivePrefix = "arXiv",
    reportNumber = "DO-TH-98-07, WUE-ITP-98-019",
    doi = "10.1007/s100520050289",
    journal = "Eur. Phys. J. C",
    volume = "5",
    pages = "461--470",
    year = "1998"
}

@article{Kretzer:2000yf,
    author = "Kretzer, S.",
    title = "{Fragmentation functions from flavor inclusive and flavor tagged e+ e- annihilations}",
    eprint = "hep-ph/0003177",
    archivePrefix = "arXiv",
    reportNumber = "DO-TH-00-04",
    doi = "10.1103/PhysRevD.62.054001",
    journal = "Phys. Rev. D",
    volume = "62",
    pages = "054001",
    year = "2000"
}

@article{Kniehl:2000fe,
    author = "Kniehl, Bernd A. and Kramer, G. and Potter, B.",
    title = "{Fragmentation functions for pions, kaons, and protons at next-to-leading order}",
    eprint = "hep-ph/0010289",
    archivePrefix = "arXiv",
    reportNumber = "DESY-00-086, MPI-PHT-2000-10",
    doi = "10.1016/S0550-3213(00)00303-5",
    journal = "Nucl. Phys. B",
    volume = "582",
    pages = "514--536",
    year = "2000"
}



\end{document}